\pdfoutput=1  

\documentclass[aps,prd,superscriptaddress,floatfix,nofootinbib,preprintnumbers,eqsecnum,twocolumn,dvipsnames]{revtex4-1}

\usepackage{amsmath,float,graphicx,placeins,amssymb,multirow,pgfplots,pgfplotstable}
\usepackage{empheq}
\usepackage{tikz}
\usepackage{comment}
\usepackage{enumerate,slashed,cancel,comment,color,bbm,graphicx,rotating,placeins,mathtools,multirow,hhline}
\usepackage[normalem]{ulem}
\usepackage{array}

\usepackage{hyperref}
\usepackage{color}
 \hypersetup
 {
   colorlinks,
   linkcolor={blue!80!black},
   citecolor={blue!70!black},
   urlcolor={blue!70!black}
 }

\usepackage{lipsum}
\usepackage{diagbox}
\usetikzlibrary{shapes.geometric,arrows}
\graphicspath{}

\def\beq{\begin{equation}}
\def\eeq{\end{equation}}
\def\beqn{\begin{eqnarray}}
\def\eeqn{\end{eqnarray}}

\def\be{\begin{equation}}
\def\ee{\end{equation}}
\def\bea{\begin{eqnarray}}
\def\eea{\end{eqnarray}}
\def\half{\mbox{\small ${\frac{1}{2}}$}}

\def\quarter{\mbox{\small ${\frac{1}{4}}$}}

\newcommand{\newc}{\newcommand}
\def\calZ{{\cal Z}}
\def\calM{{\cal M}}
\def\calV{{\cal V}}
\def\calF{{\cal F}}

\def\bQ{{\bf Q}}
\def\bT{{\bf T}}

\def\Qs{{\bf q}}

\def\barOmega{{\overline{\Omega}}}

\def\barkappa{{\overline{\kappa}}}

\def\half{{\textstyle{1\over 2}}}
\def\quarter{{\textstyle{1\over 4}}}
\def\ie{{\it i.e.}\/}
\def\eg{{\it e.g.}\/}
\def\etc{{\it etc}.\/}
\def\inbar{\,\vrule height1.5ex width.4pt depth0pt}
\def\IR{\relax{\rm I\kern-.18em R}}
 \font\cmss=cmss10 \font\cmsss=cmss10 at 7pt
\def\IQ{\relax{\rm I\kern-.18em Q}}
\def\IZ{\relax\ifmmode\mathchoice
 {\hbox{\cmss Z\kern-.4em Z}}{\hbox{\cmss Z\kern-.4em Z}}
 {\lower.9pt\hbox{\cmsss Z\kern-.4em Z}}
 {\lower1.2pt\hbox{\cmsss Z\kern-.4em Z}}\else{\cmss Z\kern-.4em Z}\fi}

\begin{document}

\title{Stasis, Stasis, Triple Stasis}

\def\andname{\hspace*{-0.5em}} 

\def\andname{\hspace*{-0.5em}}

\author{Keith R. Dienes}
 \email[Email address: ]{dienes@arizona.edu}
 \affiliation{Department of Physics, University of Arizona, Tucson, AZ 85721 USA}
 \affiliation{Department of Physics, University of Maryland, College Park, MD 20742 USA}
  \author{Lucien Heurtier}
\email[Email address: ]{lucien.heurtier@durham.ac.uk}
\affiliation{IPPP, Durham University, Durham, DH1 3LE, United Kingdom}
\author{Fei Huang}
\email[Email address: ]{fei.huang@weizmann.ac.il}
 \affiliation{Department of Particle Physics and Astrophysics, Weizmann Institute of Science, Rehovot 7610001, Israel}
\author{Tim M.P. Tait}
 \email[Email address: ]{ttait@uci.edu}
 \affiliation{Department of Physics and Astronomy, University of California, Irvine, CA  92697  USA}
 \author{Brooks Thomas}
 \email[Email address: ]{thomasbd@lafayette.edu}
 \affiliation{Department of Physics, Lafayette College, Easton, PA  18042 USA}

\begin{abstract}
Many theories of BSM physics predict the existence of large or infinite towers of 
decaying states.   In a previous paper~\cite{Dienes:2021woi} we pointed out that this 
can give rise to a surprising cosmological phenomenon that we dubbed ``stasis'' during 
which the relative abundances of  matter and radiation remain constant across extended
cosmological eras even though the universe is expanding.   Indeed, such stasis epochs 
are universal attractors, with the universe necessarily entering (and later exiting) 
such epochs for a wide variety of initial conditions.  Matter/radiation stasis is 
therefore an important and potentially unavoidable feature of many BSM cosmologies.
In this paper we extend our arguments to universes containing significant amounts of 
vacuum energy, and demonstrate that such universes also give rise to various forms of stasis 
between vacuum energy and either matter or radiation.    We also demonstrate the existence of 
several forms of ``triple stasis'' during which the abundances of matter, radiation, and 
vacuum energy all simultaneously remain fixed
despite cosmological expansion.  We further describe several close variants of stasis which 
we call ``quasi-stasis'' and ``oscillatory stasis'', and discuss the circumstances under 
which each of these can arise.   Finally, we develop a general formalism for understanding 
the emergence of stasis within BSM cosmologies irrespective of the number or type of different 
energy components involved.   Taken together, these results greatly expand the range of 
theoretical and phenomenological possibilities for the physics of the early universe, 
introducing new types of cosmological eras which may play an intrinsic and 
potentially inevitable role within numerous BSM cosmologies.
\end{abstract}
\maketitle

\tableofcontents

\def\ie{{\it i.e.}\/}
\def\eg{{\it e.g.}\/}
\def\etc{{\it etc}.\/}
\def\taubar{{\overline{\tau}}}
\def\qbar{{\overline{q}}}
\def\kbar{{\overline{k}}}
\def\bQ{{\bf Q}}
\def\calT{{\cal T}}
\def\calN{{\cal N}}
\def\calF{{\cal F}}
\def\calM{{\cal M}}
\def\calZ{{\cal Z}}

\def\beq{\begin{equation}}
\def\eeq{\end{equation}}
\def\beqn{\begin{eqnarray}}
\def\eeqn{\end{eqnarray}}
\def\apo{\mbox{\small ${\frac{\alpha'}{2}}$}}
\def\half{\mbox{\small ${\frac{1}{2}}$}}
\def\sqapo{\mbox{\tiny $\sqrt{\frac{\alpha'}{2}}$}}
\def\sqap{\mbox{\tiny $\sqrt{{\alpha'}}$}}
\def\sqapxtwo{\mbox{\tiny $\sqrt{2{\alpha'}}$}}
\def\aptwo{\mbox{\tiny ${\frac{\alpha'}{2}}$}}
\def\apofour{\mbox{\tiny ${\frac{\alpha'}{4}}$}}
\def\bosqtwo{\mbox{\tiny ${\frac{\beta}{\sqrt{2}}}$}}
\def\btosqtwo{\mbox{\tiny ${\frac{\tilde{\beta}}{\sqrt{2}}}$}}
\def\apofour{\mbox{\tiny ${\frac{\alpha'}{4}}$}}
\def\sqaptwo{\mbox{\tiny $\sqrt{\frac{\alpha'}{2}}$}  }
\def\apoeight{\mbox{\tiny ${\frac{\alpha'}{8}}$}}
\def\sapoeight{\mbox{\tiny ${\frac{\sqrt{\alpha'}}{8}}$}}

\newc{\gsim}{\lower.7ex\hbox{{\mbox{$\;\stackrel{\textstyle>}{\sim}\;$}}}}
\newc{\lsim}{\lower.7ex\hbox{{\mbox{$\;\stackrel{\textstyle<}{\sim}\;$}}}}
\def\calM{{\cal M}}
\def\calV{{\cal V}}
\def\calF{{\cal F}}
\def\bQ{{\bf Q}}
\def\bT{{\bf T}}
\def\Qs{{\bf q}}

\def\half{{\textstyle{1\over 2}}}
\def\quarter{{\textstyle{1\over 4}}}
\def\ie{{\it i.e.}\/}
\def\eg{{\it e.g.}\/}
\def\etc{{\it etc}.\/}
\def\inbar{\,\vrule height1.5ex width.4pt depth0pt}
\def\IR{\relax{\rm I\kern-.18em R}}
 \font\cmss=cmss10 \font\cmsss=cmss10 at 7pt
\def\IQ{\relax{\rm I\kern-.18em Q}}
\def\IZ{\relax\ifmmode\mathchoice
 {\hbox{\cmss Z\kern-.4em Z}}{\hbox{\cmss Z\kern-.4em Z}}
 {\lower.9pt\hbox{\cmsss Z\kern-.4em Z}}
 {\lower1.2pt\hbox{\cmsss Z\kern-.4em Z}}\else{\cmss Z\kern-.4em Z}\fi}
\def\trho{\tilde{\rho}}
\def\hatt{\hat{t}}
\def\tt{\tilde{t}}
\def\nn{\nonumber}
\def\tdiff{\Delta t}

\section{Introduction, motivation, and basic idea}

Many theories of physics beyond the Standard Model (BSM) predict the existence of infinite 
towers of unstable states.  In theories involving extra spacetime dimensions, such states 
might be the Kaluza-Klein (KK) states associated with the spectra  of  
quantized momenta in the compactified dimensions.  Alternatively, in theories with 
dark sectors consisting of strongly coupled gauge theories, such states might be the 
infinite towers of increasingly heavy bound-state resonances.  Likewise, in string theory, 
such towers of states can take the form of not only the Kaluza-Klein and winding-mode states 
associated with the compactification geometry but also the infinite towers of fundamental 
string resonances which represent the quantized excitations of the fluctuating strings 
and 
branes themselves.  In fact, some BSM models can contain mixtures of all of these states, 
with mass scales that depend on the particular BSM model under study.

In general, such states are likely to be unstable.  As a result, they 
will decay, 
either to lighter states within the same tower or directly to Standard-Model
 states.  
The heavier states will generally decay first since they are likely to have the 
largest decay widths for a given final state.   Likewise, for a given initial state, 
the largest decay widths will generically arise for decays to the lightest available 
final states, thereby endowing such states with considerable kinetic energies and 
rendering them relativistic.  Such decay products may therefore be considered as 
functionally equivalent to radiation. Of course, the detailed properties  of such 
decays will depend on the particular BSM model under study.  However, as a 
general feature, we can expect the different states in our tower to decay 
{\it sequentially}\/ to very light states, with the heavier states decaying first, 
then the next-heaviest states, and so forth down the tower.  With exceedingly 
large (or infinite) towers of states, this decay sequence
may extend over a significant period of time before finally terminating once the 
lightest states have decayed. 

In Ref.~\cite{Dienes:2021woi}, we considered the cosmological implications of such 
extended decay sequences occurring in the early universe and found that such extended 
decay sequences can lead to a surprising cosmological phenomenon which we 
called ``stasis''.  {\it During this form of stasis, the abundances 
of matter and radiation in the universe remain constant across extended cosmological 
epochs even though the universe continues to expand}\/.   At first glance, it might 
seem that such a phenomenon is impossible.
After all, any cosmological epoch consisting of both radiation and matter will 
transition from radiation-dominated to matter-dominated, purely as a result of 
cosmological expansion.
This simple observation is a consequence of the fact that the energy density 
associated with matter scales as $a^{-3}$ where $a$ is the cosmological scale factor, 
while that of radiation scales as $a^{-4}$. 
Thus, as the universe expands, an increasing fraction of the total energy density 
takes the form of matter rather than radiation, thereby causing the matter abundance 
to rise and the radiation abundance to fall.  
However, such a transition from radiation domination to matter domination need not 
occur if the BSM model in question gives rise to 
counterbalancing effects which convert matter back into 
radiation~\cite{Barrow:1991dn,Dienes:2021woi,Dienes:2022zgd}. 
   
In Ref.~\cite{Dienes:2021woi}, we demonstrated that such an extended decay sequence
can furnish precisely the sort of counterbalancing effect that is needed, converting 
matter (in the form of the original infinite towers of heavy states)  into radiation 
(in the form of the decay products, either photons or other highly energetic 
light states).  Indeed, as demonstrated in Ref.~\cite{Dienes:2021woi}, this 
counterbalancing process can persist across many $e$-folds of cosmological expansion 
as the decays work their way down the tower. Moreover, this counterbalancing effect can
precisely compensate for the effects of cosmological expansion, so that a period 
of {\it bona fide}\/ stasis ensues during which the abundances of matter and radiation 
remain constant throughout this entire epoch.  Quite remarkably, this possibility 
requires no fine-tuning and actually emerges as a {\it global attractor}\/ within 
the relevant cosmological framework~\cite{Barrow:1991dn,Dienes:2021woi, Dienes:2022zgd}.
Moreover, despite this attactor behavior, we demonstrated in Ref.~\cite{Dienes:2021woi} 
that such matter/radiation stasis epochs ultimately have a finite duration, ending 
naturally when the lightest states at the bottom of the tower decay.  Thus, the universe  
not only {\it enters}\/ into a stasis epoch but also {\it emerges}\/ from it in a 
natural way.  Indeed, a similar kind of stasis between matter and radiation 
can also arise through the decays of primordial black 
holes~\cite{Barrow:1991dn,Dienes:2022zgd}.

The primary goal of this paper is to extend this discussion to include universes 
which contain significant amounts of {\it vacuum energy}\/.   First, we will 
investigate the extent to which BSM physics can give rise to a two-component stasis 
within a universe consisting of vacuum energy and matter.
We will then seek to understand whether we can similarly achieve a two-component stasis between 
vacuum energy and radiation.  Finally, we will go for broke and 
ask whether it is possible to have a ``triple stasis'' in which vacuum energy, matter, 
and radiation can all simultaneously be in stasis with each other.   Equally importantly, 
we will also investigate whether such forms of stasis require fine-tuning, or whether 
they follow the example of matter/radiation stasis and also emerge as cosmological 
attractors. We will also construct a ``phase diagram'' for stasis which will enable us to
understand how all of these different forms of stasis can merge and transition between 
each other as we change the underlying parameters within our BSM models.
 
Ultimately, we shall find that all of these new forms of stasis can indeed exist and 
emerge naturally from BSM physics. 
This in turn reinforces our belief that the stasis phenomenon is in fact a fairly 
generic and robust feature in certain types of cosmologies involving BSM physics. 

One important subtlety in our work concerns the manner in which we incorporate vacuum 
energy into our discussion of stasis.  At first glance, it might seem that vacuum 
energy can be treated simply as just another cosmological fluid whose pressure $p$ and 
energy density $\rho$ are related via the equation of state $p = -\rho$ (\ie, with an 
equation-of-state parameter $w= -1$).   However, as we shall discuss, 
na\"{i}vely taking $w= -1$ leads to a slew of  important mathematical and physical
complications.  For this reason, one important aspect of our work is to develop a 
method in which we might successfully model vacuum energy.   However, as we shall 
demonstrate here and in Ref.~\cite{toappear}, stasis is possible and emerges naturally 
regardless of the particular manner in which vacuum energy is modeled.

How to model vacuum energy is not the only subtlety we shall encounter.   For example, 
we shall find that there are often multiple ways of performing certain critical calculations.   
While some methods work best in certain contexts, other methods work best in other contexts.  
Accordingly, with an eye towards potential future applications of our work, in this paper we 
shall outline all relevant methods of performing certain calculations and demonstrate how they 
relate to each other.

This paper is organized as follows.  First, in Sect.~\ref{subsect:2A}, we review the 
results of Ref.~\cite{Dienes:2021woi} which focused on stasis between matter and 
radiation.   We do this not only as review, but also in order to establish our overall 
notation and calculational procedures.  In Sect.~\ref{subsect:general_lessons}, we then 
extract certain general lessons from this example --- general lessons which will prove 
critical later in this paper as we expand the scope of our analysis.
Then, in Sect.~\ref{sec:LambdaMatter}, we begin our discussion of how vacuum energy 
may be introduced into this picture.   It is here that we discuss the subtleties 
associated with the introduction of vacuum energy into the stasis framework, but we 
ultimately demonstrate that a similar stasis can be achieved between vacuum energy 
and matter once these subtleties are satisfactorily addressed.  
In Sect.~\ref{sec:LambdaGamma} we then demonstrate that a similar pairwise stasis 
can exist between vacuum energy and radiation.  Finally, we conclude our discussion 
of pairwise stases in Sect.~\ref{sec:Generalpairwise} by outlining various elements 
of their common algebraic structure.

Sect.~\ref{sec:TripleStasis} in some sense serves as the central focal point  
of this paper.  In this section, pulling together our results from previous sections 
and extending them in certain critical ways, we demonstrate that we can even achieve
a {\it triple stasis}\/ involving vacuum energy, 
matter, and radiation simultaneously.  As we shall see, this is a highly non-trivial 
result because the existence of three different pairwise stases between three different 
energy components does not generally imply the existence of a triple stasis amongst 
them all simultaneously.  Indeed, we shall find that several additional constraints 
must be satisfied in order to allow such a triple stasis to exist.   Fortunately, these
constraints are not severe, and many BSM cosmologies give rise to triple stasis as well.

In Sect.~\ref{sec:attractor}, we then turn our attention to the attractor behavior 
associated with these new forms of stasis.  We  ultimately find that all of the stasis 
solutions we examine in this paper are indeed global attractors.  Likewise, in 
Sect.~\ref{sec:phase_diagram}, we present a ``phase diagram'' for the stasis phenomenon and 
demonstrate how the different types of stasis we have examined in this paper are actually
different ``phases'' of the same stasis 
phenomenon in different limits of the underlying parameter space.

Along the way we also develop a general formalism which enables a general study of stasis 
regardless of the number and types of different energy components involved.   In this 
way we also isolate the underlying ingredients that allow stasis to exist.  With these 
insights in hand, in Sect.~\ref{sec:variants} we then investigate what happens when some, 
but not all, of these ingredients are present.   In this way, we discover several new
``variants'' of the original stasis idea.  One of these, discussed in 
Sect.~\ref{subsect:Quasi-stasis}, is a new theoretical possibility in which the 
cosmological abundances of different energy components are not strictly constant but 
instead exhibit highly suppressed time-evolution.  
This too is not seen in the standard cosmological timelines, but may also 
represent a valid possibility for early-universe physics in certain situations.
We shall refer to this phenomenon as {\it quasi-stasis}\/.   We also discover a different 
variant of stasis --- one in which the abundances are again not constant but instead 
{\it oscillate}\/ around their central stasis values.   This phenomenon, which we shall call  
{\it oscillatory stasis}\/, is discussed in Sect.~\ref{subsect:OscillatoryStasis}.~ 
Yet another possibility is discussed in Sect.~\ref{subsect:stasis_unrealized}.~   
We then make concluding remarks and present ideas for further research in
Sect.~\ref{sec:conclusions}.

Just as in Ref.~\cite{Dienes:2021woi}, our main interest in this paper is the stasis 
phenomenon itself --- \ie, the existence of stable mixed-component cosmological 
eras --- and the manner in which such cosmologies emerge from BSM physics.  
Needless to say, phenomenological constraints may make it difficult for an epoch 
of stasis to appear within certain portions of the standard $\Lambda {\rm CDM}$ 
cosmological timeline (particularly those near or after Big-Bang Nucleosynthesis).  
Indeed, such phenomenological constraints in turn might be used in order to constrain 
the range of possible BSM theories governing physics at higher energy scales.
However, other regions of parameter space may be able to accommodate stasis epochs 
without difficulty.  In this paper we shall therefore study our BSM-inspired 
realizations of stasis as general theoretical phenomena, and defer discussion of 
their various phenomenological implications and constraints to future work.   
That said, the emergence of these different forms of stasis within BSM cosmologies 
gives rise to a host of new theoretical and phenomenological possibilities for 
early-universe model-building across the entire cosmological timeline.  Such 
possibilities are therefore ripe for future exploration.

\FloatBarrier

\section{Matter/radiation stasis\label{sec:MatterGamma}}

In this section we begin by reviewing the results of Ref.~\cite{Dienes:2021woi} 
concerning the possibility of stasis between matter and radiation. As discussed 
in the Introduction, our purpose for including this review is two-fold.   First, 
our analyses of each kind of stasis that we shall be discussing in subsequent 
sections can be patterned after our discussion of this case. This section will 
therefore introduce the main ideas and establish our notation.  But second, we 
shall find that many of the ingredients of this matter/radiation stasis will become 
part of the larger triple-stasis structure that we shall eventually construct in
Sect.~\ref{sec:TripleStasis}.~  Thus it will be important to recall the details 
of this case before proceeding further.  

\subsection{Algebraic analysis\label{subsect:2A}}

We begin, as in Ref.~\cite{Dienes:2021woi}, by assuming a flat 
Friedmann-Robertson-Walker (FRW) universe containing two components:
a tower of matter states $\phi_\ell$ where the indices $\ell=0,1,2,...,N-1$ are assigned
in order of increasing mass; and radiation (collectively denoted $\gamma$) into which 
the $\phi_\ell$ can decay. This radiation may consist of photons or other highly 
relativistic particles.  We shall let $\rho_\ell$ and $\rho_\gamma$ denote the 
corresponding energy densities and $\Omega_\ell$ and $\Omega_\gamma$ the corresponding
abundances.  We shall also assume that the dominant decay mode of the $\phi_\ell$ is 
into radiation, and let $\Gamma_\ell$ denote the corresponding decay rates.  Note that 
in this section we shall not require that the $\phi_\ell$ components be scalars, and 
indeed any (non-relativistic) matter fields are acceptable.   We shall also implicitly 
assume that $N$ is large (or potentially even infinite).

Recall that for any energy density $\rho_i$ (where $i$ denotes matter, radiation, 
or vacuum energy), the corresponding abundance $\Omega_i$ is given by 
\beq
  \Omega_i   ~\equiv~ \frac{8\pi G}{3H^2} \,\rho_i~,~
\label{MGOmegadef}
\eeq
where $H$ is the Hubble parameter and $G$ is Newton's constant.
From this it follows that
\beq
  \frac{d\Omega_i}{dt} ~=~ \frac{8\pi G}{3} 
  \left( \frac{1}{H^2} \frac{d\rho_i}{dt} - 2 \frac{\rho_i}{H^3} 
  \frac{dH}{dt} \right)~.
\label{MGstepone}
\eeq
We can simplify this expression 
through the use of the Friedmann ``acceleration'' equation for $dH/dt$, which in a 
universe consisting of only matter and radiation takes the form
\beqn
  \frac{dH}{dt}  
  ~&=&~ -H^2 - \frac{4\pi G}{3} \left(  \sum_i \rho_i + 3 \sum_i p_i\right)\nonumber\\
  ~&=&~ - \half H^2 \left( 2+ \Omega_M + 2 \Omega_\gamma\right)\nonumber\\
  ~&=&~ - \half H^2 \left( 4-\Omega_M \right)~.
\label{MGacceleq}
\eeqn
Note that in passing to the second line we have defined the total matter abundance
$\Omega_M \equiv \sum_\ell \Omega_\ell$, and in passing to the third line we have 
imposed the constraint $\Omega_M + \Omega_\gamma=1$, as suitable for a universe 
containing only these two energy components.  Substituting Eq.~(\ref{MGacceleq}) 
into Eq.~(\ref{MGstepone}) we then obtain
\beq
    \frac{d\Omega_i}{dt} 
     ~=~ \frac{8\pi G}{3H^2} \frac{d\rho_i}{dt} + H \Omega_i \left( 4-\Omega_M\right)~,
\label{MGconvert}
\eeq
whereupon taking $i=M$ (for the total matter abundance) or $i=\gamma$ (for the 
total radiation abundance) yields
\beqn
    \frac{d\Omega_M}{dt} 
     ~&=&~ \frac{8\pi G}{3H^2} \sum_\ell \frac{d\rho_\ell}{dt} 
     + H \Omega_M \left( 4-\Omega_M \right) ~ \nonumber\\
    \frac{d\Omega_\gamma}{dt} 
     ~&=&~ \frac{8\pi G}{3H^2} \frac{d\rho_\gamma}{dt}  
    + H \Omega_\gamma \left( 4-\Omega_M\right)~.
\label{MGconvert2}
\eeqn
These are thus general relations for the time-evolution of $\Omega_M$ and 
$\Omega_\gamma$ in terms of $d\rho_\ell/dt$ and $d\rho_\gamma/dt$ in a universe 
consisting of matter and radiation.

Given the relations in Eq.~(\ref{MGconvert2}), our final step is to insert 
appropriate ``equations of motion'' for $d\rho_\ell/dt$ and $d\rho_\gamma/dt$.  
Since each $\phi_\ell$ is assumed to decay into radiation $\gamma$ with
rate $\Gamma_\ell$, and given that each decay process conserves energy, these 
equations of motion are given by
\beqn
  \frac{d\rho_\ell}{dt} ~&=&~ -3 H \rho_\ell - \Gamma_\ell \rho_\ell~ \nonumber\\
  \frac{d\rho_\gamma}{dt} ~&=&~ -4 H \rho_\gamma + \sum_\ell \Gamma_\ell \rho_\ell~ .
\label{MGeoms}
\eeqn
While the final term on each line reflects the effects of the decays, the first term 
on the right side of each line reflects the redshifting effects of cosmological 
expansion for matter and radiation respectively.  Given these expressions for 
$d\rho_\ell/dt$ and $d\rho_\gamma/dt$, we find that Eq.~(\ref{MGconvert2}) then takes 
the form
\beq
    \frac{d\Omega_M}{dt} 
     ~=~ -\sum_\ell \Gamma_\ell \Omega_\ell  
     + H \left( \Omega_M - \Omega_M^2\right)  
\label{MGconvert3}
\eeq
with $d\Omega_\gamma/dt = - d\Omega_M/dt$.

We are seeking a 
steady-state ``stasis'' solution in which $\Omega_M$ and $\Omega_\gamma$ are constant.
Clearly such a solution will arise if the effects of the $\phi_\ell$ decays are 
precisely counterbalanced by the cosmological expansion.  We therefore wish to impose, 
at the very minimum, the condition that $d\Omega_M/dt=0$, 
which from Eq.~(\ref{MGconvert3}) yields the constraint
\beq
   \sum_\ell^{\phantom{x}} \Gamma_\ell \Omega_\ell  ~=~ H (\Omega_M - \Omega_M^2)~.
\label{MGcondition}
\eeq
However, it is not sufficient for this condition to hold only for an instant of 
time --- we want this condition to hold over an extended {\it interval}\/ of time.
In order to achieve this, we shall actually demand something stronger, namely that 
 this condition hold for {\it all}\/ times $t$.
In imposing this latter constraint we are actually implicitly demanding an 
{\it eternal}\/ stasis, one without beginning or end.  However, once we understand the
conditions that characterize such an eternal stasis state, we shall then discuss the 
physics that will actually restrict this stasis state to finite duration, essentially
introducing  not only a natural {\it entrance}\/ into the stasis state but also a 
natural {\it exit}\/ from it.

Demanding that Eq.~(\ref{MGcondition}) hold for all time requires not only that 
this equation hold at one given time, but also that both sides of this equation 
have precisely the same time-dependence when the abundances are held fixed 
(which is the defining property of the stasis configuration we are hoping to understand).  
Let us therefore assume that $\Omega_M$ and $\Omega_\gamma$ are fixed to their stasis 
values $\barOmega_M$ and $\barOmega_\gamma$.   Under these stasis conditions, we can 
actually solve for the Hubble parameter directly via Eq.~(\ref{MGacceleq}), obtaining 
the exact solution 
\beq
  H(t) = \left(\frac{2}{4-\barOmega_M}\right) \,\frac{1}{t}~
    ~~~\Longrightarrow~~~~
    \barkappa =  \frac{6}{4-\barOmega_M}~ 
\label{MGHubble}
\eeq
where $\barkappa$ generally corresponds to the parametrization $H(t)= \barkappa/(3t)$ during stasis.
Indeed, from Eq.~(\ref{MGHubble}) we verify the standard results that $H(t) = 2/(3t)$
for $\barOmega_M=1$ (\ie, for a matter-dominated universe),
while $H(t) = 1/(2t)$ for $\barOmega_M=0$ (\ie, for a radiation-dominated universe).
Substituting Eq.~(\ref{MGHubble}) into Eq.~(\ref{MGcondition}) then yields our 
matter/radiation stasis condition
\begin{empheq}[box=\fbox]{align}
~~\sum_\ell^{\phantom{X}} \Gamma_\ell\Omega_\ell(t)  ~&=~ 
\frac{\barkappa}{3} ~
     \barOmega_M (1-\barOmega_M) \, \frac{1}{t} 
 ~~\nonumber\\
   ~&=~  \biggl(
    2-\barkappa\biggr) \, \barOmega_M\, 
\frac{1}{t} ~.~~
\label{MGmasterconstraints}
\end{empheq}
where the individual matter abundances $\Omega_\ell(t)$ within this  eternal-stasis
cosmology
are given by
\beqn
    \Omega_\ell(t) ~=~ \Omega_\ell^{\ast}  
                 \left( \frac{t}{t_\ast}\right)^{2-\barkappa}  e^{-\Gamma_\ell (t-t^{(0)})}~.
\label{MGOmegal}
\eeqn
Here $t_\ast$ is some fiducial time within this stasis epoch, while $\Omega_\ell^{\ast}\equiv  \Omega_\ell(t_\ast)$ and 
$\barOmega_M\equiv \sum_\ell \Omega_\ell(t)$ for all $t$.  By contrast, the quantity 
$t^{(0)}$ within Eq.~(\ref{MGOmegal}) denotes the (presumed common) production time 
for the $\phi_\ell$ states, and thus serves as the zero of the clock according 
to which the decay lifetimes of these states are measured.

Of course, given our tower of components $\phi_\ell$, the condition in 
Eq.~(\ref{MGmasterconstraints}) cannot be strictly satisfied for all times.  Thus, we cannot truly have an eternal stasis.
For example, regardless of whether the tower of $\phi_\ell$ states is finite or infinite, 
there is an early time immediately after these states are produced at $t^{(0)}$ during which the decay 
process is just beginning and thus will not yet have grown sufficient to counterbalance 
cosmological expansion.  Indeed, the very presence of a production time $t^{(0)}$ within Eq.~(\ref{MGOmegal}) in some sense invalidates the assumption of a truly eternal stasis.   Likewise, there will eventually come a time at which all of the 
decays will have essentially concluded, at which point we expect our period of stasis to end.  
However, the critical issue is whether there exist solutions for the spectrum of decay widths 
$\lbrace \Gamma_\ell\rbrace$ and abundances $\lbrace \Omega_\ell\rbrace$ across our tower of states
which will lead to an extended period of stasis {\it during}\/ the sequential decay process.  

Our assertion in this paper is that many well-motivated theories of physics beyond the 
Standard Model 
give rise to towers of states with exactly this property.
In order to study this question in a general way, we observe  --- as discussed in the 
Introduction and in Ref.~\cite{Dienes:2021woi} --- that many well-motivated BSM theories 
give rise to infinite towers of states $\phi_\ell$ 
whose decay widths and initial abundances either exactly or approximately 
satisfy scaling relations of the forms
\beq
       \Gamma_\ell = \Gamma_0\left(\frac{m_\ell}{m_0}\right)^\gamma~,~~~~~
       \Omega_\ell^{(0)} = \Omega_0^{(0)} \left(\frac{m_\ell}{m_0}\right)^\alpha~,~~
\label{MGscalings}
\eeq
where $\alpha$ and $\gamma$ are general scaling exponents, 
where the $\phi_\ell$ mass spectrum takes the form 
\beq
          m_\ell ~=~ m_0 + (\Delta m) \ell^\delta~
\label{MGmassform}
\eeq
with $m_0\geq 0$, $\Delta m>0$, and $\delta>0$ all
treated as general free parameters,
and where
the superscript `$(0)$' within Eq.~(\ref{MGscalings}) 
denotes the time $t=t^{(0)}$ at which the $\phi_\ell$ states are initially produced
(thereby setting a common clock for the subsequent $\phi_\ell$ decays). 
For example, if the $\phi_\ell$ are the KK excitations of a five-dimensional scalar
field compactified on a circle of radius $R$ (or a $\mathbb{Z}_2$ orbifold thereof), we will have
either $\lbrace m_0,\Delta m,\delta\rbrace = \lbrace m, 1/R, 1\rbrace$
or $\lbrace m_0,\Delta m,\delta\rbrace =\lbrace m, 1/(2 m R^2), 2\rbrace$,
depending on whether $m R \ll1$ or $mR\gg 1$, respectively, where $m$ denotes the 
four-dimensional scalar mass~\cite{Dienes:2011ja, Dienes:2011sa}.  Alternatively, 
if the $\phi_\ell$ are the bound states of a strongly-coupled gauge theory, we have 
$\delta = 1/2$, where $\Delta m$ and $m_0$ are determined by the Regge slope and intercept 
of the strongly-coupled theory, respectively~\cite{Dienes:2016vei}.  The same values also 
describe the excitations of a fundamental string.  Thus $\delta=\lbrace 1/2, 1,2\rbrace$ can 
serve as compelling ``benchmark'' values.  Likewise, the exponent $\gamma$ is ultimately
governed by the particular $\phi_\ell$ decay mode.  For example, we will have $\gamma= 2d - 7$ 
if each $\phi_\ell$ state decays to photons through a dimension-$d$ contact operator of the 
form ${\cal O}_\ell \sim c_\ell \phi_\ell {\cal F}/\Lambda^{d-4}$ where $\Lambda$ is an 
appropriate mass scale and where ${\cal F}$ is an operator built from photon fields.  Thus,
values such as $\gamma=\lbrace 3,5,7\rbrace$ serve as relevant benchmarks.  
Indeed, $\gamma=1$ is also relevant in cases in which $\phi_\ell$ are scalars decaying 
into fermions.  Finally, $\alpha$ is determined by the original production mechanism 
for the $\phi_\ell$ fields.  For example, it is easy to see that $\alpha<0$ for 
misalignment production~\cite{Dienes:2011ja, Dienes:2011sa}, while $\alpha$ can generally 
be of either sign for thermal freeze-out~\cite{Dienes:2017zjq}.

Our goal will then be to determine for which combinations of these scaling exponents 
$(\alpha,\gamma,\delta)$ and other dimensionful parameters 
$(m_0, \Delta m, \Gamma_0, \Omega_0^{(0)})$ an extended stasis state may arise.
In this way, through a general study of these scaling exponents and dimensionful parameters, 
we can survey the effects of many different BSM theories at once.
Of course, within the context of this model, we shall assume that $\Omega_M=1$ at $t=t^{(0)}$ 
before the decay process has begun.  This reflects the fact that no radiation has yet been 
generated at  $t=t^{(0)}$, and that our universe at that time consists of only the initial 
$\phi_\ell$ states.  This in turn requires that we choose the overall normalization
$\Omega_0^{(0)} = \left[\sum_{\ell=0}^{N-1} (m_\ell/m_0)^\alpha\right]^{-1}$.

For future use, it will prove convenient to define two different dimensionless 
combinations of the above parameters:
\beq 
   \eta~\equiv~\alpha+ \frac{1}{\delta} ~,
  ~~~~~~~ 
   \xi~\equiv ~
   \frac{2}{\barkappa}\,
   \frac{m_0}{\Gamma_0}~.
\label{eq:etaxidef}
\eeq
Roughly speaking, we shall find that $\eta$ describes how the energy density scales per unit mass 
across the tower, while $\xi$ describes the decay rates of our tower components (as parametrized through $\Gamma_0$) relative to the 
overall rate of cosmological expansion during stasis (as parametrized through $\barkappa$).
Indeed many of our future results will depend directly on these two quantities.  Unlike $\eta$, 
however, we shall find that $\xi$ makes its appearance in our equations only after our system has 
reached stasis.  It is for this reason that $\xi$ has been defined directly in terms of 
$\barkappa$ rather than $\kappa$ itself.   Thus $\xi$ has a fixed value during stasis.

Given the scaling relations in Eqs.~(\ref{MGscalings}) and
(\ref{MGmassform}), we can now evaluate the conditions for stasis by evaluating the sum which appears on the left side of 
our constraint equation in Eq.~(\ref{MGmasterconstraints}).  Let us first focus on the
behavior of $\Omega_\ell(t)$.
In Eq.~(\ref{MGOmegal}), we provided an expression for $\Omega_\ell(t)$ in terms of a fiducial time $t_\ast$ already within stasis, but in order to connect with the abundance-scaling relation in Eq.~(\ref{MGscalings}) we would like to replace $t_\ast$ with the production time $t^{(0)}$.
However, the expression in Eq.~(\ref{MGOmegal}) assumed an eternal stasis that was already in effect at the fiducial time $t_\ast$.   
Indeed, it is for this reason that we were able to assume within Eq.~(\ref{MGOmegal}) that $\Omega_\ell(t)$ accrues a net 
gravitational 
redshift factor $(t/t_\ast)^{2-\barkappa}$ between $t_\ast$ and $t$.
However, as discussed above, these assumptions will no longer be true if $t_\ast$ is replaced by $t^{(0)}$, since we now expect
that stasis will not emerge until some time after $t^{(0)}$. 

For simplicity and generality, we shall therefore let $h(t_i,t_f)$ denote the net gravitational redshift 
factor that accrues between any two times $t_i$ and $t_f$. 
We thus have
\beq
    \Omega_\ell(t) ~=~ \Omega_\ell^{(0)}\,   h(t^{(0)},t)  \, e^{-\Gamma_\ell (t-t^{(0)})}~.
\label{MGOmegalnew}
\eeq
Note that this $h$-factor is necessarily $\ell$-independent since the gravitational redshift affects all 
components equally.  It turns out that there are important subtleties associated with our use of an $h$-factor in this way, but these subtleties will not affect our results.  We shall therefore defer a more detailed discussion of this $h$-factor until Sect.~\ref{subsect:h-section}.~ However, given Eq.~(\ref{MGOmegalnew}), we then find that
\beqn
 \sum_\ell \Gamma_\ell \Omega_\ell(t) ~&=&~ 
 \Gamma_0 \Omega_0^{(0)} \, h(t^{(0)},t) \nonumber\\
 \times && \!\! \sum_\ell \left( \frac{m_\ell}{m_0} \right)^{\alpha+\gamma} \!
         e^{ - \Gamma_0 \left( \frac{m_\ell}{m_0}\right)^\gamma (t-t^{(0)})}
         ~,~~~ ~~~~~~
\label{MGbigsum1}
\eeqn
where we have used the scaling relations in Eq.~(\ref{MGscalings}).

\begin{figure*}[t!]
\centering
\includegraphics[keepaspectratio, width=0.49\textwidth]{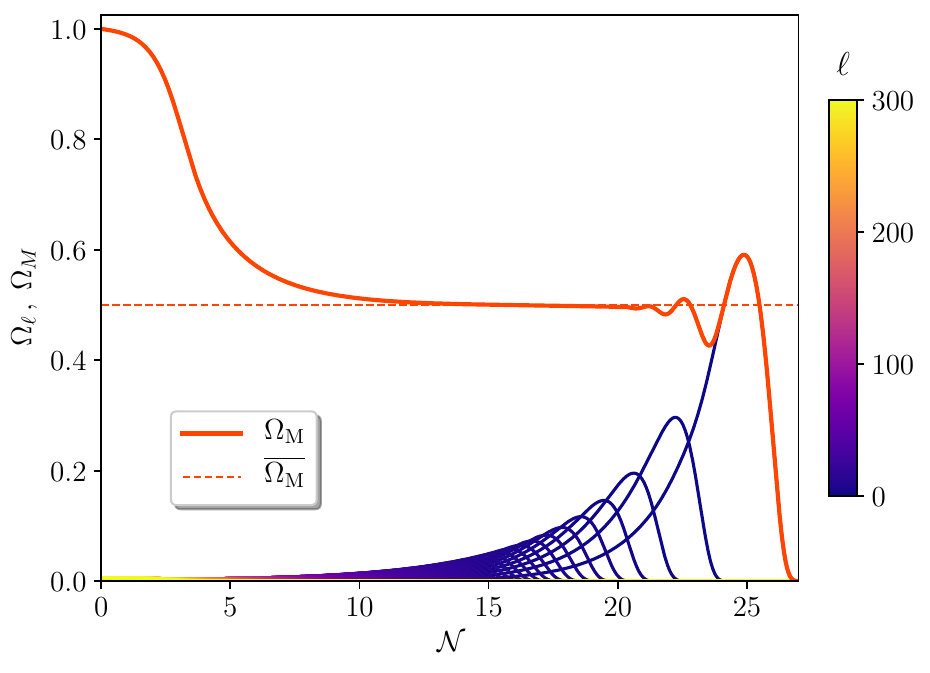}
\hfill
\includegraphics[keepaspectratio, width=0.49\textwidth]{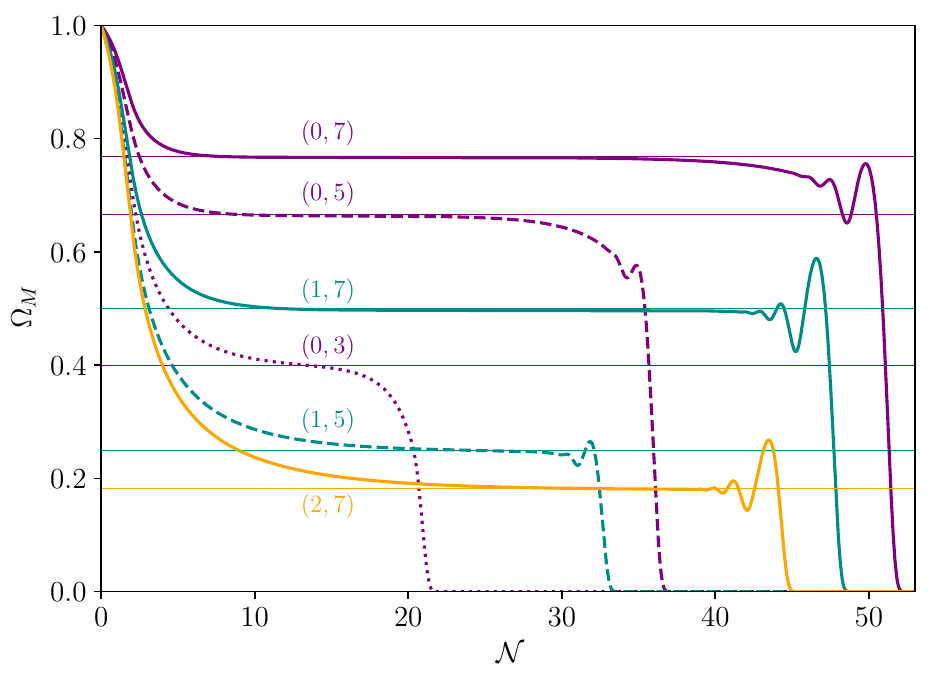}
\caption{ Matter/radiation stasis. {\it Left panel}\/:  The individual matter abundances 
$\Omega_\ell$ (shown with colors ranging from orange to blue) and the corresponding total matter abundance $\Omega_M$ (red), 
plotted as functions of the number $\calN$ of $e$-folds since the initial time of production. 
Even though the individual abundances $\Omega_\ell$ exhibit complex behaviors which are affected 
by cosmological expansion as well as $\phi_\ell$ decay, the system quickly evolves into a stasis 
state in which their sum $\Omega_M$ becomes constant.  These curves were generated through a 
direct numerical solution of the Boltzmann equations corresponding to our discrete tower of 
decaying states without invoking any approximations, and correspond to the parameter choices
$(\alpha,\gamma,\delta)=(1,7,1)$ --- for which $\barOmega_M=1/2$ ---  with 
$\Delta m = m_0$, $N=300$, and $\Gamma_{N-1}/H^{(0)}=0.01$.
{\it Right panel}\/:   
The total matter abundances $\Omega_M$, plotted as  functions of $\calN$  for different values 
of $(\alpha,\gamma)$ satisfying the constraints in Eq.~(\ref{MGrange}).  For each plot we have 
taken $\delta=1$, $\Delta m=m_0$, $N=10^5$, and $H^{(0)}/\Gamma_{N-1} = 0.1$.  In each case we 
see that the system settles into a prolonged stasis epoch lasting many $e$-folds, with a 
corresponding stasis abundance $\barOmega_M$ predicted by Eq.~(\ref{MGstasisOmegaM}).
Ultimately, in all cases, the stasis epoch  ends as we approach the final decays of the lightest 
modes. Both figures are taken from Ref.~\cite{Dienes:2021woi}.
\label{fig:review}}
\end{figure*}

In order to evaluate this sum, we can pass to a continuum limit in which we have a 
large number of $\phi_\ell$ states.  We can therefore imagine that the spectrum
of decay times $\tau_\ell\equiv \Gamma_\ell^{-1}$ is nearly continuous, merging 
to form a continuous variable ${\tau}$.    We can likewise view the discrete 
spectrum of energy densities $\rho_\ell$ and abundances $\Omega_\ell$ as continuous 
functions $\rho(\tau)$ and $\Omega(\tau)$ where the states are now indexed by the 
continuous ${\tau}$-variable corresponding to their decay times.  This allows us to rewrite 
our expressions such that $\ell$ is eliminated in favor of $\tau$.   For example, we 
replace $m_\ell/m_0$ with $(\Gamma_0 \tau)^{-1/\gamma}$. 
We can then convert the $\ell$-sum over states to a ${\tau}$-integral, {\it i.e.}\/,
\beq 
    \sum_\ell ~\Longleftrightarrow~ 
    \int d{\tau}  \, n_{\tau}(\tau)~,
\label{sum-to-integral_tau}
\eeq
where $n_{\tau}\equiv |d\ell/d{\tau}|$ is the density of states per unit ${\tau}$.
 
Of course, this passage from the sum to the integral involves a number of approximations whose 
effects are discussed in Ref.~\cite{Dienes:2021woi}.   One important observation is that this 
approximation becomes increasingly accurate for times $t$ which are far from the ``boundary'' 
(or ``edge'') effects associated with the initial entry into or exit from the stasis epoch.  
For a finite tower $\phi_\ell$ with $\ell=0,1,...,N-1$ with $N\gg 1$, this implies that our 
integral approximation will be especially valid for times $t$ within the range
\beq
    \tau_{N-1} ~\ll~ t ~\ll~ \tau_{0} 
\label{latertime}
\eeq
where of course we recall our original assumption that $t^{(0)} < \tau_{N-1}$, 
where $t^{(0)}$ is the initial production time for the $\phi_\ell$ tower.  The requirement that $t\gg \tau_{N-1}$ implies that we are focusing on a time interval after which a sizable number of states at the top of the tower have already decayed from matter to radiation.  This allows our integral approximation to capture the general behavior of our system after any initial transient effects have dissipated and the flow of energy density from matter to radiation is well underway.

Evaluating the sums in Eq.~(\ref{MGbigsum1}) in this way and assuming $\gamma>0$ and $\eta>0$ leads 
to the result
\beqn
 && \sum_\ell \Gamma_\ell \Omega_\ell(t) ~=~ \frac{ \Gamma_0 \Omega_0^{(0)}}{\gamma\delta}
     \left( \frac{m_0}{\Delta m}\right)^{1/\delta} \, \Gamma\left( \frac{\eta}{\gamma}+1\right)
          \nonumber\\
  && ~~~~~~~~~~ \times~
     h(t^{(0)},t) \,
     \left[ \Gamma_0 (t-t^{(0)}) \right]^{ -1-\eta/\gamma}~,~~\nonumber\\
\label{MGresult2}
\eeqn
where $\Gamma(z)$ is the Euler gamma function.   Likewise, letting $t_\ast$
continue to denote a fiducial time at which stasis has already developed, we can write~\cite{Dienes:2021woi}
\beqn
  h(t^{(0)},t) ~&=&~  h(t^{(0)},t_\ast)\, h(t_\ast,t) \nonumber\\
    ~&=&~ h(t^{(0)},t_\ast)\, \left( \frac{t}{t_\ast} \right)^{2-\barkappa}~.~~~~
\label{MGfactorizedh}
\eeqn
Substituting this result into Eq.~(\ref{MGresult2}), we see that 
$\sum_\ell \Gamma_\ell \Omega_\ell$ will scale as $1/t$ --- as required by 
Eq.~(\ref{MGmasterconstraints}) --- only if we take $t\gg t^{(0)}$ and require
\beq
\boxed{
   ~~\frac{\eta}{\gamma} ~=~ 2-\barkappa
      ~.~~
}
\label{MGalphagammaconstraint}
\eeq
Equivalently, through Eq.~(\ref{MGHubble}), this yields 
\beq
  \boxed{
  ~~\barOmega_M ~=~ \frac{  2 \gamma - 4\eta }{2 \gamma - \eta }~.~~
}
\label{MGstasisOmegaM}
\eeq 
Thus, for $t\gg t^{(0)}$ and for any values of $\eta$ and $\gamma$ within the ranges
\beq
 \gamma~>~0 ~,~~~~~~~
0 ~< ~ \eta ~\leq~ \frac{\gamma}{2}~,
\label{MGrange}
\eeq
our system has a stasis configuration during which 
$\barOmega_M$ is given in
Eq.~(\ref{MGstasisOmegaM}).  
Indeed, it is further shown in Ref.~\cite{Dienes:2021woi} 
that this state is a {\it global attractor}\/, and thus our system will evolve 
into the stasis state even if it does not begin in stasis at $t=t^{(0)}$.   This attractor 
behavior is discussed further in Sect.~\ref{sec:attractor}.   The stasis 
state then ends when the last $\phi_\ell$ has decayed.  Indeed, it is shown in
Ref.~\cite{Dienes:2021woi} that this stasis state will generally last for $\calN_s$ 
$e$-folds, where
\beq
     \calN_s ~\approx  ~  \frac{2\gamma \delta}{4-\barOmega_M} \log N~.
\label{MGNN}
\eeq
Here $N$ is the number of $\phi_\ell$ states in the tower.  We thus see that we can 
adjust the number $\calN_s$ of $e$-folds associated with the stasis epoch simply by
adjusting $N$.

In Fig.~\ref{fig:review} we illustrate this matter/radiation stasis by plotting the numerical results that emerge from an exact Boltzmann evolution of our individual abundances 
within the cosmology defined by the successive decays of our $\phi_\ell$ states.
In the left panel we plot the individual abundances $\Omega_\ell$ (shown in orange/blue) as well as the corresponding total matter abundance $\Omega_M$ (red) as functions of the number $\calN$ of $e$-folds since the initial time of production. The individual abundances $\Omega_\ell(t)$ exhibit complex behaviors, first   
rising due to cosmological expansion and then falling due to $\phi_\ell$ decay.
However, as time advances, this happens
in such a way that {\it the identity of the $\phi_\ell$ state with the largest abundance keeps changing}\/ as the decays work their way down the $\phi_\ell$ tower.  This causes the individual $\Omega_\ell(t)$ abundances to keep crossing each other in an interleaving,
cross-hatched fashion, as shown. 
This cross-hatched behavior for the individual abundances is a hallmark of the stasis state.
Indeed, despite the complex behaviors exhibited by the individual abundances $\Omega_\ell(t)$, we see that the system quickly evolves into a stasis state in which their sum $\Omega_M$ becomes constant.

In the right panel of Fig.~\ref{fig:review} we plot the total matter abundances $\Omega_M$ as functions of $\calN$ for a variety of different values of $(\alpha,\gamma)$ satisfying
the constraints in Eq.~(\ref{MGrange}). 
In each case we see that the system settles into a prolonged stasis epoch lasting many $e$-folds, with a corresponding stasis abundance $\barOmega_M$ predicted by Eq.~(\ref{MGstasisOmegaM}).
Ultimately, in all cases, the stasis epoch  ends as we approach the final decays of the lightest modes.

\subsection{General lessons going forward\label{subsect:general_lessons}}

Having reviewed the main results of Ref.~\cite{Dienes:2021woi}, let us now extract several lessons from this analysis which will serve as critical guideposts for our subsequent work in this paper.

First, we note that our 
main stasis constraint in Eq.~(\ref{MGmasterconstraints}) has been formulated by demanding that $d\Omega_M/dt=0$ for all times $t$ during stasis.   It is this which led to the constraint equation in Eq.~(\ref{MGalphagammaconstraint}) and to the result for $\barOmega_M$ in Eq.~(\ref{MGstasisOmegaM}).
However, an alternative approach would have been to 
proceed by demanding the condition
\beq 
        \sum_\ell \Omega_\ell(t) ~=~ \barOmega_M
\label{MGintegralform}
\eeq 
where each $\Omega_\ell(t)$ is given in Eq.~(\ref{MGOmegal}).  This simple condition would then replace the condition in Eq.~(\ref{MGmasterconstraints}).
Evaluating the sum within Eq.~(\ref{MGintegralform}) would then lead to an integral that is similar to the one we encountered between
Eqs.~(\ref{MGbigsum1})
and (\ref{MGresult2}),
and demanding that the resulting quantity be independent of time would then produce the same results as in Eqs.~(\ref{MGalphagammaconstraint}) and (\ref{MGstasisOmegaM}).

Of course, these two approaches are not distinct:  the former is simply the time-derivative of the latter.  Indeed,
the requirement that $\sum_\ell \Omega_\ell(t)$ be time-independent during stasis is ultimately tantamount to the constraint that the quantities in Eq.~(\ref{MGmasterconstraints}) scale as $1/t$ during stasis.  We shall therefore refer to these two possible formulations of our stasis constraints as employing either the {\it differential}\/ approach or the {\it integral}\/ approach.
However, these approaches generally have different advantages.   The integral approach is  more direct, but it requires having an explicit solution for the abundances $\Omega_\ell(t)$ such as in Eq.~(\ref{MGOmegal}) or Eq.~(\ref{MGOmegalnew}).
For complicated cosmologies, these solutions may not always be easily determined.  By contrast, the differential approach does not require this information.   Moreover, because the differential approach is the time-derivative of the integral approach, it explicitly describes the {\it flow}\/ of energy densities and abundances between our different energy components.  Indeed, as we shall see more explicitly in Sect.~\ref{sec:TripleStasis}, the left side of Eq.~(\ref{MGmasterconstraints}) functions as the driving ``pump'' of energy density from matter to radiation, while the right side depends on the Hubble parameter and thus captures the redshifting effects of cosmological expansion.

Thus, going forward, we shall  utilize either
the derivative forms of our stasis constraints, as in Eq.~(\ref{MGmasterconstraints}), or the integral forms of such stasis constraints, as in
Eq.~(\ref{MGintegralform}), 
choosing whichever form is calculationally cleaner given the particular stasis scenario under study.  Indeed, we shall even occasionally find that a mixture of both approaches is needed in order to obtain certain results.

A second relevant issue concerns whether there are any {\it further}\/ constraints that should be imposed in order to achieve a stasis epoch.
After all, our analysis in Sect.~\ref{subsect:2A} 
merely ensured that the left side of the stasis constraint in Eq.~(\ref{MGmasterconstraints}) has the required $1/t$ scaling dependence.  Indeed, this is what gave rise to the constraint in Eq.~(\ref{MGalphagammaconstraint}).
We shall refer to such constraints 
as  {\it scaling constraints}\/.
However, it might seem that 
there is an additional constraint that we should also impose, namely that which ensures the correct {\it prefactor}\/
in Eq.~(\ref{MGmasterconstraints}).  
In general such prefactor constraints may contain additional information beyond that which emerges from the scaling constraints. 

This issue is more subtle than it may at first appear.   
Given the result in Eq.~(\ref{MGresult2}), we see that we can write 
$\sum_\ell \Gamma_\ell \Omega_\ell(t)= 
C' t^{-1}$ where the prefactor $C'$ is given by
\beqn 
   C' ~&\equiv&~ \frac{1}{\gamma\delta}
     \left( \frac{m_0}{\Delta m}\right)^{1/\delta}  
     \Gamma\left( \frac{\eta}{\gamma}+1\right)  h(t^{(0)},t_\ast)
      \,\frac{\Omega_0^{(0)} }{(\Gamma_0 t_\ast)^{\eta/\gamma}} \nonumber\\
      ~&=&~
      \frac{ 1}{\gamma\delta}
     \left( \frac{m_0}{\Delta m}\right)^{1/\delta} 
     \Gamma\left( \frac{\eta}{\gamma}+1\right)   
     \,\Omega_0(\tau_0) \,e^{\Gamma_0 (\tau_0-t^{(0)})}~.~~~
             \nonumber\\
\label{MGCdef}
\eeqn
In writing Eq.~(\ref{MGCdef}) we have used Eq.~(\ref{MGalphagammaconstraint}) in order 
to simplify our expressions, and we have likewise defined  $\tau_0\equiv \Gamma_0^{-1}$. 
In this connection, we note that in passing to the second line of Eq.~(\ref{MGCdef}) the 
value of $C'$ has become independent of the choice of $t_\ast$, as it must, since $t_\ast$ 
is only a fiducial time with no physical significance.  

In principle the result for $C'$ in Eq.~(\ref{MGCdef}) bears no relation to the 
desired prefactor $(2-\barkappa)\barOmega_M$ in Eq.~(\ref{MGmasterconstraints}).    
However, if we calculate $\barOmega_M$ by directly evaluating the sum in Eq.~(\ref{MGintegralform}) by similarly 
utilizing our integral approximation and imposing the constraint in 
Eq.~(\ref{MGalphagammaconstraint}), we find that $\sum_\ell \Omega_\ell (t) = C$ where
$C$ is the same expression as in Eq.~(\ref{MGCdef}) except that the argument of the 
Euler gamma function is now given by $\eta/\gamma$ rather than $\eta/\gamma+1$.
Using the identity $\Gamma(1+z)=z\Gamma(z)$, we can therefore bundle these results 
together in order to find that our two prefactors are related to each other via
\beq
    C' ~=~ \left(\frac{\eta}{\gamma}\right)
     \,C  
     ~=~ \left(2-\barkappa\right)\,C~. 
\label{bundled}
\eeq
Of course, this is only a {\it relative}\/ relation between $C$ and $C'$ --- one which 
is independent of their individual absolute sizes.  However, it eliminates all of the 
complicated factors such as those in Eq.~(\ref{MGCdef}) which arose from our conversion 
of the discrete sum to an integral, and it is also consistent with Eq.~(\ref{MGmasterconstraints}).

Given these results, the final remaining step 
is to demonstrate that either of our two prefactors $C$ or $C'$ takes the correct {\it absolute}\/ size ---
\ie, that
\beq
C ~=~ \barOmega_M~,
\label{eq:abs1}
\eeq
or that
\beq
C'~=~ \left(\frac{\eta}{\gamma}\right)\barOmega_M~=~ (2-\barkappa)\, 
\barOmega_M~,
\label{eq:abs2}
\eeq
 where $\barOmega_M$ is the matter abundance that ultimately emerges during stasis. 
In conjunction with Eq.~(\ref{bundled}), this would then guarantee the correct prefactors in Eq.~(\ref{MGmasterconstraints}).
Indeed, from Eq.~(\ref{eq:abs2}) and the definition of $C'$ above Eq.~(\ref{MGCdef}), we find that 
\beq
 \sum_\ell \Gamma_\ell \Omega_\ell(t) ~=~
     \left(\frac{\eta}{\gamma}\right) \barOmega_M \, \frac{1}{t}~
\label{pumpMatterGamma}
\eeq
during stasis.
However, this final absolute-prefactor constraint in Eq.~(\ref{eq:abs1}) or (\ref{eq:abs2})  is unlike the others.  
Whereas our overall scaling constraint and {\it relative}\/-prefactor constraint are independent of the specific approximations
involved in passing from Eq.~(\ref{MGbigsum1}) 
to Eq.~(\ref{MGresult2}) [or equivalently in obtaining a precise value for the discrete sum $\sum_\ell \Omega_\ell(t)$],
the {\it absolute}\/-prefactor constraint in 
Eq.~(\ref{eq:abs1}) or (\ref{eq:abs2})
is highly sensitive to the details of these approximations.

Fortunately, as discussed in Ref.~\cite{Dienes:2021woi}, we need 
 not worry about this absolute-prefactor constraint because it basically functions as an overall normalization constraint, and 
  the proper normalizations of these sums are ultimately guaranteed by the attractor behavior of this system.
Indeed, even if this overall normalization constraint is not initially satisfied, the system will inevitably flow towards the stasis attractor solution in which the proper overall normalization comes into balance.  This feature is illustrated explicitly in Fig.~\ref{fig:review} and in Ref.~\cite{Dienes:2021woi}.

It is important to understand how this balancing occurs.  {\it A priori}\/, the individual 
abundances are given in Eq.~(\ref{MGOmegalnew}) where $h(t^{(0)},t)$ represents the net 
gravitational redshift factor that accrues between the original production time $t^{(0)}$ 
and any later time $t$.  Thus, $h(t^{(0)},t)$ carries with it an explicit time-dependence.   
Indeed, because this factor is $\ell$-independent, this same factor also appears within the 
total sum $\sum_\ell \Omega_\ell(t)$.  However, once our system settles into a stasis 
configuration, $h(t^{(0)}, t)$ factorizes as in Eq.~(\ref{MGfactorizedh}) where $t_\ast$ 
is any fiducial time during stasis.   The factor $(t/t_\ast)^{2-\barkappa}$  within 
Eq.~(\ref{MGfactorizedh}) is then cancelled as part of the overall scaling constraints, 
leaving behind the time-independent factor $h(t^{(0)},t_\ast)$ which appears in 
Eq.~(\ref{MGCdef}).   This factor thus represents the part of the net redshift that occurs 
between $t^{(0)}$ and $t_\ast$, and its value has already been dynamically adjusted during the 
pre-stasis epoch so as to ensure that $C=\barOmega_M$.   

Phrased somewhat differently, we may define 
\beq
  h(t) ~\equiv~ h(t^{(0)},t) \left( \frac{t}{t_\ast} \right)^{-2+\barkappa}~.
\eeq
In general, this quantity is time dependent and evolves significantly during the time interval 
from $t^{(0)}$ to $t_\ast$ prior to the emergence of stasis.  However, upon 
reaching $t_\ast$, we know that our system has entered the stasis 
regime.  The quantity $h(t)$ is then a constant, and from Eq.~(\ref{MGfactorizedh}) we see that 
this constant is what we have been calling $h(t^{(0)}, t_\ast)$.   
Indeed, during the pre-stasis epoch when $h(t)$ was still evolving, this quantity was heading towards 
(and ultimately assumed) precisely the value $h(t^{(0)}, t_\ast)$ needed to ensure that 
$C=\barOmega_M$.   We will discuss these $h$-factors in further detail in Sect.~\ref{subsect:h-section}.

There is also one additional constraint that is worthy of note.   As indicated below Eq.~(\ref{MGbigsum1}), it was necessary to assume in passing from Eq.~(\ref{MGbigsum1}) 
to Eq.~(\ref{MGresult2}) that
$\gamma>0$ and that 
\beq
   \eta ~ > ~ 0~.
\label{logavoidance}
\eeq
 Indeed, if our initial parameters had satisfied ${\eta=0}$, we would have obtained a logarithmic (rather than power-law) dependence on $t$, and there would have been no hope of achieving a true stasis in such a case because of the resulting logarithmic drift.
 
Of course $\gamma>0$ is a perfectly natural assumption for our underlying model, since this 
implies that our decay widths grow with $\ell$ --- \ie, that the more massive $\phi_\ell$ states 
decay more rapidly than do the lighter $\phi_\ell$ states.  By contrast, it is 
not {\it a priori}\/ required that ${\eta>0}$ (or equivalently that $\alpha+1/\delta>0$).  
In principle, this therefore becomes an additional constraint that must be imposed on our model in order 
to avoid logarithmic drift and achieve stasis.  However, it turns out that this constraint 
is already guaranteed by Eq.~(\ref{MGalphagammaconstraint}): since purely matter-dominated 
and radiation-dominated universes have $\barkappa=2$ and $\barkappa=3/2$ respectively, any 
universe  exhibiting a two-component stasis between matter and radiation must necessarily 
have $3/2<\barkappa<2$.  Eq.~(\ref{MGalphagammaconstraint}) then implies that $\eta>0$.
The constraint ${\eta>0}$ is also already implicit within 
Eq.~(\ref{MGrange}).

Thus, to summarize, we see that in general there are several kinds of constraints that must be 
satisfied in order to achieve stasis:
\begin{itemize}
  \item  overall scaling constraints such as that in Eq.~(\ref{MGalphagammaconstraint});
  \item {\it relative}\/ prefactor constraints such as that in Eq.~(\ref{bundled});
  \item {\it absolute}\/ prefactor constraints such as that in Eq.~(\ref{eq:abs1}) 
    or (\ref{eq:abs2}) 
    which ensure the correct overall normalizations;  and
  \item 
    constraints  such as that in Eq.~(\ref{logavoidance}) which ensure that there is no 
    logarithmic drift, and that a true power-law time-dependence emerges from our sums over states, 
    ultimately to be cancelled through cosmological redshifting effects.
\end{itemize}
In general, the absolute-prefactor constraints will be satisfied as a consequence of the attractor 
behavior associated with our stasis solutions when these solutions are indeed attractors.  
However, {\it a priori}\/, each of the other constraints is generally capable of yielding new restrictions on our model or 
independent information concerning the properties of the resulting stasis.  Of course, for the 
case of the pairwise matter/radiation stasis we have examined here, we have found that the 
relative-prefactor and logarithm-avoidance constraints are already satisfied whenever the 
overall scaling constraint is satisfied.  In other words, matter/radiation stasis
may be 
viewed as an over-constrained system which nevertheless gives rise to stasis solutions because 
these different constraints happen to be redundant (or subsumed within each other) without 
providing additional information.  However, as we shall shortly see, this will not always be 
the case.

We close this section with one final comment.  Throughout this section, we have been referring to Eq.~(\ref{MGalphagammaconstraint}) as a {\it constraint}\/ equation.   In reality, however, 
this equation is not a constraint on our scaling exponents $(\alpha,\gamma,\delta)$ so much 
as a {\it prediction} for the value of $\kappa$ during the resulting stasis.
Indeed, so long as our scaling exponents are chosen to lie within the ranges specified in 
Eq.~(\ref{MGrange}), we see that Eq.~(\ref{MGalphagammaconstraint}) simply allows us to calculate 
the corresponding value of $\barkappa$, and this in turn allows us [via Eq.~(\ref{MGHubble})] 
to determine the stasis abundance $\barOmega_M$, as in Eq.~(\ref{MGstasisOmegaM}).
Thus, so long as our scaling exponents are chosen to lie within the ranges specified in 
Eq.~(\ref{MGrange}), we will {\it always}\/ obtain a stasis.  We see, then, that our desire to achieve matter/radiation 
stasis does not impose any actual constraints on our model beyond those in Eq.~(\ref{MGrange});   indeed, stasis
emerges quite robustly for all values of the relevant parameters within these ranges.
We shall nevertheless often refer to Eq.~(\ref{MGalphagammaconstraint}) and its cousins as constraint 
equations in what follows.

\FloatBarrier

\section{Vacuum-energy/matter stasis\label{sec:LambdaMatter}}

We now turn our attention to one of the main tasks of this paper:  the inclusion of {\it vacuum energy}\/ into the stasis discussion. 
In this section we begin by exploring the possibility of a stasis between vacuum energy and matter, and the ways in which this might arise from BSM physics.  Of course, we know that any epoch in which both matter and vacuum energy are present and which is initially matter-dominated will evolve --- simply as the result of cosmological expansion --- into a vacuum-energy-dominated epoch.  This is because  the matter energy density falls like $\rho_M\sim a^{-3}$ as the universe expands,
while the vacuum-energy density $\rho_\Lambda$ remains constant.  Achieving stasis between matter and vacuum energy therefore requires a mechanism to counterbalance this effect and convert vacuum energy back to matter.

\FloatBarrier

\subsection{Modeling vacuum energy and the transition to matter}

To study this, we begin with a discussion of how we might introduce vacuum energy into our analysis.  As we shall see, this issue turns out to be surprisingly subtle and requires some care.

By definition, vacuum energy has equation-of-state parameter $w= -1$.   Indeed, such a state is pure potential energy (this is the ``vacuum energy'') and no kinetic energy.   One natural approach towards studying a stasis involving vacuum energy and matter would therefore be to repeat the analysis in Sect.~\ref{sec:MatterGamma} for matter/radiation stasis, only replacing the equation of state of the initial $\phi_\ell$ components from $w=0$ to $w= -1$ until the moment they each undergo some sort of transition to matter.   We would then replace the post-transition equation of state in Sect.~\ref{sec:MatterGamma} from $w= 1/3$  (radiation) to $w=0$.  

Of course, it would also be necessary to identify a mechanism for this ``transition'' --- \ie, to 
identify a
general process through which vacuum energy can be converted into matter, in much the same way as the  process of particle decay converted matter into radiation in Sect.~\ref{sec:MatterGamma}.~ 
However, it is not difficult to identify such a process.
Let us consider the
  coherent state consisting of the zero-momentum modes of a scalar field $\phi$ of mass $m$.
   At early times, when the Hubble parameter is large (with $3H \gg 2m$), this field is severely overdamped 
   and thus has little kinetic energy. This is therefore a situation in which the energy of the field can be considered pure potential energy (vacuum energy), with equation-of-state parameter $w\approx -1$. However, as the universe expands, the Hubble parameter generally drops.   As a result, the field eventually becomes underdamped (with $3H\lsim 2m$) and begins to experience damped oscillations. In general, these oscillations quickly virialize, whereupon the energy of this field is split equally between potential and kinetic energy.  The corresponding equation-of-state parameter is then $w=0$, and the corresponding energy density  behaves as matter as far as issues pertaining to cosmological expansion are concerned.  

We thus see that the overdamped/underdamped transition at $3H(t)=2m$ provides a natural mechanism for converting vacuum energy to matter, in exactly the same way as particle decay at $t=1/\Gamma$ provided a natural mechanism in Sect.~\ref{sec:MatterGamma} for converting matter into radiation.   Of course, near the transition time, our field has a non-zero kinetic energy and thus has neither $w=-1$ nor $w=0$. 
Indeed, during such a transition period,  the energy density of our field can be interpreted as a mixture of   vacuum energy and matter.
However, our main point is that an overdamped/underdamped transition has the net effect of converting vacuum energy to matter.
In the following discussion, for the purpose of calculational simplicity, we shall idealize this transition by disregarding the ``transient'' effects that arise near $3H(t)\approx 2m$, and instead approximate our scalar field as having $w= -1$ whenever
$3H(t)> 2m$ and $w=0$ otherwise.

Given this understanding, we might attempt to 
generate a long-lived vacuum-energy/matter stasis by initially assuming a tower of coherent overdamped scalar fields $\phi_\ell$ with masses $m_\ell$ and equations of state $w_\ell =-1$, in complete analogy with the initial configuration in Sect.~\ref{sec:MatterGamma}.~  
We would then allow these $\phi_\ell$ states to undergo successive transitions to an underdamped phase as the falling Hubble parameter $H(t)$ crosses the successive critical transition points $2m_\ell/3$.    Such underdamping transitions would then proceed down the $\phi_\ell$ tower, just as before, and potentially establish a stasis epoch along the way.  Indeed, at any moment, the lighter fields would still be in the overdamped phase while the heavier fields will have already transitioned to the underdamped phase. 

This approach would clearly be the most straightforward analogue of the scenario discussed in Sect.~\ref{sec:MatterGamma}.~ 
Ultimately, however, this approach does not work.  The reason is simple:  such a system begins as pure vacuum energy, with a total equation-of-state parameter $w_{\rm tot} = -1$.   Such a universe therefore has a Hubble parameter which is constant and never falls.   As a result, there is no possibility of any fields becoming underdamped, and likewise no possibility of any subsequent transitions from vacuum energy to matter.  Indeed, such a  system simply remains ``stuck'' in its initial state, with no subsequent dynamics at all. There is thus no way in which a true stasis can develop in such a theory --- we simply have the total initial abundances $\Omega_\Lambda=1$ and $\Omega_M=0$ 
in perpetuity.   Of course, this is itself a kind of degenerate ``stasis'', but it is uninteresting for our purposes.

We shall therefore need to modify this na\"\i ve picture in such a way that our system will actually evolve away from its initial state, with ensuing cascading overdamped/underdamped transitions that convert vacuum energy to matter.
There are several approaches we might follow in order to achieve this:

\begin{itemize}
\item  We could begin by positing that the initial state of our system also includes some additional non-vacuum-energy component.  Such an additional energy component would then introduce a non-trivial time-dependence for the Hubble parameter, and this would in turn eventually trigger the cascading overdamped/underdamped transitions that we require.   Unfortunately, doing this requires that our model now include extra components beyond our initial tower of states $\phi_\ell$.   While there is nothing wrong with this (and indeed such additional components may ultimately be well-motivated on phenomenological grounds), the introduction of such extra components beyond our tower of $\phi_\ell$ states is not in the spirit of our previous analysis and would introduce arbitrary new features and parameters into our model.

\item  A second option --- indeed, one which is more ``minimal'' and which does not introduce new fields --- would be to {\it deform}\/ our model slightly by imagining that our overdamped $\phi_\ell$ fields actually have a small kinetic energy (\ie, a ``slow roll'') in addition to their potential energies. 
For algebraic simplicity, we can incorporate this kinetic energy into our model by imagining that each of our $\phi_\ell$ fields has an arbitrary common fixed equation-of-state parameter $w$ in the range $ -1<w<0$ prior to becoming underdamped (after which we can assume that each field transitions to having $w=0$, as above). For example, we might consider $w$ to be extremely close to (but slightly greater than) $-1$.  Our model would then proceed exactly as before, with sharp overdamped/underdamped transitions occurring when $3H(t) = 2 m_\ell$.   Indeed, 
in such a model, non-zero values of the quantity $w+1$ would essentially function as a ``regulator'' which allows us to avoid the difficulties associated with taking $w=-1$.   We could then define our case of interest --- namely that with ``vacuum energy'' --- as the $w\to -1$ limit of this model.   Indeed, as we shall find, the $w\to -1$ limit is typically well-behaved even if the precise $w= -1$ endpoint is not. 
  
\item  Finally, we could consider the full equations of motion for the scalar fields $\phi_\ell$ without any approximations.   
These equations are those of a damped driven
oscillator in which the Hubble damping terms carry a non-trivial time-dependence.   
Within an arbitrary universe that has not yet reached stasis, such equations of 
motion lack analytical solutions, so this approach would necessarily be numerical.  
\end{itemize}

The latter two approaches have complementary strengths and weaknesses.  Within the $w>-1$ approach we are positing a relatively simple behavior for the $\rho_\ell$ energy densities, and thus this model can be understood and solved analytically.  Moreover, taking $w> -1$ as a regulator successfully allows us to avoid having an initial Hubble parameter which remains constant, thereby allowing the dynamics of our system to ``start'' on its own.  Indeed, we shall find that this model yields results in the $w\to -1$ limit which match much of what we expect 
from a na\"\i ve treatment 
in which we simply allow the vacuum-energy component to have $w= -1$ at the outset but in which the dynamics is somehow ``started'' in other ways.  Finally, this model has the side benefit of allowing us to study the prospects for achieving stasis for arbitrary $w$.  In this way we could thereby understand how the properties of the resulting stasis, if any, depend on $w$.   

Unfortunately, this model, while suitable for understanding overall cosmological features associated with stasis, lacks a microscopic (particle-physics) Lagrangian description.   By contrast, the full scalar-dynamics model
has a {\it bona fide}\/ realization in terms of the physics of a scalar field evolving in an external cosmology (\ie, subject to Hubble friction).  Such a model is thus the rigorous setting for the $3H=2m$ transition that allows us to convert from overdamped to underdamped behavior, or equivalently from vacuum energy to matter. 
Although this model cannot be solved analytically outside the stasis regime, 
a numerical analysis is possible.  Of course,
within this approach, the equation-of-state parameter $w$ for each field during the overdamped phase is not an input parameter over which we have direct control, but is instead an {\it output}\/ of the numerical simulation. Indeed, $w$ may not even be constant, nor will it necessarily be the same for each field.   In a similar way, the value of $w$ for each field {\it after}\/ the field becomes underdamped will not be strictly zero, but will also continue to have a non-trivial time-dependence.   Thus,  within this model it is only an approximation to assert that the underdamped phase results in pure ``matter'', just as it is an approximation to state that the overdamped phase is pure ``vacuum energy'' with fixed $w= -1$.

In this paper we shall adopt the general-$w$ model (as described in the second bullet above), allowing $w$ to lie anywhere within the range $-1<w<0$.   As we have discussed, this will enable us to study the stasis phenomenon analytically.  This in turn will also allow us to determine the conditions for stasis and ultimately study the behavior of this model as $w\to -1$.   
However, we shall study the full dynamical-scalar model in Ref.~\cite{toappear}, and we shall find that stasis emerges within the dynamical-scalar model as well.  Indeed, our results here and in Ref.~\cite{toappear} will together allow us to  verify that these two models, despite their differences, yield similar results. We shall therefore regard both models as demonstrating that vacuum energy can be successfully introduced into the overall stasis framework.

\FloatBarrier 
\subsection{Stasis analysis for general $w$\label{subsect:genw}}

As discussed above, we shall begin our analysis of a possible vacuum-energy/matter stasis in the same way as we did in Sect.~\ref{sec:MatterGamma}, specifically by assuming
a flat Friedmann-Robertson-Walker (FRW) universe containing a tower of scalar fields $\phi_\ell$ with masses $m_\ell$, where the indices $\ell=0,1,2,...,N-1$ are assigned in order of increasing mass.   At any time $t$, the fields for which $3H(t)> 2m_\ell$ will be assumed to be overdamped with a fixed equation-of-state parameter $w$ which we can imagine is close to (but greater than) $-1$. By contrast, the fields for which $3H(t)<2m_\ell$ will be assumed to be underdamped, with a fixed equation-of-state parameter $w=0$.   Thus the overdamped/underdamped transition at $3H=2 m_\ell$ converts vacuum energy (or its close approximation) to matter and proceeds down the tower as time advances.  We shall let $\rho_\Lambda$ and $\rho_M$ denote the total energy densities of this system attributable to vacuum and matter respectively, while $\rho_\ell$ will denote the energy density associated with the individual scalar field $\phi_\ell$ while it is still overdamped.  We shall also let $\Omega_\Lambda$, $\Omega_M$, and $\Omega_\ell$ represent the corresponding abundances.

Our analysis begins just as for matter/radiation stasis in Sect.~\ref{sec:MatterGamma}.~
Indeed, for any energy density $\rho_i$ (where $i=\Lambda, M,\ell$), 
the corresponding abundance $\Omega_i$ continues to be given by Eq.~(\ref{MGOmegadef}), from which Eq.~(\ref{MGstepone}) continues to follow.
However, the Friedmann ``acceleration'' equation now takes the form 
\beqn
  \frac{dH}{dt}  
  ~&=&~ -H^2 - \frac{4\pi G}{3} \left(  \sum_i \rho_i + 3 \sum_i p_i\right)\nonumber\\
  ~&=&~ -H^2 - \frac{4\pi G}{3} \left[  (1+ 3w)\rho_\Lambda + \rho_M \right]\nonumber\\
  ~&=&~ - \textstyle\frac{3}{2} \,H^2\, \left( 1+w\Omega_\Lambda \right)~,
\label{acceleq}
\eeqn
or equivalently 
\beq 
       \kappa ~=~ \frac{2}{1+w \Omega_\Lambda}~
\label{LMHubble}
\eeq
where we generally identify $\kappa$ through the parametrization 
\beq
   \frac{dH}{dt} ~=~ - \frac{3}{\kappa} H^2~.
\label{eq:kappagendef}
\eeq
Substituting Eq.~(\ref{acceleq}) into Eq.~(\ref{MGstepone}) we then obtain
\beq
    \frac{d\Omega_i}{dt} 
     ~=~ \frac{8\pi G}{3H^2} \frac{d\rho_i}{dt} + 3 H \Omega_i \left( 1+w\Omega_\Lambda \right)~,
\label{convert}
\eeq
yielding
\beqn
    \frac{d\Omega_\Lambda}{dt} 
     ~&=&~ \frac{8\pi G}{3H^2}  \frac{d\rho_\Lambda}{dt} 
     + 3H \Omega_\Lambda \left( 1+w\Omega_\Lambda \right) ~ \nonumber\\
    \frac{d\Omega_M}{dt} 
     ~&=&~ \frac{8\pi G}{3H^2} \frac{d\rho_M}{dt}  
    + 3H \Omega_M \left(1+w\Omega_\Lambda\right)~.
\label{convert2}
\eeqn
These are thus general relations for the time-evolution of $\Omega_\Lambda$ and $\Omega_M$ in terms of 
$d\rho_\Lambda/dt$ and $d\rho_M/dt$.

Given the relations in Eq.~(\ref{convert2}), our final step is to insert an appropriate equation of motion for 
$d\rho_\Lambda/dt$ (with the understanding that $d\rho_M/dt= -d\rho_\Lambda/dt$ by conservation of energy).  It is here that we introduce the idea that vacuum energy is converted to matter when the individual states $\phi_\ell$ become underdamped and begin oscillating.   For each field, this is presumed to occur precisely at the time $t_\ell$ when $3H(t_\ell)=2m_\ell$. In this paper, we shall refer to $t_\ell$ as a critical ``underdamping'' time.  Thus,
whereas the equations of motion given in Eq.~(\ref{MGeoms}) corresponded to the case in which the transition from matter to radiation occurred 
through an exponential decay term $\rho_\ell(t)\sim e^{-\Gamma_\ell (t-t^{(0)})}$, we shall now model  the corresponding vacuum-energy/matter transition term as $\rho_\ell \sim \Theta(t_\ell-t)$ where $\Theta(x)$ denotes the Heaviside $\Theta$-function [for which $\Theta(x) = 1$ for $x\geq 0$ and $\Theta(x)=0$ otherwise].  This enforces our expectation that $\rho_\ell$ is non-zero (and can thus be attributed to vacuum energy) only for $t\leq t_\ell$. 
  Likewise, the energy density for a fluid with equation-of-state parameter $w$ generally scales as $a^{-3(1+w)}$.
The corresponding equation of motion for each individual (vacuum) energy density $\rho_\ell$ is therefore given by
\beqn
  \frac{ d \rho_\ell}{dt} ~&=&~ 
  \rho_\ell  \frac{d}{dt}\Theta(t_\ell-t)
     -3 (1+w) H \rho_\ell ~\nonumber\\
 ~&=&~ 
 - \rho_\ell \delta(t_\ell-t)
     -3 (1+w) H \rho_\ell~,~~~~
\label{eq:drhoelldt}
\eeqn
whereupon we see that the equation of motion for the total vacuum-energy density $\rho_\Lambda$ in this system is given by
 \beq
 \frac{d\rho_\Lambda}{dt} ~=~ 
 - \sum_\ell \rho_\ell\, \delta(t_\ell-t) 
      - 3 (1+w) H \rho_\Lambda~.
\label{intermedsum}
 \eeq

Just as in Sect.~\ref{sec:MatterGamma}, let us now evaluate this sum by passing to a continuum limit in
which we truly have a large number
of $\phi_\ell$ states.  However, in this case it will prove more useful to imagine that it is the spectrum
of {\it underdamping}\/ times $t_\ell$ which is nearly continuous, merging to form a continuous variable ${\hat t}$.    We can likewise view the discrete spectrum of energy densities $\rho_\ell$ and abundances $\Omega_\ell$ as continuous functions $\rho({\hat t})$ and $\Omega(\hat t)$ where the states are now indexed by the continuous ${\hat t}$-variable corresponding to their underdamping times.  
We can then convert the $\ell$-sum over states to a ${\hat t}$-integral, {\it i.e.}\/,
\beq 
    \sum_\ell ~\Longleftrightarrow~ 
    \int d{\hat t}  \, n_{\hat t}({\hat t})~,
\label{sum-to-integral_that}
\eeq
where $n_{\hat t}\equiv |d\ell/d{\hat t}|$ is the density of states per unit ${\hat t}$.
  Of course, this passage from the sum to the integral involves a number of approximations whose effects are similar to those discussed discussed in Sect.~\ref{sec:MatterGamma} and in Ref.~\cite{Dienes:2021woi}.  
  In particular, our integral approximation will be especially valid for times $t$ within the range
\beq
    t_{N-1} ~\ll~ t ~\ll~ t_{0} 
\eeq
where we are of course assuming $t^{(0)} < t_{N-1}$. 
In other words, we are focusing on a time interval after which a sizable number of states at the top of the tower have already transitioned from vacuum energy to matter. 

Within this integral approximation, 
Eq.~(\ref{intermedsum}) then becomes
\beqn
 \frac{d\rho_\Lambda}{dt} ~&=&~ 
    - \int d{\hat t} \, n_{\hat t}({\hat t})
     \, \rho({\hat t}) \, \delta({\hat t}-t)~ -3(1+w) H \rho_\Lambda
      \nonumber\\
&=& ~ - n_{\hat t}(t) \, \rho(t) - 3(1+w)H\rho_\Lambda~.
\label{eom1}
 \eeqn
In a similar way we also find that
\beq 
  \frac{d \rho_M}{dt} ~=~ 
  +n_{\hat t}(t) \, \rho(t) - 3 H \rho_M~
\label{eom2}
\eeq
where the first term results from the conservation of energy that governs the  process of converting  vacuum energy to matter while the second term incorporates the gravitational redshifting that is experienced by matter under cosmological expansion.
Substituting these results into Eq.~(\ref{convert2}) we then find 
\beq
   \frac{d \Omega_\Lambda}{dt} ~=~
       - n_{\hat t}(t) \,\Omega(t)
              -3w H \Omega_\Lambda (1-\Omega_\Lambda)~
\label{convert3}
\eeq
with $d\Omega_M/dt = - d\Omega_\Lambda/dt$.
Note that $ \Omega_\Lambda (1-\Omega_\Lambda) = \Omega_\Lambda \Omega_M = \Omega_M (1-\Omega_M)$.

The differential equation for $\Omega_\Lambda$ in Eq.~(\ref{convert3}) is completely general, describing the
 time-evolution of $\Omega_\Lambda$ and $\Omega_M$.
One necessary (but not sufficient) condition for stasis is that
$d\Omega_\Lambda/dt=0$.   
This then yields the constraint
\beq
    \phantom{\sum_\ell^\infty}\!\!\!\!
    n_{\hat t}(t)\, \Omega(t)  ~=~ -3 w H\, \Omega_\Lambda\,(1- \Omega_\Lambda)~.
\label{condition}
\eeq
Of course, for stasis we wish to have situations in which Eq.~(\ref{condition}) holds not only instantaneously, but also over an extended period of time.  This requires not only that Eq.~(\ref{condition}) hold instantaneously, but that both sides of Eq.~(\ref{condition}) have the same time-dependence.   

However, it is  straightforward to determine the time-dependence of the Hubble parameter $H$ during an assumed period of stasis during which $\Omega_\Lambda$ and $\Omega_M$ are fixed at stasis values $\barOmega_\Lambda$ and $\barOmega_M$, respectively.
Under such conditions, we can solve
Eq.~(\ref{acceleq}) directly to find the exact solution
\beq
    H(t) = \left[\frac{2}{3(1+w\barOmega_\Lambda)}\right] \,\frac{1}{t} ~~\Longrightarrow~~
     \barkappa= \frac{2}{1+w \barOmega_\Lambda}~,~~
\label{Hubble}
\eeq
in agreement with the result in Eq.~(\ref{LMHubble}).
Note that this result holds 
for all $\barOmega_\Lambda$, {\it including}\/ $\barOmega_\Lambda = 1$, so long as $w> -1$.
Indeed, for $\barOmega_\Lambda=0$, we verify from Eq.~(\ref{Hubble}) the standard result that $H(t)= 2/(3t)$ for a matter-dominated universe.
The solution for $H(t)$ in Eq.~(\ref{Hubble}) in turn implies that our underdamping times $t_\ell$ during stasis are given by 
\beq 
         t_\ell ~=~ \frac{\barkappa}{2 m_\ell}
         ~=~  \frac{1}{1+w \barOmega_\Lambda}  \frac{1}{m_\ell}     ~.
\label{t-to-ell}
\eeq 
Likewise, this solution for $H(t)$ implies that 
 the scale factor grows during stasis as 
\beq
         a(t) ~=~
         a_\ast \left( \frac{t}{t_\ast}\right)^{\barkappa/3}
         ~=~ 
            a_\ast \left( \frac{t}{t_\ast}\right)^{2/(3 +3w \barOmega_\Lambda)}~
\label{scalefactor}
\eeq 
with $t_\ast$ representing an arbitrary early fiducial time during stasis, as in Sect.~\ref{sec:MatterGamma}, and 
 the `$\ast$' subscript indicating that the relevant quantity is evaluated at $t=t_\ast$. 
It then follows from Eq.~(\ref{eq:drhoelldt}) that
\beqn 
 \Omega_\ell(t) ~&=&~ \Omega_\ell^\ast \left( \frac{t}{t_\ast}\right)^2 \, 
 \left[\frac{a(t)}{a_\ast}\right]^{-3(1+w)}  \Theta\left(
  \frac{\barkappa}{2 m_\ell} -t \right)~ ~~~\nonumber\\
  ~&=&~ \Omega_\ell^\ast \left( \frac{t}{t_\ast}\right)^{2- (1+w) \barkappa }  
\Theta\left(
  \frac{\barkappa}{2 m_\ell} -t \right)~ ~~~
  \label{Omegal}
\eeqn
where we have further assumed that the fiducial time $t_\ast$ is prior to the transition of $\phi_\ell$ from vacuum energy to matter at $t_\ell$.   This final assumption will be discussed further and justified in Sect.~\ref{subsect:h-section}. 

Given the result in Eq.~(\ref{Hubble}), we find that Eq.~(\ref{condition}) will be
satisfied for an extended period of time
so long as 
\begin{empheq}[box=\fbox]{align}
~~n_{\hat t}(t) \, \Omega(t)  ~&=~ 
 \biggl\lbrack
 -w\,\barkappa\, \barOmega_\Lambda 
 (1-\barOmega_\Lambda)\biggr\rbrack \, \frac{1}{t} ~~\nonumber\\
   ~&=~  \biggl\lbrack
    2-(1+w)\barkappa\biggr\rbrack \, \barOmega_\Lambda\, 
\frac{1}{t} ~.~~
\label{masterconstraints}
\end{empheq}
This result, which is the vacuum-energy/matter analogue of Eq.~(\ref{MGmasterconstraints}), thus becomes our condition for vacuum-energy/matter stasis within the general-$w$ model.
Of course, as discussed in Sect.~\ref{subsect:general_lessons},  this condition is ultimately equivalent to the constraint that  $\sum_\ell \Omega_\ell (t)= \barOmega_\Lambda$ where $\Omega_\ell(t)$ is given in Eq.~(\ref{Omegal}), only expressed in differential form (and in the continuum limit) in order to expose the details of how the stasis is explicitly maintained.
We shall demonstrate this equivalence explicitly below.  

In parallel with our analysis in 
Sect.~\ref{sec:MatterGamma}, our goal is now to demonstrate that generic models of BSM physics that 
give rise to such towers of scalar states $\phi_\ell$ 
experiencing such overdamping/underdamping transitions will satisfy the stasis constraint in Eq.~(\ref{masterconstraints})
as exactly as possible over an extended time interval.    For this purpose, we can adopt the same generic parametrization that was discussed in Sect.~\ref{sec:MatterGamma}.~
Specifically, we shall imagine a tower of scalars $\phi_\ell$ 
 whose abundances and masses satisfy the scaling relations in 
 Eqs.~(\ref{MGscalings}) and (\ref{MGmassform}) respectively.
 We shall also assume that the initial production time $t^{(0)}$ for these scalar fields has occurred during a period 
in which none of the $\phi_\ell$ has yet become underdamped.   As a result, in complete analogy with 
 the model in Sect.~\ref{sec:MatterGamma}, we are implicitly assuming that  $\Omega_\Lambda=1$ during this initial period.

Within the context of this general model, we can now evaluate the quantities which appear on the left side of 
our constraint equation in Eq.~(\ref{masterconstraints}).   
By demanding that Eq.~(\ref{masterconstraints}) holds, we will then obtain the conditions on our model parameters that are required for stasis.   Of course, the question of whether and how these conditions may come to be satisfied for arbitrary initial conditions is a separate one which requires a different (dynamical) analysis.   We shall defer such a dynamical analysis to Sect.~\ref{sec:attractor}.

We begin by calculating $n_{\hat t}(t)$. 
Recall that
 $n_{\hat t}(t)$ is the
density of states per unit transition time $\hat t$, evaluated for precisely that part of the $\phi_\ell$ tower for which $t_\ell =t$.  
From Eqs.~(\ref{MGmassform}) and
(\ref{t-to-ell}) 
we have
\beq
    \ell ~=~ \left[ 
       \frac{\barkappa}{2}\,
       \frac{1}{ (\Delta m) t_\ell}\right]^{1/\delta} 
\label{elldef}
\eeq
where we have adopted the simplifying approximation $m_0\ll (\Delta m) \ell^\delta$.
We thus find
\beq
n_{\hat t}(t) ~\equiv~ \left| 
\frac{d\ell}{dt_\ell}\right|_{t_\ell=t}
 \!=~ 
 \frac{1}{\delta}  
\left( \frac{\barkappa}{2 \Delta m }\right)^{1/\delta} 
t^{-1-1/\delta} ~.~
\label{ndensity}
\eeq

Likewise, $\Omega(t)$ is the
 abundance that is disappearing from our tally of vacuum-energy abundances at time $t$ [\ie, the abundance $\Omega_\ell(t)$ evaluated at time $t$ and for that value of $\ell$ for which $t_\ell=t$].   However, just as in Sect.~\ref{sec:MatterGamma},
 we recognize that our model provides 
 specific scaling relations in Eq.~(\ref{MGscalings}) for the energy densities $\rho_\ell$ and corresponding abundances $\Omega_\ell$ which hold only at the production time $t^{(0)}$.
 In order to calculate $\Omega(t)$ within the context of our model, we therefore can no longer  use Eq.~(\ref{Omegal}), which assumed the existence of an eternal stasis.  Instead, we must express our abundances at time $t$ in terms of the corresponding abundances at $t^{(0)}$.
To do this we can follow our analysis from Sect.~\ref{sec:MatterGamma}.~ In particular, from Eqs.~(\ref{t-to-ell}) and (\ref{Omegal})
we have 
\beqn
\Omega(t) ~&=&~ \Omega_\ell(t)\bigg|_{t_\ell=t} \nonumber\\
 &=&~ \Omega_0^{(0)}\left(
  \frac{m_\ell}{m_0} \right)^\alpha h(t^{(0)},t_\ast) \left( \frac{t}{t_\ast}\right)^{2-(1+w)\barkappa} \nonumber\\
&=&~
\Omega_0^{(0)}
\left( \frac{\barkappa}{2m_0 t}\right)^\alpha 
   h(t^{(0)},t_\ast)
    \left( \frac{t}{t_\ast}\right)^{2-(1+w)\barkappa} \!\!\!~~~~~~~~~
\label{Omegadensity}
\eeqn
where $h(t^{(0)},t_\ast)$, as in Sect.~\ref{sec:MatterGamma}, denotes the net gravitational redshift factor that accrues between the initial production time $t^{(0)}$ and the fiducial time $t_\ast$ at which our system has reached stasis. Once again, the presence of such an $h$-factor will be discussed in more detail and justified in Sect.~\ref{subsect:h-section}.

Putting the pieces together, we thus find that requiring 
$n_{\hat t}(t) \Omega(t)\sim t^{-1}$ 
yields the constraint 
\beq
\boxed{
~~\eta ~=~ 2 - (1+w)\,\barkappa~,~~
}
\label{alphadeltatwo_w}
\eeq
whereupon we have
$n_{\hat t}(t) \Omega(t)= C' t^{-1}$
with
\beq
C' ~\equiv~ \frac{\Omega_0^{(0)}}{\delta} 
 \left( \frac{m_0}{\Delta m}\right)^{1/\delta} 
\left( \frac{\barkappa}{2 m_0 t_\ast} \right)^{\eta} h(t^{(0)},t_\ast) ~.
\label{Cdef}
\eeq

We thus see that within our model specified by Eqs.~(\ref{MGscalings}) and (\ref{MGmassform}), 
stasis will emerge only if Eq.~(\ref{alphadeltatwo_w}) is satisfied.  However, just as with Eq.~(\ref{MGalphagammaconstraint}), we may view this not as a constraint on our original model parameters $(\alpha,\delta,w)$ so much as a prediction for the resulting stasis value $\barkappa$.  
We thus see that stasis is realized for all 
$\eta$ within the range
\beq
   0 ~<~ \eta ~<~ -2w~,
\label{rangge}
\eeq
 with the resulting stasis abundance $\barOmega_\Lambda$  given by
\beq 
  \barOmega_\Lambda ~=~ 
    \frac{ \eta + 2 w}
     {(2-\eta) w}~
\label{LambdaMstasisabundance}
\eeq 
and $\barOmega_M=1-\barOmega_\Lambda$. Indeed, we see that $\barOmega_\Lambda$ is always within the range
$0< \barOmega_\Lambda < 1$. 

The limit as $w\to -1$ is particularly interesting. 
If we were to take $w= -1$ directly, we would find from Eq.~(\ref{LambdaMstasisabundance}) that 
$\barOmega_\Lambda|_{w=-1}=0/0$ is
indeterminate.
This is why we introduced $w$, treating $1+w$ as a regulator.  However, evaluating our stasis abundance $\barOmega_\Lambda$ for general $w$ as in Eq.~(\ref{LambdaMstasisabundance}) and then taking the $w\to -1$ limit, we find that
$\lim_{w\to -1} \barOmega_\Lambda = 1$.
Indeed, this limiting value for $\barOmega_\Lambda$ is consistent with our expectation for $w= -1$ that $\Omega_\Lambda$ will never depart from its initial value if that initial value is $1$, since in that case the Hubble parameter $H(t)$ remains constant and there is no dynamics within this model.  We thus obtain a sensible result even as $w\to -1$.

It is illustrative to verify that Eq.~(\ref{alphadeltatwo_w})
 also emerges from  the ``integral'' form of the stasis condition in Eq.~(\ref{masterconstraints}), namely the defining requirement  $\sum_\ell\Omega_\ell(t)= \barOmega_\Lambda$, where the abundances $\Omega_\ell(t)$ are given in Eq.~(\ref{Omegal}).  
 As we shall see, the same calculational ingredients are involved in both calculations, only slightly reshuffled.   
 In order to evaluate $\sum_\ell \Omega_\ell(t)$, we begin by noting that the Heaviside $\Theta$-functions within Eq.~(\ref{Omegal}) imply  that at any time $t$ there will be a maximum $\ell$-value $\ell_{\rm max}(t)$ for which the corresponding abundances 
$\Omega_\ell(t)$
are still non-zero and thus contributing to the total vacuum-energy abundance $\Omega_\Lambda$ of the system.  Indeed, for any $t$, no values of $\Omega_\ell(t)$ with $\ell> \ell_{\rm max}(t)$ can contribute to $\Omega_\Lambda$.
We can therefore eliminate all of the Heaviside $\Theta$-functions from the $\Omega_\ell(t)$ expressions within the sum by introducing a time-dependent maximum value $\ell_{\rm max}(t)$ on the states that should be included in the sum.  We therefore have
\beqn
\sum_\ell \Omega_\ell(t) ~&=&~
 \sum_{\ell=0}^{\ell_{\rm max}(t)}
    \Omega_\ell^\ast \left(
        \frac{t}{t_\ast}\right)^{2-(1+w)\barkappa} \nonumber\\
  ~&=&~
\sum_{\ell=0}^{\ell_{\rm max}(t)}
    \Omega_0^\ast 
    \left( \frac{m_\ell}{m_0}\right)^\alpha
     \left(
        \frac{t}{t_\ast}\right)^{2-(1+w)\barkappa} \nonumber\\
&\approx&~ 
  \int_t^\infty d\hat t ~ n_{\hat t}(\hat t)~
\Omega_0^\ast \,
 \left( \frac{\barkappa}{2 m_0 \hat{t}} \right)^\alpha
\left( \frac{t}{t_\ast}\right)^{2-(1+w)\barkappa} \nonumber\\
&=& ~
\frac{\Omega_0^\ast}{\delta }
\left( \frac{\barkappa}{2 \Delta m}\right)^{1/\delta } 
\left( \frac{\barkappa}{2 m_0}\right)^\alpha 
\left( \frac{t}{t_\ast}\right)^{2-(1+w)\barkappa}  \nonumber\\
&& ~~~~~~~~~~~~~~~~\times \, 
 \int_t^\infty d\hat t ~ \hat{t}^{
  -1-\eta }
  \nonumber\\
&=& ~
\frac{\Omega_0^\ast}{\delta }
\left( \frac{\barkappa}{2 \Delta m}\right)^{1/\delta } 
\left( \frac{\barkappa}{2 m_0}\right)^\alpha 
\left( \frac{t}{t_\ast}\right)^{2-(1+w)\barkappa}  \nonumber\\
&& ~~~~~~~~~~~~~~~~\times \, 
  \frac{1}{\eta} ~ t^{-\eta}  ~,
\label{intromax}
\eeqn 
where we have adopted the shorthand
\beq
\Omega_\ell^\ast ~\equiv~ \Omega_\ell(t_\ast)~=~ \Omega_\ell^{(0)}h(t^{(0)},t_\ast)
\label{shorthand}
\eeq
for all $\ell$ (including $\ell=0$).
In the first two lines of Eq.~(\ref{intromax}) we have used 
Eqs.~(\ref{MGscalings}) and (\ref{t-to-ell}),
while in passing to the third line we have
utilized the same integral approximation discussed above in terms of the continuous
$\hat t$-variable.   Similarly, in passing to the fourth line of Eq.~(\ref{intromax}) we have utilized Eq.~(\ref{ndensity}) 
for the density of states per unit $\hat t$, and in passing to the final line we have recognized from Eq.~(\ref{rangge}) that $\eta>0$.
However, demanding that the final result in Eq.~(\ref{intromax}) be independent of $t$ leads to the same constraint as we obtained previously in Eq.~(\ref{alphadeltatwo_w}).
Thus the integral form of our stasis condition leads to the same overall scaling constraint as we obtained through the differential form.

Thus far in our analysis of vacuum-energy/matter stasis we have concentrated on the overall scaling constraint.   Indeed, as discussed in Sect.~\ref{subsect:general_lessons}, we must also consider the associated prefactor constraints (both relative and absolute) as well as the associated logarithm-avoidance constraint.
However, it is straightforward to see that --- just as for matter/radiation stasis --- the relative-prefactor and logarithm-avoidance constraints are redundant with the overall scaling constraint and therefore provide no new restrictions.   In particular, writing $\sum_\ell \Omega_\ell(t) = C$, we find from Eq.~(\ref{intromax})   
that
$C$ is given by the same expression as in
Eq.~(\ref{Cdef}) except divided by $\eta$, thereby enabling us to extract the relative-prefactor relation
\beq 
     C' ~=~ \eta \, C ~=~ \left[ 2-(1+w)\barkappa\right]\,C~.
\eeq
This is the vacuum-energy/matter analogue of the relation in Eq.~(\ref{bundled}).   We thus find that our desired prefactor relation in Eq.~(\ref{masterconstraints}) will be satisfied as long as 
\beq
             C ~=~ \barOmega_\Lambda
\eeq
or equivalently
\beq
      C' ~=~ \eta \,\barOmega_\Lambda ~=~ \left[ 2-(1+w)\barkappa\right]\,\barOmega_\Lambda~.
\eeq
 Likewise, from this result we have the stasis result
\beq
    n_{\hat t}(t)\, \Omega(t) ~=~
           \eta \,\barOmega_M\, \frac{1}{t}~,
\label{pumpLambdaMatter}
\eeq
as required by Eq.~(\ref{masterconstraints}).
However, these last absolute-prefactor constraints will naturally come into balance dynamically, since the attractor behavior of this solution (which we will discuss in Sect.~\ref{sec:attractor}) automatically adjusts $h(t^{(0)},t_\ast)$ so as to ensure that $C = \barOmega_\Lambda$.    We thus see that our model naturally gives rise to the correct prefactor constraints in Eq.~(\ref{masterconstraints}) as well.

\begin{figure*}[t!]
\centering
~\hskip -0.2 truein \includegraphics[keepaspectratio, width=0.54\textwidth]{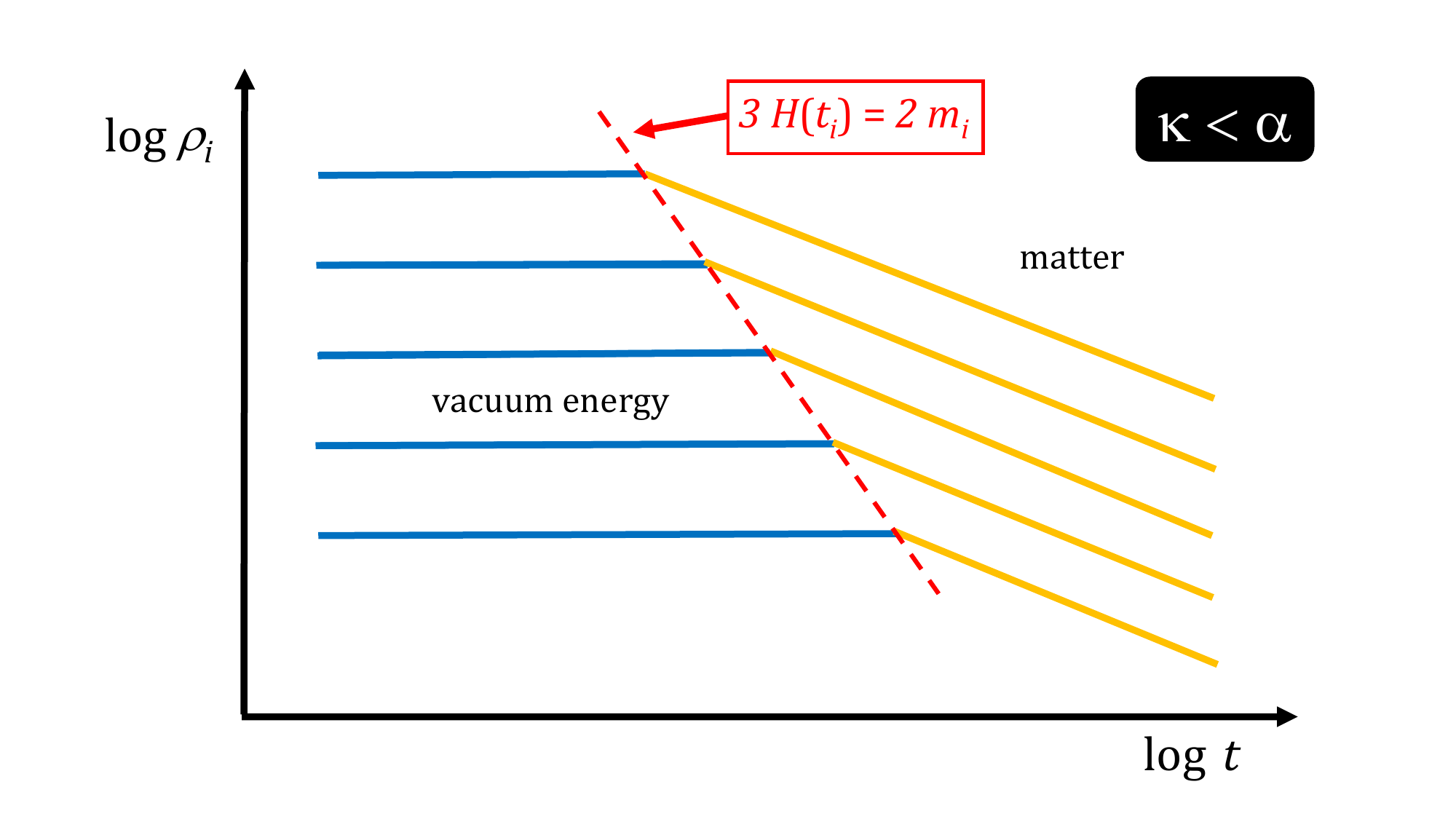}
\hskip -.45 truein
\includegraphics[keepaspectratio, width=0.54
\textwidth]{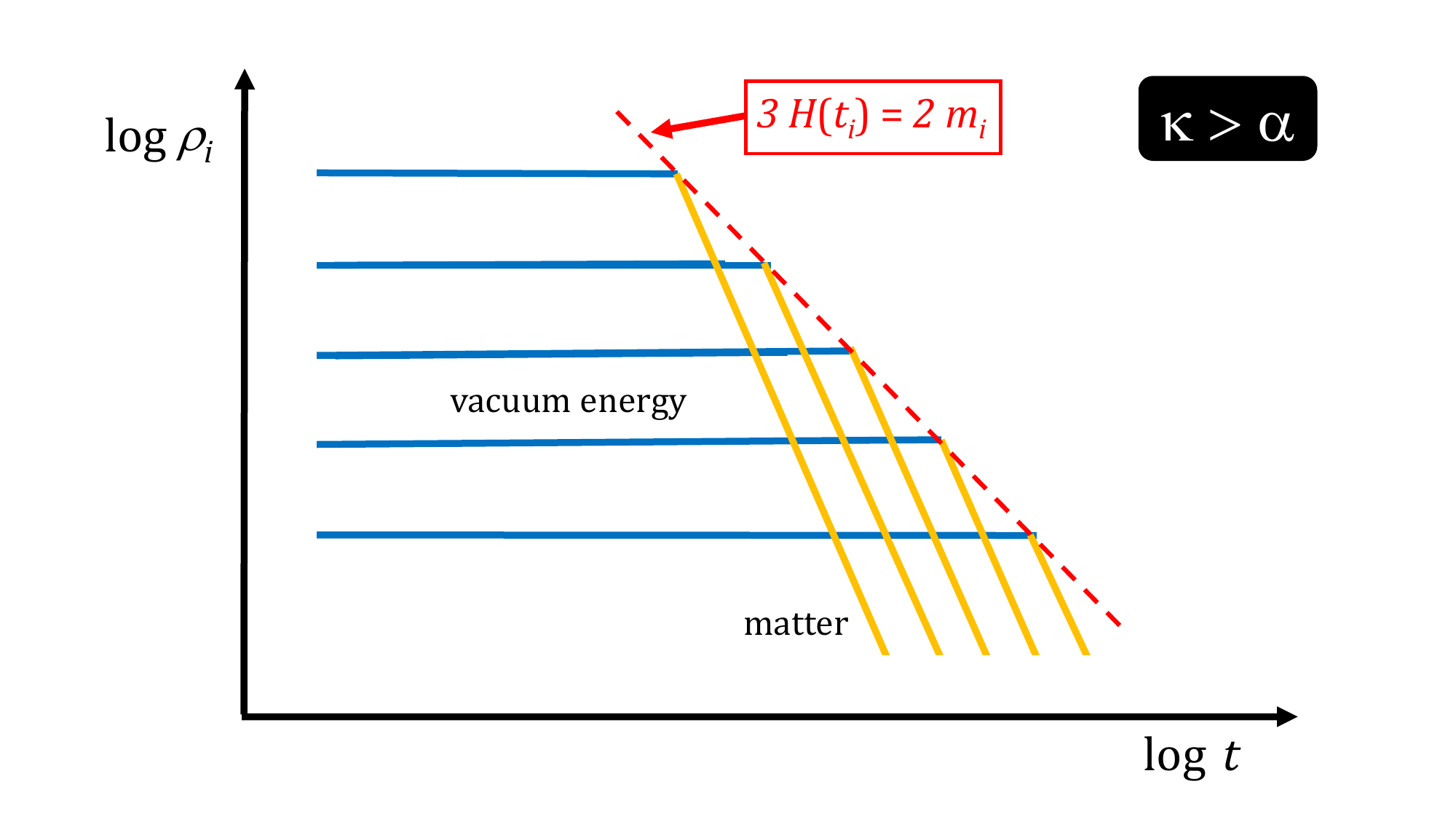}
\caption{ 
Two different scenarios illustrating possible behaviors for the individual energy densities $\rho_\ell(t)$ (blue/orange), 
sketched as functions of time,
as our corresponding tower of states $\phi_\ell$  undergoes sequential transitions from an overdamped phase to an underdamped phase at times $t_\ell$ for which $3H(t_\ell)=2m_\ell$.  These energy densities $\rho_\ell(t)$ are interpreted as corresponding to vacuum energy (blue) or matter (orange) when the corresponding states $\phi_\ell$ are overdamped or underdamped, respectively, with the dashed line (red) indicating the moments of transition between these two regimes.  
For this purposes of these sketches we have assumed that states with greater energy densities have greater masses (\ie, $\alpha>0$).  We have also chosen to sketch the case in which $w \to -1$ for the vacuum energy, causing the blue lines to be almost exactly horizontal.  The left and right panels show the behaviors that emerge for $\barkappa<\alpha$ and
$\barkappa>\alpha$, respectively.
In the latter case, the energy densities ``reflect'' off the red transition line, thereby giving rise to the  sequential ``inter-leaved'' energy-density crossings that are the hallmark of the stasis phenomenon.
}
\label{fig:turnon}
\end{figure*}

As discussed in Sect.~\ref{subsect:general_lessons}, we must also satisfy the logarithm-avoidance constraint.
Indeed, in the present case this is nothing but the constraint $\eta>0$ that enabled us to avoid obtaining a logarithmic time-dependence when passing to the final line of Eq.~(\ref{intromax}).    However, this constraint is already subsumed into our overall scaling constraint, as we see from the allowed ranges in Eq.~(\ref{rangge}).    Moreover, we see that Eq.~(\ref{alphadeltatwo_w})   
is not really a constraint on the input parameters of our model so much as a prediction for the resulting stasis value $\barkappa$
and therefore $\barOmega_\Lambda$.
Thus, so long as the input parameters of our model satisfy Eq.~(\ref{rangge}),
a stasis state will necessarily emerge.

It is also instructive to understand in a qualitative way the behavior of the energy densities $\rho_\ell$ during the stasis epoch as the decays of our individual $\phi_\ell$ states proceed down the tower.   As we have seen in Eq.~(\ref{MGscalings}),
the energy densities $\phi_\ell$ have initial
values $\sim (m_\ell/m_0)^\alpha$ at $t=t^{(0)}$. In the $w\to -1$ limit, these quantities then remain time-independent until the time $t=t_\ell$, defined by the constraint $3H(t_\ell)=2m_\ell$, after which they scale as $\rho_\ell(t) \sim t^{-\barkappa}$.
Moreover, we have seen in Eq.~(\ref{t-to-ell})
that $t_\ell\sim m_\ell^{-1}$, implying that $\rho_\ell\sim t_\ell^{-\alpha}$.
Given these observations, rough sketches of two possible time-evolutions for each $\rho_\ell(t)$ during stasis appear in Fig.~\ref{fig:turnon}. 
These energy densities are sketched in blue when they correspond to vacuum energy, and yellow when they correspond to matter. The dividing line between these two phases is indicated in red, and given our result $\rho_\ell\sim t_\ell^{-\alpha}$ we see that this line has slope $-\alpha$ on this log-log plot.
For simplicity these sketches are drawn with equally spaced initial values of $\log \phi_\ell$, but this property is chosen for graphical simplicity and will play no role in our analysis.

The primary difference between the two panels of Fig.~\ref{fig:turnon} 
concerns the value of $\barkappa$ that governs the logarithmic slope of each $\rho_\ell(t)$ after $t=t_\ell$.  Indeed, the left panel of Fig.~\ref{fig:turnon} corresponds to the case with $\barkappa<\alpha$,
while the right panel of Fig.~\ref{fig:turnon} corresponds the case with $\barkappa>\alpha$.   However,  it is immediately apparent that the change in the sign of this inequality has a profound effect on the behavior of the corresponding energy densities.  
 For $\barkappa < \alpha$, the energy densities $\rho_\ell$ necessarily remain in the same relative order in which they began, with $\rho_{\ell'}(t)> \rho_\ell(t)$ for all $t$ as long as $\ell'>\ell$.   For $\barkappa>\alpha$, by contrast, the energy densities $\rho_\ell(t)$ undergo successive pairwise crossings as time evolves.   Thus while the energy density associated with the top component $\ell_{\rm max}=N-1$ begins as the largest, eventually the energy density associated with $\ell_{\rm max}-1$ becomes the largest, then that with $\ell_{\rm max}-2$, and so forth.

\begin{figure}
    \includegraphics[width=0.99\linewidth]{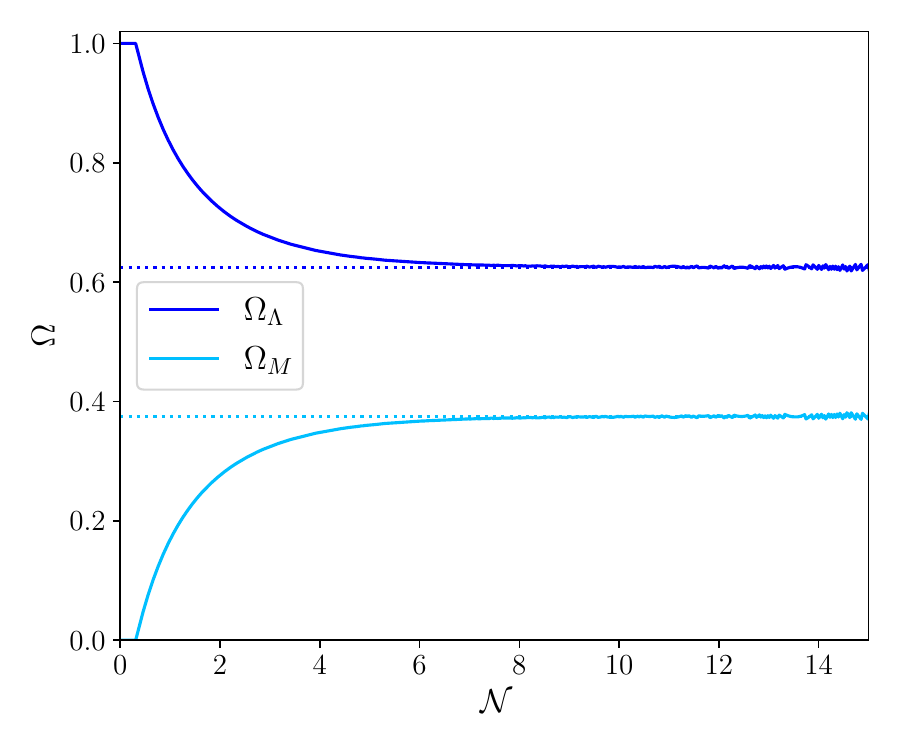}
    \caption{Vacuum-energy/matter stasis.
 Here the total vacuum-energy and matter abundances, $\Omega_\Lambda(t)$ and $\Omega_M(t)$ respectively, are plotted as functions of the number $\calN$ of $e$-folds since the initial production time $t^{(0)}$, taking
$\alpha=0.7$, $\delta=2$, and $w=-0.8$
 as reference values.  We have also taken $H^{(0)}/m_{N-1} = 0.75$.
 As we see, the system eventually evolves into a stasis state with $\barkappa=4$,
 $\barOmega_\Lambda = 5/8$,
 and $\barOmega_M=  3/8$.
 This stasis state ends only when the last component of the tower transitions to matter.   As with matter/radiation stasis, the total number of $e$-folds of stasis ${\cal N}_s$ is determined by the total number $N$ of states in the tower.   In analogy with Eq.~(\ref{MGNN})   for matter/radiation stasis, we expect 
 $\calN_s\sim\log N$.   Similar stasis behavior emerges for all values of $(\alpha,\delta,w)$ within the ranges in Eq.~(\ref{rangge}), with corresponding stasis abundances $\barOmega_\Lambda$ determined by Eq.~(\ref{LambdaMstasisabundance}) and $\barOmega_M=1-\barOmega_\Lambda$. }
 \label{fig:vac_M} 
\end{figure}

 It is this latter behavior involving successive pairwise energy-density crossings which underlies the stasis phenomenon.   This is particularly evident from the left panel of Fig.~\ref{fig:review}, which shows the analogous situation with matter/radiation stasis.
  Thus, just from consideration of these sorts of figures, we can immediately see that stasis requires $\barkappa > \alpha$.   Of course, this result is entirely consistent with the full stasis condition in Eq.~(\ref{alphadeltatwo_w}).  Indeed, we see from Eq.~(\ref{LMHubble}) that $\barkappa>2$, whereupon we see from Eq.~(\ref{alphadeltatwo_w}) that $\alpha<2$.   Thus stasis necessarily requires $\alpha<\barkappa$, consistent with the right panel of Fig.~\ref{fig:turnon} but not the left panel. 
 
 In general, these results will apply to all of the stasis situations we shall consider in this paper.   In each case, we will require successive energy inversions proceeding down the $\phi_\ell$ tower as time evolves.  As we have seen, this requires a very particular behavior as our energy densities $\rho_\ell(t)$ approach the red lines that indicate our transitions, as in Fig.~\ref{fig:turnon}.   Speaking qualitatively, we may regard our energy densities $\phi_\ell(t)$ as either passing {\it through}\/ these red lines, as in the left panel of Fig.~\ref{fig:turnon}, or being {\it reflected}\/ by these red lines, as in 
 the right panel of Fig.~\ref{fig:turnon}.  {\it It is the case of reflection that induces the behavior that underlies the stasis phenomenon.}
 
In Fig.~\ref{fig:vac_M} we show the emergence of vacuum-energy/matter stasis for a system in which we take 
$\alpha=0.7$, 
$\delta=2$, and $w=-0.8$ as reference values.
As we see, this eventually results in a stasis with $\barkappa=4$, consistent with Eq.~(\ref{alphadeltatwo_w}), which in turn implies that
$\barOmega_\Lambda=  5/8$ and
$\barOmega_M=  3/8$.

\FloatBarrier 
\section{Vacuum-energy/radiation stasis\label{sec:LambdaGamma}}

In Sect.~\ref{sec:LambdaMatter}, we demonstrated  
that vacuum energy and matter can be in stasis with each other.  Our analysis took place 
within a cosmology containing a tower of scalar fields $\phi_\ell$ with masses $m_\ell$ which 
sequentially transition from an overdamped phase (during which their energies are identified 
as vacuum energy) to an underdamped phase (during which their energies are identified as 
those of matter).  Indeed, these transitions occur at the times $t_\ell$ for which 
$3 H(t_\ell)=2 m_\ell$.

However, such $\phi_\ell$ fields can also experience decays into radiation, with non-zero 
decay widths $\Gamma_\ell$.  Indeed, within certain regions of parameter space, it may even 
happen that the lifetimes $\tau_\ell\equiv 1/\Gamma_\ell$ of the components $\phi_\ell$ are 
all smaller than the critical underdamping times $t_\ell$ at which these fields would have transitioned to 
an underdamped state.  In such cases, we can then have decays directly from vacuum energy to 
radiation.

In general, a universe composed entirely of vacuum energy and radiation will evolve under 
cosmological expansion from a radiation-dominated epoch to a vacuum energy-dominated epoch.
The above decays from vacuum energy back to radiation can thus provide a counterbalancing 
effect that could potentially lead to a stasis between vacuum energy and radiation. 

\subsection{Theoretical subtleties \label{subsect:killjoy}}

In order to study this phenomenon, we must first discuss several additional theoretical 
subtleties that arise in modeling the transfer of energy density from vacuum 
energy to radiation.  The rate at which this transfer of energy density occurs depends 
on the properties of the vacuum-energy component.

For example, as discussed in Sect.~\ref{sec:LambdaMatter}, one natural realization
of $\rho_\Lambda$ in a particle-physics context is the energy density associated with one 
or more overdamped scalar fields $\phi_\ell$ which are displaced from their potential minima.  
The equation of motion for each such field is 
$\ddot{\phi}_\ell + (3H +\Gamma_\ell)\dot{\phi}_\ell + \partial V/\partial\phi_\ell = 0$, 
where $V$ is the scalar potential and where a dot denotes a time derivative.  
At late times, when $3H(t) \ll 2m_\ell$ and the field is 
well within the underdamped regime, $\Gamma_\ell$ is often approximated as constant 
and identified with the proper decay width of $\phi_\ell$ in Minkowski 
space~\cite{Albrecht:1982mp,Abbott:1982hn}.  However, this heuristic treatment of the 
dissipation rate is not appropriate while the field is in the overdamped regime~\cite{Kofman:1997yn}.
Rather, $\Gamma_\ell$ must be calculated using methodologies appropriate for analyzing the 
non-equilibrium dynamics of a quantum field interacting with its environment, such as the 
closed-time path (\ie, Schwinger-Keldysh) formalism (for reviews, see, \eg, Ref.~\cite{Bellac:2011kqa}) 
or the inference time formalism (for reviews, see, \eg, Refs.~\cite{Calzetta:2008iqa,Berges:2015kfa}).
A number of subtleties arise in these calculations as a result of the background cosmology. 
For example, since there is no global time-like Killing vector in an FRW universe, 
particle energy is not manifestly conserved.  A variety of processes which are 
forbidden in Minkowski space --- including the decay of a field 
into its own quanta~\cite{Boyanovsky:1996ab,Boyanovsky:2004gq,Bros:2010rku,Boyanovsky:2011xn} ---
can therefore contribute to the dissipation rate.  The form which $\Gamma_\ell$ takes is highly 
model dependent and in general depends non-trivially both on the temperature of the radiation 
bath --- or, equivalently, on the value of $\rho_\gamma$ --- and on the time-varying expectation 
value of $\phi$ (for reviews and discussion, see, \eg, 
Refs.~\cite{Berera:2008ar,Bastero-Gil:2009sdq,Kamali:2023lzq,Berera:2023liv}). 

We note that there is another natural mechanism via which the energy density associated 
with an overdamped scalar field can be transferred directly to a radiation-like energy component with $w=1/3$.  This mechanism is similar to the mechanism discussed in 
Sect.~\ref{sec:LambdaMatter} for transferring energy density from vacuum energy to 
matter, which involved an overdamped/underdamped transition, but operates
in scenarios in which the quadratic term for $\phi_\ell$ in $V$ 
is negligible or vanishing, and $\phi_\ell$ is instead dominated by a higher-order
polynomial $V_{\ell}\sim \phi_\ell^{2n}$ with $n>1$.
At early times, while $3H \gg 2(\partial^2 V/\partial\phi_\ell^2)^{1/2}$ and $\phi_\ell$ is 
effectively stationary, the energy density $\rho_\ell$ associated with that field scales 
with $a$ like vacuum energy.  However, once $3H \approx 2(\partial^2 V/\partial\phi_\ell^2)^{1/2}$, 
the field $\phi_\ell$ begins oscillating around its potential minimum.  The effective 
equation-of-state parameter for $\phi_\ell$ during this oscillatory phase, time-averaged 
over many cycles of oscillation, is $w_\ell \approx (n-1)/(n+1)$~\cite{Turner:1983he}.  
Thus, for $n=2$ --- \ie, for a quartic potential --- the equation-of-state parameter for 
such an oscillating scalar field is identical to that for radiation.  As with the 
overdamped/underdamped transition discussed in Sect.~\ref{sec:LambdaMatter}, we can 
obtain some insight into the cosmological dynamics of scenarios involving such scalars by 
idealizing this transition as an instantaneous one in which $\phi_\ell$ is approximated 
as having $w = -1$ whenever $3H > 2(\partial^2 V/\partial\phi_\ell^2)^{1/2}$ 
and $w = 1/3$ otherwise.

For simplicity, we shall focus in what follows  on 
the case in which $\Gamma_\ell$ is effectively constant and independent of $\rho_\gamma$.  
While an analysis based on this form of $\Gamma_\ell$
does not have a straightforward motivation in terms of a top-down model, it can nevertheless serve as 
a convenient starting point for the analysis of specific models with more complicated, time- and 
temperature-dependent dissipation rates.  We also note that the results obtained for
this form of $\Gamma_\ell$ in the instantaneous-decay approximation turn out to be applicable, 
through a straightforward mapping, to the case of a scalar field $\phi_\ell$ with a quartic 
potential whose energy density scales like that of radiation once it begins 
oscillating~\cite{toappear}.  Moreover, as we shall see, the results we obtain for a constant 
$\Gamma_\ell$ will provide some mathematical insight into certain limiting cases of the 
three-component cosmological system on which we shall focus in Sect.~\ref{sec:TripleStasis}.

\subsection{Algebraic analysis \label{subsect:VacEnergyRadiationStasis}}

Given these understandings and assumptions, we shall now 
repeat the algebraic steps in the previous sections
in order to investigate the possibility of achieving a two-component stasis between vacuum energy and radiation. 
Towards this end, our analysis will essentially be a hybrid of the analyses in Sects.~\ref{sec:MatterGamma} and \ref{sec:LambdaMatter}:
we shall treat the vacuum energy according to the general-$w$ approach of Sect.~\ref{subsect:genw}, assuming that each of our $\phi_\ell$ fields evolves with a fixed equation-of-state parameter $w> -1$, but we shall also assume that this vacuum energy simultaneously experiences an exponential decay with lifetime $\tau_\ell\equiv \Gamma_\ell^{-1}  \ll t_\ell$.

Within a cosmology consisting of only vacuum energy and radiation, the Hubble parameter evolves as
\beq
  \frac{dH}{dt}  
  ~=~  -\textstyle\frac{1}{2} \,H^2\, \bigl[
    4+ (3w-1) \Omega_\Lambda \bigr]~,
\eeq
or equivalently 
\beq
 \kappa ~=~ \frac{6}{4 + (3w-1) \Omega_\Lambda}~
\eeq
where $\kappa$ continues to be defined through the parametrization in Eq.~(\ref{eq:kappagendef}).
Likewise, the vacuum-energy density $\rho_\ell$ associated with each $\phi_\ell$ field evolves as
\beq
\frac{d\rho_\ell}{dt} ~=~ 
 - 3 (1+w) H \rho_\ell - \Gamma_\ell \rho_\ell~,
 \eeq
 whereupon we find
 \beq
 \frac{d\Omega_\Lambda}{dt} ~=~
 -\sum_\ell \Gamma_\ell \Omega_\ell + (1-3 w) H \Omega_\Lambda (1-\Omega_\Lambda)~.
 \eeq
Setting $d\Omega_\Lambda/dt=0$ and inserting the stasis Hubble parameter $H(t)= \barkappa/(3t)$ then yields the corresponding stasis condition
\begin{empheq}[box=\fbox]{align}
~~\sum_\ell^{\phantom{A}} \Gamma_\ell \Omega_\ell(t)   ~&=~ 
 \biggl\lbrack
 \left( \frac{1}{3}-w\right) \barkappa\, \barOmega_\Lambda 
 (1-\barOmega_\Lambda)\biggr\rbrack \, \frac{1}{t} ~~\nonumber\\
   ~&=~  \biggl\lbrack
    2-(1+w)\barkappa\biggr\rbrack \, \barOmega_\Lambda\, 
\frac{1}{t} ~.~~
\label{masterconstraints2}
\end{empheq}

This condition for vacuum-energy/radiation stasis can also be satisfied within the model of Eqs.~(\ref{MGscalings}) and (\ref{MGmassform}).   In order to determine the resulting constraints on our model parameters $(\alpha,\gamma,\delta,w)$, we can follow the steps in 
Sect.~\ref{sec:MatterGamma}.~  
Indeed, for this purpose we adopt Sect.~\ref{sec:MatterGamma} rather  than Sect.~\ref{sec:LambdaMatter} as our guide because the former 
also involved an exponential decay from one energy component to another.
Repeating the steps in Sect.~\ref{sec:MatterGamma}, we can convert the sum in Eq.~(\ref{masterconstraints2})
to an integral.  Moreover, rather than introduce a continuous variable $\hat t$ of underdamping times as in Sect.~\ref{sec:LambdaMatter}, 
we follow Sect.~\ref{sec:MatterGamma} and work in terms of a continuous 
variable $\tau$ of {\it decay}\/ times $\tau$.
Assuming an eternal stasis,
we then find that our parameters must satisfy the relation
\beq 
   \boxed{
~~\frac{\eta}{\gamma} ~=~ 2 - (1+w)\barkappa~.~~
}
\label{LambdaRadiationconstraint}
\eeq
This is thus the analogue of Eqs.~(\ref{MGalphagammaconstraint}) and (\ref{alphadeltatwo_w}).
Once again, this is not 
a constraint on $(\alpha,\gamma,\delta,w)$ so much as a prediction for $\barkappa$.
Indeed, so long as
\beq
       0 ~<~ \frac{\eta}{\gamma} 
       ~<~ \frac{1-3w}{2}~,
\label{LGrange}
\eeq
we find that the resulting value of $\barOmega_\Lambda$ during stasis is given by 
\beq
  \barOmega_\Lambda ~=~ \frac{ 2 (2\eta +3w\gamma-\gamma)}{(1-3w)(\eta-2\gamma)}~.
\label{LamRadOmega}
\eeq
Note that
$0< \barOmega_\Lambda < 1$ so long as 
Eq.~(\ref{LGrange}) is satisfied.

Interestingly, for $w= -1$ we find that $\barOmega_\Lambda=1$ for any $\eta$ and $\gamma$.   
This indicates that the only ``stasis'' that develops in the $w= -1$ case is that with 
which we started, namely a universe containing nothing but vacuum energy.
However, for $w> -1$, we find that a non-trivial stasis develops with $\barOmega_\Lambda <1$.

Following the results in Sect.~\ref{sec:MatterGamma}, we may also 
compare the overall coefficients that enter into our stasis constraints.
For this purpose, let us define $C$ and $C'$ via
\beqn 
   \sum_\ell \Omega_\ell(t) ~&=&~ C \nonumber\\
   \sum_\ell \Gamma_\ell \Omega_\ell(t) ~&=&~ C'\, t^{-1}~
\label{CCprimedefsexponential}
\eeqn
where we have imposed the relation in Eq.~(\ref{LambdaRadiationconstraint}). 
We can explicitly evaluate the sums on the left sides of Eq.~(\ref{CCprimedefsexponential}) 
by converting to integrals over a continuous $\tau$-variable, as discussed above.   
We then find that $C$ and $C'$ are related via
\beq
  C'~=~ \frac{\eta}{\gamma}\,C ~=~
     \left[ 2-(1+w)\barkappa\right]\,C~
\label{CCprimeLambdaGamma}
\eeq
where
\beq 
   C ~\equiv~ \frac{1}{\gamma\delta}
     \left( \frac{m_0}{\Delta m}\right)^{1/\delta}  
   \!   h(t^{(0)},t_\ast)
      \,\frac{\Omega_0^{(0)} }{(\Gamma_0 t_\ast)^{\eta/\gamma}} \,
 \Gamma\left( \frac{\eta}{\gamma}\right)  ~.~~~
\label{LambdaGCdef}
\eeq
These solutions for $C$ and $C'$ exactly match those in Sect.~\ref{sec:MatterGamma}.~
As in previous cases, the attractor behavior of the stasis solution (to be discussed in 
Sect.~\ref{sec:attractor}) then inevitably ensures that $C=\barOmega_\Lambda$.  Indeed, 
this happens in the same manner as described in Sect.~\ref{subsect:general_lessons}.~
The result in Eq.~(\ref{CCprimeLambdaGamma}) then ensures that the prefactors within 
Eq.~(\ref{masterconstraints2}) match precisely.  This also ensures that Eq.~(\ref{pumpMatterGamma}) holds for vacuum-energy/radiation stasis as well. 

Although we have treated our decay process as a {\it bona fide}\/ exponential decay, it will 
prove both instructive and useful to repeat this calculation within the framework of an 
{\it instantaneous-decay approximation}\/ in which our $\phi_\ell$ states decay suddenly and 
completely at $t=\tau_\ell=1•/\Gamma_\ell$.   This is tantamount to approximating the exponential decays 
as sharp cutoffs by replacing
\beq
e^{-t/\tau_\ell} ~\to~ \Theta( \tau_\ell-t)~.
\label{instantaneousdecay}
\eeq
Implementing this substitution, we find that our calculations now more closely resemble the 
calculations in Sect.~\ref{sec:LambdaMatter} (wherein the underdamping times $t_\ell$ played 
the role of $\tau_\ell$) rather than those in Sect.~\ref{sec:MatterGamma}.~  In this case,
the constraint equations for stasis in Eq.~(\ref{masterconstraints2}) are replaced by
\begin{empheq}[box=\fbox]{align}
~~n_\tau(t)\, \Omega(t)   ~&=~ 
 \biggl\lbrack
 \left( \frac{1}{3}-w\right) \barkappa\, \barOmega_\Lambda 
 (1-\barOmega_\Lambda)\biggr\rbrack \, \frac{1}{t} ~~\nonumber\\
   ~&=~  \biggl\lbrack
    2-(1+w)\barkappa\biggr\rbrack \, \barOmega_\Lambda\, 
\frac{1}{t} ~.~~
\label{masterconstraints2_instantaneous}
\end{empheq}

This condition for vacuum-energy/radiation stasis can also be satisfied within the model of 
Eqs.~(\ref{MGscalings}) and (\ref{MGmassform}).
A calculation similar to that in Eq.~(\ref{intromax})
tells us that $n_\tau(t)\Omega(t)$ will scale as $t^{-1}$,
as required, only if Eq.~(\ref{LambdaRadiationconstraint}) is satisfied.   We thus see that the instantaneous-decay approximation leads to precisely the same scaling relation for the $(\alpha,\gamma,\delta,w)$ parameters as full exponential decay.  Likewise, defining our $C$-coefficients via the relations
\beqn 
   \sum_\ell \Omega_\ell(t) ~&=&~ C \nonumber\\
   n_\tau(t) \, \Omega(t) ~&=&~ C'\, t^{-1}~,
\label{CCprimedefsinstantaneous}
\eeqn
we again obtain the relation between $C$ and $C'$ given in 
Eq.~(\ref{CCprimeLambdaGamma}), where now $C$ is given by
\beq 
   C ~\equiv~ \frac{1}{\gamma\delta}
     \left( \frac{m_0}{\Delta m}\right)^{1/\delta}  
   \!   h(t^{(0)},t_\ast)
      \,\frac{\Omega_0^{(0)} }{(\Gamma_0 t_\ast)^{\eta/\gamma}} 
 ~.~~~
\label{LambdaGCdef2}
\eeq
The attractor behavior of the stasis solution then guarantees that $C= \barOmega_\Lambda$, as before, and likewise yields Eq.~(\ref{pumpLambdaMatter}).

Thus, even within the instantaneous-decay approximation, we find that stasis is achieved. 
This is a rather remarkable result, demonstrating that full exponential decay and the instantaneous-decay approximation are equally valid as far as stasis calculations are concerned.   Indeed, both full exponential decay and the instantaneous-decay approximation lead to the identical scaling relation in Eq.~(\ref{LambdaRadiationconstraint}) and the identical relative and absolute prefactor constraints in Eqs.~(\ref{CCprimeLambdaGamma}).
(The same is also true for the logarithmic-avoidance constraint, which now takes the form $\eta/\gamma>0$.)
Indeed, the only difference between the two formalisms for treating the decay process is in the precise absolute value of $C$, but this information is ultimately washed away due to the attractor nature of the stasis solution.
These observations thus suggest a certain robustness to the existence of the stasis phenomenon, indicating that our conclusions regarding the emergence of stasis are largely independent of the precise details concerning how the decays of the $\phi_\ell$ states are ultimately modeled.

Interestingly, we observe that the effective equation-of-state parameter for this vacuum-energy/radiation system during stasis is given by
\beq 
     \langle w\rangle
     ~=~  w\barOmega_\Lambda+ \frac{\barOmega_\gamma}{3}~.
\label{eq:weffdef}
\eeq
Thus, when 
$\barOmega_\gamma = -3w\barOmega_\Lambda$, our system has $\langle w\rangle=0$.  In other words, for the purposes of cosmological expansion, our universe behaves as if it were effectively matter-dominated despite the lack of an actual matter component.
Following the arguments in Ref.~\cite{Dienes:2021woi}, this means that this system can even co-exist with a ``spectator'' (non-interacting) additional {\it matter}\/ energy component with abundance $\Omega_M$ because the introduction of such an additional spectator matter component will not disturb the stasis value $\langle w\rangle$ that has already been realized for the vacuum energy and radiation components.  This then becomes an example of a system in which vacuum energy, radiation, and matter can all co-exist in a stasis configuration.   Unfortunately, in such a system the matter must not have any energy-transferring interactions with either the vacuum energy or the radiation.  It is for this reason that we refer to the matter as a ``spectator'' component of the total energy.     Indeed, in such a system the matter abundance $\Omega_M$ remains fixed only because the vacuum energy and radiation in their stasis configuration with $\barOmega_\gamma= -3w\barOmega_\Lambda$ conspire to produce a universe whose expansion rate is effectively that of a matter-dominated universe.   As a result, $\Omega_M$ neither grows nor shrinks as a result of cosmological expansion, and thus remains constant without receiving energy from (or losing energy into) the other components.

\begin{figure}[htb]
\includegraphics[width=0.99\linewidth]{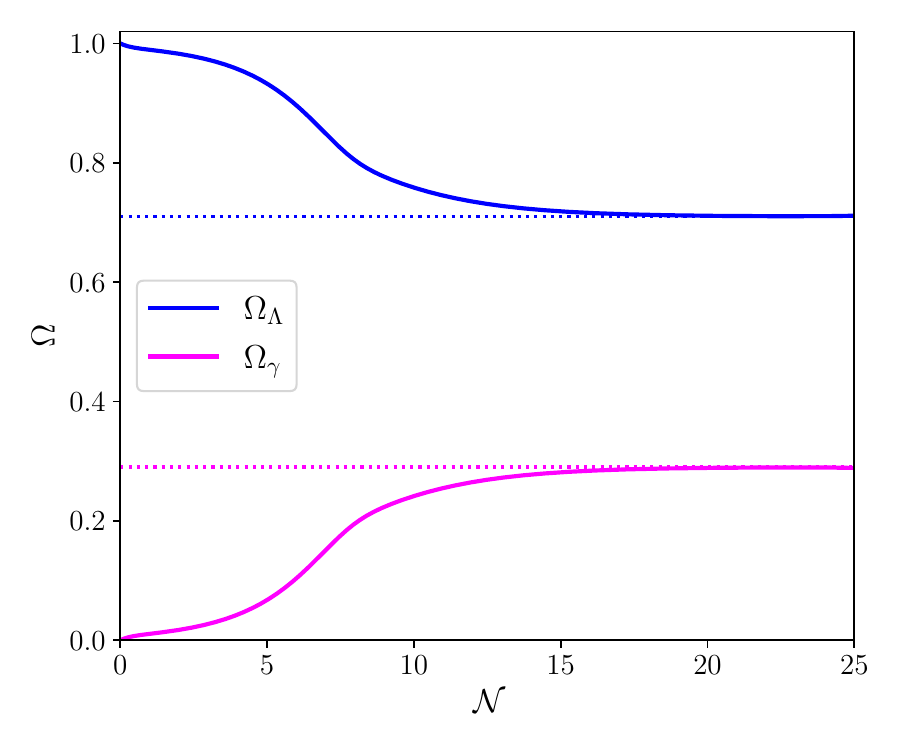}
\caption{
Vacuum-energy/radiation stasis.
 Here we plot the total vacuum-energy and radiation abundances, $\Omega_\Lambda$ and $\Omega_\gamma$ respectively, as functions of time, taking  $\alpha=0.5$,
 $\delta=2$, $\gamma=1$, $w= -0.7$,
 and $H^{(0)}/\Gamma_{N-1}=20$ as reference values and working with the exponential-decay formalism. 
 As we see, these benchmark values lead to 
 a stasis with $\barkappa = 10/3$,
 implying $\barOmega_\Lambda = 22/31\approx 0.71$
 and $\barOmega_\gamma= 9/31\approx 0.29$.
 This stasis state ends only when the last component of the tower decays to radiation.    Similar stasis behavior emerges for all values of $(\alpha,\delta,\gamma,w)$ within the range in Eq.~(\ref{LGrange}), with corresponding stasis abundances $\barOmega_\Lambda$ determined by Eq.~(\ref{LamRadOmega}).
\label{fig:vac_gamma}} 
\end{figure}

In Fig.~\ref{fig:vac_gamma}, we show
the emergence of vacuum-energy/radiation stasis for a system in which we take $\alpha=0.5$,
 $\delta=2$, $\gamma=1$, and $w= -0.7$ as reference values. 
 As we see, these lead to 
 a stasis with $\barkappa = 10/3$,
 implying $\barOmega_\Lambda = 22/31$
 and $\barOmega_\gamma= 9/31$.  Indeed, as with the stases in Sect.~\ref{sec:MatterGamma} and \ref{sec:LambdaMatter}, this is a pairwise stasis involving only two components in which energy flows from one directly into the other, bypassing the third completely.
 A similar stasis would emerge for any parameters within the range in Eq.~(\ref{LGrange}).

\FloatBarrier

\section{Algebraic structure of pairwise stases\label{sec:Generalpairwise}}

In general, the analyses of Sects.~\ref{sec:MatterGamma}, \ref{sec:LambdaMatter}, and 
\ref{sec:LambdaGamma} share an underlying algebraic structure which applies to any ``pairwise'' 
stasis (\ie, any stasis between two energy components).  
To demonstrate this, we shall maintain full generality by assuming that our two components 
have constant equation-of-state parameters $(w_1,w_2)$ with corresponding abundances 
$(\Omega_1,\Omega_2)$  such that $\Omega_1+\Omega_2=1$.  For concreteness we shall also 
assume that $w_1<w_2$.   This means that cosmological expansion will tend to convert 
an $\Omega_2$-dominated universe into an $\Omega_1$-dominated universe.   Thus stasis 
can be achieved only in the presence of some method  of converting $\Omega_1$ back 
into $\Omega_2$.  Because this conversion operates in a manner opposite to the natural 
effects of cosmological expansion, we shall refer to such a method of converting 
$\Omega_1$ back into $\Omega_2$ as a ``pump''.

Given this setup, and following our previous steps, it is straightforward to verify that 
our abundances $\Omega_{1,2}$ evolve according to the differential equations 
\beqn \label{eqGeneralSystem}
  \frac{d\Omega_1}{dt} ~&=&~ 3H (w_2-w_1) \Omega_1 \Omega_2 - P_{12}\nonumber\\
  \frac{d\Omega_2}{dt} ~&=&~ 3H (w_1-w_2) 
  \Omega_1\Omega_2 + P_{12} ~~~
\label{gendiffeqs}
\eeqn
where $P_{12}$ schematically denotes the pump term which transfers abundance from 
$\Omega_1$ to $\Omega_2$.  Because we are interested in studying only the 
algebraic structure of the pairwise stasis, we shall leave this pump term unspecified.
We verify from Eq.~(\ref{gendiffeqs}) that the gravitational redshifting effects 
indeed tend to increase $\Omega_1$ and deplete $\Omega_2$ if $w_2> w_1$. 
We also verify, 
as required, that $d\Omega_2/dt= -d\Omega_1/dt$.  Our condition for stasis is therefore
\beq
  P_{12} ~=~ 3H(w_2-w_1) \barOmega_1\barOmega_2~.
\eeq
In this system we generally have
\beq
  \kappa ~=~ \frac{2}{1+ w_1\Omega_1 + w_2\Omega_2}~.
\label{kappaseesaw}
\eeq
Indeed this result holds regardless of whether we are in a stasis epoch.
Thus our stasis condition takes the general form
\begin{empheq}[box=\fbox]{align}
  ~~P_{12} ~&=~ \barkappa \,(w_2-w_1) \,\barOmega_1\barOmega_2 \,\frac{1}{t} ~\nonumber\\
  &=~ \biggl[ 2- (1+w_1)\barkappa\biggr] \,\barOmega_1 \frac{1}{t}~.~~
\label{w1w2pump}
\end{empheq}

This result is the $(w_1,w_2)$  generalization of Eqs.~(\ref{MGmasterconstraints}), 
(\ref{masterconstraints}), and (\ref{masterconstraints2}).
However, we now see that this constraint is  independent of the particular realization of the pump term $P_{12}$ in terms of an underlying BSM physics model.  
Indeed, as we have stressed,  this result reflects the general algebraic structure underlying all pairwise stases. 

To proceed further we may assume that during stasis, our pumps have a general time-dependence of the form
\beq
P_{12}(t)~\sim~ t^{-1-p+2 - (1+w_1)\barkappa}
\eeq
where $p$ is a general constant.
Indeed, all of the pumps we have considered in this paper have this time-dependence, 
with $p=\eta= \alpha+1/\delta$ for the pump in Sect.~\ref{sec:LambdaMatter} 
and $p=\eta/\gamma$ for the pump in Sect.~\ref{sec:MatterGamma}.~
Our pump will then have the required $1/t$ time-dependence only if
\beq
     p ~=~ 2 - (1+w_1)\barkappa~.
\label{barkappap}
\eeq
This serves as our general overall scaling constraint.   This in turn implies that during 
stasis our pump must generally take the form
\beq
  P_{12} ~=~ p \,\barOmega_1 \frac{1}{t}~.
\label{pumpsoln}
\eeq
This is consistent with our prior results in Eqs.~(\ref{pumpMatterGamma}) and (\ref{pumpLambdaMatter}).

Given these results, the structure of the pairwise stasis solution is clear.  In general we learn from Eq.~(\ref{barkappap}) that
\beq
 \barkappa ~=~ \frac{2-p}{1+w_1}~.
\eeq
We also observe from Eq.~(\ref{kappaseesaw}) that
the abundance-weighted average $\langle w\rangle$ of $w$-values during stasis,
\ie,
\beq
  \langle w\rangle ~\equiv~ w_1 \barOmega_1 + w_2
   \barOmega_2~,
\label{waverage}
\eeq
is given by
$\langle w\rangle = 2/\barkappa-1$, or equivalently
\beq
 \langle w\rangle ~=~
  2\left( \frac{1+w_1}{2-p} \right) - 1~.
\label{wtrisoln}
\eeq
In conjunction with the constraint $\barOmega_1+\barOmega_2=1$, Eqs.~(\ref{waverage}) and (\ref{wtrisoln}) then allow us to solve directly for $\barOmega_{1,2}$, 
yielding
\beq
  \barOmega_1 ~=~ 
  \frac{2(w_2-w_1)-p(1+w_2)}{(2-p)(w_2-w_1)}~.
\label{eq:GeneralwsOmega1}
\eeq
This of course agrees with our previous results in 
Eqs.~(\ref{MGstasisOmegaM}), (\ref{LambdaMstasisabundance}), and (\ref{LamRadOmega}).
We likewise have
\beq 
 \barOmega_2 ~=~
    \frac{p(1+w_1)}{(2-p)(w_2-w_1)}~.~~
\label{eq:GeneralwsOmega2}
\eeq
In this connection, we note that we always must have $p<2$.   More specifically, for this system we have $w_1< \langle w\rangle<w_2$, or equivalently
\beq
    \frac{2}{1+w_2} ~<~
     \barkappa ~<~ \frac{2}{1+w_1}~.
\eeq
This implies via Eq.~(\ref{barkappap}) that all values of $p$ lead to a consistent stasis solution so long as 
\beq
  0 ~<~ p ~<~  \frac{2(w_2-w_1)}{1+w_2}~.
\eeq

We conclude with three important comments.   First, we see from the above analysis that the details of the pump are relevant {\it only for determining the abundance-weighted ``center''  of our system in $w$-space}\/, as in Eq.~(\ref{wtrisoln}).  By contrast, once this $w$-average $\langle w\rangle$ is determined, the two stasis abundances $\barOmega_{1,2}$ are situated relative to this average in a pump-independent way, as in Eq.~(\ref{waverage}).
Indeed, we shall more fully exploit this way of thinking in Sect.~\ref{subsect:seesaw}
when discussing triple stasis.

Second, we note that the solution for $\langle w\rangle$ in Eq.~(\ref{wtrisoln}) can also be written in a form that more manifestly respects the $1\leftrightarrow 2$ symmetry between our two components.
This can be done by extracting $\barkappa$ from the first line of Eq.~(\ref{w1w2pump}) [in conjunction with Eq.~(\ref{pumpsoln})]
rather than from Eq.~(\ref{barkappap}).
We thereby obtain the stasis solution
\beq
\langle w\rangle ~=~ 
\frac{4 (w_2-w_1) \barOmega_1\barOmega_2}{ (1+ \barOmega_1-\barOmega_2) p} - 1~.
\eeq
This, in conjunction with the constraint in Eq.~(\ref{waverage}), also permits an evaluation of 
the stasis abundances $\barOmega_{1,2}$.
    
Finally, we note that
in this section we have limited our discussion to the algebraic structure of pairwise stases --- namely the required relations between  pumps and cosmological expansion.
However, the critical remaining issue is to realize such  pumps in terms of
actual underlying BSM particle-physics models.  For example, in Sects.~\ref{sec:MatterGamma},
\ref{sec:LambdaMatter}, and \ref{sec:LambdaGamma}
we worked within the context of models involving
large towers of states and demonstrated that we could realize the pumps needed for stasis in terms of natural particle-physics processes such as particle decays and/or underdamping transitions.
It is this success that we consider to be the primary achievement of our previous analyses.
 
\section{Triple stasis\label{sec:TripleStasis}}

In our previous paper~\cite{Dienes:2021woi} (and as reviewed in 
Sect.~\ref{sec:MatterGamma}), we demonstrated that matter can exist in stasis 
with radiation.   Likewise, in Sects.~\ref{sec:LambdaMatter} and~\ref{sec:LambdaGamma}, 
we further demonstrated that vacuum energy can exist in stasis with matter or with radiation, respectively.  Each of these configurations represents 
a pairwise stasis between two different types 
of energy components.    Given this, the obvious next question is to determine 
whether it is possible to have a {\it triple stasis}\/ 
in which vacuum energy, matter, and radiation all co-exist in stasis with each other.   

At the end of Sect.~\ref{sec:LambdaGamma}, we noted that this can occur in a universe 
in which the matter energy component is merely a non-interacting ``spectator''.   
However, the question we now wish to investigate concerns whether we can have a 
true triple stasis in which all three energy components are interacting non-trivially 
with each other.

Of course, in a universe that contains all three energy components, cosmological 
expansion inevitably shifts the identity of the dominant component along the chain
\beq 
       \gamma ~\to~ M ~\to~ \Lambda~,
\eeq
\ie, in the direction of decreasing $w$.
In other words, a mixed-component universe that starts in a radiation-dominated configuration will eventually tend to become matter-dominated and then vacuum-dominated, simply as a result of cosmological expansion.   However, we are now seeking to determine if this entire process can be simultaneously counterbalanced by
\beqn
     {\rm underdamping~transition:}&~~~& \Lambda \to M ~~~~ \nonumber\\
      {\rm decay:}&~~~& M\to \gamma ~.~~~~~~~
\eeqn
Indeed, as we have discussed in Sect.~\ref{sec:Generalpairwise}, each of these effects essentially serves as a ``pump'' which counterbalances the natural 
tendencies induced by cosmological expansion by transferring energy back up towards components 
with larger values of $w$.  While we already know that this counterbalancing can occur 
successfully for each step individually, the question is to determine whether (and to what extent) both 
counterbalancings can co-exist within the same overall cosmology.

We emphasize that this is, {\it a priori}\/, a highly non-trivial question.
Even though a given energy component $A$ might come into stasis with another energy component $B$ in an $A/B$ universe,
and even though $A$ might also come into stasis with $C$ in an $A/C$ universe,
and even though $B$ might come into stasis with $C$ in a $B/C$ universe, it does not necessarily follow that
$A$, $B$, and $C$ can all simultaneously come into stasis in an $A/B/C$ universe.
This is because each of our previous energy-transfer processes (underdamping and decay) would now need to operate in a universe which also contains a
{\it third}\/ energy component. This third component affects the Hubble parameter and thus the
overall expansion rates whose effects would need to cancel for a triple stasis.

Phrased slightly differently, a true triple stasis can arise only if both processes (underdamping and decay) can occur
simultaneously {\it while embedded within a common cosmology}\/.
Indeed, triple stasis requires that these processes be capable of co-existing with each other within the same cosmological setting, and  this co-existence requirement may (and ultimately will) place new mathematical constraints on each.

To study this, we shall proceed in several steps.
We shall begin, as in Sect.~\ref{sec:Generalpairwise}, by studying the general algebraic structure of triple stasis in a model-independent way, deriving constraints that any simultaneous pumping transitions must satisfy in order 
to achieve triple stasis.   We shall then explore the different possible configurations that a 
triple stasis may exhibit in terms of our underlying particle-physics model involving large towers of states with particle decays and underdamping transitions. After this, we shall perform a general analysis of triple stasis and 
derive the overall scaling equations that must be satisfied in order for triple stasis to exist.
In so doing, we shall discover an additional constraint which must be satisfied in order for the overdamping and decay transitions to co-exist within the same cosmology.

With these results in hand, we  shall then proceed to consider the corresponding prefactor constraints.  
Unlike the situations that arose in previous sections for pairwise stases, we shall now find that our prefactor constraints are no longer redundant with our scaling constraints.   Instead, we shall see that they actually supply additional information.  Finally, we shall pull all the pieces together in a graphical, intuitive way which demonstrates how triple stasis ultimately operates. 

As might be imagined, this section is in some sense the 
central core
 of this paper.    We shall therefore attempt to provide as many different perspectives on our results as possible --- general and specific, algebraic and intuitive.  
All of these perspectives will be useful in subsequent sections when we extend the results of this section in order to consider the attractor nature of all of our stases, when we develop a phase-space understanding of the stasis phenomenon as a whole, and when we extend our analysis to consider various close variants of stasis. 

\FloatBarrier
\subsection{Algebraic structure of triple stasis}

To analyze the algebraic structure of stasis, we shall repeat our previous steps, only now within 
a completely general cosmology simultaneously comprising all three components (vacuum energy, matter, 
and radiation), all treated dynamically.  Following the previous analyses, we find in all generality that
\beq 
 \kappa  ~\equiv~   \frac{6}{ 2 +(1+3w) \Omega_\Lambda + \Omega_M + 2 \Omega_\gamma}~
\label{eq:triplekappa2}
\eeq 
where $\kappa$, as always, is related to the rate of change of the Hubble parameter via Eq.~(\ref{eq:kappagendef}).
We then have
\beq 
  \frac{d\Omega_i}{dt} ~=~ \frac{8\pi G}{3H^2}  \frac{d\rho_i}{dt} + \frac{6}{\kappa} H \Omega_i~.
\label{eq:preoms}
\eeq
We now must insert the equations of motion $d\rho_i/dt$ for our system.
In general, our system will have equations of motion with the algebraic structure
\beqn
\frac{d\rho_\Lambda}{dt}~&=&~ -P^{(\rho)}_{\Lambda M} - P^{(\rho)}_{\Lambda \gamma}  - 3(1+w) H \rho_\Lambda \nonumber\\
\frac{d\rho_M}{dt}~&=&~ +P^{(\rho)}_{\Lambda M} - P^{(\rho)}_{M \gamma}  -3 H \rho_M  \nonumber\\
\frac{d\rho_\gamma}{dt}~&=&~ +P^{(\rho)}_{\Lambda \gamma} + P^{(\rho)}_{M \gamma}  -4 H \rho_\gamma
\label{pumpterms}
\eeqn
where $P^{(\rho)}_{ij}$ denotes the ``pump'' term that describes the conversion of energy density 
$\rho$ from type $i$ to type $j$, and where the signs preceding these terms indicate whether this 
pump is acting as a sink ($-$) or source ($+$).  Inserting this into Eq.~(\ref{eq:preoms}) we obtain
\beqn \label{eq:TrueSystem}
\frac{d\Omega_\Lambda}{dt} ~&=&~ 
  -P_{\Lambda M} - P_{\Lambda \gamma}  + \frac{6}{\kappa} H \Omega_\Lambda 
  -3(1+w) H \Omega_\Lambda\nonumber\\
 \frac{d\Omega_M}{dt} ~&=&~ 
  P_{\Lambda M} - P_{M \gamma}  + \frac{6}{\kappa} H \Omega_M- 3H\Omega_M\nonumber\\
  \frac{d\Omega_\gamma}{dt} ~&=&~ 
 P_{\Lambda \gamma} + P_{M \gamma}  + \frac{6}{\kappa} H \Omega_\gamma -  4H\Omega_\gamma~\nonumber\\
 \eeqn 
where 
\beq
P_{ij}(t) ~\equiv~ \frac{8\pi G}{3H(t)^2} \,P^{(\rho)}_{ij}(t)~.
\eeq 
Thus $P_{ij}(t)$ denotes a pump for {\it abundances}\/, while $P_{ij}^{(\rho)}(t)$ denotes the 
corresponding pump for {\it energy densities}\/.  As a self-consistency check, we observe that 
$d\Omega_\Lambda/dt +d\Omega_M/dt +d\Omega_\gamma/dt$ indeed vanishes.

Let us now investigate the conditions under which our system can be in an (eternal) stasis epoch.  
During such a period, we must certainly have $d\Omega_\Lambda/dt= d\Omega_M/dt= d\Omega_\gamma/dt=0$.   
This gives rise to the conditions
\beqn 
 P_{\Lambda M} + P_{\Lambda \gamma}  ~&=&~ \frac{6}{\barkappa}  H \barOmega_\Lambda -3(1+w) H  \barOmega_\Lambda \nonumber\\
  -P_{\Lambda M} + P_{M \gamma} ~&=&~  \frac{6}{\barkappa}  H \barOmega_M- 3 H\barOmega_M\nonumber\\
 -P_{\Lambda \gamma} - P_{M \gamma} ~&=&~ \frac{6}{\barkappa} H \barOmega_\gamma -  4 H\barOmega_\gamma~
\label{inttt}
\eeqn 
where the pump terms $P_{ij}$ are to be evaluated during a period of stasis.   Likewise, from 
Eqs.~(\ref{eq:kappagendef}) and (\ref{eq:triplekappa2}) we can solve for the Hubble parameter $H$ during stasis, obtaining
\beq
         H(t) ~=~  \frac{\barkappa}{3 t}~.
\label{Hubbletimestasis}
\eeq
Inserting this into Eq.~(\ref{inttt}) we obtain
\begin{empheq}[box=\fbox]{align}
  P_{\Lambda M} + P_{\Lambda \gamma}  ~&=~  \Bigl[2 - (1+w)\barkappa\Bigr]~ 
    \barOmega_\Lambda \, \frac{1}{t} ~~\nonumber\\
  ~ -P_{\Lambda M} + P_{M \gamma}  ~&=~ \Bigl[ 2-\barkappa \Bigr] ~\barOmega_M\, \frac{1}{t}
    \nonumber\\
 -P_{\Lambda \gamma} - P_{M \gamma} ~&=~ \!\left[ 2-\frac{4\barkappa}{3} \right]~ 
   \barOmega_\gamma \, \frac{1}{t}~.~
 \label{eq:alstruc}
 \end{empheq}

These, then, are the most general conditions for a triple stasis involving vacuum energy, matter, 
and radiation.  
(Similar conditions can likewise be derived for any three energy components, or even for more than three components.)
Indeed, any pump terms $P_{ij}$ satisfying these equations as functions of time will 
lead to an extended (eternal) stasis epoch.  With $\barkappa$ given in Eq.~(\ref{eq:triplekappa2}) 
and with $\barOmega_\Lambda+
\barOmega_M + \barOmega_\gamma=1$ we immediately verify that   
the sum of these 
equations vanishes, implying that only two of these equations are truly independent of each other, as 
expected.  We also note that $2-4\barkappa/3<0$ for all $(\barOmega_\Lambda,\barOmega_M,\barOmega_\gamma)$ 
within the ranges $0\leq \Omega_i\leq 1$ with $\sum_i \Omega_i=1$.  Thus while both sides of the first 
equation in Eq.~(\ref{eq:alstruc}) are necessarily positive, both sides of the third equation are 
necessarily negative.   By contrast, the sign of both sides of the second equation ultimately depends 
on whether the pumping action produces a net flow of energy into or out of matter. Of course, during 
stasis, the sign of this net flow will be exactly as needed in order to compensate for the effects of 
cosmological expansion, the latter also having either a positive or negative sign depending on the 
particular values of $(\barOmega_\Lambda,\barOmega_M,\barOmega_\gamma)$.  

We have already remarked that Eq.~(\ref{w1w2pump}) describes the algebraic structure of pairwise stases.
From this perspective, Eq.~(\ref{eq:alstruc}) is the triple-stasis
analogue of Eq.~(\ref{w1w2pump}) 
and likewise describes the algebraic structure of triple stasis, once again in terms of general pumps but focused on vacuum energy, matter, and radiation.

\FloatBarrier
\subsection{Surveying the possible configurations
\label{subsec:surveying}}

Our job is now to 
construct a particle-physics model of stasis in which the $P_{ij}$ pump terms are consistent with these equations.
To do this, let us return to our model consisting of a tower of zero-mode scalar fields $\phi_\ell$, 
with $\ell = 0, 1,\ldots, N-1$.   However we shall now take into account not only the underdamping 
transition of Sect.~\ref{sec:LambdaMatter} but also the possibility of particle decay, as 
discussed in Sects.~\ref{sec:MatterGamma} and \ref{sec:LambdaGamma}.~
In general, a given field $\phi_\ell$ of mass $m_\ell$ and decay width $\Gamma_\ell$ will experience 
an underdamping transition at the time $t_\ell$ for which $3H(t_\ell)=2m_\ell$, while this same field 
will also decay with lifetime $\tau_\ell=1/\Gamma_\ell$.   For simplicity we shall assume that $t_\ell<\tau_\ell$ for all $\ell$  --- an assumption which will be discussed in more detail below --- and we shall adopt the 
``instantaneous decay'' approximation in Eq.~(\ref{instantaneousdecay})  in which a given state 
$\phi_\ell$ is presumed to decay instantaneously and fully at $t=\tau_\ell$.
As discussed below Eq.~(\ref{instantaneousdecay}), this instantaneous-decay assumption will not be critical for any of our 
important results.  However, this assumption allows the decay transition to more closely resemble the 
underdamping 
transition, with each treated  as occurring 
 instantaneously and completely at a specified time.
Later in this section we shall also consider the case with full exponential decay and verify that our basic results remain intact.

In general, these fields $\phi_\ell$ will have time-dependent energy densities $\rho_\ell(t)$.  
As in previous sections, we shall interpret this energy density as vacuum energy for times $t<t_\ell$,  
but as matter if $t_\ell<t<\tau_\ell$ and as radiation if 
$t>\tau_\ell$.   We shall let $\rho_{\rm tot}^{(\Lambda)}(t)$, $\rho_{\rm tot}^{\rm (M)}(t)$, and
$\rho_{\rm tot}^{(\gamma)}(t)$ represent the corresponding total energy densities of each type at any given time $t$, and let
$\Omega_\Lambda(t)$, $\Omega_M(t)$, and $\Omega_\gamma(t)$ represent the corresponding total abundances. 

Within a period of stasis, the underdamping time $t_\ell$ associated with underdamping 
transitions is given by the condition $3H(t_\ell)=2m_\ell$, or equivalently $t_\ell = \barkappa/(2m_\ell)$ where 
we have used Eq.~(\ref{Hubbletimestasis}).
This  provides a relation between 
$t_\ell$ and the corresponding mass $m_\ell$.  Likewise, from Eq.~(\ref{MGscalings}) 
we see that the lifetimes $\tau_\ell$ are given by
\beq
  \tau_\ell~=~ \frac{1}{\Gamma_\ell} ~=~ \frac{1}{\Gamma_0}
  \left( \frac{m_\ell}{m_0}\right)^{-\gamma}~.
\label{taulifetime}
\eeq
This provides a relation between $\tau_\ell$ and the mass $m_\ell$.   It therefore 
follows that when $\gamma\not=1$ there is a critical mass for which $t_\ell=\tau_\ell$.  
Indeed, letting the subscript `$X$\/' denote this critical point, we find
\beq
  t_X^{\gamma-1} = 
    \xi^{-\gamma} \,\Gamma_0^{1-\gamma} ~,~~~~~
  m_X^{\gamma-1} = \xi\, m_0^{\gamma-1}~,~
\label{crossingvalues}
\eeq
where $\xi$ is defined in Eq.~(\ref{eq:etaxidef}).
Of course, for $\gamma=1$, there is no single point $X$ at which $t_X=\tau_X$. 

It also follows from Eqs.~(\ref{t-to-ell})
and (\ref{taulifetime}) that
 \beq
\frac{\tau_\ell}{t_\ell} ~=~ 
 \xi\, \left(\Gamma_0 \tau_\ell\right)^{1-1/\gamma}~
\label{eq:similartriangles}
\eeq
This result holds for all values of $\gamma$. 
  However, for $\gamma\not=1$ we find
that $\xi = (\Gamma_0 t_X)^{1/\gamma-1}$, whereupon we have $\tau_\ell/t_\ell= (\tau_\ell/t_X)^{1-1/\gamma}$.

In complete analogy with the sketch in the right panel of Fig.~\ref{fig:turnon}, we may now sketch the anticipated behavior of the energy densities $\rho_\ell$ in this system as a function of time during a potential period of triple stasis.   As we shall demonstrate, there are a number of distinct possibilities for how these energy densities might behave.   We shall therefore begin by discussing these different possibilities.

For convenience, following the discussion in Sect.~\ref{sec:LambdaMatter}, we shall continue to assume for the purpose of such energy-density sketches that $w\approx -1$, so that $\rho_\ell(t)$ is approximately constant for all $t<t_\ell$.   Of course, the critical new feature for triple stasis is the existence of {\it two}\/ independent transitions, the first between vacuum energy and matter occurring at the critical underdamping time $t_\ell$ for which $3H(t_\ell)=2m_\ell$, and the second between matter into radiation occurring at the decay time $\tau_\ell=1/\Gamma_\ell$. 
At any time $t_\ell$, the original energy density $\rho_\ell^{(0)}$ which experiences the underdamping transition from vacuum energy to matter scales as
\beq
   \rho_\ell^{(0)}~\sim~ m_\ell^\alpha~\sim~
     t_\ell^{-\alpha}~.
\label{redlineturnon}
\eeq
On a log-log plot of energy density versus time, this underdamping ``$3H=2m$'' transition line would thus appear with slope $-\alpha$, precisely as in the right panel of Fig.~\ref{fig:turnon}.
Likewise, at any time $\tau_\ell$, the original energy density $\rho^{(0)}_\ell$ which decays from matter to radiation scales as
\beq
\rho^{(0)}_\ell ~\sim~ m_\ell^\alpha ~\sim~
   \tau_\ell^{-\alpha/\gamma}~.
\label{redlinedecay}
\eeq
On a log-log plot of energy density versus time, this decay transition ``$t=1/\Gamma$'' line would thus appear with slope $-\alpha/\gamma$.   

In principle, these two transition lines govern the behavior of the system.   However, for the purposes of our energy-density sketches it will prove useful to introduce a third transition line.
This line is motivated by the observation that each energy density $\rho_\ell$ no longer has its initial value $\rho_\ell^{(0)}$ at the time when the corresponding $\phi_\ell$ field decays because of the existence of the earlier underdamping transition at $t=t_\ell$ which converted the corresponding $\rho_\ell$ energy density from vacuum energy to matter.  As a result, by the time $t=\tau_\ell$ is reached, this energy density has now fallen to
$\rho_\ell \sim \rho_\ell^{(0)} (\tau_\ell/t_\ell)^{-\barkappa}$ because it has spent the time interval between $t_\ell$ and $\tau_\ell$ behaving as matter rather than vacuum energy.
We thus have
\beq
    \rho_\ell(\tau_\ell) ~\sim~
  m_\ell^\alpha \left( \frac{\tau_\ell}{t_\ell}\right)^{-\barkappa} ~\sim~ \tau_\ell^{-\alpha/\gamma - \barkappa(\gamma-1)/\gamma}~.
\label{redlineeffdecay}
\eeq
On a log-log plot of energy density versus time, this ``effective decay'' transition line would thus appear with slope 
$-\alpha/\gamma - \barkappa (\gamma-1)/\gamma$.
Because this line corresponds to the actual values of the energy densities $\rho_\ell$ at the times when the corresponding states $\phi_\ell$ are  decaying, it is this third line which in some sense represents the true decay transition, relating the energy density at the decay time to the time at which the decay transition occurs.  By contrast, the ``$\tau=1/\Gamma$'' line discussed above corresponds to the decay line that would have been relevant {\it if no prior transition to matter had occurred}\/. 

Putting these observations together, we see that on a log-log plot of energy density versus time, we now expect to have  {\it three}\/ transition lines with different slopes:
\beqn
  3H=2m~ {\rm line}: && ~~~~{\rm slope=} 
          -\alpha~,\nonumber\\
  \tau=1/\Gamma~ {\rm line}: && ~~~~{\rm slope=} 
          -\alpha/\gamma~,\nonumber\\
{\rm effective~decay~line:} && ~~~~{\rm slope=} 
          -\alpha/\gamma -\barkappa(\gamma-1)/\gamma~.\nonumber\\
\label{slopes}
\eeqn
Of course, for $\gamma\not=1$, all three of these lines  intersect at the critical point $X$ described in Eq.~(\ref{crossingvalues}).  By contrast, for $\gamma=1$, these lines are all parallel.

There is one final point that we must also consider before sketching the possible behaviors of the energy densities during triple stasis.  This is the fact that the magnitudes of our energy densities $\rho_\ell$ should continue to experience successive pairwise crossings as time evolves,
just as we observed in the right panel of Fig.~\ref{fig:turnon}, so that the identity of the $\phi_\ell$ component with the largest energy density is continually changing as our transitions proceed down the tower, just as each energy density changes from vacuum energy to matter to radiation.  As we discussed in relation to the right panel of Fig.~\ref{fig:turnon},
it is this behavior involving successive energy-density crossings which underlies the pairwise stasis phenomenon, and we expect the same to be true for triple stasis as well.
In Sect.~\ref{sec:LambdaMatter}, this crossing inter-leaved behavior was the result of the energy densities $\rho_\ell(t)$ ``reflecting'' off the transition line;   indeed, this property is ultimately what distinguished the behavior sketched in the right panel of Fig.~\ref{fig:turnon} from 
that sketched in the left panel, the latter of which does not lead to stasis.  Towards this end, we shall similarly expect a reflecting inter-leaved behavior for triple stasis.   However, we now have {\it two}\/ transition lines relative to which such reflections could potentially occur.   As a result, we can consider situations in which our energy densities reflect off the first transition line, but we can alternatively consider situations in which our energy densities pass through the first transition line and reflect off the second line instead.

\begin{figure*}[t!]
\centering
~\hskip -0.2 truein \includegraphics[keepaspectratio, width=0.54\textwidth]{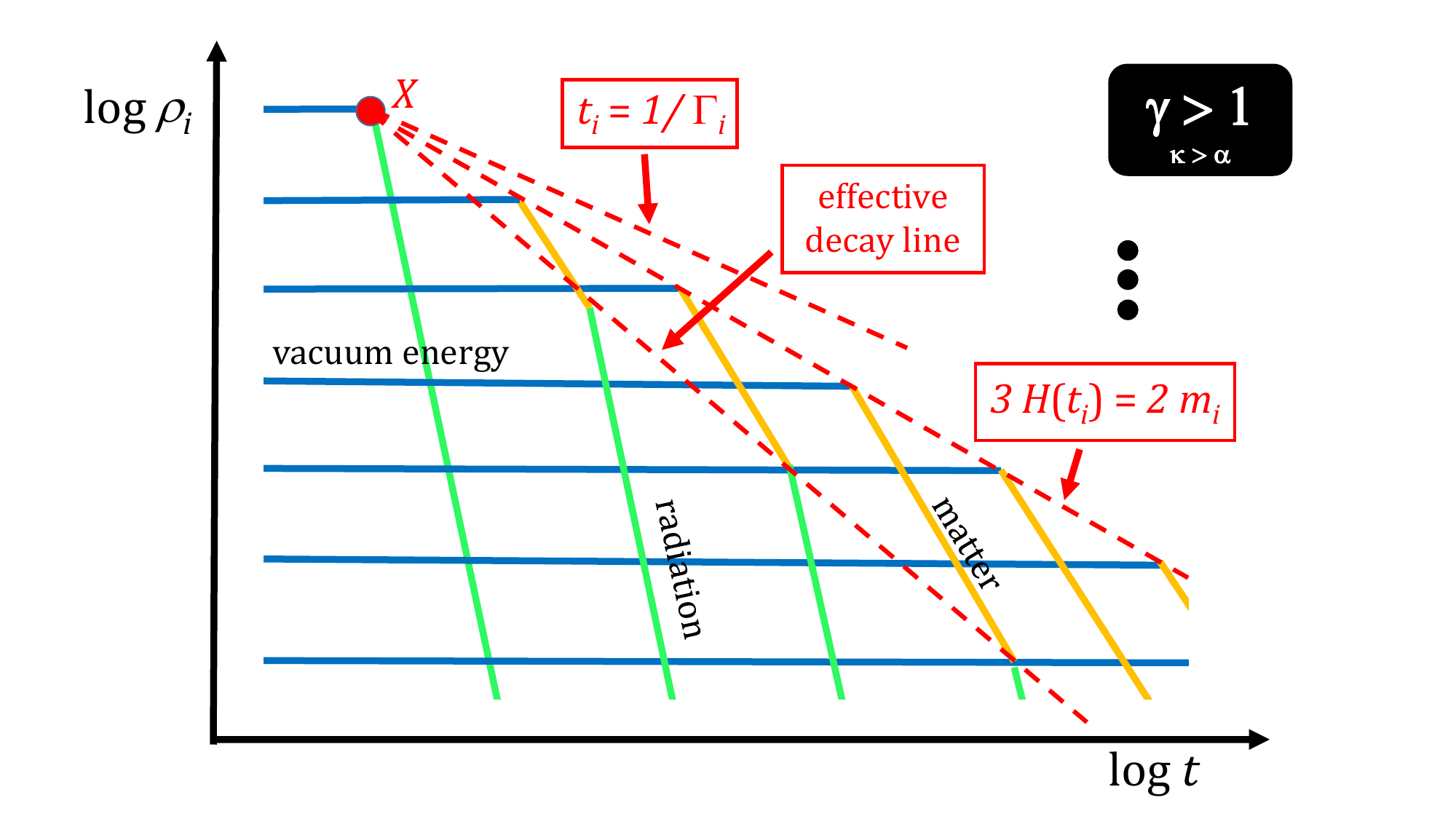}
\hskip -.45 truein
\includegraphics[keepaspectratio, width=0.54
\textwidth]{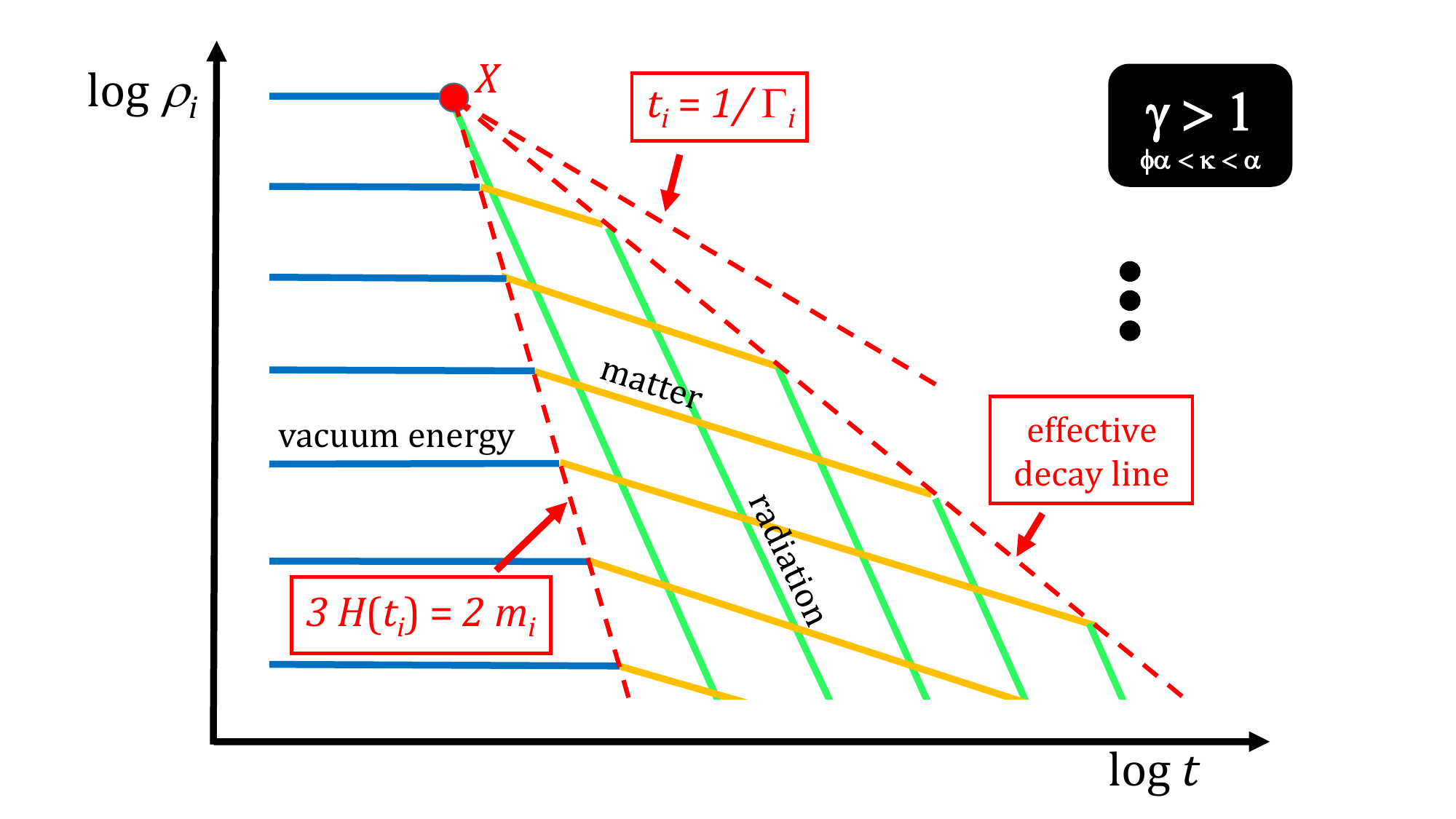}
\caption{ {\it Left panel:}\/ The individual energy densities $\rho_\ell(t)$ for the states $\phi_\ell$ with $m_\ell<m_X$, sketched as functions of time during a stasis epoch with $\gamma>1$ and $\barkappa>\alpha$. Each state undergoes a transition from an overdamped phase (blue) to an underdamped phase (yellow) at a time $t_\ell$ for which $3H(t_\ell)=2m_\ell$ before subsequently decaying to radiation (green) at the time $\tau_\ell \equiv 1/\Gamma_\ell$. 
The dashed lines (red) indicate the moments of transition between these different behaviors  and have slopes given in Eq.~(\protect\ref{slopes}).
For the purposes of this sketch we have assumed that states with greater energy densities have greater masses (\ie, that $\alpha>0$).  We have also assumed that $H(t)\sim 1/t$ with a fixed constant of proportionality, as appropriate during stasis, and we have taken $w\approx -1$ so that our vacuum-energy lines are essentially horizontal.    We see by considering vertical time-slices through this figure that the identity of the state with the greatest energy density is time-dependent, with lighter and lighter states carrying increasing shares of the total energy density as time evolves.
{\it Right panel:}\/   Same as left panel, but now sketched for $\barkappa$ within the range $\phi \alpha < \barkappa < \alpha$ where $\phi\equiv (1+\gamma/3)^{-1}$.}
\label{fig:ffig1}
 ~\hskip -0.2 truein \includegraphics[keepaspectratio, width=0.54\textwidth]{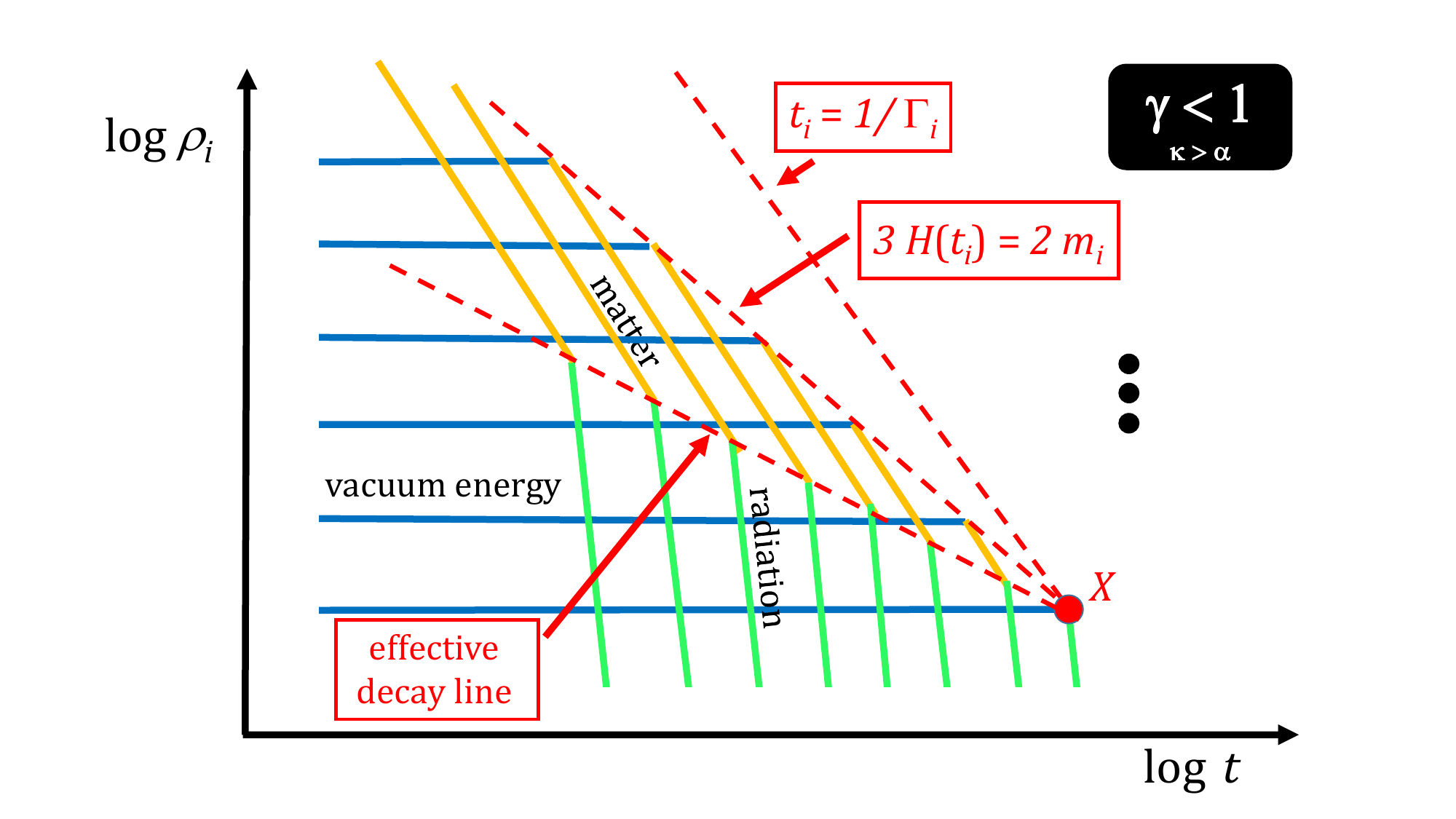}
\hskip -0.45 truein
\includegraphics[keepaspectratio, width=0.54\textwidth]{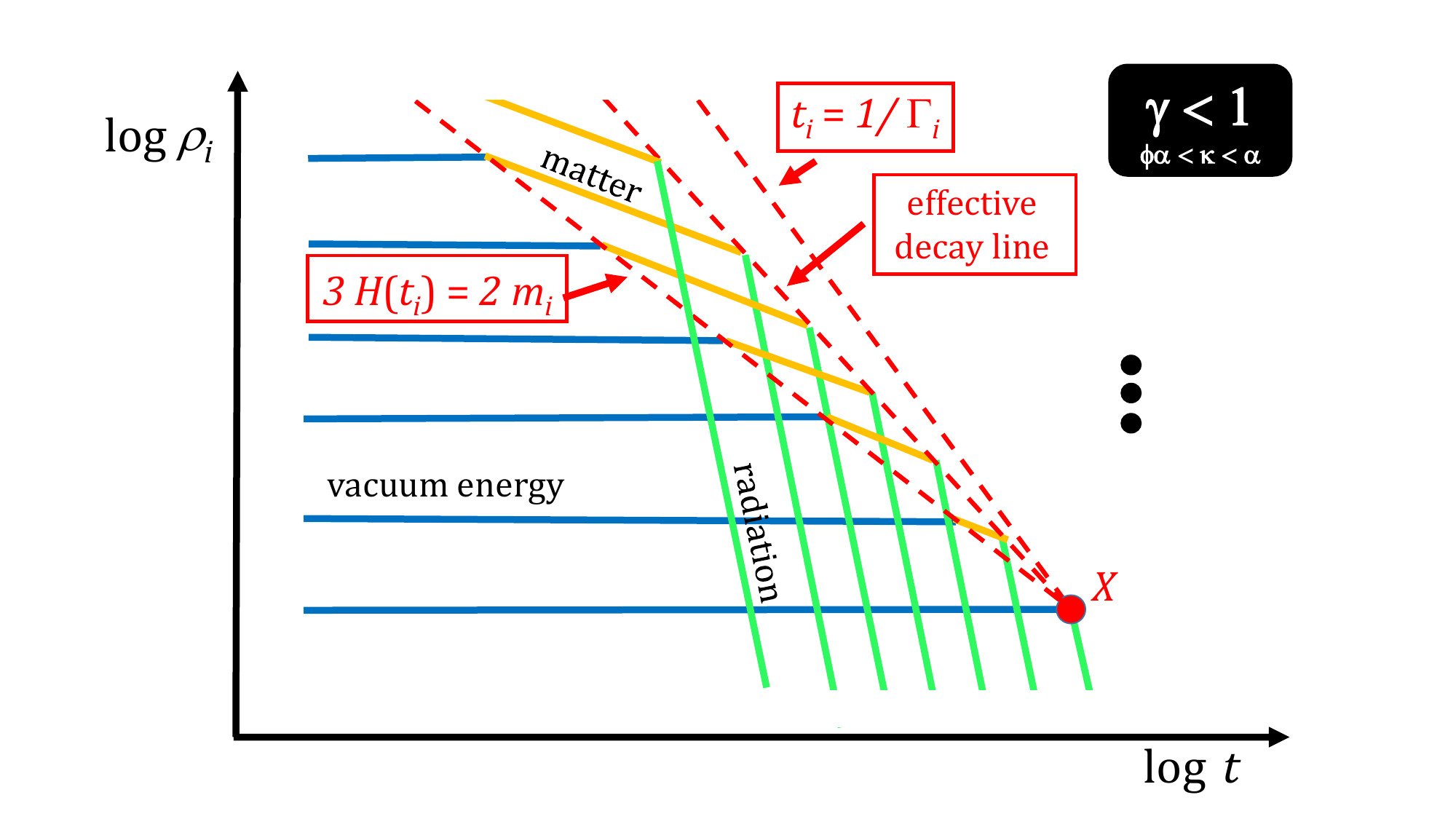}
\caption{ {\it Left panel:}\/
Same as Fig.~\protect\ref{fig:ffig1}, but now sketched for $\gamma<1$.  Here it is the {\it heavier}\/ portion of the tower with $m>m_X$ for which the states experience  transitions from overdamped to underdamped behavior before decaying to radiation.
{\it Right panel:}\/ Same as left panel, but for $\barkappa$ within the range $\phi \alpha < \barkappa< \alpha$ where $\phi\equiv (1+\gamma/3)^{-1}$.}
\label{fig:ffig3}
 ~\hskip -0.2 truein \includegraphics[keepaspectratio, width=0.54\textwidth]{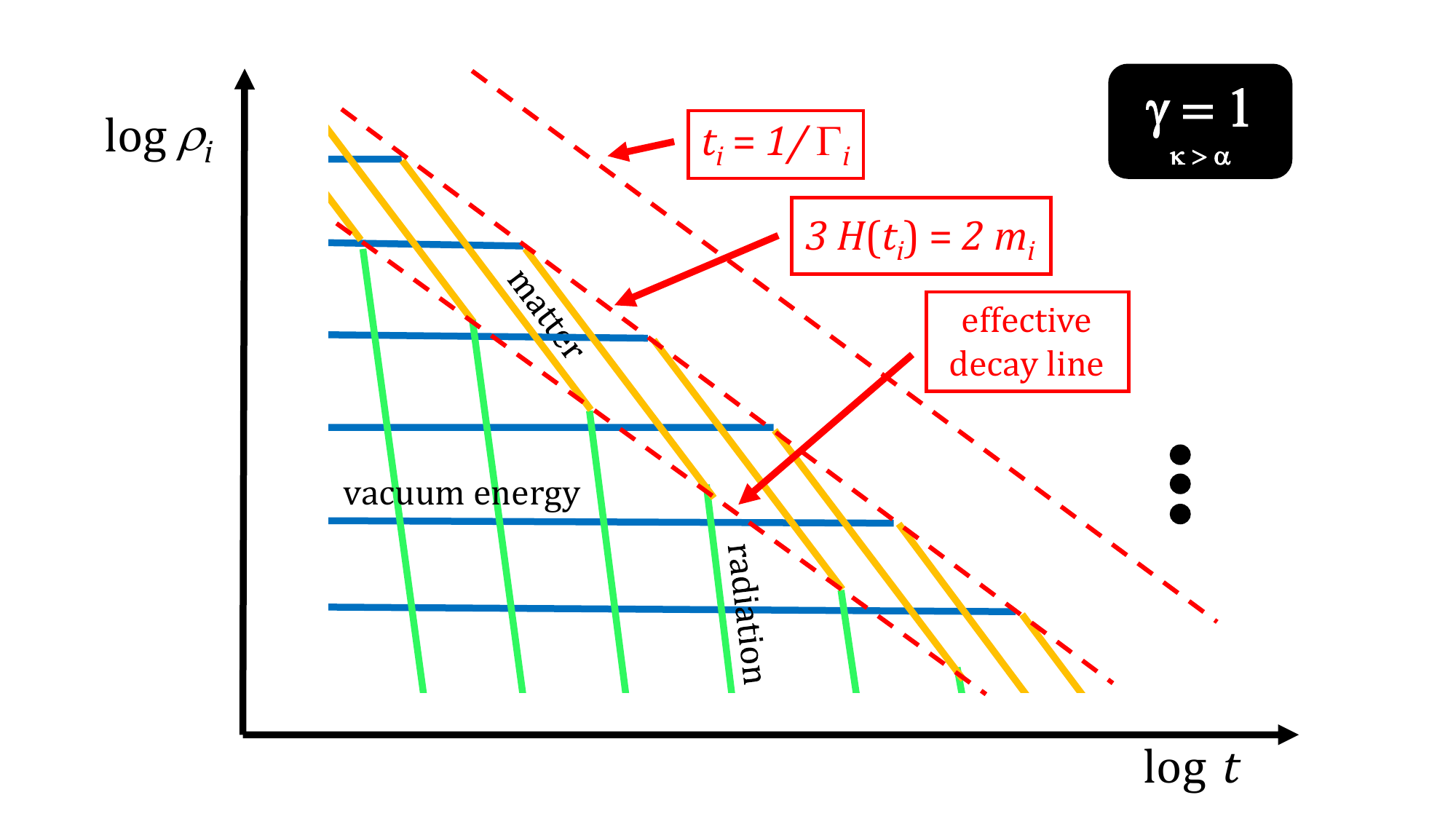}
\hskip -0.45 truein
\includegraphics[keepaspectratio, width=0.54\textwidth]{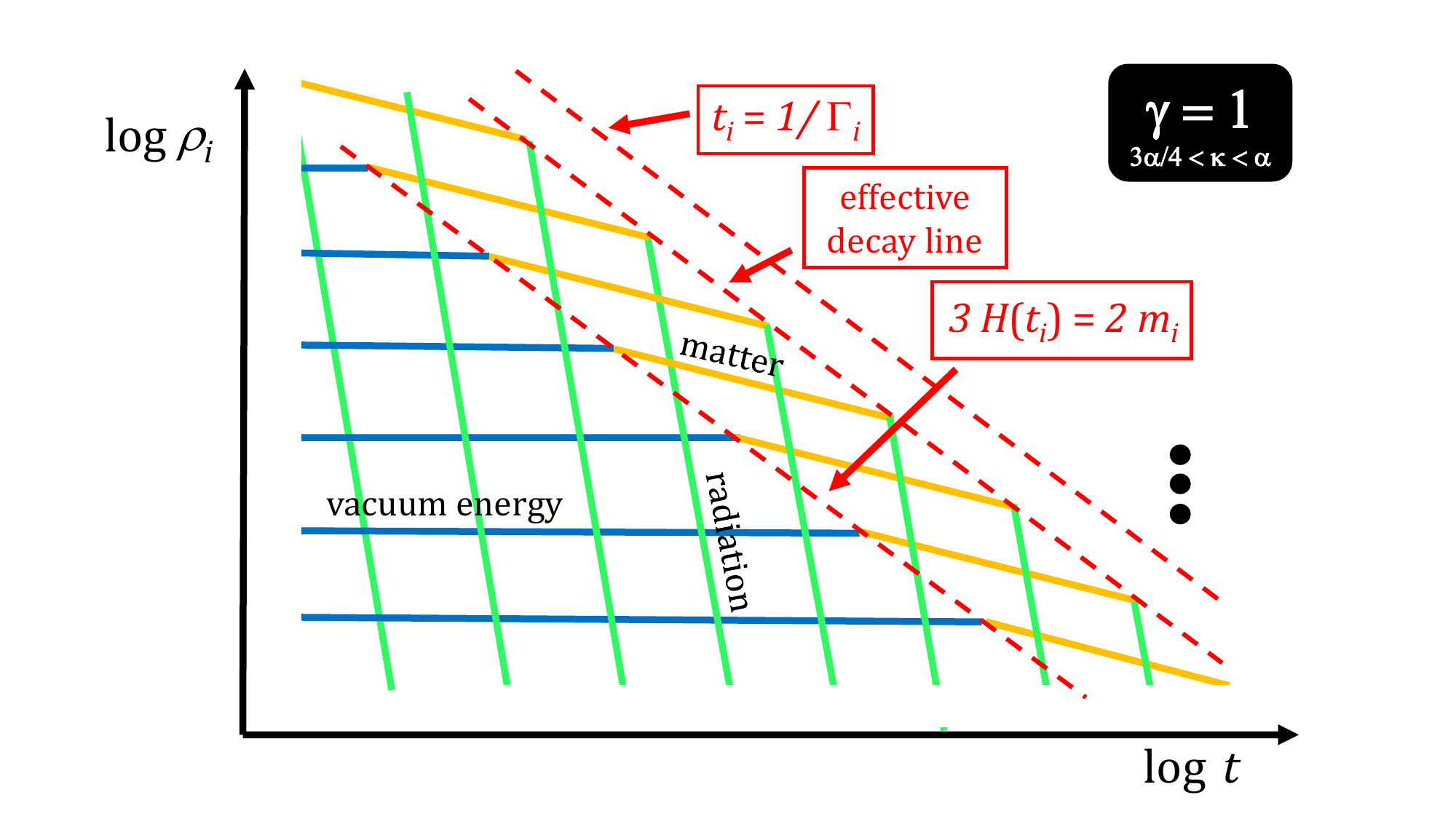}
\caption{ {\it Left panel:}\/
Same as Figs.~\protect\ref{fig:ffig1} and \protect\ref{fig:ffig3}, but now sketched for $\gamma=1$.  As long as $\xi>1$, each component across the {\it entire}\/ tower experiences a transition from overdamped to underdamped behavior before decaying to radiation.  {\it Right panel:}\/  Same as left panel, but now sketched for $\barkappa$ within the range $3\alpha/4 < \barkappa<\alpha$.  }
\label{fig:ffig5}
\end{figure*}

Putting all of these ingredients together, we see that there are in principle {\it six}\/
different classes of possible behaviors for the energy densities during triple stasis. 
These six resulting possible behaviors correspond to taking $\gamma>1$, $\gamma<1$, or $\gamma=1$,
and then in each case considering either reflection off the $3H=2m$ transition line or reflection off the effective decay line.  These six resulting possible behaviors are sketched in Figs.~\ref{fig:ffig1} through \ref{fig:ffig5}. 
Given that the slopes of the yellow (matter) and green (radiation) $\rho_\ell(t)$ lines are respectively given by $-\barkappa$ and $-4\barkappa/3$, we see that the condition for reflecting off the $3H=2m$ transition line is simply $\barkappa>\alpha$.  By contrast,
the condition for {\it not}\/ reflecting off this line but instead reflecting off the effective decay line is given by $\phi \alpha <\barkappa<\alpha$ where $\phi\equiv (1+\gamma/3)^{-1}$.

\begin{figure*}
\centering
\includegraphics[keepaspectratio, width=1.0\textwidth]{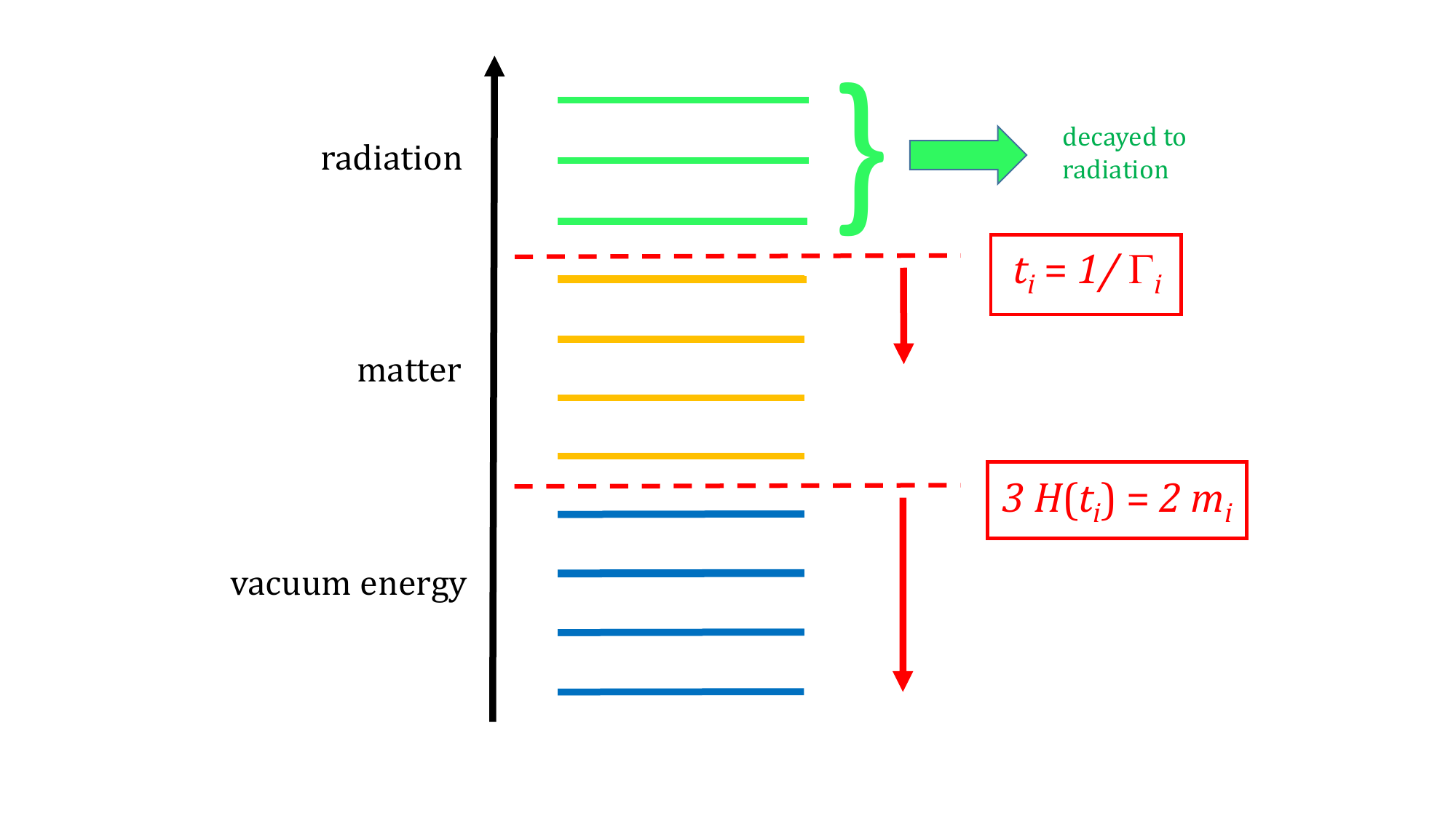}
~\vskip -0.3 truein 
\caption{{\it Triple stasis in action:}\/  The structure of the tower $\phi_\ell$ during a potential triple-stasis epoch.   The individual states are shown in order of increasing $\ell$, just as they are at the initial time in Figs.~\ref{fig:ffig1} through \ref{fig:ffig5}, with colors determined by considering appropriate vertical time slices through these figures as time evolves. At any moment two separate transitions are occurring at different locations within the tower:  an underdamping transition from vacuum energy (blue) to matter (yellow) occurring at a lower location within the tower, and a decay transition from matter (yellow) to radiation (green) simultaneously occurring at a higher location.   As time evolves, both transitions work their way down the tower, with the color of a given state changing from blue to yellow and then from yellow to green as each of the two transitions sweeps past it.   
For $\gamma>1$ the decay transition proceeds down the tower more slowly
than the underdamping transition, thereby allowing the matter region of the tower to continually increase in size.  By contrast, for $\gamma<1$ the decay transition proceeds more quickly down the tower than the underdamping transition, ultimately catching up with it at $m=m_X$.  For $\gamma=1$ both transitions move down the tower at the same rate.   All of this occurs within an expanding universe, thereby potentially supporting a triple-stasis epoch. }
\label{fig:triple_stasis_in_action}
\end{figure*}

As evident from Figs.~\ref{fig:ffig1} through \ref{fig:ffig5},
the behavior of our system is highly sensitive to whether
$\gamma>1$ (as in Fig.~\ref{fig:ffig1}),
$\gamma<1$ (as in Fig.~\ref{fig:ffig3}),
or $\gamma=1$ (as in Fig.~\ref{fig:ffig5}).
For $\gamma>1$, we find that $\tau_\ell > t_\ell$ for all $m_\ell < m_X$.   Thus the portion of the tower with $m< m_X$ corresponds to a region in which we can expect the underdamping transition to {\it precede}\/ the eventual decay, consistent with our original assumptions.
Indeed, as time evolves, the states in this region each experience a transition from vacuum energy to matter and then ultimately from matter to radiation.   Thus, within this region, we might imagine a triple stasis emerging if $m_0\ll m_{N-1} < m_X$.   By contrast, for $\gamma <1$, we find that $\tau_\ell> t_\ell$ only for $m_\ell > m_X$.   It is therefore within the portion of the tower with $m> m_X$ that we might imagine a triple stasis emerging.  This is especially true  if $m_{N-1}\gg m_X$, so that a triple stasis has time to develop within this region.  As the transitions proceed down from $m_{N-1}$  and approach $m_X$,  the ``matter'' phase experienced by the corresponding states prior to decay has shorter and shorter duration until it disappears entirely at $m=m_X$, thereby extinguishing this part of the triple stasis.
Finally, for $\gamma=1$, no critical point $X$ exists.    We then find that $\tau_\ell/t_\ell=  
2m_0/(\barkappa\Gamma_0)=\xi$
for {\it all}\/ masses $m_\ell$.   Thus in this case our {\it  entire}\/ tower could potentially support a triple stasis if $\tau_\ell/t_\ell>1$ (\ie, if $\xi>1$), but could never support a triple stasis otherwise.

Thus, to summarize, in our model of triple stasis we shall restrict our attention to those portions of our tower --- and those times $t$ --- that satisfy the conditions
\beq
\begin{cases}
  ~ m< m_X, ~t>t_X  & {\rm for}~~ \gamma>1\\
  ~ m> m_X, ~t<t_X & {\rm for}~~ \gamma<1\\
  ~ \xi >1,~
  ~{\rm any}\,(m,t)~ & 
   {\rm for}~~ \gamma=1~.
\end{cases}
\label{stasisregions}
\eeq
 Of course, we stress that the sharp 
inequalities within each of the different cases in Eq.~(\ref{stasisregions}) exist 
only because we have assumed the instantaneous-decay approximation. 

The regions of our $\phi_\ell$ towers given in Eq.~(\ref{stasisregions}) are precisely those for which $t_\ell< \tau_\ell$, thereby allowing a matter phase to emerge for each $\ell$.
Thus, within the regions outlined in Eq.~(\ref{stasisregions}), 
 we are assured that our tower simultaneously gives rise to vacuum energy, matter, and radiation. Likewise, within the times indicated in Eq.~(\ref{stasisregions}), there are two independent transitions occurring simultaneously:   the damping transition from vacuum energy to matter, and the decay transition from matter to radiation.  Taking vertical time-slices through the relevant portions of Figs.~\ref{fig:ffig1} through \ref{fig:ffig5}, we then obtain the situation illustrated in Fig.~\ref{fig:triple_stasis_in_action}.  
At any moment,  the states within the lowest portion of the tower have not yet experienced any transitions and can thus be interpreted as contributing to vacuum energy, while the states within an intermediate middle portion can be interpreted as matter
 and the states within the upper portion have already decayed to radiation.  Although the underdamping and decay transitions at any fixed time are occurring at different locations within the tower  --- the former occurring for lighter states and the latter occurring for heavier states --- they are each independently making their way down the tower. 
For $\gamma>1$, the decay transition makes its way down the tower more slowly than the underdamping transition, implying that  as time evolves an increasingly large portion of the tower behaves as matter.   By contrast, for $\gamma<1$ the reverse is true:  the decay transition makes its way down the tower more {\it rapidly}\/ than the underdamping transition, and ultimately catches up to it when $m_\ell=m_X$ (which signals the boundary of our region of interest as far as triple stasis is concerned).   Finally, for $\gamma=1$, these transitions make their way down the tower at exactly the same rate.

In the rest of this section 
we shall be interested in situations in which both transitions are still occurring within
the relevant portions of our tower, far from any ``edge'' effects either at the top or bottom of the regions of interest indicated within Eq.~(\ref{stasisregions}).  This in turn implies that we shall focus our attention on situations in which our states $\phi_\ell$, $\ell=0,1,...,N-1$, have a maximum mass $m_{N-1}$ satisfying
\beq
\begin{cases}
  ~m_X > m_{N-1} \gg 0~~  & {\rm for}~~
     \gamma > 1\\
  ~m_{N-1} \gg m_X & {\rm for}~~ \gamma<1\\
  ~m_{N-1} \gg 0  & {\rm for}~~\gamma=1~.
\end{cases}
\label{stasisregions2}
\eeq

Finally, before concluding this discussion, let us briefly add further context to our assumption that
$t_\ell< \tau_\ell$ for all $\ell$.   In making this assumption, we are explicitly disregarding the possibility that $t_\ell > \tau_\ell$ --- \ie, that the vacuum energy can decay directly to radiation even within the instantaneous-decay approximation.  Of course, as discussed at the beginning of Sect.~\ref{sec:LambdaGamma}, it is indeed possible for vacuum energy to decay or dissipate directly to radiation (or to a component that functions cosmologically as radiation, with $w=1/3$).   However, for our current purposes, such transitions are of less interest because they would completely bypass the matter phase which is needed in order to have a true triple stasis. 
Moreover, the point $X$ described in Eq.~(\ref{crossingvalues}) signifies the critical location within the $\phi_\ell$ tower which separates $t_\ell<\tau_\ell$ behavior from $t_\ell > \tau_\ell$ behavior.  Thus, by 
choosing to focus our attention on only one side of $X$ but not the other, we are implicitly enforcing the restriction that $t_\ell<\tau_\ell$ so that an intermediate matter phase appears.

These observations hold only within the instantaneous-decay approximation.  By contrast, a treatment involving a full exponential decay (replete with exponential tails in both directions) would continue to allow vacuum energy to be converted directly to our $w=1/3$ component even when $t_\ell<\tau_\ell$.  Indeed, this possibility can occasionally be quite significant, and becomes extremely remote only when $\tau_\ell \gg t_\ell$.   From Eq.~(\ref{eq:similartriangles}), we see that the latter situation arises only if
$\xi\gg 1$.
Thus, while it is certainly a mathematically self-consistent choice to adopt the instantaneous-decay approximation and restrict our attention to $\phi_\ell$ towers for which $t_\ell<\tau_\ell$ for all $\ell$, we expect such a treatment to match the results of a fully physical exponential decay only when $\xi\gg 1$.  
This issue will be discussed in more detail later in this section and in Sect.~\ref{sec:phase_diagram}.

Given this understanding, we shall proceed
by adopting the instantaneous-decay approximation and assuming that $t_\ell<\tau_\ell$ for all $\ell$.  We shall do this with the understanding that such a model represents what we may consider to be a faithful representation of the full exponential decay within the  $\xi\gg 1$ regime.
With this model in hand, our goal is then to determine
whether the resulting system can host a true triple stasis in which the total abundances $\Omega_\Lambda$, $\Omega_M$, and $\Omega_\gamma$ each remain fixed despite cosmological expansion.  

\FloatBarrier
\subsection{Overall scaling constraints
\label{subsect:overall_scaling_constraints}}

Within this system, our first step is to evaluate 
the pump terms in Eq.~(\ref{eq:alstruc}) during a period of stasis.   
This in turn 
requires that we determine
$d\rho^{(i)}_{\rm tot}/dt$ for $i=\Lambda, M,\gamma$ within this model, since the pump terms $P_{ij}$ are defined through their appearance in Eq.~(\ref{pumpterms}).
In order to derive these equations, we shall proceed in stages. 
First,  
within each of the ranges specified above in Eqs.~(\ref{stasisregions}) and (\ref{stasisregions2}) we  can approximate the corresponding energy-density contributions from each $\phi_\ell$ component during (eternal) stasis as 
\beqn
\rho^{(\Lambda)}_\ell(t) ~&=&~ \rho^\ast_\ell  \,
\left( \frac{t}{t_\ast} \right)^{-(1+w)\barkappa}
\Theta(t_\ell-t) \nonumber\\
\rho^{(M)}_\ell(t) ~&=&~ \rho^\ast_\ell \,
\left(  \frac{t_\ell}{t_\ast} \right)^{-(1+w)\barkappa}
\left(  \frac{t}{t_\ell}\right)^{-\barkappa} \, \nonumber\\
&&~~~~~~~~~~~~ \times \, \Theta(t-t_\ell) \,
\Theta(\tau_\ell -t) 
  \nonumber\\
\rho^{(\gamma)}_\ell(t) ~&=&~ \rho^\ast_\ell \,
\left( \frac{t_\ell}{t_\ast} \right)^{-(1+w)\barkappa}
 \left( \frac{\tau_\ell}{t_\ell}\right)^{-\barkappa} 
 \left( \frac{t}{\tau_\ell} \right)^{-4\barkappa/3} ~\nonumber\\
 &&~~~~~~~~~~~~ \times \,
 \Theta(t-\tau_\ell) ~~~~\nonumber\\
 \label{trip-solns}
 \eeqn
where $t_\ast$ is a fiducial early time within our eternal stasis 
prior to $t_\ell$.
Indeed, these are the relations whose $w\to -1$ limits are sketched in Figs.~\ref{fig:ffig1} through \ref{fig:ffig5}, but we shall keep $w$ arbitrary for our algebraic analysis.
The Heaviside functions within these equations capture the manner in which the original energy density $\rho_\ell$ of each $\phi_\ell$ field is transferred between the vacuum-energy, matter, and radiation  components at times $t_\ell$ and $\tau_\ell$ as the universe expands.

Of course, strictly speaking, our assumption that $t_\ast<t_\ell$ for all $\ell$ is inconsistent with our assumption that our stasis is eternal.
Indeed, Eq.~(\ref{trip-solns}) may be viewed as the triple-stasis analogue of Eq.~(\ref{MGOmegal}), for which similar issues arose.
However, just as in Sect.~\ref{sec:MatterGamma}, we shall  temporarily proceed with this assumption since it will not affect our eventual stasis constraints, deferring a formal justification of our adoption of this assumption to Sect.~\ref{subsect:h-section}. 

Given the energy densities in Eq.~(\ref{trip-solns}), we immediately find
\beqn 
 \frac{d\rho_\ell^{(\Lambda)}}{dt}~&=&~ 
    -\rho^\ast_\ell \left( \frac{t_\ell}{t_\ast} \right)^{-(1+w)\barkappa} \delta(t-t_\ell) \nonumber\\
      && ~~~~~~~~~~ -3 (1+w) H \rho_\ell^{(\Lambda)} \nonumber\\
  \frac{d\rho_\ell^{(M)}}{dt}~&=&~ 
    \rho^\ast_\ell \left( \frac{t_\ell}{t_\ast} \right)^{-(1+w)\barkappa} \delta(t-t_\ell) \nonumber\\
      && -  \rho^\ast_\ell \left( \frac{t_\ell}{t_\ast} \right)^{-(1+w)\barkappa} \left( \frac{\tau_\ell}{t_\ell}\right)^{-\barkappa} \delta(t-\tau_\ell)\nonumber\\
       &&~~~~~~~~~~~~~~~~~~~~~~~~~~~~~~
    - 3 H \rho^{(M)}_\ell  \nonumber\\
  \frac{d\rho_\ell^{(\gamma)}}{dt}~&=&~ 
     \rho^\ast_\ell \left( \frac{t_\ell}{t_\ast} \right)^{-(1+w)\barkappa}
     \left( \frac{\tau_\ell}{t_\ell} \right)^{-\barkappa}\, \delta(t-\tau_\ell) 
     \nonumber\\
  &&~~~~~~~~~~~~~~~~~~~~~~~~~~~~~~
     - 4 H \rho_\ell^{(\gamma)}   ~.~~~~\nonumber\\
\label{trimid2}
\eeqn

In order to calculate the total energy densities 
$\rho_i =\sum_\ell \rho_\ell^{(i)}$, we now wish to sum these equations over all of 
the $\phi_\ell$ states in our theory.   To do this, we shall pass to the continuum limit 
in which we express our $\ell$-parameter in terms of the corresponding underdamping time 
$t_\ell$ (as in Sect.~\ref{sec:LambdaMatter}) or lifetime $\tau_\ell$ 
(as in Sect.~\ref{sec:MatterGamma})
and then treat $t_\ell$ or $\tau_\ell$ as a continuous parameter $\hat t$ or $\tau$ respectively.  In other words, we shall replace
\beq
     \sum_\ell \to \int d \hat t \, n_{\hat t}(\hat t) ~~~~{\rm or}~~~~
\sum_\ell \to \int d\tau \, n_{\tau}(\tau)~,~~
\label{choices}
\eeq
where $n_{\hat t}(\hat t)$ and $n_\tau(\tau)$ are respectively the densities of states per unit $\hat t$ and $\tau$, evaluated at the locations within our tower for which the underdamping time or lifetime is given by $\hat t$ or $\tau$, respectively. Indeed, for each term within Eq.~(\ref{trimid2}),
we shall choose the first option within Eq.~(\ref{choices}) if the relevant term in Eq.~(\ref{trimid2}) contains a $\delta$-function involving $t_\ell$, and the second option if the relevant term contains a $\delta$-function involving $\tau_\ell$.

We shall likewise consider the energy densities $\rho_\ell(t)$ to be labeled
not by the discrete variable $\ell$ but by the continuous variables $\hat t$ or $\tau$.   However, each of these energy densities itself evolves as a function of time.   Thus for each energy density we actually have two kinds of time variables in play, one telling us {\it which}\/ energy density we are talking about (\ie, corresponding to {\it which}\/ $\phi_\ell$ within the tower) and the other telling us {\it when}\/ during the evolution of the universe that energy density should be evaluated. 
Towards this end,  for absolute clarity, we shall let
$\rho_{\hat t}(t_1;t_2)$ denote the energy density ---
evaluated at time $t_2$ ---
of that particular $\phi$-field which {\it becomes underdamped}\/ precisely at $t_1$,  and likewise let $\rho_{\tau} (t_1;t_2)$ denote the energy density --- evaluated at time $t_2$ --- of that particular $\phi$-field which {\it decays}\/ at $t_1$.
We can then replace
\beq
 \rho_\ell (t) ~\to~ \rho_{\hat t} (t_\ell ;t) ~~~{\rm or}\/~~~
 \rho_\ell (t) ~\to~ \rho_{\tau} (\tau_\ell;t)~.
 \label{replacements}
 \eeq
 Of course, the fiducial energy density $\rho_\ell^{\ast}$ now becomes either
 $\rho_{\hat t}(t_\ell;t_\ast)$ or
 $\rho_{\tau}(\tau_\ell;t_\ast)$, depending on the chosen integration variable.
 
Given these substitutions, it becomes relatively straightforward to evaluate 
the $\ell$-summations of the terms appearing in Eq.~(\ref{trimid2}).
For example, we find
\beqn
&&  \sum_\ell \rho^\ast_\ell 
  \left( \frac{t_\ell}{t_\ast} \right)^{-(1+w)\barkappa} 
\delta(t-t_\ell) \nonumber\\
&& ~~~~\to~ 
  \int d\hat t \, n_{\hat t}(\hat t)\, \rho_{\hat t}(\hat t; t_\ast)\,
\left( \frac{\hat t}{t_\ast} \right)^{-(1+w)\barkappa} 
\delta(t-\hat t)\nonumber\\ 
&& ~~~~~~~~~~=~ 
 n_{\hat t}(t)\, 
 \rho_{\hat t}(t; t_\ast)\,
\left( \frac{t}{t_\ast} \right)^{-(1+w)\barkappa} ~.
\eeqn

In a similar vein, we also have
\beqn
&& \sum_\ell \rho^\ast_\ell 
\left( \frac{t_\ell}{t_\ast} \right)^{-(1+w)\barkappa}
     \left( \frac{\tau_\ell}{t_\ell} \right)^{-\barkappa}\, \delta(t-\tau_\ell) \nonumber\\
&& ~=~ \sum_\ell \rho^\ast_\ell 
  \left(\xi \Gamma_0 t_\ast\right)^{w\barkappa} 
  \left( \frac{\tau_\ell }{t_\ast}\right)^{-\barkappa} 
      (\Gamma_0\tau_\ell)^{-w \barkappa/\gamma} \, 
     \delta(t-\tau_\ell) \nonumber\\
 && ~\to ~  \int d\tau \, n_\tau(\tau)\,
    \rho_\tau( \tau; t_\ast) \,
  \left(\xi \Gamma_0 t_\ast\right)^{w\barkappa} 
  \left( \frac{\tau }{t_\ast}\right)^{-\barkappa} \nonumber\\
    && ~~~~~~~~~~~~~~~~~~~~~~~~~~~~~~~~~~~~~
      \times (\Gamma_0\tau)^{-w \barkappa/\gamma}  
      \delta(t-\tau)\nonumber\\
  && ~ = ~ 
  \left(\xi\Gamma_0 t_\ast\right)^{w\barkappa} 
  \left( \frac{t }{t_\ast}\right)^{-\barkappa} 
      (\Gamma_0 t)^{-w \barkappa/\gamma} \,
 n_\tau(t)\,
 \rho_\tau( t; t_\ast) \,
~\nonumber\\
\label{newinsertion}
\eeqn
where in passing to the second line we have 
used Eq.~(\ref{eq:similartriangles}).

We thus find that Eq.~(\ref{trimid2}) takes the anticipated form in Eq.~(\ref{pumpterms}), 
where can now identify the pump terms for this model:
\beqn
  P^{(\rho)}_{\Lambda M} ~&=&~ \left( \frac{t}{t_\ast} \right)^{-(1+w)\barkappa} 
    n_{\hat t}(t)\, 
\rho_{\hat t}(t; t_\ast)
\nonumber\\
  P^{(\rho)}_{\Lambda \gamma} ~&=&~ 0 \nonumber\\
  P^{(\rho)}_{M\gamma} ~&=&~ 
      \left(\xi\Gamma_0 t_\ast\right)^{w\barkappa} 
      \left( \frac{t }{t_\ast}\right)^{-\barkappa} 
      (\Gamma_0 t)^{-w \barkappa/\gamma}\nonumber\\
    && ~~~~~~~~~~~~~ \times ~
      n_{\tau}(t)\, 
      \rho_\tau( t; t_\ast)~.      
\eeqn
As discussed at the end of Sect.~\ref{subsec:surveying},
the vanishing of the $P_{\Lambda\gamma}^{(\rho)}$ pump is a consequence 
of our adoption of the instantaneous-decay approximation and our assumption that 
$t_\ell< \tau_\ell$ for all $\ell$ (or equivalently that we are working within the 
$\xi\gg 1$ limit of a treatment based on taking a adopting a full exponential decay).
These results in turn yield
\beqn 
  P_{\Lambda M} ~&=&~ 
    \left( \frac{t}{t_\ast} \right)^{-(1+w)\barkappa} n_{\hat t}(t)\,
    \Omega_{\hat t}(t; t_\ast) \nonumber\\
  P_{\Lambda \gamma} ~&=&~ 0 \nonumber\\
  P_{M\gamma} ~&=&~ 
      \left(\xi\Gamma_0 t_\ast\right)^{w\barkappa} 
      \left( \frac{t }{t_\ast}\right)^{-\barkappa} 
      (\Gamma_0 t)^{-w \barkappa/\gamma} \nonumber\\
    && ~~~~~~~~~~~~~ \times ~
      n_{\tau}(t)\, 
      \Omega_\tau( t; t_\ast)
\eeqn
where 
$\Omega_{\hat t,\tau}(t_1;t_2)\equiv 
\frac{8\pi G}{3 H(t_2)^2} 
\rho_{\hat t,\tau}(t_1;t_2)$.

Of course, these results make intuitive sense.
Indeed, within a period of stasis we see that
 \beq
 \left( \frac{t}{t_\ast} \right)^{-(1+w)\barkappa}
\, \Omega_{\hat t}(t; t_\ast)   ~=~
 \Omega_{\hat t}(t; t) ~, 
 \label{tstargone}
 \eeq 
and this is nothing but the abundance of the field which is 
becoming underdamped precisely 
at the time $t$, evaluated at the moment of the underdamping transition.  
Indeed, the disappearance of $t_\ast$ within Eq.~(\ref{tstargone}) is consistent with our original adoption of $t_\ast$ as a mere fiducial time within our eternal stasis.
Likewise, we find
\beq 
  \left(\xi\Gamma_0 t_\ast\right)^{w\barkappa} 
  \left( \frac{t }{t_\ast}\right)^{-\barkappa} 
      (\Gamma_0 t)^{-w \barkappa/\gamma}\,
 \Omega_\tau( t; t_\ast) ~=~   
 \Omega_\tau( t; t)~,
\eeq
and this is nothing but the abundance of the field which is decaying precisely at the time $t$, evaluated at the moment of decay.
We can thus write our pumping terms in the final forms
\begin{empheq}[box=\fbox]{align}
         ~~ P_{\Lambda M} ~&=~ n_{\hat t}(t)\, \Omega^{(\Lambda)}_{\hat t}(t;t)
          \nonumber\\
          P_{\Lambda \gamma} ~&=~ 0 \nonumber\\
          P_{M\gamma} ~&=~
    n_{\tau}(t)\, 
    \Omega_\tau^{(M)}(t;t)
           ~
\label{pumpings2}
\end{empheq} 
where we have added the superscripts $(\Lambda)$ and $(M)$
as a reminder that the associated abundances can be interpreted as corresponding to vacuum energy and matter, respectively, at times $t=\hat t$ and $t=\tau$.
Indeed, these pumping terms describe the rates at which  abundances are being transferred between our different energy components at the moments they pass across the underdamping and decay thresholds, respectively.  

We now proceed to 
evaluate these pumping terms 
within the framework of the model introduced in Sect.~\ref{sec:MatterGamma}.~ 
In this way we shall be able to 
determine the conditions under which these pumps might simultaneously satisfy the triple-stasis constraints in Eq.~(\ref{eq:alstruc}).
Recall that $n_{\hat t} (t)$ is the density of states per unit underdamping time $\hat t$, evaluated for that portion of the tower which is becoming underdamped at time $t$, while $n_\tau(t)$ is the density of states per unit decay time $\tau$ evaluated for that portion of the tower  that is decaying at time $t$.
A straightforward calculation then yields
\beqn
  n_{\hat t} (t) \equiv \left| \frac{d\ell}{d\hat t}\right|_{\hat t=t} ~&=&~
  \frac{1}{\delta} \left( \frac{\barkappa}{2\, \Delta m\,t}\right)^{1/\delta} \frac{1}{t} ~\nonumber\\
   n_\tau(t) \equiv \left| \frac{d\ell}{d\tau} \right|_{\tau=t} ~&=&~\frac{1}{\gamma\delta}  
   \left( \frac{m_0}{\Delta m}\right)^{1/\delta} 
   (\Gamma_0 t)^{-1/(\gamma\delta)} 
  \, \frac{1}{t}~,~~~\nonumber\\
\label{ns}
\eeqn
where we have again taken $m_0\ll (\Delta m) \ell^\delta$.
Similarly, we can evaluate the abundances in Eq.~(\ref{pumpings2}), obtaining
\beqn
 && \Omega^{(\Lambda)}_{\hat t}(t;t) ~=~\Omega_0^\ast
  \left(\frac{\barkappa}{2m_0t}\right)^\alpha 
          \left( \frac{t}{t_\ast} \right)^{2-(1+w)\barkappa}
  ~\nonumber\\
&&\Omega^{(M)}_\tau (t;t) ~=~\Omega_0^\ast
 \left( \frac{m_\ell}{m_0}\right)^\alpha
 \left( \frac{t_\ell}{t_\ast} \right)^{2-(1+w)\barkappa}
  \left( \frac{t}{t_\ell} \right)^{2-\barkappa}
 \nonumber\\
 &&~~=~\Omega_0^\ast
 \left(\xi \Gamma_0 t_\ast\right)^{w\barkappa} 
 \left(\frac{t}{t_\ast}\right)^{2-\barkappa} 
 (\Gamma_0 t)^{-\alpha/\gamma-w\barkappa/\gamma} ~.\nonumber\\
\label{Omegas}
\eeqn 
Given these results, we see that
\begin{eqnarray}
P_{\Lambda M} ~&=&~ \frac{\Omega_0^\ast}{\delta t}\left(\frac{m_0}{\Delta m}\right)^{1/\delta}
  \!  \left(\xi\Gamma_0t_\ast\right)^{-\eta} \left(\frac{t}{t_\ast}\right)^{2-\barkappa - (\eta + w\barkappa)} 
    \nonumber \\
P_{M\gamma} ~&=&~ \frac{\Omega_0^\ast}{\gamma\delta t }
    \left(\frac{m_0}{\Delta m}\right)^{1/\delta}
    \left(\Gamma_0t_\ast\right)^{-\eta/\gamma}
     \nonumber \\
  & & ~~~\times \left[\xi (\Gamma_0t_\ast)^{1-1/\gamma}\right]^{w\barkappa}
    \left(\frac{t}{t_\ast}\right)^{2-\barkappa - (\eta+w\barkappa)/\gamma}\,.
    \nonumber \\
\label{eq:PumpScalingswPrefs}
\end{eqnarray}
Thus, the scaling for our first pump $P_{\Lambda M}$ will be consistent with the 
scaling $t^{-1}$ required for triple stasis only if 
\beq
\boxed{
        ~~\eta ~=~2 - (1+w)\barkappa~.~~
}
\label{constw1}
\eeq
This is precisely the constraint that we already obtained for the single pairwise $(\Lambda, M)$ stasis in Sect.~\ref{sec:LambdaMatter}.~
Likewise, demanding that our second pump $P_{M\gamma}$ scale as $t^{-1}$ yields the constraint
\beq
\boxed{
        ~~\eta ~=~2\gamma - (\gamma+w)\barkappa~.~~
}
\label{constw1prime}
\eeq
Subtracting Eqs.~(\ref{constw1}) and (\ref{constw1prime}) then yields the constraint
\beq
\boxed{
      ~~(2-\barkappa) \left( 1-\frac{1}{\gamma}\right) ~=~0~.~~
} 
\label{consteq}
\eeq
Indeed, this last constraint ensures that our two pumps $P_{\Lambda M}$ and $P_{M\gamma}$ are compatible with each other within the same background cosmology.
 
\begin{figure}[htb]
\centering
\includegraphics[width=\linewidth]{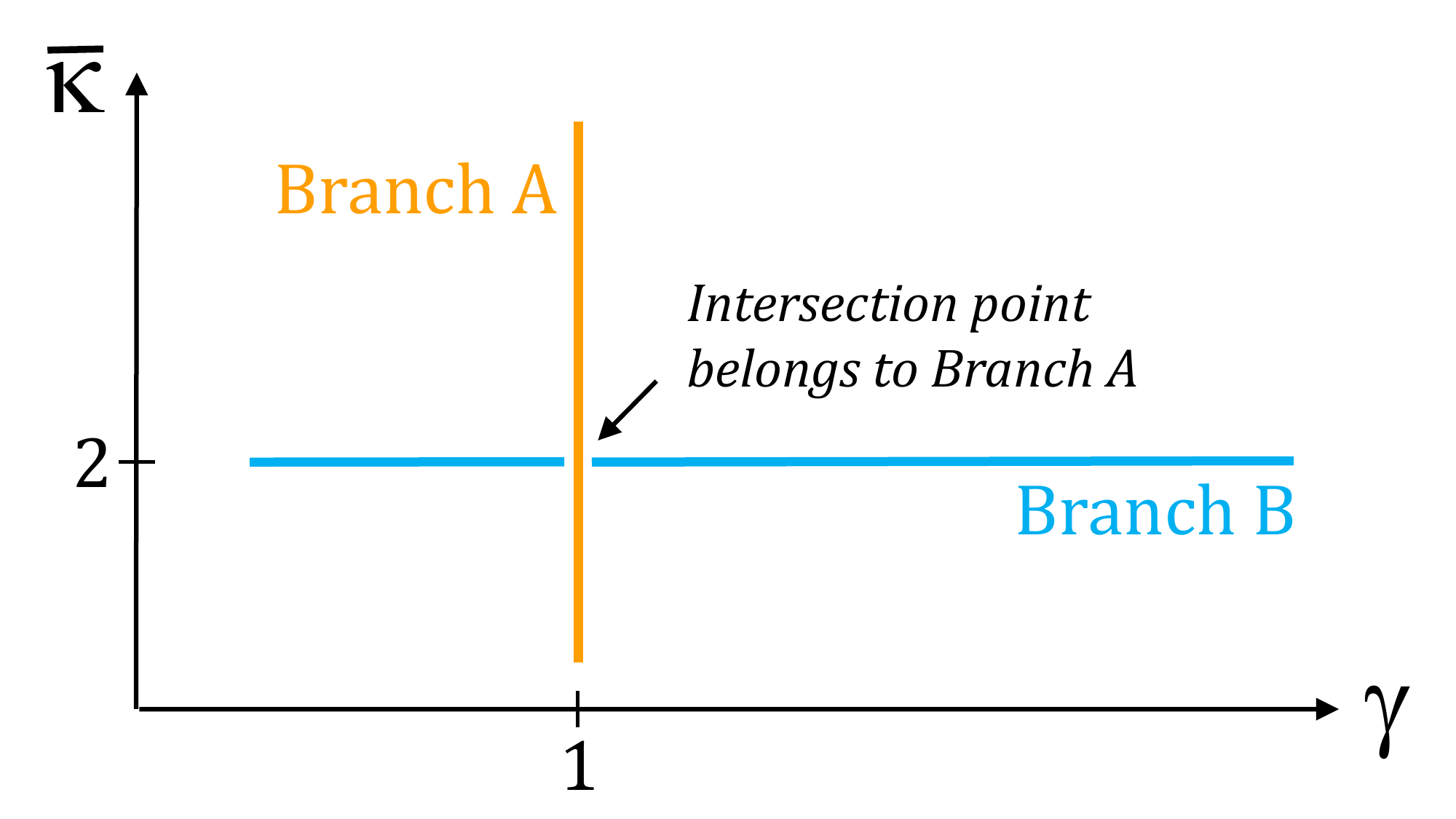}
\caption{Two branches of solutions to Eqs.~(\ref{consteq}).   These branches are defined as in Eq.~(\ref{twobranches}) with the convention that the point with $(\gamma,\barkappa)=(1,2)$ is considered to be part of Branch~A but not Branch~B.~   With this convention the two branches are mutually exclusive.}
\label{fig:branches}
\end{figure}

The result in Eq.~(\ref{consteq}) indicates that there are two mutually disjoint ``branches'' of our theory that are potentially capable of yielding triple stasis:
one has $\gamma=1$ and any value of $\barkappa$, while the other has $\barkappa=2$ and any value of $\gamma$.   Such branches have a common intersection point with $\gamma=1$ and $\barkappa=2$.  However, for reasons to become clear, it will prove useful to define these two branches to be mutually exclusive by assigning the $(\gamma,\barkappa)=(1,2)$ point to be a member of the first of these branches but not the second.
We thus define Branches~A and B to consist of potential stasis solutions satisfying not only Eq.~(\ref{constw1}) but also the additional constraints
\beqn
 {\rm Branch~A}: ~~~~ && \gamma=1,~~
 {\rm any} ~\barkappa~  \nonumber\\
{\rm Branch~B}: ~~~~ &&  \barkappa=2,~~
 {\rm any} ~\gamma\not=1 ~.~~~~~~~ 
\label{twobranches}
\eeqn
The $(\gamma,\barkappa)$ parameter-space domains corresponding to these 
additional constraints 
are sketched in Fig.~\ref{fig:branches}.
We shall soon find that these two branches have very different behaviors.

At this stage, we have learned that any possible triple-stasis solutions are limited to Branch~A or Branch~B.~
Moreover, given that we have already satisfied the constraint in Eq.~(\ref{consteq}) by limiting our attention to Branch~A or Branch~B, 
Eq.~(\ref{constw1}) becomes the 
only additional relation that we have thus far governing such potential triple stases.
However, this relation is no different from that in Sect.~\ref{sec:LambdaMatter}, with any choice of input parameters $(\alpha,\delta,w)$ leading to a unique value of $\barkappa$.
Of course, if we also choose $\gamma=1$ we find ourselves on the Branch~A line.
By contrast,  leaving $\gamma$ arbitrary, we find that only those choices of $(\alpha,\delta,w)$ that lead to $\barkappa=2$ correspond to Branch~B.

In either case, however, we see that $\barkappa$ is uniquely determined from our choice of the initial parameters $(\alpha,\delta, w)$.
In the previous pairwise cases which involved only two 
stasis abundances, such predictions for  $\barkappa$ 
(along with the usual normalization constraint $\sum_i \barOmega_i=1$) 
were sufficient to permit us to uniquely determine the corresponding stasis 
abundances.    However, for {\it triple}\/ stasis, the
specification of $\barkappa$ only restricts us to a {\it line}\/ of solutions for 
the stasis abundances $\barOmega_i$.
Thus, at first glance, it would no longer appear possible to obtain firm predictions for the stasis abundances. 

\subsection{Prefactor constraints, log-avoidance constraints, and the emergence of triple stasis}  
\label{subsect:abundance_calculation}

Fortunately, we have two further classes of constraints at our disposal:   these are the log-avoidance constraints as well as the 
prefactor constraints.   
These general types of constraints were discussed 
in some detail in Sect.~\ref{subsect:general_lessons}.~
In the case of the pairwise stases in Sects.~\ref{sec:MatterGamma}, \ref{sec:LambdaMatter}, and \ref{sec:LambdaGamma}, we found that
these prefactor 
constraints were redundant with our scaling constraints.  Indeed, this redundancy 
is what allowed stasis to be a self-consistent phenomenon in such pairwise situations.   
{\it However, for triple stasis, we shall find that these prefactor constraints are no longer redundant with 
our scaling constraints and therefore contain further information.}\/
Indeed, we shall find that
this further information will play two roles.
First, it will provide an additional condition for stasis, one which further restricts the potential solutions that we have thus far obtained.  In particular, this additional information will allow us to distinguish between Branches~A and B and thereby demonstrate that only one of these branches gives rise to a full triple stasis.
However,  this additional information will 
also allow us 
to determine the stasis abundances $\barOmega_{\Lambda,M,\gamma}$ uniquely.  

In order to derive these results, we follow previous sections and shift from the differential to the integral form of our calculations and  directly evaluate the abundances $\barOmega_{\Lambda,M,\gamma}$ during stasis.  In order to expose the common algebraic structure of these calculations, we shall evaluate these abundances in parallel.
Our calculation begins, as before, with Eq.~(\ref{trip-solns}). Even though this equation gives expressions for the energy densities $\rho_\ell(t)$,  the corresponding abundances $\Omega_\ell(t)$ take precisely the same forms except multiplied by
$(t/t^{(0)})^2$.
The Heaviside $\Theta$-function structure within Eq.~(\ref{trip-solns}) tells us that at any time $t$ we can interpret our abundance as corresponding to
 vacuum energy, matter, or radiation
 according to
\beqn
\Lambda:&~~~~~~~& t< t_\ell\nonumber\\
M:&~~~~~~~& t_\ell<t<\tau_\ell\nonumber\\
\gamma:&~~~~~~~& t> \tau_\ell~.
\eeqn
At any time $t$, this in turn implies that the lower parts of our $\phi_\ell$ tower are generally still vacuum energy while the middle parts of our tower have already transitioned to matter and the upper parts have already decayed to radiation.  Indeed, this configuration is consistent with Fig.~\ref{fig:triple_stasis_in_action}. 

Our goal, of course, is to sum over the contributions from each relevant part of the tower in order to obtain our total abundances $\barOmega_{\Lambda,M,\gamma}$.
In order to perform these sums in a parallel fashion, we shall express each state at level $\ell$ in terms of its decay lifetime $\tau_\ell$ and then treat these decay lifetimes as forming a continuous parameter $\tau$.
We can thus rewrite
\beqn
      m_\ell/m_0 ~&\rightarrow&~
      \left( \Gamma_0 \tau\right)^{-1/\gamma} \nonumber\\
      t_\ell ~&\rightarrow&~
     \left(\Gamma_0 \tau\right)^{1/\gamma}/(\xi \Gamma_0)~
\label{eq:subst}
\eeqn
where in the last line we have made use of Eq.~(\ref{eq:similartriangles}).  Finally, at any moment $t$, we can calculate the critical $\tau$-values that demarcate the boundaries between
the regions of the $\phi_\ell$ tower corresponding to vacuum energy, matter, and radiation.  Indeed, the boundary between vacuum energy and matter occurs where $t=t_\ell$, while that between matter and radiation occurs where $t=\tau_\ell$.   We thus have
\beqn
\Lambda:&~~~~~~~&  (\xi \Gamma_0 t)^\gamma \tau_0 < \tau < \tau_0 \nonumber\\
M:&~~~~~~~& t < \tau< (\xi \Gamma_0 t)^\gamma \tau_0 \nonumber\\
\gamma:&~~~~~~~&  \tau_{N-1} < \tau < t ~
\label{eq:int_limits}
\eeqn
where $\tau_0\equiv 1/\Gamma_0$.
Note that strictly speaking, our range of $\tau$-values stretches from $\tau_{N-1}$ (the smallest value, corresponding to the top state $\phi_{N-1}$ within the tower) all the way to $\tau_0$ (the largest value, corresponding to the bottom state $\phi_0$ within the tower).  In previous sections, we have noted that we are ultimately interested in times $t$ for which  $\tau_{N-1} \ll t \ll \tau_0$, so that we are far from any ``edge'' effects.    We therefore made the approximations $\tau_{N-1}\approx 0$
and $\tau_0 \approx \infty$.   However, in order to be absolutely rigorous, we have retained these precise values within our integration limits for $\barOmega_\Lambda$ and $\barOmega_\gamma$ in Eq.~(\ref{eq:int_limits}).   

Given these ingredients, our calculation of the total abundances $\barOmega_{\Lambda,M,\gamma}$ is relatively straightforward. 
Just as in previous sections, we shall no longer assume an eternal stasis, but instead recognize that stasis will only begin at times $t\gg t^{(0)}$ where $t^{(0)}$ is the production time for our $\phi_\ell$ states. 
We shall also now seek to evaluate these abundances in terms of their values $\Omega_{\Lambda,M,\gamma}^{(0)}$ at $t^{(0)}$ rather than at a fiducial time $t_\ast$ during an eternal stasis. 
Towards this end, we shall again introduce an $h$-function $h(t^{(0)},t_\ast)$ which describes the evolution of the abundances due to gravitational redshifting between $t^{(0)}$ and $t_\ast$.
We shall further assume that our abundances all share this common factor
even though there are now {\it three}\/ abundances in play, rather than only two.  At first glance, the presence of three abundances would seem to allow room for a second $h$-factor, even while demanding that the sum of the abundances remain at $1$.  However, as we shall demonstrate in Sect.~\ref{subsect:h-section}, our assumption of a common $h$-function across all three abundances is indeed appropriate for our stasis calculations.  We shall therefore proceed with only one $h$-function, deferring further discussion of this issue to Sect.~\ref{subsect:h-section}.

In the continuum limit with $\tau$ chosen as our continuous variable, each of these abundances can be ultimately expressed in the form
\begin{equation}
  \barOmega_i ~=~ \int d\tau \, n_{\tau}(\tau) \, \Omega^{(i)}_{\tau}(\tau;t) 
\end{equation}
for $i=\Lambda,M,\gamma$.
Here $n_\tau(\tau)$ is the density of states per unit $\tau$, while $\barOmega_i(\tau;t)$ represents the contribution to the
abundance $\barOmega_i$, evaluated at time $t$, from the field which decays at time $\tau$.  
Of course, in each case the integral will be delimited according to Eq.~(\ref{eq:int_limits}).

The density of states $n_\tau(\tau)$ is given in Eq.~(\ref{ns}). Likewise, the abundances
 $\Omega_\tau^{(i)}(\tau;t)$
follow readily from the energy densities in Eq.~(\ref{trip-solns})
upon making the substitutions listed in Eqs.~(\ref{eq:similartriangles}) and (\ref{eq:subst}):
\begin{eqnarray}
\Omega^{(\Lambda)}_{\tau}(\tau;t) 
   \,&=&\, \Omega_{\tau}(\tau;t^{(0)}) \, h(t^{(0)},t_\ast)\,
      \left( \frac{t}{t_\ast} \right)^{2-(1+w)\barkappa} \, \nonumber\\
    && \times \,
      \Theta\left(\frac{(\Gamma_0\tau)^{1/\gamma}}{\xi\Gamma_0} - t\right) \nonumber\\
\Omega^{(M)}_{\tau}(\tau;t) \,&=&\, 
     \Omega_{\tau}(\tau;t^{(0)}) \, 
       h(t^{(0)},t_\ast)\, 
      \left(\frac{t}{t_\ast}\right)^{2-\barkappa} \nonumber\\
      && \times \, \left[\frac{(\Gamma_0\tau)^{1/\gamma}}{\xi\Gamma_0 t_\ast} \right]^{-w\barkappa}
      \,\Theta\left(t-\frac{(\Gamma_0\tau)^{1/\gamma}}{\xi\Gamma_0}\right)\,
      \nonumber\\
      && \times \, 
      \Theta(\tau - t) 
      \nonumber\\
\Omega^{(\gamma)}_{\tau}(\tau;t) \,&=&\, \Omega_{\tau}(\tau;t^{(0)})\, h(t^{(0)},t_\ast)\, 
      \left(\frac{t}{t_\ast}\right)^{2-\barkappa} \nonumber\\
      && \times \, 
      \left[\frac{(\Gamma_0\tau)^{1/\gamma}}{\xi\Gamma_0 t_\ast} \right]^{-w\barkappa}
        \,\left(\frac{t}{\tau}\right)^{-\barkappa/3}
       \nonumber\\
    && \times \, 
       \Theta(t-\tau)~ .~~~~
   \label{trip-solns-Omegas}
\end{eqnarray} 
However, we know that $\Omega_\tau(\tau;t^{(0)})= \Omega_0^{(0)} (\Gamma_0\tau)^{-\alpha/\gamma}$.  [This is the continuum limit of the relation $\Omega_\ell^{(0)} = \Omega_0^{(0)} (m_\ell/m_0)^\alpha$.]
It therefore follows that we can write all three stasis abundances in the compact form
\begin{eqnarray}
   \barOmega_\Lambda  ~&=&~ A(t) ~
       \int_{\tau_0(\xi\Gamma_0 t)^\gamma}^{\tau_0} d\tau \, 
       (\Gamma_0 \tau)^{-1-\eta/\gamma}
       \nonumber \\
  \barOmega_M ~&=&~ A(t)\,  (\xi\Gamma_0 t)^{w\barkappa} \nonumber\\
    & & ~~~~~~ \times~ 
      \int_t^{\tau_0(\xi\Gamma_0 t)^\gamma} d\tau \,
      (\Gamma_0\tau)^{-1-(\eta+w\barkappa)/\gamma} 
      \nonumber \\
  \barOmega_\gamma  ~&=&~ A(t) \,
       (\xi\Gamma_0 t)^{w\barkappa} \, (\Gamma_0 t)^{-\barkappa/3}   
       \nonumber \\ 
    & & ~~~~~~ \times~ 
    \int_{\tau_{N-1}}^t d\tau \, 
       (\Gamma_0\tau)^{-1-(\eta + w\barkappa)/\gamma + \barkappa/3}~,
       \nonumber\\
\label{trip-Omegas-ints}
\end{eqnarray}
where each integral is delimited according to Eq.~(\ref{eq:int_limits})
and where the common $t$-dependent prefactor in each expression is
\begin{equation}
    A(t) ~\equiv~ \frac{\Omega_0^{(0)}\Gamma_0}{\gamma\delta}  
       \left(\frac{m_0}{\Delta m}\right)^{1/\delta} 
       h(t^{(0)},t_\ast)\,
       \left(\frac{t}{t_\ast}\right)^{2-(1+w)\barkappa}~.
\end{equation}

Our first interest is in the $t$-dependence of each expression within Eq.~(\ref{trip-Omegas-ints}).
This in turn depends in part on the exponent of $\tau$ 
in the corresponding integrand.  Since we are evaluating these total abundances
during an epoch of triple stasis --- \ie, at a time both well after the heaviest
$\phi$-field decays and well before the lightest $\phi$-field begins to 
oscillate --- it follows that $\tau_{N-1} \ll t \ll \hat{t}_0 = \tau_0/\xi$.
Thus, to a very good approximation, we find that $\barOmega_\Lambda$ has the $t$-scaling behavior
\begin{equation}
  \barOmega_{\Lambda} ~\sim~ \begin{cases} 
      \displaystyle t^{2-(1+w)\barkappa} & ~~\eta < 0 \\
      \displaystyle t^{2-(1+w)\barkappa}\,\log
        \left(\xi\Gamma_0 t\right) & ~~\eta = 0 \\
      \displaystyle t^{2-(1+w)\barkappa - \eta}  & ~~\eta > 0~.
    \end{cases}
\label{eq:OLscaling}
\end{equation}
Likewise, we find that $\barOmega_M$ has the $t$-scaling behavior
\beq
\barOmega_M \! ~
\sim ~\! \begin{cases} 
      \displaystyle t^{2-\barkappa}\Biggl[t^{-(\eta+w\barkappa)/\gamma} \\[-10pt]
        ~~~~~~~~~~~ 
        \displaystyle - \frac{\tau_0^{-(\eta+w\barkappa)/\gamma}}
        {(\xi\Gamma_0 t)^{\eta+w\barkappa}}\Biggr] 
       &  ~~ \eta + w\barkappa \neq 0 \\       
      \displaystyle t^{2-\barkappa}
        \,\log\left[\xi(\Gamma_0 t)^{1-1/\gamma}\right]
        & ~~\eta +w\barkappa = 0~.\\
    \end{cases} 
\label{eq:OMscaling}
\eeq
Finally, we find that $\barOmega_\gamma$ scales with time according to
\begin{equation}
  \barOmega_{\gamma} ~\sim ~ \begin{cases} 
      \displaystyle t^{2-\barkappa - (\eta + w\barkappa)/\gamma}  & 
     ~~ \eta +w\barkappa < \barkappa\gamma/3 \\       
      \displaystyle t^{2-4\barkappa/3}
        \,\log\left(\frac{t}{\tau_{N-1}}\right) 
        & ~~\eta +w\barkappa = \barkappa\gamma/3 \\
        \displaystyle t^{2-4\barkappa/3} 
        & ~~\eta + w\barkappa > \barkappa\gamma/3~.
    \end{cases}
\label{eq:Ogscaling}
\end{equation}

Given these scaling behaviors, we now seek to determine the conditions under which all three of these abundances will be independent of time.
It will prove simplest to begin by considering $\barOmega_M$.  Close inspection of our results for $\barOmega_M$ in Eq.~(\ref{eq:OMscaling}) indicates that $\barOmega_M$ will be a constant only if
\beq
\eta= 2-(1+w)\barkappa~~~~{\rm and}~~~~ \gamma=1~.
\label{eq:OMconds}
\eeq
Indeed, the upper line of Eq.~(\ref{eq:OMscaling}) will be a constant only if we additionally impose the requirement that $\barkappa\not =2$ (so that the relevant condition for that case applies), but the lower line applies if $\barkappa=2$ and also  yields a constant so long as $\gamma=1$.    We thus find that any value of $\barkappa$ is allowed in Eq.~(\ref{eq:OMconds}).

With Eq.~(\ref{eq:OMconds}) in hand, 
let us now consider our results for $\barOmega_\gamma$ in Eq.~(\ref{eq:Ogscaling}).
From Eq.~(\ref{eq:Ogscaling}), we learn that the logarithmic case can never yield a constant, while the lower case within Eq.~(\ref{eq:Ogscaling}) is fundamentally inconsistent with Eq.~(\ref{eq:OMconds}).   By contrast, the upper case within Eq.~(\ref{eq:Ogscaling}) yields a constant so long as
\beq
   \barkappa ~>~ 3/2~.
\label{eq:kappalimit}
\eeq      
However, this relation is not a surprise:  {\it any}\/ system consisting of a non-trivial mixture of vacuum energy, matter, and radiation must have $\barkappa> 3/2$ simply because radiation alone has $\kappa=3/2$ while matter and vacuum energy necessarily have $\kappa$-values that are larger than $3/2$.    Indeed, this constraint is always satisfied (essentially by construction).  It therefore does not represent an additional constraint that needs to be imposed on our system, and we need not consider it further.

Finally, we must demand that $\barOmega_\Lambda$ also be a constant.
From our results for $\barOmega_\Lambda$ in Eq.~(\ref{eq:OLscaling}) we immediately observe that the middle line can never yield a constant (thanks to the logarithm), while the top line is fundamentally inconsistent with the first relation in Eq.~(\ref{eq:OMconds}) since the power of $t$ is consistent with Eq.~(\ref{eq:OMconds}) only if $\eta=0$, while the relevant condition for that case requires $\eta<0$.    Indeed, the only case within
Eq.~(\ref{eq:OLscaling}) for which $\barOmega_\Lambda$ can be a constant is that on the final line, whereupon we obtain another condition for constant abundances, namely
\beq
   \eta ~>~ 0~.
\label{eq:poseta}
\eeq    
However, this relation is entirely subsumed within the first relation in Eq.~(\ref{eq:OMconds}).    To see this, we recognize that a universe consisting only of vacuum energy within the general-$w$ model would have $\kappa= 2/(1+w)$, while matter- or radiation-dominated universes would necessarily have smaller $\kappa$-values, namely $\kappa=2$ or $\kappa=3/2$ respectively.
Thus, by construction, triple stasis can only give rise to universes satisfying $\barkappa<2/(1+w)$, whereupon the first relation in Eq.~(\ref{eq:OMconds}) immediately yields Eq.~(\ref{eq:poseta}).
Thus Eq.~(\ref{eq:poseta}) --- like Eq.~(\ref{eq:kappalimit}) --- does not provide an additional constraint on our system and can henceforth be disregarded.

We thus conclude that the conditions for triple stasis are simply those listed in 
Eqs.~(\ref{eq:OMconds}).
Indeed, because our analysis has followed the integral form of our constraints and has therefore involved evaluating the abundances $\barOmega_{\Lambda,M,\gamma}$ directly,
these are the {\it complete set of constraints needed for triple stasis}\/.
From this observation we learn several important things:
\begin{itemize}
    \item  The scaling relation in Eq.~(\ref{constw1}) that emerged from the differential form of our constraints now appears within the integral form as well, as the first constraint within Eq.~(\ref{eq:OMconds}).
    \item   The additional constraint that we now have within Eq.~(\ref{eq:OMconds}), namely $\gamma=1$, can be identified as a {\it log-avoidance constraint}\/.   We shall discuss this additional constraint below.  
    \item   Finally, we have not yet examined our prefactor constraints.  However, we already see that any such prefactor constraints cannot possibly be needed as preconditions for stasis, since stasis has already been assured through the above constraints alone.  Indeed, as we shall see, our prefactor constraints will serve another purpose entirely, ultimately allowing us to solve for the abundances $\barOmega_{\Lambda,M,\gamma}$ during stasis.
\end{itemize}

In Sect.~\ref{subsect:overall_scaling_constraints}, we found that the requirements of triple stasis --- as obtained through the differential form of our constraints --- led uniquely to two ``branches'' of potential solutions.  These branches were indicated in Eq.~(\ref{twobranches})
and sketched in Fig.~\ref{fig:branches}.~
However, comparing with our current constraints obtained via the integral form of our stasis relations, we now see that Branch~B violates the $\gamma=1$ constraint and thus does not lead to triple stasis.   
It is easy to see what is going wrong within Branch~B.~   Recall that Branch~B is defined by the overall scaling constraint in Eq.~(\ref{constw1})
along with the extra conditions $\barkappa=2$ and $\gamma\not=1$.  Under these circumstances, we find from Eqs.~(\ref{eq:OLscaling}) and (\ref{eq:Ogscaling}) that both $\barOmega_\Lambda$ and $\barOmega_\gamma$ nominally continue to remain constant.    However we find from Eq.~(\ref{eq:OMscaling}) that $\barOmega_M$ now accrues a nominally logarithmic time dependence.  
It is for this reason that we may regard $\gamma=1$ as a logarithm-avoidance constraint.
Indeed, this nominal logarithmic time dependence for $\barOmega_M$ emerges for all points along Branch~B (with the understanding that 
the case with $\barkappa=2$ and $\gamma=1$ belongs to Branch~A).~  
{\it Thus, we conclude that Branch~B fails to yield a true triple stasis.}\/

In this context, one important comment is in order --- one which explains the word ``nominally'' which we have used above.  Even though we occasionally obtain what look like logarithmic time dependences within Eqs.~(\ref{eq:OLscaling}), (\ref{eq:OMscaling}), and (\ref{eq:Ogscaling}), this does not imply that our corresponding abundances actually exhibit time dependences which are precisely logarithmic. 
This is because the derivation leading to these equations assumed the existence of an eternal stasis.   Indeed, this was the underpinning of Eq.~(\ref{trip-solns}) with which we started our calculations.  For example, within Eq.~(\ref{trip-solns}) we implicitly assumed that our FRW scale factor $a$ evolves  as a fixed power of time, yet such behavior emerges only if the corresponding $\kappa$ is a constant.
Thus the results in Eqs.~(\ref{eq:OLscaling}),
(\ref{eq:OMscaling}), and
(\ref{eq:Ogscaling}) are absolutely 
trustworthy only in situations in which all three abundances are constants.

By contrast, in order to determine the true time dependence of our abundances in situations in which 
these abundances vary with time, we would ultimately need a more general derivation in which such initial stasis assumptions are not made. 
This has the potential to deform our results away from those obtained 
from Eqs.~(\ref{eq:OLscaling}), (\ref{eq:OMscaling}), and (\ref{eq:Ogscaling}) above.
Indeed, the fastest way to see that deformations    are needed in such cases is to  recognize from the above analysis that Branch~B appears to lead to only one abundance (specifically $\barOmega_M$) which has a non-zero time dependence, yet this cannot be consistent with our overall normalization constraint $\sum_i\barOmega_i=1$.   Thus even $\barOmega_\Lambda$ and $\barOmega_\gamma$ must accrue a suppressed time dependence as well.
However, even if the actual time dependences that emerge within such an analysis are not precisely logarithmic, they will continue to be non-vanishing.  Thus our conclusion that Branch~B does not lead to stasis remains unchanged.   The resulting time dependences are nevertheless likely to be  highly suppressed compared with our usual $\Lambda$CDM expectations.    

These observations are relevant for understanding the properties of Branch~B.~ Because Branch~B involves abundances which are not truly static,  it may therefore seem that Branch~B is uninteresting.  However,
although Branch~B does not lead to stasis, it does lead to a new phenomenon:
abundances which evolve unexpectedly slowly as  functions of time.
Indeed, although Branch~B fails to satisfy the $\gamma=1$ constraint in
Eq.~(\ref{eq:OMconds}), it does satisfy our overall scaling constraint
and thus gives rise to abundances exhibiting extremely suppressed 
time-evolutions.
This in and of itself is also an 
interesting new feature which does not emerge in the standard $\Lambda$CDM cosmologies
but which may nevertheless have important phenomenological implications.  
We shall refer 
to this phenomenon involving abundances with extremely suppressed time dependences as a {\it quasi-stasis}\/.
 We shall return to this issue in
Sect.~\ref{sec:variants}.
   
By contrast, Branch~A satisfies all of our stasis constraints, even when 
$\barkappa=2$, by virtue of the fact that $\gamma=1$ along the full length 
of Branch~A.~   Thus Branch~A exhibits a full triple stasis, thereby verifying that triple stasis is possible!    Even though the case with $\barkappa=2$ corresponds to the logarithmic case within Eq.~(\ref{eq:OMscaling}),
the fact that $\gamma=1$ assures us that the argument of the logarithm is itself time-independent.   Thus the stasis behavior is preserved even in this case.

Now that we have determined the conditions for (eternal) stasis in Eq.~(\ref{eq:OMconds}), we can proceed
to evaluate the corresponding stasis
abundances $\barOmega_\Lambda$,
$\barOmega_M$, and $\barOmega_\gamma$.
Indeed, from the bottom line of Eq.~(\ref{eq:OLscaling}), both lines of Eq.~(\ref{eq:OMscaling}),
and the top line of Eq.~(\ref{eq:Ogscaling})
we find
\beqn
  \barOmega_\Lambda ~&=&~ \frac
  {\Omega_0^{(0)}}{\delta\eta}  
    \left(\frac{m_0}{\Delta m}\right)^{1/\delta}  
    h(t^{(0)},t_\ast)\,
    \left( \xi \Gamma_0 t_\ast\right)^{-\eta}~ \nonumber\\    
\barOmega_M ~&=&~ 
    \displaystyle \frac{\Omega_0^{(0)}}{\delta}  
    \left(\frac{m_0}{\Delta m}\right)^{1/\delta} 
    h(t^{(0)},t_\ast) \,
    \xi^{w\barkappa}
    \left(\Gamma_0 t_\ast\right)^{-\eta} X^{-1}~\nonumber\\
\barOmega_\gamma ~&=&~ \frac{\Omega_0^{(0)}}{(4\barkappa/3 -2) \delta}  
    \left(\frac{m_0}{\Delta m}\right)^{1/\delta} 
      h(t^{(0)},t_\ast) \,
    \xi^{w\barkappa}
    \left(\Gamma_0 t_\ast\right)^{-\eta}~\nonumber\\
\label{eq:BarOmegawPrefs} 
\eeqn
where the quantity $X$ within $\barOmega_M$ is given by
\begin{equation}
  X ~\equiv~ \begin{cases}
      \displaystyle \frac{2-\barkappa}{1-\xi^{\barkappa-2}} 
        & ~~\barkappa  \neq 2 \\ 
     \left( \log \,\xi \right)^{-1} 
        & ~~\barkappa = 2~.~~
    \end{cases}
  \label{eq:XMDef} 
\end{equation}
Note that L'H\^opital's Rule guarantees that $X$ is continuous across $\eta+w\barkappa=0$.

Likewise, from Eq.~(\ref{eq:PumpScalingswPrefs}), we find that our pumps during stasis take the form
\begin{eqnarray}
P_{\Lambda M} ~&=&~ \frac{\Omega_0^{(0)}}{\delta }\left(\frac{m_0}{\Delta m}\right)^{1/\delta}
 h(t^{(0)},t_\ast)\,
    \left(\xi\Gamma_0 t_\ast\right)^{-\eta} 
   \, \frac{1}{t} \nonumber \\
P_{M\gamma} ~&=&~ \frac{\Omega_0^{(0)}}{\delta  }
    \left(\frac{m_0}{\Delta m}\right)^{1/\delta}
  h(t^{(0)},t_\ast)\,
  \xi^{w\barkappa}
    \left(\Gamma_0 t_\ast\right)^{-\eta}
  \frac{1}{t}~.~\nonumber\\
\label{eq:PumpScalingswPrefs2}
\end{eqnarray}
For $\eta+w\barkappa=0$ (\ie, for $\barkappa=2$), we thus immediately observe that 
\beq 
    P_{\Lambda M} ~=~ P_{M\gamma}~.
\eeq
This case will be discussed extensively below. 
  However in all other cases we find that our pumps are unequal.

It is now straightforward to derive the relative prefactor constraints for triple stasis.  Following the same path as in previous sections (but bypassing the $C$- and $C'$-coefficients since our only interest is in the {\it relative}\/ prefactor constraints), we can simply write our pumps in 
 Eq.~(\ref{eq:PumpScalingswPrefs2}) directly in terms of our abundances in Eq.~(\ref{eq:BarOmegawPrefs}):
\beqn
P_{\Lambda M} ~&=&~ 
     \eta\, \barOmega_\Lambda
     \frac{1}{t}~
     \nonumber\\
P_{M\gamma} ~&=&~
    X \barOmega_M \frac{1}{t}
    ~=~
    \left( \frac{4\barkappa}{3}  -2\right) \barOmega_\gamma \frac{1}{t}~,~~~~
\label{firsttwopumps}
\eeqn
where the second expression for $P_{M\Lambda}$ follows from the relation
\beq
    X \barOmega_M ~=~ \left(
    \frac{4\barkappa}{3}-2\right) 
    \barOmega_\gamma~
\label{eq:firsteq}
\eeq
which emerges from a direct comparison between the final two lines in Eq.~(\ref{eq:BarOmegawPrefs}).
Happily, the results in Eq.~(\ref{firsttwopumps}) 
immediately satisfy the 
first and third pump constraints in Eq.~(\ref{eq:alstruc}), providing further evidence that our model successfully yields stasis.     Indeed, these are the constraints governing the independent flows of energy density out of $\Omega_\Lambda$ and into $\Omega_\gamma$, respectively.   
However, just as in previous sections, 
we learn nothing new from these two relative prefactor constraints.

\begin{figure*}[htb]
\centering
  \includegraphics[width=0.32\linewidth]{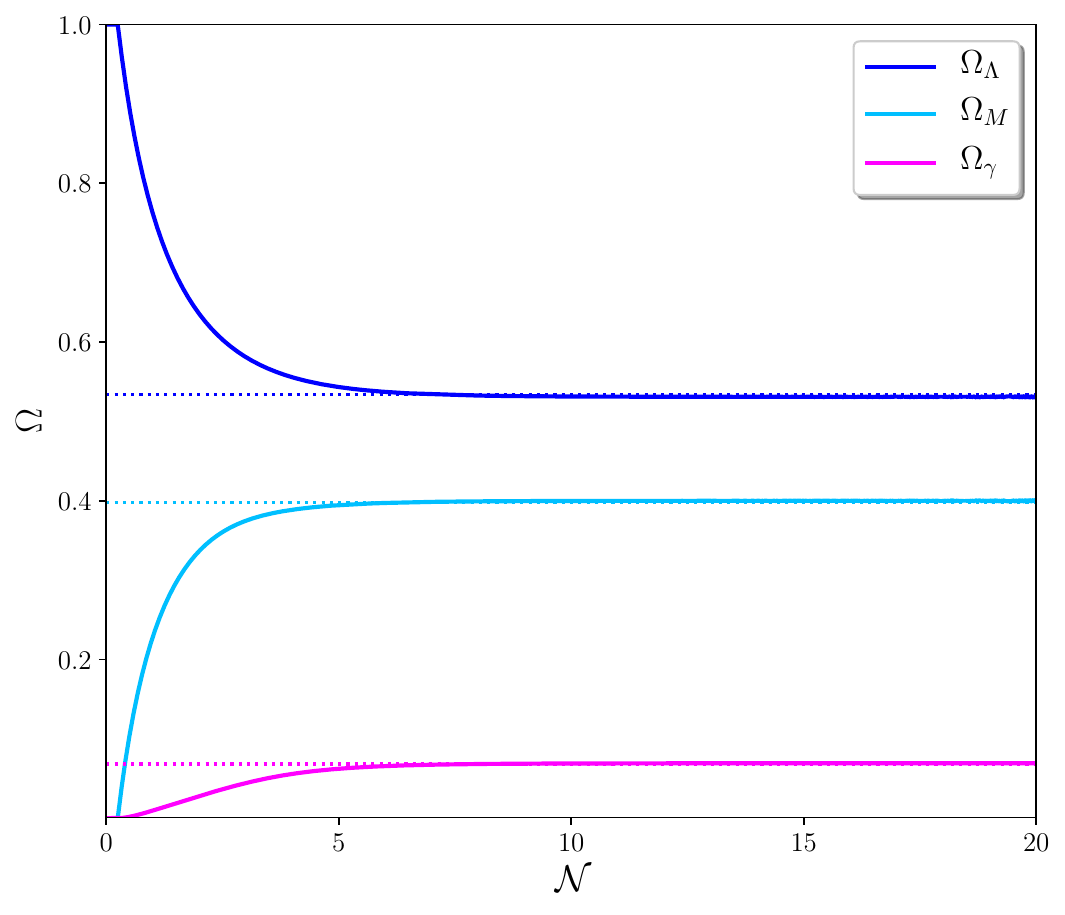}
  \includegraphics[width=0.32\linewidth]{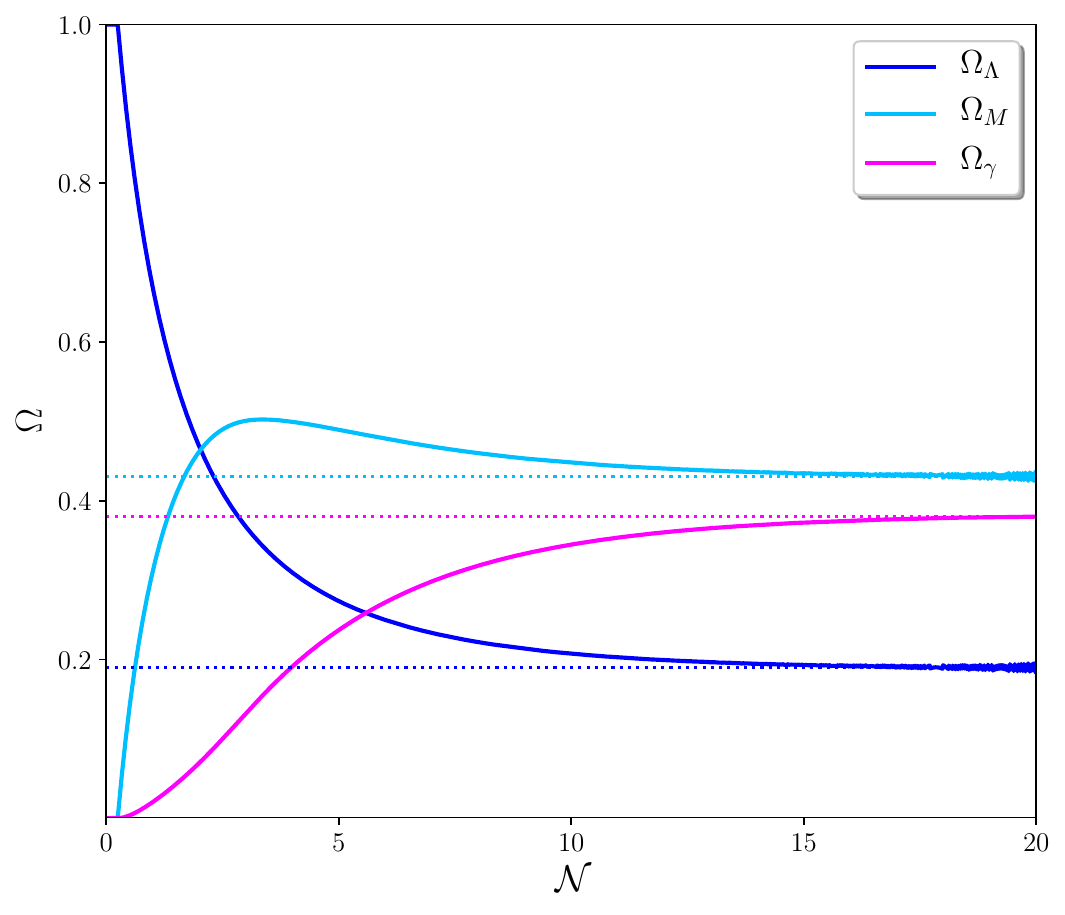}
  \includegraphics[width=0.32\linewidth]{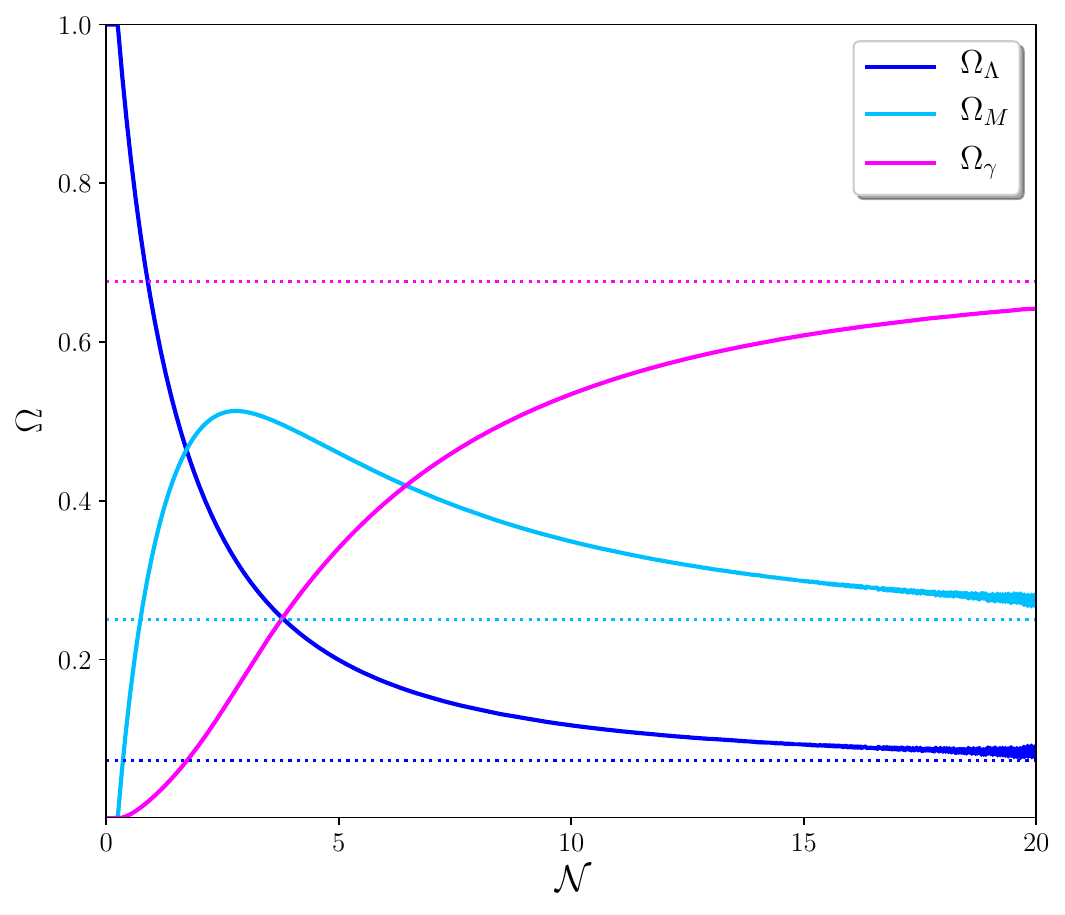}
\caption{Triple stasis!   The abundances $\Omega_\Lambda$ (solid dark blue curves), 
  $\Omega_M$ (solid cyan curves), and $\Omega_\gamma$ (solid magenta curves), plotted as 
  functions of the number $\calN$ of $e$-folds of cosmic expansion since the production 
  time $t^{(0)}$. In each case these abundances approach and then remain fixed at the 
  corresponding stasis values (dotted lines) predicted in Eqs.~(\ref{finalstasisabundances}) 
  and (\ref{abundance_avalues}), with $X$ given in Eq.~(\ref{eq:XDefImproved}).
  These curves are calculated for reference values 
  $\alpha= 2/3$ (left panel), $1$ (middle panel), and $1.1$ (right panel), holding 
  $\delta=3$, $\gamma=1$, and $w= -2/3$ fixed across all panels.  
  The full exponential nature of the decay process was taken into account in modeling the 
  transition of energy density from matter to radiation.
  These choices lead to stasis values $\barkappa = 3$, $2$, and $1.7$, respectively, 
  and we have adjusted $\Gamma_0/m_0$ in each case in order to hold $\xi=20$ fixed for 
  all panels.  These figures confirm that we can achieve triple stasis for $\barkappa>2$, 
  $\barkappa=2$, and $\barkappa<2$, respectively, and that it takes less time to achieve 
  triple stasis as $\barkappa$ increases. 
\label{fig:BranchA}}
\end{figure*}

There is, however, another prefactor constraint --- one which is non-trivial.  
This is the {\it middle}\/ constraint in Eq.~(\ref{eq:alstruc}) which links the 
two pumps $P_{\Lambda M}$ and $P_{M\gamma}$ to each other and thereby ensures that 
they are compatible with each other within a single cosmology in which the sum of all 
three abundances is restricted to remain at $1$.  [Alternatively, as discussed below
Eq.~(\ref{eq:alstruc}), this constraint is not new if we impose the relation in
Eq.~(\ref{eq:triplekappa2}), which likewise ensures that our cosmology simultaneously 
includes all three energy components.]
Comparing our results in Eq.~(\ref{firsttwopumps}) with the middle equation in 
Eq.~(\ref{eq:alstruc}) we obtain the additional constraint
\beq
   \eta\, \barOmega_\Lambda ~=~
   (X + \barkappa -2) \,\barOmega_M~.
\label{eq:secondeq}   
\eeq
Taking this constraint together with Eq.~(\ref{eq:firsteq}) and the constraint that 
\hbox{$\sum_i \barOmega_i=1$}, we then find that our relative prefactor constraints
can be satisfied only if our stasis abundances are given by
\beq
\boxed{
~~\barOmega_i ~=~ r_i \, \big/ \, \textstyle{\sum_{j} r_j~~ }
}
\label{finalstasisabundances}
\eeq
where 
\begin{empheq}[box=\fbox]{align}
    ~~ r_\Lambda ~&=~ (4\barkappa-6) (X+\barkappa-2)~~ \nonumber\\
     r_M ~&=~ (4\barkappa-6)\,\eta~~\nonumber\\
     r_\gamma~&=~ 3 \eta X~.~~
\label{abundance_avalues}
\end{empheq}
Of course, we know from Eq.~(\ref{constw1}) that
\beq
   \barkappa ~=~ \frac{2-\eta}{1+w}~.
\label{kappasolution}
\eeq
With Eq.~(\ref{kappasolution}) inserted into Eq.~(\ref{abundance_avalues}), we therefore 
find that our final stasis abundances in Eq.~(\ref{finalstasisabundances}) can be written 
directly in terms of the input variables $(\alpha,\gamma,\delta,w)$.
Indeed, only for these values of our three abundances is the middle pump equation 
in Eq.~(\ref{eq:PumpScalingswPrefs}) 
consistent with the other two [or equivalently is the value of $\barkappa$ in these 
constraint equations consistent with its original definition in Eq.~(\ref{eq:triplekappa2})].

The above analysis, which employed the instantaneous-decay approximation, may 
be refined in a straightforward manner in order to account for the full exponential nature of the 
$\phi_\ell$ decays.  After making the replacements
$\Theta(\tau -t)\rightarrow e^{-t/\tau}$ 
and 
$\Theta(t-\tau)\rightarrow 1-e^{-t/\tau}$ in Eqs.~(\ref{trip-solns}) and~(\ref{trip-solns-Omegas}),
we find that Eq.~(\ref{firsttwopumps})
still holds, but with the expression for $X$
in Eq.~(\ref{eq:XMDef}) replaced by
\beq
X~\equiv~ 
  \frac{\Gamma(3-\barkappa,\xi^{-1})}{\Gamma(2-\barkappa,\xi^{-1})}~.~~
\label{eq:XDefImproved}
\eeq
Here $\Gamma(a,z)$ denotes the
upper incomplete gamma function, with the usual domain of the 
argument $a$ extended to include all non-positive real values:
\begin{equation}
    \Gamma(a,z)~\equiv~ \int_{z}^{\infty} dy~y^{a-1}e^{-y}~.
\end{equation}
It therefore 
follows that the stasis abundances $\barOmega_i$ have the same forms as in
Eqs.~(\ref{finalstasisabundances})
and (\ref{abundance_avalues}), but with $X$ defined in
Eq.~(\ref{eq:XDefImproved}).

\begin{figure*}[htb]
\centering
\includegraphics[width=\linewidth]{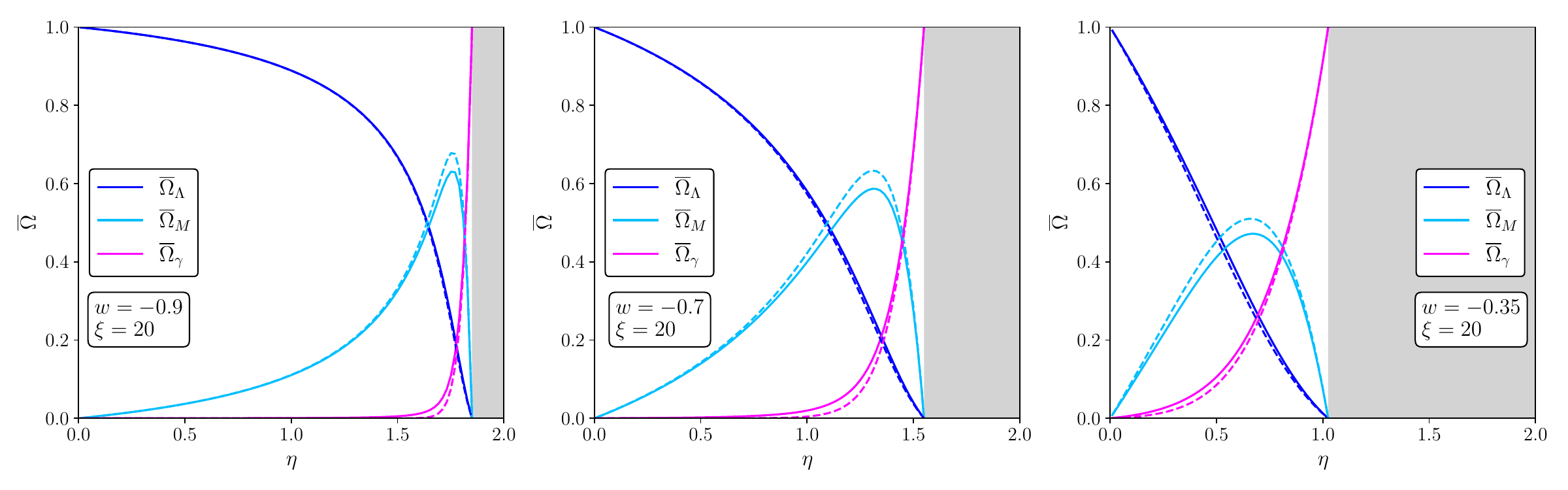}\\
\includegraphics[width=\linewidth]{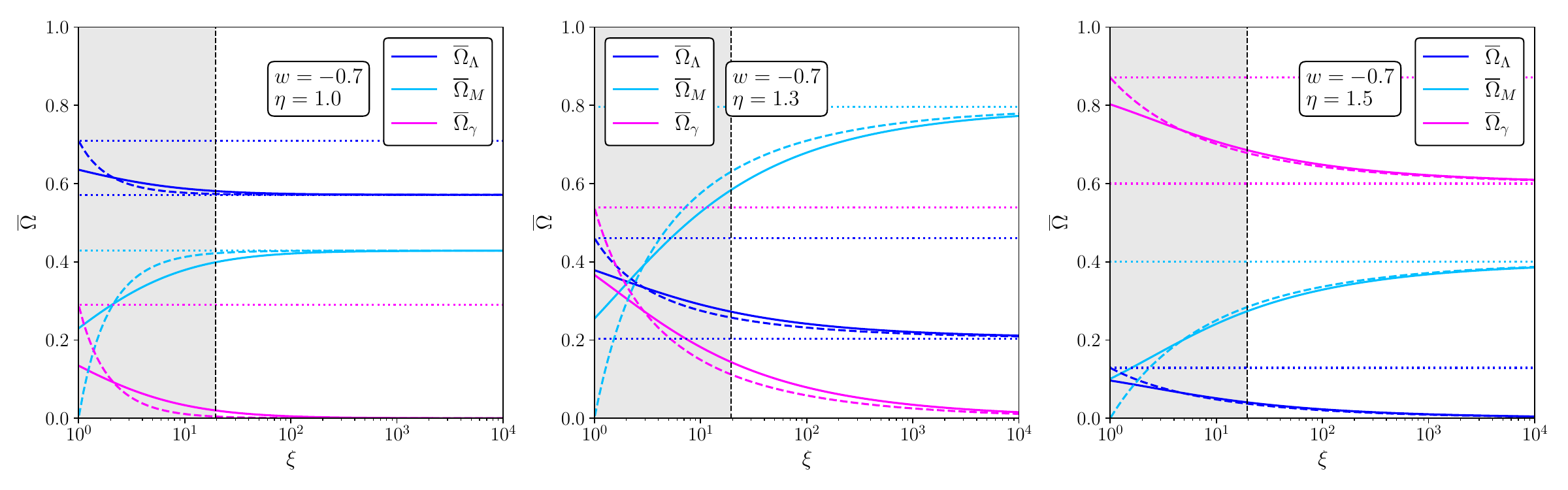}
\caption{{\it Top panels}\/: The stasis abundances $\barOmega_\Lambda$ (blue), 
$\barOmega_M$ (cyan), and $\barOmega_\gamma$ (magenta), plotted as functions 
of $\eta$ for different values of $w$ with $\xi$ held fixed.   
The solid curves are calculated assuming full exponential decay,
while the dashed curves are calculated within the instantaneous-decay approximation.  The gray region
on the right side of each panel is excluded by the constraint $0<\eta\leq (1-3w)/2$,
which follows from the constraint $\barkappa > 3/2$ in 
Eq.~(\protect\ref{eq:kappalimit}).
{\it Bottom panels}\/: The stasis abundances $\barOmega_\Lambda$ (blue), 
$\barOmega_M$ (cyan), and $\barOmega_\gamma$ (magenta), plotted as 
functions of $\xi$ for different values of $\eta$ 
with $w$ held fixed.  
The gray regions on the left sides of these panels correspond to $\xi < \xi_{\rm min}$ and 
are thus outside our regime of validity.
In the instantaneous-decay approximation, we see that $\barOmega_M \rightarrow 0$ as  
$\xi\to 0$ irrespective of the value of $\eta$, whereupon our triple stasis 
reduces to a pairwise stasis involving vacuum energy and radiation.  By contrast, 
the behavior of the abundances as $\xi\to\infty$ limit depends on the 
relationship between $\eta$ and $w$.  For $\eta < 2w$, as illustrated in the
left and middle panels, $\barOmega_\gamma\rightarrow 0$ and the triple 
stasis reduces to a pairwise stasis involving vacuum energy and matter.  By contrast, for 
$\eta > 2w$, as illustrated in the right panel, $\barOmega_\Lambda\rightarrow 0$ 
and the triple stasis reduces to a pairwise stasis involving matter and radiation.
\label{fig:Omegas}}
\end{figure*}

We emphasize once again that the model of triple stasis that we have presented in this section
represents a complete and self-consistent picture of the underlying dynamics only when certain 
conditions are satisfied.  In particular, since we are neglecting the direct transfer of energy 
density from vacuum energy to radiation, this model is valid only within the regime wherein this 
transfer of energy density has a negligible effect on the cosmological dynamics.  Thus, in what 
follows, we shall restrict our attention to regions of our parameter space wherein the ratio 
$\tau_\ell/t_\ell$ is sufficiently large for all $\phi_\ell$ that only a negligible fraction of 
the comoving number density of each particle species would have decayed at times $t < t_\ell$.  
In other words, we shall require that 
\begin{equation}
    1 - e^{-\Gamma_\ell t_\ell} ~<~ \epsilon_{\rm dec}~,
  \label{eq:LargexiCondit}
\end{equation}
for all $\phi_\ell$, where $\epsilon_{\rm dec}$ is an arbitrary small number.  However, since 
triple stasis requires that $\gamma = 1$, Eq.~(\ref{eq:similartriangles}) implies that 
$\Gamma_\ell t_\ell = \xi^{-1}$ for all $\phi_\ell$.  Thus, the condition in 
Eq.~(\ref{eq:LargexiCondit}) is tantamount to imposing an lower bound on $\xi$ of the form
\begin{equation}
    \xi ~>~ \xi_{\rm min} ~\equiv~ \frac{-1}{\log (1-\epsilon_{\rm dec})}~.
  \label{eq:ximin}
\end{equation}
In what follows, we shall take $\epsilon_{\rm dec} = 0.05$, which yields 
$\xi_{\rm min} \approx 19.5$.  That said, we note that since no energy density is transferred to 
radiation by the decay of each $\phi_\ell$ at times $t<\tau_\ell$ in the instantaneous-decay 
approximation, our analysis is formally valid in this approximation for all $\xi > 1$.

In Fig.~\ref{fig:BranchA}, we illustrate the emergence of triple stasis along Branch~A.~  
In each panel of Fig.~\ref{fig:BranchA}, we
plot the three abundances $\Omega_\Lambda$, $\Omega_M$, and $\Omega_\gamma$ as functions 
of the number of $e$-folds since the original production time within the framework of a 
full exponential decay.   These three cases correspond to sections of Branch~A with 
$\barkappa>2$, $\barkappa=2$, and $\barkappa<2$ respectively.   In all cases we find 
that a triple stasis is reached, with the stasis abundances precisely matching the 
predictions in Eqs.~(\ref{finalstasisabundances}) and (\ref{abundance_avalues})
with $X$ given in Eq.~(\ref{eq:XDefImproved}).  This then provides numerical confirmation 
of the existence of triple stasis.  We also note that it takes longer to reach triple 
stasis as $\barkappa$ is decreased.

In Fig.~\ref{fig:Omegas}, we illustrate how the expressions for the 
stasis abundances $\barOmega_\Lambda$, $\barOmega_M$, and $\barOmega_\gamma$ in 
Eqs.~(\ref{finalstasisabundances}) and (\ref{abundance_avalues})
depend on our model parameters $\eta$, $w$, and $\xi$.  In the panels appearing along 
the top row of the figure, we plot these abundances as functions of $\eta$ for three 
different values of $w$, with $\xi$ fixed.  We note that since $\xi$ 
depends on $\barkappa$, and hence on $\eta$, the ratio $m_0/\Gamma_0$ also varies 
with $\eta$ in each panel such that $\xi$ remains constant within each panel.  The results shown
in the left, middle, and right panels of this row correspond to the choices $w=-0.9$, $w=-0.7$, 
and $w=-0.35$, respectively, and in all three panels we have taken $\xi = 20$ --- a
value which exceeds $\xi_{\rm min}$ for our chosen value of $\epsilon_{\rm dec}$.
The solid curves in each panel indicate the values of $\barOmega_\Lambda$ (blue), 
$\barOmega_M$ (cyan), and $\barOmega_\gamma$ (magenta) obtained using 
the form for $X$ in Eq.~(\ref{eq:XDefImproved}) corresponding to full exponential decay, 
while the dashed curves indicate the corresponding abundances obtained using the form 
for $X$ in Eq.~(\ref{eq:XMDef}) corresponding to the instantaneous-decay approximation.  
The gray region on the right side of each of these panels is excluded by the constraint 
$0<\eta\leq (1-3w)/2$, which follows from the constraint $\barkappa > 3/2$ in 
Eq.~(\protect\ref{eq:kappalimit}).

\begin{figure*}[htb]
\centering
\includegraphics[width=0.8\linewidth]{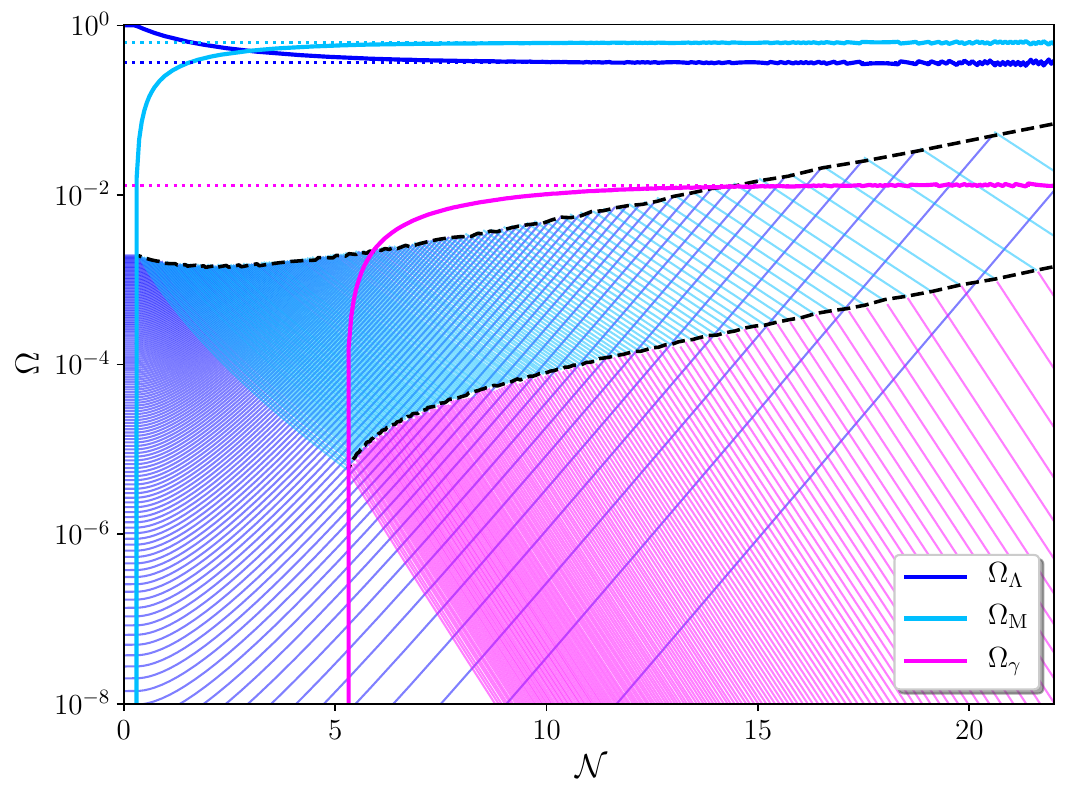}
\caption{Triple stasis ``under the hood'', illustrating how the individual abundances 
$\Omega_\ell(t)$ conspire to bring our system into --- and then sustain --- triple stasis.  
Working within the framework of the instantaneous-decay approximation
and choosing $(\alpha,\gamma,\delta,w)=(1,1,5,-0.7)$ in order to highlight critical features, 
we plot the three total abundances $\Omega_\Lambda$ (thick dark-blue solid line), 
$\Omega_M$ (thick cyan solid line), and $\Omega_\gamma$ (thick magenta solid line) 
as functions of the number $\calN$ of $e$-folds since the initial production time.
We also plot the individual abundances $\Omega_\ell(t)$ (thin lines) as they evolve in time, 
starting as vacuum energy (dark blue) before transitioning to matter (cyan) and eventually 
to radiation (magenta).  For the sake of graphical clarity 
we have  only shown every 20th individual abundance $\Omega_\ell(t)$, starting from the 
top of the tower, and adjusted $\Gamma_0/m_0$ such that $\xi=300$. 
The underdamping transition between the vacuum-energy and matter phases occurs along the 
upper dashed black line, while the instantaneous-decay transition between the matter and 
radiation phases occurs along the lower dashed black line.   This figure can be viewed 
as an explicit numerical realization of Fig.~\ref{fig:ffig5} except that it is plotted 
for individual abundances $\Omega_\ell(t)$ rather than individual energy densities 
$\rho_\ell(t)$ and also includes  the initial transient behavior of our system as it 
evolves from the production time $t^{(0)}$ into stasis.
The red transition lines in Fig.~\ref{fig:ffig5} correspond to the dashed black lines here,
asymptotically becoming parallel once stasis is reached.  This figure can also be 
viewed as a triple-stasis analogue of the left panel of Fig.~\ref{fig:review}.} 
\label{fig:underhood}
\end{figure*}

From these panels we observe that the universe during stasis is effectively 
vacuum-energy dominated in the $\eta\to 0$ limit, with
$\barOmega_\Lambda \rightarrow 1$ and $\barOmega_M,\barOmega_\gamma \rightarrow 0$.
Indeed, this holds true regardless of the value of $w$.
However, as $\eta$ increases,  we see that $\barOmega_\gamma$ rises monotonically 
and reaches unity at the 
point at which $\eta$ reaches its maximum value, while $\barOmega_\Lambda$ falls 
monotonically to zero over the same interval.  By contrast, $\barOmega_M$
initially increases with $\eta$, reaching a maximum (the value of which depends 
non-trivially on $\eta$, $w$, and $\xi$), and then falling to zero as $\eta$ increases 
toward its maximum value.  
Thus, in each case, we see that our triple stasis as a function of $\eta$ 
{\it interpolates}\/ between a fully vacuum-energy  dominated universe and a 
radiation-dominated one.

In the panels along the bottom row of Fig.~\ref{fig:Omegas}, 
we instead plot $\barOmega_\Lambda$, $\barOmega_M$, and $\barOmega_\gamma$ as functions of 
$\xi$ for different values of $\eta$, with $w$ held fixed.  The results in the left, 
middle, and right panels of this row correspond to the choices $\eta=1.0$, $\eta=1.3$, and 
$\eta=1.5$, respectively, and for all three panels we have taken $w = -0.7$.  
The gray region on the left side of each of these three panels indicates the region 
wherein $\xi < \xi_{\rm min}$, which lies outside our regime of validity.  

We observe from these panels that the behavior 
of the abundances in the $\xi\rightarrow \infty$ limit depends on the relationship between 
$\eta$ and $w$.  For $\eta < 2w$, as illustrated in the left and middle panels, 
$\barOmega_\gamma\rightarrow 0$ and the triple stasis reduces to a pairwise stasis 
involving vacuum energy and matter alone in both the exponential-decay treatment and 
the instantaneous-decay approximation.  By contrast, for $\eta > 2w$, as illustrated in 
the right panel, $\barOmega_\Lambda\rightarrow 0$ and the triple stasis reduces to a 
pairwise stasis involving matter and radiation alone in both the exponential-decay model and 
the instantaneous-decay approximation.  In the opposite limit, as $\xi$ becomes small,
the exponential-decay treatment becomes increasingly unreliable, since the direct transfer of 
vacuum-energy to radiation, which we are neglecting, becomes important when $\xi < \xi_{\rm min}$.
However, the instantaneous-decay approximation is formally valid for all $\xi < 1$.  In this 
approximation, $\barOmega_M\rightarrow 0$ in the $\xi\rightarrow 1$ limit irrespective of 
the relationship between $\eta$ and $w$.  Thus, in this limit, the triple stasis reduces to a 
pairwise stasis involving vacuum energy and radiation.

It is also instructive to ``look under the hood'' and examine how these triple stasis configurations are realized in terms of the behaviors of the abundances $\Omega_\ell(t)$ of the individual constituents within the towers.
In Fig.~\ref{fig:underhood}, we have chosen a triple stasis corresponding to the input parameters $(\alpha,\gamma,\delta,w)=(1,1,5,-0.7)$
and plotted the total abundances $\Omega_\Lambda$ (thick dark-blue solid line), $\Omega_M$ (thick cyan solid line), and $\Omega_\gamma$ (thick magenta solid line) as functions of the number $\calN$ of $e$-folds since the initial production time.  We have also plotted the individual abundances $\Omega_\ell(t)$ corresponding to the uppermost states within the tower that give rise to this stasis over the time interval shown.  These latter abundances are shown  as thin lines as they transition between vacuum energy (blue), then matter (cyan), and ultimately radiation (magenta).   The underdamping transition between the vacuum-energy and matter phases occurs along the upper dashed black line, while the instantaneous-decay transition between the matter and radiation phases occurs along the lower dashed black line.

This figure can be viewed as an explicit numerical realization of the left panel of Fig.~\ref{fig:ffig5} except that it is plotted for individual {\it abundances}\/ $\Omega_\ell(t)$ rather than individual {\it energy densities}\/ $\rho_\ell(t)$.
Indeed, for each line this change from energy density to abundance introduces an extra overall factor of $H^2$ (which scales as $t^2$ during stasis), thereby tilting all slopes upward relative to those shown in Fig.~\ref{fig:ffig5}. 
However, this figure also includes  the initial transient behavior of our system as it evolves from the production time $t^{(0)}$ into stasis.
 The red transition lines in Fig.~\ref{fig:ffig5} correspond to the dashed black lines here, asymptotically becoming parallel once stasis is reached.
 This figure can also be viewed as a triple-stasis analogue of the left panel of Fig.~\ref{fig:review}.

As we see from Fig.~\ref{fig:underhood}, each abundance begins immediately after production with a horizontal slope, as appropriate for a highly vacuum-energy-dominated universe.  However, these curves all begin to curve upwards as increasing amounts of matter are produced near the top of the tower, thereby affecting the Hubble parameter for the overall cosmology.  However, this behavior is part of the overall transition to stasis.   Indeed, we see that within several $e$-folds our abundances begin to exhibit the ``cross-hatched'' behavior that is the hallmark of stasis, with the identity of the most abundant state continually shifting down the tower as time evolves.   The fact that these abundances ``reflect'' off the underdamping transition line rather than the effective decay line (not shown) allows us to identify this figure as the analogue of the left panel, rather than the right panel, within Fig.~\ref{fig:ffig5}.   This is consistent with the fact that Fig.~\ref{fig:ffig5} was evaluated numerically for parameters corresponding to $\barkappa=8/3$, whereupon we  see that indeed $\barkappa>\alpha$.

\FloatBarrier
\subsection{More about $h$-factors\label{subsect:h-section}}

At long last, we are now in a position 
to circle back and address one of the assumptions 
 with which we started in
Sect.~\ref{subsect:overall_scaling_constraints}, namely our assumption 
discussed below Eq.~(\ref{trip-solns})
that $t_\ast < t_\ell$ for all $\ell$, where $t_\ast$ is a fiducial time during stasis.  
Strictly speaking, such an assumption cannot hold throughout our tower:  since $t_\ell$ gets smaller and smaller as we proceed up the tower, we must inevitably reach a point at which $t_\ell$ becomes less than $t_\ast$.   The states above this point therefore violate our assumption.  Yet this assumption has been made at many points throughout this paper --- not only below Eq.~(\ref{trip-solns}), but also in Sect.~\ref{sec:LambdaMatter} [see below Eq.~(\ref{Omegal})]  and implicitly  in Sects.~\ref{sec:MatterGamma}
and \ref{sec:LambdaGamma}.~  Although this issue has been relevant in each of these earlier cases, it is within the case of triple stasis that this issue becomes the most critical.  It is for this reason that we have deferred this discussion until now.

There is another related concern that might also seem to cast doubt on our previous analyses.  If we were to go back to Eq.~(\ref{trip-solns}) and describe the time-evolution of the individual constituent energy densities $\rho_\ell^{(\Lambda,M,\gamma)}$ since the initial {\it production}\/ time $t^{(0)}$ --- as needed in order to make contact with the scaling relations in Eq.~(\ref{MGscalings}) that define our BSM model ---
we might attempt to follow our results from previous sections such as those in Eq.~(\ref{MGOmegalnew}) and write these energy densities in terms of an appropriate $h(t^{(0)},t_\ast)$ function which describes the net ({\it a priori}\/ unknown) gravitational redshifting that occurs between $t^{(0)}$ and $t_\ast$.
However, as we proceed toward the heavier states in the tower, both $t_\ell$ and $\tau_\ell$ become increasingly small.   As a result, it is possible to reach a point at which  the relative ordering between $t_\ast$ and  $t_\ell$,  and potentially even  between $t_\ast$ and $\tau_\ell$, will change.  However, if either $t_\ell$ or $\tau_\ell$ becomes smaller than $t_\ast$, then the corresponding field will already begin to behave as matter or radiation before reaching stasis.   This means that the 
time-evolution of such a field will be {\it different}\/ than it would have been for fields with $t_\ell$ larger than $t_\ast$, since the latter fields will behave as vacuum energy all the way until reaching stasis.
It is therefore possible  that a single undetermined $h$-function may not be sufficient to describe all the states in our tower prior to reaching stasis.  

This argument can also be phrased in terms of the {\it total}\/ abundances $\Omega_{\Lambda,M,\gamma}$.
Within our  pairwise stasis analyses only one $h$-function  
was needed because there was only one independent total abundance whose behavior we needed to describe between $t^{(0)}$ and $t_\ast$.  Indeed, the second abundance was not an independent degree of freedom because the sum of our two abundances was fixed at $1$.   However, for a triple stasis, we now have {\it three}\/ total abundances which must sum to $1$, implying that there could be {\it two}\/ independent $h$-functions describing our total abundances.  

The presence of multiple  $h$-functions is exceedingly dangerous for the analysis we have been performing.  In this paper we have been deriving not only overall scaling constraints and  log-avoidance constraints, but also relative prefactor constraints.  For example, in the case of triple stasis, such relative prefactor constraints are none other than the relations in  Eq.~(\ref{firsttwopumps})
which express our pumps directly in terms of our abundances, thereby bypassing all of the leading coefficients that are common to the pumps and the abundances.   When there is only a single $h$-function, 
it cancels from both sides of these {\it relative}\/ prefactor constraints.   
This unknown $h$-function will therefore not affect  the nature of such constraints. 
However, with multiple $h$-functions, it is possible that different $h$-functions will appear on each side of our relative prefactor constraints.   Such $h$-functions would no longer cancel, thereby bringing our previous results into doubt.

 These are all serious concerns.
 However, we will now demonstrate that none of these worries are ultimately realized.   In particular, we will explain why we can indeed assume that $t_\ast < t_\ell$ in our analysis, and why only
 a single $h$-function is relevant for calculations of stasis quantities such as the stasis abundances $\barOmega_i$.
It therefore follows that all of our previous results remain valid.

To understand why only a single $h$-function is relevant --- and likewise to understand why we may assume $t_\ast < t_\ell$ in our analysis --- 
let us go back to 
Eq.~(\ref{trip-solns}) and attempt to write our different energy densities $\rho_\ell(t)$ in terms of the initial energy densities $\rho_\ell^{(\Lambda,M,\gamma)}(t)$ at the production time $t=t^{(0)}$.    We know, of course, that $t_\ell<\tau_\ell$, and likewise we know that vacuum energy, matter, and radiation respectively correspond to $t<t_\ell$, $t_\ell < t < \tau_\ell$, and $t> \tau_\ell$.   We also seek to be describing these energy densities during stasis, and therefore we know $t> t_\ast$. However, since different parts of the tower will have different relative orderings of $t_\ast$, $t_\ell$, and $\tau_\ell$, we will make no assumption regarding this ordering.
We then find that there are only six different potential orderings of the relevant timescales in our model:
\beqn
     \Lambda:~~~~~    &&   t^{(0)} < t_\ast < t      < t_\ell < \tau_\ell \nonumber\\
          M1:~~~~~    &&   t^{(0)} < t_\ast < t_\ell < t      < \tau_\ell \nonumber\\
          M2:~~~~~    &&   t^{(0)} < t_\ell < t_\ast < t      < \tau_\ell \nonumber\\
    \gamma 1:~~~~~    &&   t^{(0)} < t_\ast < t_\ell < \tau_\ell < t  \nonumber\\
    \gamma 2:~~~~~    &&   t^{(0)} < t_\ell < t_\ast < \tau_\ell < t  \nonumber\\
    \gamma 3:~~~~~    &&   t^{(0)} < t_\ell < \tau_\ell < t_\ast < t ~.~~~
\label{diffcases}
\eeqn 
  Given these orderings, we can then write down the corresponding energy densities $\rho_\ell$ at time $t$ 
 in terms of their values $\rho_\ell^{(0)}$ at $t^{(0)}$.   Each takes the general form
 \beq
       \rho_\ell(t) ~=~ \rho_\ell^{(0)}\, \cdot \, G_\ell(t)\, \cdot \, {\rm Heaviside}~~
\eeq
where `Heaviside' denotes the specific $\Theta$-function combinations in Eq.~(\ref{trip-solns}) for vacuum energy, matter, and radiation, and where   
$G_\ell(t)$ are the net gravitational redshift factors given by
\beqn
   G_\ell^{(\Lambda)} (t) \,&=&\, h_\Lambda(t^{(0)},t_\ast) 
               \left(  \frac{t}{t_\ast}   \right)^{-(1+w)\barkappa} 
               \nonumber\\
   G_\ell^{(M1)}(t) \,&=&\, h_\Lambda(t^{(0)},t_\ast) 
      \left(  \frac{t_\ell}{t_\ast}   \right)^{-(1+w)\barkappa} 
      \left(  \frac{t}{t_\ell}   \right)^{-\barkappa} 
             \nonumber\\
    G_\ell^{(M2)}(t) \,&=&\, 
         h_\Lambda(t^{(0)},t_\ell) \,
       h_M(t_\ell,t_\ast) 
      \left(  \frac{t}{t_\ast}   \right)^{-\barkappa} 
             \nonumber\\
    G_\ell^{(\gamma 1)}(t) \,&=&\, 
        h_\Lambda(t^{(0)},t_\ast) 
      \left(  \frac{t_\ell}{t_\ast}   \right)^{-(1+w)\barkappa} \!\!
      \left(  \frac{\tau_\ell}{t_\ell}   \right)^{-\barkappa} \!\!
      \left(   \frac{t}{\tau_\ell}\right)^{-4\barkappa/3} 
             \nonumber\\
  G_\ell^{(\gamma 2)}(t) \,&=&\, 
         h_\Lambda(t^{(0)},t_\ell) \,
         h_M      (t_\ell,t_\ast) 
      \left(  \frac{\tau_\ell}{t_\ast}   \right)^{-\barkappa} 
      \left(  \frac{t}{\tau_\ell}   \right)^{-4\barkappa/3} 
             \nonumber\\            
 G_\ell^{(\gamma 3)}(t) &=&~ 
         h_\Lambda(t^{(0)},t_\ell) \,
         h_M      (t_\ell,\tau_\ell)\, 
         h_\gamma      (\tau_\ell,t_\ast) 
      \left(  \frac{t}{t_\ast}   \right)^{-4\barkappa/3}.   
      \nonumber\\
\label{manyhfunctions}
 \eeqn      
Here $h_\Lambda$, $h_M$, and $h_\gamma$ represent the unknown but distinct gravitational redshift factors that are accrued by vacuum energy, matter, and radiation prior to stasis (\ie, prior to $t_\ast$).  
It is the appearance of all three redshift factors in these expressions which is our primary concern.  In general, all three redshift factors will propagate throughout our subsequent calculations, potentially leading to the difficulties discussed above.

It is important to note that all of these difficulties arise only for cases in which $t_\ell > t_\ast$.   If it were possible to impose the constraint that $t_\ast< t_\ell$, then cases $M2$, $\gamma2$, and $\gamma3$ in Eq.~(\ref{diffcases}) would be eliminated, 
and multiple $h$-factors would no longer arise --- even in situations in which vacuum energy, matter, and radiation are all present.    Moreover, the only remaining redshift factor would be $h_\Lambda(t^{(0)}, t_\ast)$, {\it and this is entirely $\ell$-independent}\/.   It would therefore be possible to form a coherent $\ell$-sum, thereby yielding a single $h$-factor for total abundances such as $\Omega_\Lambda$, $\Omega_M$, and $\Omega_\gamma$.

\begin{figure}[t]
\centering
\includegraphics[width=1.0\linewidth]{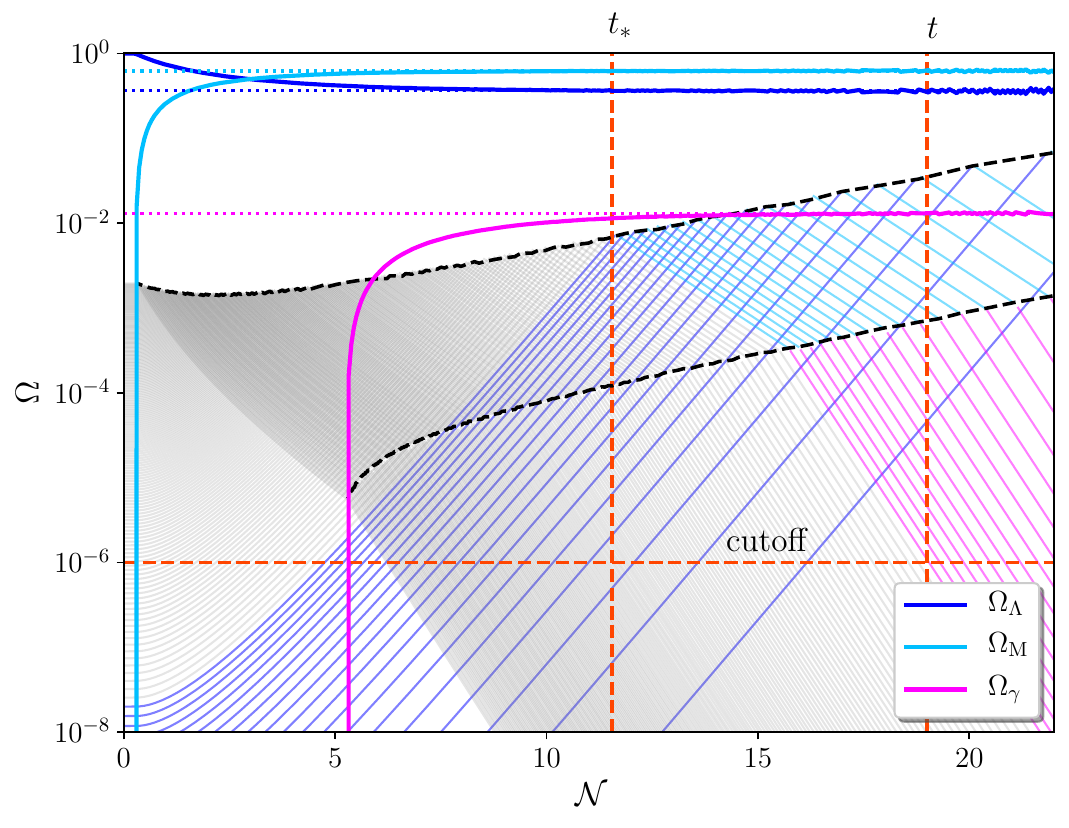}
\caption{A version of Fig.~\protect\ref{fig:underhood} illustrating that at any sufficiently late time $t$ during triple stasis it is possible to choose an earlier fiducial time $t_\ast$ --- also during triple stasis --- such that all states $\phi_\ell$ with significant abundances $\Omega_\ell$ at time $t$ are still overdamped at $t_\ast$.   Thus, when studying the properties of stasis at any sufficiently late time $t$ during triple stasis, it is a legitimate approximation to assume that all relevant states $\phi_\ell$ have $t_\ell > t_\ast$, with all other states from
Fig.~\protect\ref{fig:underhood}
disregarded and hence shown in gray.} 
\label{fig:underhood2}
\end{figure}

The issue, then, boils down to a simple question:   what justifies the assumption that $t_\ast <t_\ell$?    We have already seen that there can exist states at the top of the tower for which this assumption is not true.    What, then, would justify disregarding these states in our analysis?

To analyze this issue, let us return to Fig.~\ref{fig:underhood} and consider the behavior of the individual abundances at some time $t$ deep within stasis.  For example, such a time $t$ is indicated as the right-most vertical orange/red dashed 
line in Fig.~\ref{fig:underhood2}.  At the time $t$, it is clear that only the states with significant abundances can possibly be important players in continuing to produce the stasis phenomenon.   By contrast, states whose abundances at time $t$ have fallen below some chosen cutoff value can no longer play a significant role in supporting the stasis at time $t$.   We have indicated such a cutoff as corresponding to the horizontal orange/red line in Fig.~\ref{fig:underhood2}, and 
states whose abundances at time $t$ have fallen below this cutoff have been colored in gray.    Such states simply do not matter for the purposes of analyzing the stasis phenomenon at time $t$.   However, tracing the remaining (colored) states backwards in Fig.~\ref{fig:underhood2}, we see that there is a lower limit to the times $t_\ell$ at which such states all became underdamped.    
Thus, so long as this lower limit is still within the stasis epoch, we can identify this lower limit as $t_\ast$.    We thus have a situation in which $t_\ast< t_\ell$ for all of the states which will ultimately play a role in supporting stasis at time $t$ --- \ie, the states whose abundances exceed a critical cutoff value at time $t$.  Indeed, the deeper into the stasis epoch we go (\ie, the larger $t$ becomes), the larger the corresponding fiducial time $t_\ast$ can be for which we may safely assume $t_\ast< t_\ell$ for all relevant states.

This argument can also be understood in the forward direction, starting with our entire tower at the production time.  As we have discussed above, it is generally the heavier states within the tower which are most in danger of becoming underdamped (or potentially even decaying) prior to our system entering stasis, thereby violating our assumption that $t_\ast<t_\ell$.  However, for sufficiently late times $t$ during stasis, we need no longer consider the contributions from such heavier states, since their abundances at time $t$ will have dropped below a critical relevance cutoff.  In other words, such states can be viewed as ``turning gray'' within the conventions of Fig.~\ref{fig:underhood2}, with increasingly many states becoming gray as the time $t$ evolves.
We can then safely ignore the contributions from such states when discussing analyzing stasis at time $t$.  Since stasis persists over many $e$-folds, we are free to choose $t$ sufficiently large so that $t_\ast$ --- which is smaller than all of the relevant $t_\ell$ --- is itself within stasis.   We can then freely assume that $t_\ast<t_\ell$ for all relevant $\ell$.

We have already noted at numerous points in this paper
that one critical property of stasis is that the identity of the most-abundant state at any given time $t$ itself progresses down the tower as  $t$ increases.  Each state ``gets its day'' supporting the stasis before yielding its dominant role to the next state in the tower and subsequently diminishing into irrelevance as time proceeds.   In this sense, old age does indeed each generation waste.    But this is how the stasis state continues to be supported, extended across a cold pastoral landscape of many $e$-folds, with the abundances $\barOmega_{\Lambda,M,\gamma}$ remaining fixed as if carved in marble.

Of course, as evident in Figs.~\ref{fig:underhood} and \ref{fig:underhood2}, the heaviest states in the tower play a critical role in shaping the initial {\it transient}\/ behavior that occurs between $t^{(0)}$ and $t_\ast$.  These states thus play an important in role in guiding the system towards stasis.  However, as we have seen, these states cease to play a role within the stasis itself.
The same interpretation may be given to the different $h$-functions in Eq.~(\ref{manyhfunctions}).  
Clearly they are relevant for describing the initial transient behavior of our system.  However, the further into stasis our system gets, the smaller the influence of the initial behavior encapsulated within $h_\gamma$ and then $h_M$. Ultimately we reach a time $t$ beyond which only the single $h$-function $h_\Lambda$ remains to play a role.
Thus only $h_\Lambda$ is ultimately needed for describing the stasis state, as discussed above.   Indeed, taking $t_\ast<t_\ell$ and the ability to eliminate $h_M$ and $h_\gamma$ from our stasis calculations go hand in hand.

We thus conclude that it is legitimate to assume $t_\ast<t_\ell$ for all relevant $\ell$, whereupon we may disregard all but the single $h$-function $h_\Lambda(t^{(0)},t_\ast)$.   Moreover, this single $h$-function will cancel from all relative prefactor constraints.  We thus conclude that all of the triple-stasis results we have derived thus far remain valid.

Having discussed the $h$-factors that are relevant for individual abundances $\Omega_\ell$, let us turn to the $h$-factors that are relevant for the {\it total}\/ abundances $\Omega_\Lambda$, $\Omega_M$, and $\Omega_\gamma$ during stasis.  However, as we have seen,
the fact that we can now restrict our contributing states to those satisfying $t_\ast< t_\ell$ during stasis means that our individual abundances during stasis depend only on $h_\Lambda(t^{(0)},t_\ast)$, and this quantity is $\ell$-independent.   This factor can therefore be pulled out of any $\ell$-sum, whereupon we see that the corresponding {\it total}\/ abundances during stasis will also share this same $h$-factor.
We thus see that the $h(t^{(0)},t_\ast)$ function that has appeared throughout this section for our abundances and pumps during stasis is nothing but $h_\Lambda(t^{(0)},t_\ast)$.

That said, it is also interesting to consider the manner in which our total abundances transition from their initial values at  $t^{(0)}$ towards their ultimate stasis values.
Of course, throughout this paper we have described the {\it net}\/ effects of this pre-stasis process on our total abundances through a single $h$-factor which we have denoted $h(t^{(0)},t_\ast)$.  By extending from $t^{(0)}$ all the way to $t_\ast$, this  factor in principle encapsulates the all of this initial transient behavior.  However, we have not examined the actual time-dependence associated with this transient behavior --- \ie, the time-dependence of our abundances $\Omega_{\Lambda,M,\gamma}(t)$ {\it as they evolve}\/  towards their stasis values.  Indeed,
this evolution is of interest since it corresponds to the initial curvatures in the plots of the total abundances in Figs.~\ref{fig:review}, \ref{fig:vac_M}, \ref{fig:vac_gamma}, \ref{fig:BranchA}, and \ref{fig:underhood}.

Ultimately, this behavior can also be written in terms of the $h_\Lambda$,
$h_M$, and $h_\gamma$ functions we have discussed above.  For this analysis we shall content ourselves with understanding the basic algebraic structural elements associated with this behavior. Once again evaluating $\Omega_{\Lambda,M,\gamma}(t)$ as functions of $t$ --- but now {\it without}\/ the assumption of stasis  ---   we obtain
\beqn
\Omega_\Lambda(t) \,&=&\,\int d\tau \, 
    n_\tau(\tau)\, \Omega_0^{(0)}
     \left( \Gamma_0 \tau\right)^{-1/\gamma}\,
       h_\Lambda(t^{(0)},t)
       \nonumber\\
\Omega_M(t) \,&=&\, \int d\tau \, 
    n_\tau(\tau)\, \Omega_0^{(0)}
     \left( \Gamma_0 \tau\right)^{-1/\gamma}\,
       h_\Lambda \!\left( t^{(0)}, f(\tau)\right)
       \,\nonumber\\
&& ~~~~~~~~\times\,
     h_M\! \left( 
     f(\tau), t\right)\nonumber\\
\Omega_\gamma(t) \,&=&\, \int d\tau \, 
    n_\tau(\tau)\, \Omega_0^{(0)}
     \left( \Gamma_0 \tau\right)^{-1/\gamma}\,
       h_\Lambda\!\left( t^{(0)},
       f(\tau) \right)\,\nonumber\\
&& ~~~~~~~~\times\,
     h_M\!\left( 
     f(\tau), \tau\right)
     \,h_\gamma(\tau,t)~.
\label{approachingstasis}
\eeqn
where these $\tau$-integrals have limits that follow the general structure
\beq
{\Lambda:}~\int_{f^{-1}(t)}^{\tau_0} ~,~~~~~
{M:}~\int_{t}^{f^{-1}(t)} ~,~~~~~
{\gamma:}~\int^{t}_{\tau_{N-1}}~.~~~
\label{eq:non-stasis-limits}
\eeq
In these equations, $f(\tau)$ denotes the underdamping time 
[\ie, the time $\hat t$ at which $3H(\hat t)=2m$]
for that part of the tower which decays at time $\tau$, while $f^{-1}(t)$ denotes the inverse of this function. 
 If we were in a stasis configuration, 
 we would have $H(t)= \barkappa/(3t)$,
 whereupon we  would identify
 \beq
  f(\tau)=\frac{ (\Gamma_0 \tau)^{1/\gamma}}{\xi\Gamma_0}~~~\Longleftrightarrow~~~
  f^{-1}(t) =
    \frac{ (\xi \Gamma_0 t)^{\gamma}}{\Gamma_0}~.~~
\label{idents}
\eeq
The limits in Eq.~(\ref{eq:non-stasis-limits}) would then be nothing other than those in Eq.~(\ref{eq:int_limits}).   However, since we are not assuming stasis for this calculation, we can no longer assert that $H(t)=\barkappa/(3t)$.  Thus the identifications in Eq.~(\ref{idents}) will no longer hold.   The function $f(\tau)$ will nevertheless depend on the values of the abundances $\Omega_\Lambda(t)$ themselves, thereby creating a highly non-linear system of equations.   

The overall algebraic structure indicated in Eqs.~(\ref{approachingstasis}) and (\ref{eq:non-stasis-limits}) is valid for all times $t$.
Indeed, given this general structure, 
we thus see that it is the integrals of products of our $h_{\Lambda,M,\gamma}$ functions 
which are responsible for producing the initial curvatures for the total abundances plotted in our figures.   

One feature which is worthy of note from this structure is that the $h_\Lambda$, $h_M$, and $h_\gamma$ functions are not independent of each other.   Instead, they must be related in such  a way that $\Omega_\Lambda(t)+ \Omega_M(t) + \Omega_\gamma(t) =1$ for all $t$. 
Of course, the overall value of this sum of abundances is determined by the choice of $\Omega_0^{(0)}$ at the initial time.   However, what is remarkable is that once this value is set, this sum of abundances remains fixed as a function of $t$.   

In order to see
how this feat is ultimately accomplished, we note that the variable $t$ appears within Eq.~(\ref{approachingstasis}) in only two sets of locations.  The first set of locations consists of those within the limits of integration that determine the boundaries between the integration domains for $\Omega_\Lambda$ and $\Omega_M$,  and between $\Omega_M$ and $\Omega_\gamma$, as indicated in Eq.~(\ref{eq:non-stasis-limits}).   By contrast, the second set of locations consists of those 
within the final $h$-factors that appear within the integrands for each of the three total abundances in Eq.~(\ref{approachingstasis}).
The appearance of $t$ within the first set of locations tells us that $t$ determines the {\it partition}\/ of the total $\tau$-range $\tau_{N-1}\leq \tau\leq \tau_0$ into the individual ranges corresponding to $\Lambda$, $M$, and $\gamma$.
Indeed, as $t$ increases, the range of $\tau$ corresponding to $\Lambda$ decreases, while that corresponding to $\gamma$ increases.  This is consistent with Fig.~\ref{fig:triple_stasis_in_action}.   Thus, different portions of the $\tau$-range pass directly from one type of energy component to another.
However, this passing of abundance between the individual components $\Omega_i(t)$ does not affect the {\it total}\/ abundance
$\sum_i\Omega_i(t)$.    By contrast, it is the appearance of $t$ within the second set of locations --- \ie, within the different {\it integrands}\/ within Eq.~(\ref{approachingstasis}), each with its own unique time-dependence ---
that can potentially affect the value of $\sum_i\Omega_i$.    
Indeed, since the time $t$ appears within $h_\Lambda$ function for $\Omega_\Lambda(t)$, within the $h_M$ function for $\Omega_M(t)$, and within the $h_\gamma$ function for $\Omega\gamma(t)$, 
the cancellation of these time dependences for the sum $\sum_i\Omega_i(t)$ implies a relation between these three $h$-functions.

These assertions can be made mathematically explicit by considering the time-derivatives of the total abundances in Eq.~(\ref{approachingstasis}). 
Considering $d\Omega_\gamma(t)/dt$ first, we find
\beqn
\frac{d\Omega_\gamma(t)}{dt} ~&=&~  
  n_\tau(t) \,
\Omega_0^{(0)}\,
     \left( \Gamma_0 t \right)^{-1/\gamma}
       h_\Lambda\!\left( t^{(0)},f(t)\right)\,\nonumber\\
&& ~~~~~~~~\times\,
     h_M\!\left( f(t), t \right)
     \,h_\gamma(t ,t)~\nonumber\\
&& ~+ \int_{\tau_{N-1}}^t  d\tau \, 
    n_\tau(\tau)\,
     \Omega_0^{(0)} \left( \Gamma_0 \tau\right)^{-1/\gamma} \nonumber\\
&& ~~~~~~~~\times\,
       h_\Lambda\! \left( t^{(0)},f(\tau)\right)\,\nonumber\\
&& ~~~~~~~~\times\,
     h_M \!\left( 
     f(\tau), \tau\right)
     \,\frac{d}{dt} h_\gamma(\tau,t)~,~~~~\nonumber\\
\label{firstwriteofderivative} 
\eeqn
where the first term [top two lines of Eq.~(\ref{firstwriteofderivative})] comes from differentiating the factor of $t$ within the integration limit, while the second term [remaining three lines of Eq.~(\ref{firstwriteofderivative})] comes from differentiating the factor of $t$ in the integrand.    Note that $h_\gamma(t,t)=1$ in the second line of Eq.~(\ref{firstwriteofderivative}).   As a result, $h_\gamma$ completely disappears from the first term of Eq.~(\ref{firstwriteofderivative}) --- a feature which allows us to identify the remaining $h$-factor structure as appropriate for a matter abundance rather than a radiation abundance.  
Indeed, taken together, this first term is nothing but $n_\tau(t) \Omega_\tau^{(M)} (t;t)$, and this product in turn is nothing but our pump $P_{M\gamma}$ in Eq.~(\ref{pumpings2}).
Eq.~(\ref{firstwriteofderivative}) thus takes the relatively simple form
\beq 
\boxed{
   ~\frac{d\Omega_\gamma(t)}{dt} ~=~
P_{M\gamma}(t) + \int_{\tau_{N-1}}^t d\tau\, n_\tau(\tau)\,\frac{d}{dt} \Omega_\tau^{(\gamma)}(\tau;t)~.~
}
\label{eq:pump3}
\eeq
This result of course makes perfect intuitive sense, asserting that the total rate of change for the total radiation abundance $\Omega_\gamma$ has two contributions:   one from the abundance being pumped into radiation through  $\phi_\ell$ decays, and the second from the natural Hubble scaling associated with this abundance itself.
{\it We stress that this result holds in complete generality, and does not assume a stasis of any sort.}   We also remark that it is not surprising that one of our pumps has made an appearance in this calculation.   Eq.~(\ref{approachingstasis}) essentially represents the integral form for our analysis, while taking the time-derivative has thrown us into the differential form in which our pumps make an appearance.   As we see, this remains true even if we are not in a stasis epoch.

Proceeding similarly for $d\Omega_\Lambda(t)/dt$, we find
\beqn
 \frac{d\Omega_\Lambda(t)}{dt} ~&=&~
    -   \frac{df^{-1}(t)}{dt}\,
           n_\tau\!\!\left( f^{-1}(t)\right)\, \Omega_0^{(0)}
    \nonumber\\
    && ~~~~~~~~ \times ~
              \left[ \Gamma_0 f^{-1}(t)\right]^{-1/\gamma }
              h_\Lambda\left( t^{(0)},t \right)~\nonumber\\
&&~+ \int_{f^{-1}(t)}^{\tau_0} d\tau \, n_\tau(\tau) \,
     \Omega_0^{(0)} \, (\Gamma_0 \tau)^{-1/\gamma} 
     \nonumber\\
     && ~~~~~~~~ \times ~  \frac{d}{dt} h_\Lambda(t^{(0)},t).
\label{intermedstep}
\eeqn
However, as discussed below Eq.~(\ref{eq:non-stasis-limits}), $f(\tau)$ is essentially $\hat t$ (the continuous variable associated with underdamping times rather than decay times).   We can then identify
\beq
 \frac{df^{-1}(t)}{dt} \,
           n_\tau~=~
           n_{\hat t} 
\eeq
whereupon we see 
that the first term in Eq.~(\ref{intermedstep}) is nothing but the product of two factors:
\begin{itemize}
    \item the density of states per unit $\hat t$ evaluated for the part of the tower with $\tau$-value $f^{-1}(t)$;   and
    \item  the vacuum-energy abundance evaluated at the time $t$ for the part of the tower with $\tau$-value $f^{-1}(t)$.  
\end{itemize}
However, the first of these factors is nothing but the density of states per unit $\hat t$ for that part of the tower with $\hat t$-value $t$, previously denoted $n_{\hat t}(t)$.   Likewise, the second factor is nothing but the vacuum-energy abundance evaluated at the time $t$ for that part of the tower with $\hat t$-value $t$, previously denoted $\Omega_{\hat t}^{(\Lambda)}(t;t)$. 
Upon  comparison with Eq.~(\ref{pumpings2}), we then see that the product of these two factors is nothing but the pump $P_{\Lambda M} (t)$.
We thus have
\begin{empheq}[box=\fbox]{align}
 ~  \frac{d\Omega_\Lambda(t)}{dt} ~&=~
-P_{\Lambda M}(t) \nonumber\\
&~+~
\int_{f^{-1}(t)}^{\tau_0} d\tau\, n_\tau(\tau)\,\frac{d}{dt} \Omega_\tau^{(\Lambda)}(\tau;t)~.~
\label{eq:pump1}
\end{empheq}

Likewise, for $d\Omega_M(t)/dt$, we have
\begin{empheq}[box=\fbox]{align}
~ \frac{d\Omega_M(t)}{dt} ~&=~
P_{\Lambda M} (t) 
- 
P_{M\gamma}(t) \nonumber\\
& ~+~
\int^{f^{-1}(t)}_{t} d\tau\, n_\tau(\tau)\,\frac{d}{dt} \Omega_\tau^{(M)}(\tau;t)~.~
\label{eq:pump2}
\end{empheq}

Collecting our results in Eqs.~(\ref{eq:pump3}), (\ref{eq:pump1}), and (\ref{eq:pump2}), we thus find
\beqn
\sum_i \frac{d\Omega_i(t)}{dt} ~&=& ~
\int_{f^{-1}(t)}^{\tau_0} d\tau\, n_\tau(\tau)\,\frac{d}{dt} \Omega_\tau^{(\Lambda)}(\tau;t)~ 
\nonumber\\
&& ~+~\int^{f^{-1}(t)}_{t} d\tau\, n_\tau(\tau)\,\frac{d}{dt} \Omega_\tau^{(M)}(\tau;t)~~~~
\nonumber\\
&& ~+~ \int_{\tau_{N-1}}^t d\tau\, n_\tau(\tau)\,\frac{d}{dt} \Omega_\tau^{(\gamma)}(\tau;t)~.\nonumber\\
\label{sumchanges}
\eeqn
Interestingly, all of the pump terms have cancelled within this sum.  As anticipated above, this reflects the fact that the pump terms describe the {\it redistributions}\/ of energy density between the different energy components {\it within}\/ our system, but do not affect the total energy density of the system itself.  Only the Hubble redshifting effects can do that.

The statement that $\sum_i \Omega_i(t)$ is a constant then boils down to the constraint that $\sum_i d\Omega_i(t)/dt=0$.   We thus find that the three terms on the right side of Eq.~(\ref{sumchanges}) must cancel directly amongst themselves, \ie,
\beqn
0 ~&=&~ \int_{f^{-1}(t)}^{\tau_0} d\tau\, n_\tau(\tau)\,\frac{d}{dt} \Omega_\tau^{(\Lambda)}(\tau;t)~ 
\nonumber\\
&& ~+~\int^{f^{-1}(t)}_{t} d\tau\, n_\tau(\tau)\,\frac{d}{dt} \Omega_\tau^{(M)}(\tau;t)~~~~
\nonumber\\
&& ~+~ \int_{\tau_{N-1}}^t d\tau\, n_\tau(\tau)\,\frac{d}{dt} \Omega_\tau^{(\gamma)}(\tau;t)~.~~~~
\label{redshiftcancellation}
\eeqn
This is then a constraint on the three $h$-functions which appear within these three integrands and which help to determine the time-derivatives therein.  As noted above, these $h$-functions also help to determine the value of $f^{-1}(\tau)$ which appears within the limits associated with two of these integrals.   This is therefore a complicated system of equations which does not have an obvious analytical solution.   These results are nevertheless the underpinnings of the initial curvatures in the plots of the total abundances in Figs.~\ref{fig:review}, \ref{fig:vac_M}, \ref{fig:vac_gamma}, \ref{fig:BranchA}, and \ref{fig:underhood}.

The terms in Eq.~(\ref{redshiftcancellation}) all reflect the time-dependences that come from 
cosmological redshifting effects.   Their cancellation when summed across all three abundances implies that while one abundance has a positive time-derivative under gravitational redshifting, another must have a negative time-derivative.
Indeed, this feature is readily apparent within the figures previously cited.
That said, we stress that {\it the cancellation of gravitational redshifting factors across all abundances has nothing whatsoever to do with stasis}\/.   Indeed, while Eq.~(\ref{redshiftcancellation}) implies a cancellation between redshift factors across different abundances, stasis is a cancellation between the redshift factor {\it and the corresponding pump}\/ within {\it each}\/ abundance individually.
Indeed, this is what is required in order to have each derivative $d\Omega_\Lambda/dt$,
$d\Omega_M/dt$, and $d\Omega_\gamma/dt$ vanish  independently.

As we have stressed, the results in  Eqs.~(\ref{eq:pump3}), (\ref{eq:pump1}), and (\ref{eq:pump2})
are completely general and make no assumption of stasis.
However, it is easy to see what becomes of these results if we make the further assumption that we are within a stasis epoch.   Within a stasis epoch, we know that the vacuum-energy, matter, and radiation abundances have gravitational redshift factors that scale as
\beqn
  h_\Lambda(t_\ast,t) ~&\sim&~
  \left(\frac{t}{t_\ast}\right)^{ 2 - (1+w)\barkappa}
  \nonumber\\
  h_M(t_\ast,t) ~&\sim&~
  \left(\frac{t}{t_\ast}\right)^{ 2 - \barkappa}
  \nonumber\\
  h_\Lambda(t_\ast,t) ~&\sim&~
  \left(\frac{t}{t_\ast}\right)^{ 2 - 4\barkappa/3}~.~~~~
\eeqn
where $t_\ast$, as always, is any time within stasis prior to $t$. 
Since these $h$-functions are the only places that carry an explicit $t$-dependence within the abundances $\Omega_\tau^{(\Lambda,M,\gamma)}(\tau;t)$ that appear within the integrands of the $\tau$-integrals on the right sides of 
Eqs.~(\ref{eq:pump3}), (\ref{eq:pump1}), and (\ref{eq:pump2}), we immediately find upon taking the $t$-derivatives of these expressions 
that
\beqn 
  \frac{d}{dt} \Omega_\tau^{(\Lambda)}(\tau;t)
  ~&=&~ \Bigl[2-(1+w)\barkappa\Bigr] \, \Omega_\tau^{(\Lambda)} (\tau;t)
  \, \frac{1}{t} \nonumber\\
  \frac{d}{dt} \Omega_\tau^{(M)}(\tau;t)
  ~&=&~ \Bigl[2-\barkappa\Bigr] \, \Omega_\tau^{(M)} (\tau;t)
  \, \frac{1}{t} \nonumber\\
\frac{d}{dt} \Omega_\tau^{(\gamma)}(\tau;t)
  ~&=&~ \Bigl[2-4\barkappa/3\Bigr] \, \Omega_\tau^{(\gamma)} (\tau;t)
  \, \frac{1}{t}~.~~~~~~~
\eeqn
Performing the $\tau$-integrals in Eqs.~(\ref{eq:pump3}), (\ref{eq:pump1}), and (\ref{eq:pump2}) and then setting 
$d \Omega_i(t)/dt=0$ for each $i=\Lambda,M,\gamma$ then yields precisely the pump equations in Eq.~(\ref{eq:alstruc})
with which we started, thereby providing a critical cross-check on our results.  Indeed,  Eqs.~(\ref{eq:pump3}), (\ref{eq:pump1}), and (\ref{eq:pump2}) may be taken as the more general underpinning behind Eq.~(\ref{eq:alstruc}), one which does not assume stasis but which yields Eq.~(\ref{eq:alstruc}) as a special case. 
On the other hand, our results within 
Eqs.~(\ref{eq:pump3}), (\ref{eq:pump1}), and (\ref{eq:pump2})
assume an instantaneous decay from matter to radiation (and likewise assume that $P_{\Lambda \gamma}=0$), while the result in Eq.~(\ref{eq:alstruc}) is more general and in principle also allows for the possibility of a direct energy transfer from vacuum energy to radiation.

\FloatBarrier
\subsection{Pumps, seesaws, and energy flows}
\label{subsect:seesaw}

Finally, in order to understand these triple-stasis solutions along Branch~A more 
intuitively, let us consider the flows of energy density that they imply.

We shall begin by analyzing the case with $\barkappa=2$, since this case 
turns out to have the greatest symmetry and simplicity.  With 
$\barkappa=2$, Eq.~(\ref{constw1}) reduces to the constraint $\eta = -2w$, which implies that $\eta<2$.   Since $\delta>0$, this means we must have $\alpha<2$, and thus
$\alpha<\barkappa$.
   This then corresponds to the left panel of 
Fig.~\ref{fig:ffig5}, but not the right panel.

Given the result in Eq.~(\ref{eq:triplekappa2}), we learn that the 
constraint $\barkappa=2$ restricts our corresponding stasis abundances 
to lie along a {\it line}\/ of solutions for which
\beq 
  \barOmega_\Lambda ~=~ -\frac{1}{3w} ~\barOmega_\gamma~
\label{abundanceseesaw}
\eeq
with $\barOmega_M= 1-\barOmega_\Lambda-\barOmega_\gamma$.
Indeed, it is only the specific choices of $(\alpha, \delta, w, \xi)$ 
that distinguish between the different abundance solutions along this 
line.  For example, as $w\to -1$, a particularly symmetric distribution 
of abundances satisfying these constraints is 
$(\barOmega_\Lambda,\barOmega_M,\barOmega_\gamma)=(1/6,1/3,1/2)$.  Indeed,
this is the three-component analogue of matter/radiation equality in the 
sense that $\Omega_M$ now carries $1/3$  of the total abundance while 
the other two abundances collectively carry $2/3$.

Finally, given the constraints in Eqs.~(\ref{eq:alstruc}), we see that our 
pump terms $P_{\Lambda M}$ and $P_{M\gamma}$ for $\barkappa=2$ become 
equal during triple stasis and are given by
\beq
 P_{\Lambda M} ~=~ 
 P_{M\gamma}~=~ -2 w\, \barOmega_\Lambda \,\frac{1}{t}~=~ \frac{2}{3} \,\barOmega_\gamma  \frac{1}{t}~.
 \label{eq:pumpreduce}
 \eeq

\begin{figure*}
\centering
\includegraphics[keepaspectratio, width=0.95\textwidth]{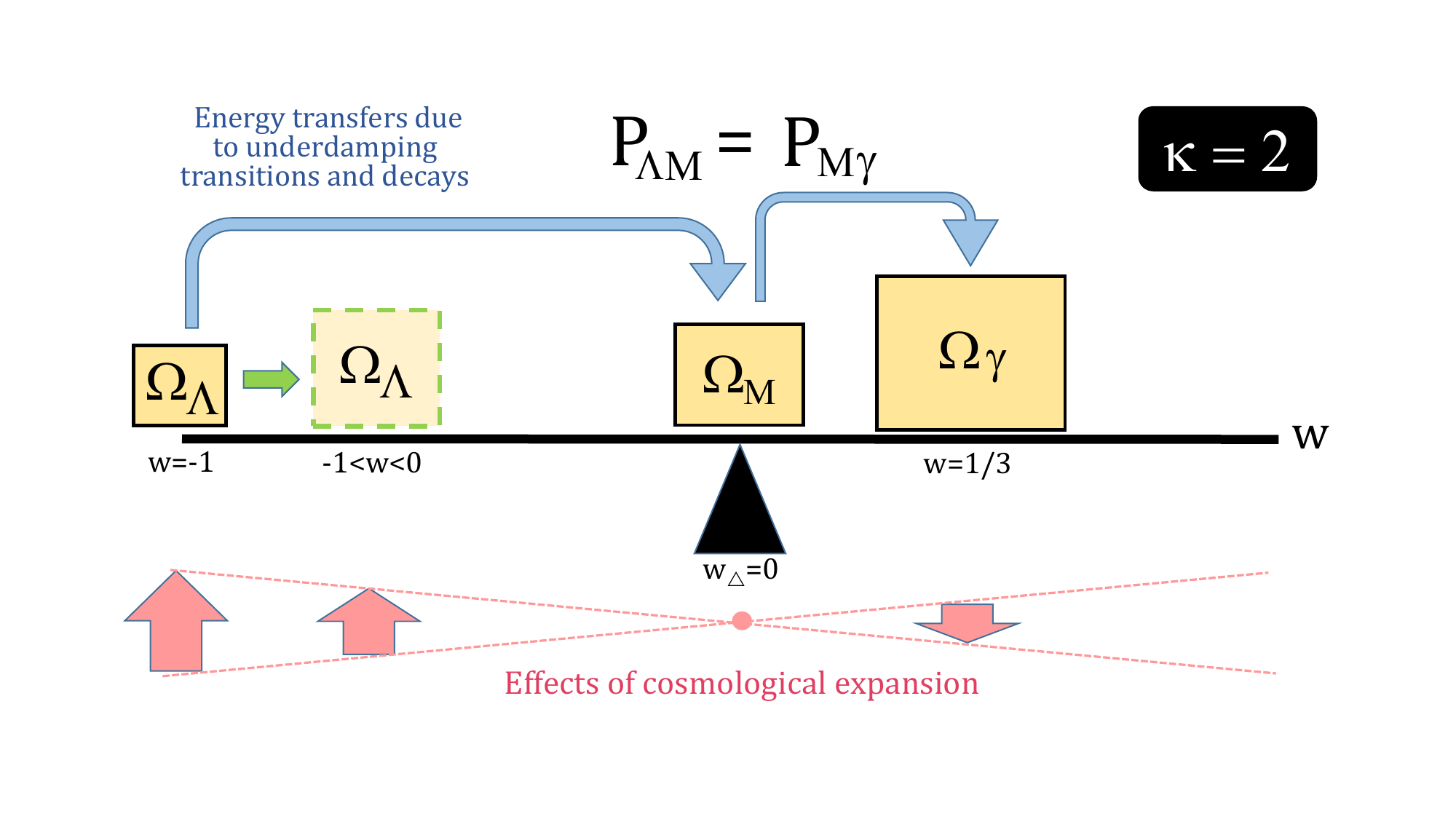}
\vskip -0.5 truein 
\caption{The ``seesaw'' structure of triple stasis with $\barkappa=2$ along Branch~A in which the 
abundances of the different energy components along the $w$-axis are balanced around a ``fulcrum'' 
located at $\langle w\rangle=w_\triangle=0$. The location of the fulcrum corresponds to a $\barkappa=2$ 
universe which is effectively matter-dominated.    The pumps $P_{\Lambda M}$ and $P_{M\gamma}$ (shown 
schematically in blue) transfer energy from vacuum energy to matter and from matter to radiation, 
respectively.   These pumps  exactly compensate for the effects of cosmological expansion wherein the 
effects due to gravitational redshifting (shown schematically in pink) either increase or decrease the 
relative abundances of these different components according to their distances from the fulcrum at 
$w_\triangle=0$.   Changing the value of $w$ for the vacuum energy component (shown schematically in green) 
corresponds to ``sliding'' the position of the corresponding abundance along the $w$-seesaw.   This abundance 
is then rescaled according to Eq.~(\ref{abundanceseesaw}) in order to maintain the balancing of the 
$w$-seesaw.   However, this increased abundance experiences a weakened gravitational redshifting  and 
thus the delicate balancing inherent in triple stasis is maintained.} 
\label{fig:seesaw}
\end{figure*}

It turns out that this $\barkappa=2$ stasis solution can be visualized 
in a particularly useful and  compelling way.
The $\barkappa=2$ constraint means that the corresponding triple-stasis universe must be effectively matter dominated, \ie, with an effective abundance-weighted equation-of-state 
parameter
\beq
  \langle w\rangle ~\equiv~ \sum_{i=\Lambda,M,\gamma}  w_i \Omega_i ~=~ 0~
  \label{seesawbalancezero}
\eeq
where $w_\Lambda = w$, $w_M=0$, and $w_\gamma=1/3$.
Indeed, this constraint is nothing but the result in Eq.~(\ref{abundanceseesaw}).
However, we may also view Eq.~(\ref{seesawbalancezero}) as the condition for a ``balancing'' along a 
$w$-{\it seesaw}\/, as illustrated in Fig.~\ref{fig:seesaw}, with the fulcrum position 
$\langle w\rangle = w_\triangle$ identified as $w_\triangle=0$.    
We emphasize in this context that having $\barkappa=2$ does not mean that all of the abundance is in the 
matter component --- it just means that the radiation abundance must be three times (or more precisely 
$-3w$ times) the vacuum-energy abundance, so that the seesaw balances.  Whatever abundance is left over 
is thus the matter abundance.   Of course, this matter abundance sits immediately above the fulcrum, so 
any common rescaling of the vacuum-energy and radiation abundances continues to maintain the seesaw balance. 
This then provides a seesaw-based explanation for the existence of a line of triple-stasis solutions, all 
of which are balanced around the fulcrum at $w_\triangle=0$ with the individual values of 
$(\alpha,\delta,w,\xi)$ selecting between them.

In this connection, we recall from the end of Sect.~\ref{sec:LambdaGamma} that we previously had a case of 
vacuum-energy/radiation abundance which allowed for a spectator matter abundance.  That universe was also 
required to be effectively matter-dominated.  The special case in Sect.~\ref{sec:LambdaGamma} can thus be 
interpreted as a variant of the current triple-stasis phenomenon -- a variant in which the two equal pumps 
$P_{\Lambda M}$ and $P_{M\gamma}$ are merged into a single pump $P_{\Lambda \gamma}$ which bypasses 
$\Omega_M$ completely, thereby rendering $\Omega_M$ a mere spectator abundance.

The manner in which this triple-stasis solution works is thus clear, and is illustrated in 
Fig.~\ref{fig:seesaw}.   Because the universe is matter-dominated, the total matter abundance 
$\Omega_M$ does not feel any tendency to evolve in either direction (rise or fall) under cosmological 
expansion.  Given this, the triple-stasis solution then operates through the following balancing act:

\begin{itemize}
    
\item  {\it Vacuum-energy abundance}\/ $\Omega_\Lambda$:  In a matter-dominated universe 
$\Omega_\Lambda$  wants to {\it rise}\/ due to cosmological expansion (as indicated through the 
pink arrows in Fig.~\ref{fig:seesaw}).  However, the pump $P_{\Lambda M}$ continually drains away 
this excess so that $\Omega_\Lambda$ stays fixed.  

\item {\it Radiation abundance}\/ $\Omega_\gamma$:  Conversely, in a matter-dominated universe the 
radiation abundance $\Omega_\gamma$ wants to {\it fall}\/ due to cosmological expansion.  However, 
here the pump $P_{M\gamma}$ keeps sourcing a fresh supply so that $\Omega_\gamma$ also stays fixed.  

\item {\it Matter abundance}\/ $\Omega_M$:  Finally, it is through the central matter abundance 
$\Omega_M$ that the ``collision'' between these two pairwise pumps occurs.  However, with 
$\barkappa=2$ this collision is non-problematic: the universe is effectively matter-dominated, and 
thus the matter abundance sits directly atop the fulcrum at $w_\triangle=0$.   In this case there is 
no tendency for the matter abundance to rise or fall, which is consistent with the fact that the pumps 
into and out of $\barOmega_M$ exactly cancel.   

\end{itemize}

Note that this is a true triple stasis, with each energy component experiencing interactions with 
others and experiencing sources and/or sinks.  Moreover, it is remarkable that this solution balances 
correctly and that our simple scalar model actually realizes it --- especially given that one of the 
pumps results from the transition from an overdamped to underdamped phase while the other pump 
involves decay, which is a completely different underlying process!  Moreover, even though this 
solution may seem trivially balanced, with equal pumps into and out of the ``central'' matter 
component $\Omega_M$, we must remember that our physical {\it realization}\/ of this triple stasis 
is actually highly non-trivial, with both pumps operating as the associated transitions make their 
way down the $\phi_\ell$ tower (as illustrated in Fig.~\ref{fig:triple_stasis_in_action}).  Finally, 
we did not need to create a contorted model with arbitrary interactions in order to realize this 
stasis --- literally {\it any}\/  coherent state of boson zero modes (such as naturally arise in 
axion physics) will necessarily experience not only a transition from an overdamped to an underdamped 
regime but also an eventual decay.  Moreover, we will naturally obtain a tower of such states if this 
boson is higher-dimensional, with the different $\phi_\ell$ fields identified as different KK modes.    
 
This seesaw picture also enables us 
to understand intuitively why this triple-stasis solution works for all $-1 <w<0$.
Towards this end, let us imagine sliding $\Omega_\Lambda$ along the seesaw from $w=-1$ to $w> -1$,
as illustrated in Fig.~\ref{fig:seesaw}, while keeping $\barOmega_M$ and $\barOmega_\gamma$ fixed.
Under these assumptions, we see from 
Eq.~(\ref{abundanceseesaw}) that 
$\barOmega_\Lambda$
grows by a factor of $w^{-1}$:
\beq 
       \barOmega_\Lambda = 3 \barOmega_\gamma
       ~~~\to~~~
       \barOmega'_\Lambda = -\frac{3}{w} \barOmega_\gamma~.
\eeq
This too  is illustrated in Fig.~\ref{fig:seesaw}.
However, following Eq.~(\ref{eq:pumpreduce}) and given that $\barkappa$ remains fixed at $\barkappa=2$ even as $w$ shifts, we see that the total pump rate $P_{\Lambda M}$ is unchanged:
\beq
  P_{\Lambda M} = 2 \barOmega_\Lambda \frac{1}{t}
   ~~~\to~~~
    P'_{\Lambda M} = -{2w} \barOmega'_\Lambda \frac{1}{t}~=~ P_{\Lambda M}~.
\eeq
Thus $P_{\Lambda M}$ remains equal to $P_{M\gamma}$.
Moreover, although the shift in $w$ causes the abundance $\barOmega_\Lambda$ to increase, it also causes the {\it rate of increase  per unit abundance}\/ due to cosmological expansion (as indicated by the wide pink arrows in Fig.~\ref{fig:seesaw}) to decrease.    Thus the net rate at which $\barOmega_\Lambda$ would tend to increase under
cosmological expansion alone is unchanged.   This too is consistent with our observation that the pump $P_{\Lambda M}$ is unchanged.

\begin{figure*}
\centering
\includegraphics[keepaspectratio, width=0.95\textwidth]{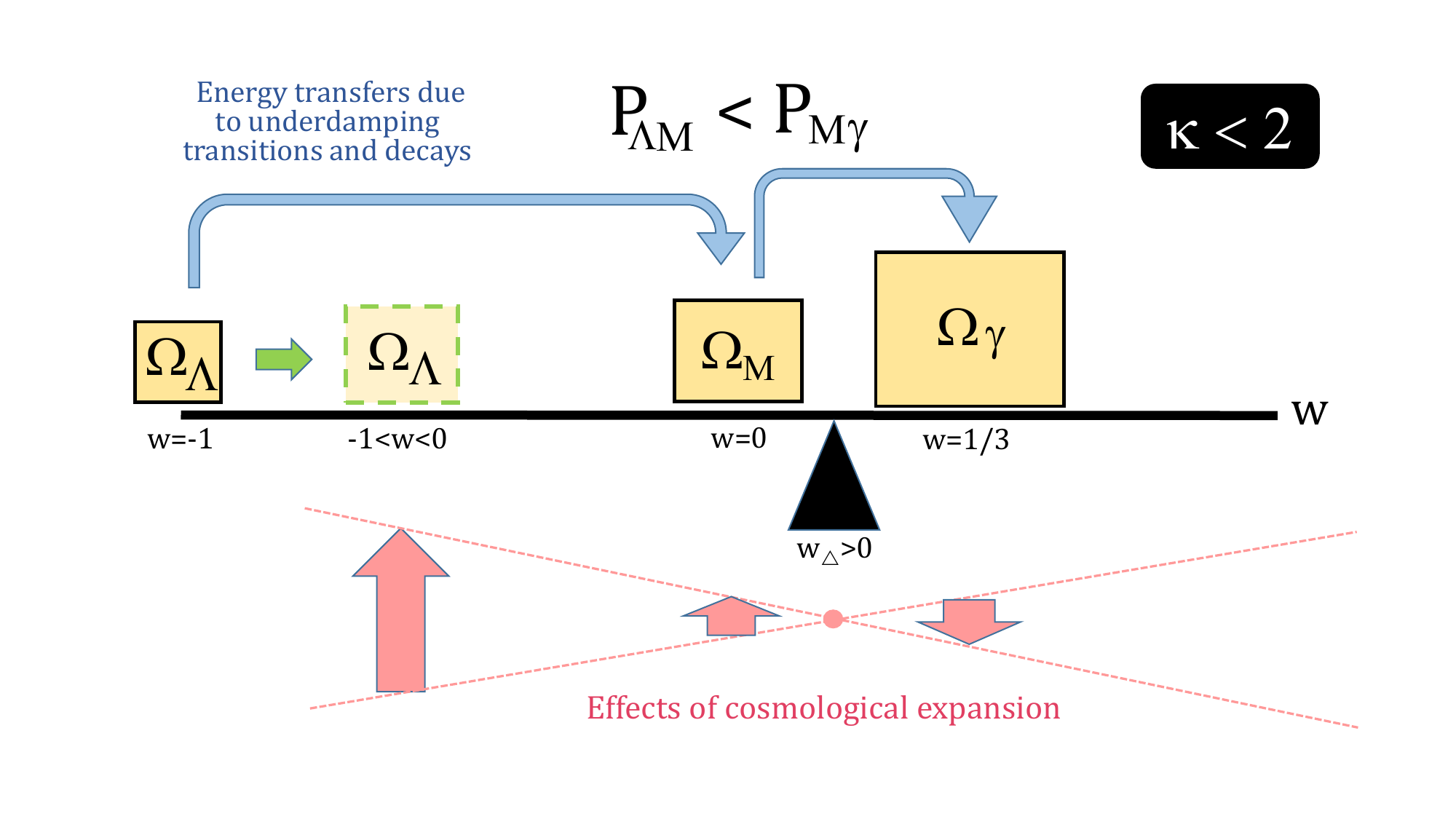}
\includegraphics[keepaspectratio, width=0.95\textwidth]{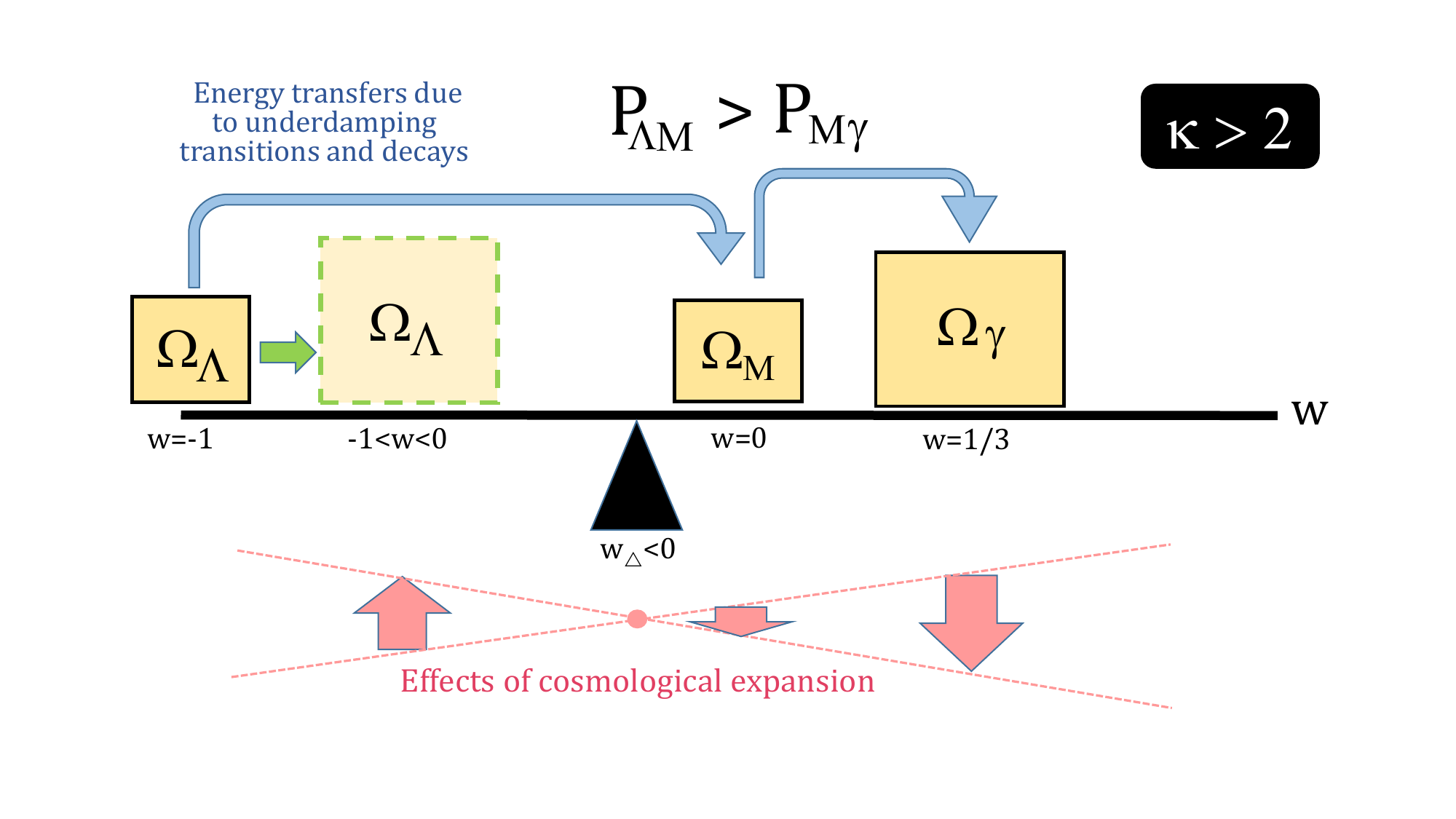}
\vskip -0.5 truein 
\caption{Same as Fig.~\ref{fig:seesaw}, except for stasis solutions with $\barkappa\not=2$.
{\it Upper panel}\/:  In this case $\barkappa<2$, implying that $w_\triangle >0$.  This causes $\Omega_M$ to tend to increase under the effects of cosmological expansion, which implies that stasis is achieved only for $P_{\Lambda M}< P_{M\gamma}$.  This inequality of the pumps is consistent with the fact that $\Omega_\Lambda/\Omega_\gamma$ must be smaller than it was in Fig.~\ref{fig:seesaw} in order to achieve a balanced ``seesaw''.   
{\it Lower panel}\/:   The same situation, but for $\barkappa>2$.  In this case the signs of all inequalities are reversed, with $P_{\Lambda M}>P_{M \gamma}$ and $\Omega_\Lambda/\Omega_\gamma$  now larger than it was in Fig.~\ref{fig:seesaw}.
}
\label{fig:seesaw2}
\end{figure*}

We now turn our attention to 
Branch~A cases with $\barkappa\not=2$.
 These situations correspond to Fig.~\ref{fig:ffig5}.   Indeed, one important feature of Branch~A when $\barkappa\not=2$ is that we can now in principle have stasis configurations with $3\alpha/4 <\barkappa<\alpha$, as illustrated in the right panel of Fig.~\ref{fig:ffig5}.   As an existence proof of such a possibility, let us consider the case with $w= -0.99$ and $(\barOmega_\Lambda,\barOmega_M,\barOmega_\gamma)=(0.05, 0.1, 0.85)$.   This configuration corresponds to $\barkappa\approx 1.621$, whereupon we see that $2-(1+w)\barkappa \approx 1.984$.   This in turn implies that $\alpha$ could potentially be as large as $1.984$, thereby exceeding $\barkappa$ while nevertheless ensuring that $3\alpha/4 < \barkappa$.
This configuration therefore meets the conditions needed in order to realize the stasis behavior in the right panel of  Fig.~\ref{fig:ffig5}.

Because we are no longer restricting our attention to $\barkappa=2$,  our resulting stasis need not be effectively matter-dominated.   This means that during stasis all three of our energy components (vacuum energy, natter, and radiation) 
can experience gravitational redshifts, implying that our two pumps $P_{\Lambda M}$ and $P_{M\gamma}$ will no longer generally be equal.   We must nevertheless continue to have a balancing as illustrated in Fig.~\ref{fig:seesaw}. 
 However, the fulcrum need no longer be located at $w_\triangle=0$. 

In general, for any system with Hubble expansion coefficient $\barkappa$, 
it is straightforward to determine the equation-of-state parameter that a 
fluid must have in order that its abundance neither rise nor fall as a 
result of cosmological expansion.  This defines the corresponding 
fulcrum location $w_\triangle$, and indeed one finds 
\beq
   w_\triangle ~\equiv ~ \frac{2}{\barkappa} -1~.
\label{fulcrumcondition}
\eeq
For a universe consisting
of only matter, radiation, and vacuum energy (the latter with 
equation-of-state parameter $w$), we then find that 
Eq.~(\ref{eq:triplekappa2}) can be rewritten in the form
\beq
    \langle w\rangle ~=~
         w \barOmega_\Lambda + \barOmega_\gamma/3 ~=~ w_\triangle~.
\label{Bline}
\eeq
This is the generalization of Eq.~(\ref{seesawbalancezero}) with arbitrary
$w_\triangle$, and tells us that our seesaw must indeed be balanced around
$w_\triangle$.  In particular we see that $w_\triangle\geq w$, with the 
inequality saturated only when $\barOmega_\Lambda=1$ and 
$\barOmega_M=\barOmega_\gamma=0$.

Given these results, the stasis constraint in Eq.~(\ref{constw1}) then 
tells us that
\beq
   w_\triangle ~=~ 
  \frac{ 2w+\eta }
         {2-\eta}~
\label{wedgevalue}
\eeq
or, equivalently, that
\beq
    w_\triangle + 1 ~=~ 
    \left( \frac{2}{2-\eta} \right)(w+1)\,.
\label{wedgevalue2}
\eeq
We thus see that $\eta$ and $w$ determine the fulcrum location 
$w_\triangle$ and thereby determine the line along which the 
corresponding abundances lie, while these same variables --- with the 
addition of $\xi$ --- determine the precise locations along this line 
for the final stasis abundances.  Thus while $\xi$ is not required in 
order to determine the fulcrum location $w_\triangle$, it is needed 
for determining the absolute magnitudes of the specific abundances 
{\it around}\/ this point.

The corresponding energy flows for $\barkappa\not=2$  can be illustrated as in Fig.~\ref{fig:seesaw2}.
In general, the results depend on whether $\barkappa<2$ or $\barkappa>2$.   In the first case (corresponding to the upper panel in Fig.~\ref{fig:seesaw2}), we illustrate the situation for $\barkappa<2$.   In this case $w_\triangle >0$, which implies that $\Omega_M$ will generally tend to increase under the effects of cosmological expansion.  Stasis is then achieved only for $P_{\Lambda M}< P_{M\gamma}$, which is consistent with the fact that $\Omega_\Lambda/\Omega_\gamma$ must be smaller than it was in Fig.~\ref{fig:seesaw} in order to achieve a balanced seesaw.   
The lower panel of Fig.~\ref{fig:seesaw2} illustrates the opposite situation with $\barkappa>2$.  

\FloatBarrier

\section{Stasis as a global attractor\label{sec:attractor}}

In this section we study the extent to which the different forms of cosmic stasis 
that we have examined in the previous sections are local or global attractors.  
As we shall see, the stasis solutions for the pairwise (two-component) systems which we 
considered in Sects.~\ref{sec:MatterGamma}
through \ref{sec:Generalpairwise}
are all global attractors.  Moreover, we shall demonstrate that the triple stasis considered 
in Sect.~\ref{sec:TripleStasis} is a global attractor as well.

\subsection{Pairwise stases}

We first analyze the attractor behavior of the pairwise stases discussed 
in Sects.~\ref{sec:MatterGamma}, \ref{sec:LambdaMatter}, and \ref{sec:LambdaGamma}.~
In each case we shall begin by demonstrating that the corresponding stasis is a local 
attractor.  With these results in hand,  we shall then proceed to demonstrate that 
these attractors are all in fact global.

Broadly speaking, in particle-physics realizations of stasis which involve 
towers of states, the physical condition which determines the time $t$ at which each 
individual such state effectively transitions from one equation of state to another can
generally be classified into one of two overall categories:

\begin{itemize}

\item \uline{Class~I}:~ transitions for which this 
transition time is {\it intrinsic}\/ to the particle in the sense that it depends 
on the time $t$ in the cosmological background frame alone and is essentially 
independent of the expansion history.  Examples of transitions within this class include
the decay transitions which underpin the 
matter/radiation and vacuum-energy/radiation stases discussed in Sects.~\ref{sec:MatterGamma} 
and~\ref{sec:LambdaGamma}, respectively.  
In these realizations of stasis, the intrinsic timescale for these transitions --- \eg, the 
proper lifetime $\tau_\ell$ of the particle $\phi_\ell$ at each level --- was assumed to scale across the tower as a function 
of mass according to a power law of the form
$\tau_\ell = \tau_0 ( m_\ell/m_0)^{-\gamma}$. Passing to the continuuum limit, we thus have
\begin{equation}
  \tau ~=~ \tau_0 \left(\frac{m}{m_0}\right)^{-\gamma}~.
\end{equation}

\item \uline{Class~II}:~ transitions for which the transition time is {\it extrinsic}\/ to 
the particle in the sense that the transition is triggered when the temperature, critical 
density, or expansion rate of the universe drops below some threshold value.  
For transitions within this second class, the time $t$ at which the transition takes place 
{\it does}\/ depend on the expansion history of the universe through the Hubble parameter 
$H(t)$.  Examples of transitions within this class include the transition
from overdamped to underdamped oscillation which underpins the vacuum-energy/matter stasis 
discussed in Sect.~\ref{sec:LambdaMatter}.~  In this realization of stasis, the transition 
of the particle species of mass $m_\ell$ is triggered at a time $t_\ell$ when $3H(t_\ell) = 2m_\ell$.  
Phrased slightly differently, this transition for each $\phi_\ell$ is triggered when the Hubble parameter $H(t)$ drops 
below the critical threshold scale $\hat H_\ell = 2 m_\ell/3$, with the corresponding time $t_\ell$ determined {\it implicitly}\/ through the condition $H(t_\ell)= \hat H_\ell$.
For complete generality, we shall 
focus in what follows on the case in which 
$\hat{H}_\ell$ scales across the tower according to a power law of the form
$\hat H_\ell= \hat H_0 (m_\ell/m_0)^{\hat{\gamma}}$
where $\hat{\gamma}$ is a general scaling exponent.
In the continuum limit this becomes
\beq
   \hat{H} ~=~ \hat{H}_0 \left(\frac{m}{m_0}\right)^{\hat{\gamma}}~.
\label{class2defn}
\eeq
Thus $\hat H$, like $\tau$, is a continuous variable that specifies a particular part of our $\phi$-tower, namely that part which has a Class~II transition occurring when $H(t)=\hat H$.  Since $H(t)$ decreases monotonically with 
$t$ in a flat universe, and since we are primarily interested in situations in which $\hat{H}$ 
increases with $m$, we shall henceforth restrict our attention to the regime in which $\hat{\gamma} > 0$.
Indeed, as we have seen, an overdamped/underdamped transition of the sort discussed in 
Sect.~\ref{sec:LambdaMatter} obeys a scaling relation of this form with $\hat{\gamma} = 1$ and $\hat H_0= 2m_0/3$.

\end{itemize}

When the universe is already deeply in stasis, the distinction between these two 
classes of transitions is not terribly important.  Indeed, since $H \approx \barkappa/(3t)$ 
within stasis, we can obtain a direct expression for the underdamping transition time $t_\ell$, namely $t_\ell= \barkappa/(2m_\ell)$, as we have used throughout this paper.
However, if we now wish to extend our analysis 
to situations in which the universe is not already in stasis,
whether a transition is intrinsic or extrinsic matters.   Thus, by rephrasing our transition times in terms of critical values $\hat H$ of the Hubble parameter, we can extend our analysis beyond stasis and allow the transition times $\hat t$ to be determined implicitly.

In what follows, we shall examine these two classes of transitions in turn.
For each class, we shall concentrate on two-component systems with abundances $\Omega_{1,2}$ and corresponding 
equation-of-state parameters $w_{1,2}$ with
$-1< w_1<w_2<1$, as in Sect.~\ref{sec:Generalpairwise}.~
Within this general framework, we shall 
 derive the coupled equations of motion for the Hubble parameter 
and for the abundance $\Omega_1$ of the component with the smaller equation-of-state parameter $w_1$.
We shall then demonstrate analytically
that the stasis solution to these equations of motion in each case is a local attractor 
for all possible such combinations of $w_1$ and $w_2$.
Finally, we shall demonstrate that 
 these stasis solutions are also  {\it global}\/ attractors
for the cases of 
physical interest discussed in Sects.~\ref{sec:MatterGamma}, \ref{sec:LambdaMatter}, 
and \ref{sec:LambdaGamma}.


\subsubsection{Local attractor behavior: Class~I transitions}

In general, for any pairwise stasis involving two components with equation-of-state parameters
$w_1$ and $w_2$, where $w_1 < w_2$, the equation of motion for $\Omega_1$ is given by 
Eq.~(\ref{gendiffeqs}).  For a stasis of this sort involving Class~I transitions, 
the pump term in this expression, 
evaluated in the continuum limit, takes the form
$P_{12} = n_\tau(t)\Omega^{(1)}_\tau(t;t)$, with $n_\tau(t)$ given by 
Eq.~(\ref{ns}).  However, since we are not assuming stasis, the expression for 
$\Omega^{(1)}(t;t)$ is given by 
\begin{equation}
  \Omega_\tau^{(1)}(\tau;t) ~=~ \Omega_0^{(0)}\,
    (\Gamma_0 \tau)^{-\alpha/\gamma}\, h_1(t^{(0)},t)\,
    \Theta(\tau -t)~~
\end{equation}
with $\tau = t$.   We caution that while the `$(1)$' superscript  on $\Omega^{(1)}_\tau(\tau;t)$ --- just like the `1' subscript on $h_1(t^{(0)},t)$ --- indicates the energy component with equation-of-state parameter $w_1$, the `$(0)$' superscript on $\Omega_0^{(0)}$ continues to indicate the value at the initial production time $t^{(0)}$.
It therefore follows that
\begin{eqnarray}
    P_{12} ~&=&~ \frac{\Omega_0^{(0)}\Gamma_0}{\gamma\delta} 
    \left(\frac{m_0}{\Delta m}\right)^{1/\delta} 
    (\Gamma_0 t)^{-1-\eta/\gamma} \,h_1(t^{(0)},t)~.\nonumber \\
  \label{eq:PumpMgammaNotInStasis}
\end{eqnarray}
Moreover, the total abundance $\Omega_1$ of the component with
equation-of-state parameter $w_1$, evaluated in the continuum limit, is 
given for $t\ll \tau_0$ by 
\begin{eqnarray}
  \Omega_1 ~&=&~ \int_{\tau_{N-1}}^{\tau_0} d\tau 
    \,n_{\tau}(\tau)\,\Omega_\tau^{(1)}(\tau;t) \nonumber \\ 
    ~&\approx&~ \frac{\Omega_0^{(0)}}{\eta\delta} 
    \left(\frac{m_0}{\Delta m}\right)^{1/\delta}
    (\Gamma_0 t)^{-\eta/\gamma} \,h_1(t^{(0)},t)~.~~
\end{eqnarray}
Comparing this expression to the expression for 
$P_{12}$ in Eq.~(\ref{eq:PumpMgammaNotInStasis}), we obtain
\begin{equation}
  P_{12} ~=~ \frac{\eta}{\gamma}\,\Omega_1\frac{1}{t}~.
\label{diffeqPumps}
\end{equation}
Interestingly, this result reduces to 
to our result
in Eq.~(\ref{pumpMatterGamma})  
during stasis.  
However, we now see that this result applies even without the assumption of stasis.
We thus find that the time-evolution of $\Omega_1$ and $H$ is 
described by the system of equations
\begin{eqnarray}
  \frac{d\Omega_1}{dt} ~&=&~ - p\,\Omega_1\frac{1}{t}
    + 3H(w_2-w_1)\Omega_1 (1 -\Omega_1) \nonumber \\
  \frac{dH}{dt} ~&=&~ -\frac{3}{2}H^2\Bigl[1+w_2+(w_1-w_2)\Omega_1\Bigr]~.
\end{eqnarray}

We may recast these equations in a more revealing form by
parametrizing the expansion rate of the universe in terms of the quantity
\begin{equation}
  \mathcal{H}(t) ~\equiv~ H(t)\left(\frac{3t}{\barkappa}\right)~
\end{equation}
which represents the ratio of the Hubble parameter $H(t)$ at any given time
$t$ to the value which it would have at that time if the universe were in stasis.
In particular, we find that 
\begin{eqnarray}
  \frac{d\Omega_1}{d\log t} ~&=&~ - p\,\Omega_1
    + \barkappa(w_2-w_1)\mathcal{H}\Omega_1 (1 -\Omega_1) \nonumber \\
  \frac{d\mathcal{H}}{d\log t} ~&=&~ \mathcal{H}
    -\frac{\barkappa}{2}\mathcal{H}^2\Bigl[1+w_2+(w_1-w_2)\Omega_1\Bigr]~.~~~
  \label{eq:StabDiffEqsGenClassIStasis}
\end{eqnarray} 
Taking $d\Omega_1/d\log t = d\mathcal{H}/d\log t = 0$, we find that indeed 
the only equilibrium solution for this system with non-vanishing $\Omega_1$
is the stasis solution in which $\mathcal{H} = 1$ and in which $\Omega_1=\barOmega_1$, 
with $\barOmega_1$ given by Eq.~(\ref{eq:GeneralwsOmega1}).

In order to determine whether or not this stasis solution is a local attractor, 
we evaluate the eigenvalues of the Jacobian matrix for the system of equations in
Eq.~(\ref{eq:StabDiffEqsGenClassIStasis}).  Using the fact that $\barkappa$ is
given by Eq.~(\ref{kappaseesaw}) during a general pairwise stasis, we can  eliminate $\barkappa$ in favor of $\barOmega_1$.
These
eigenvalues can then be written in the form
\makeatletter
\newcommand{\vast}{\bBigg@{3.3}}
\makeatother
\begin{eqnarray}
  \lambda_\pm &~=~& -\frac{1}{2}\vast[ 
    \frac{1+w_2-(w_1-w_2)\barOmega_1}{1+w_2+(w_1-w_2)\barOmega_1} 
    \nonumber \\ & &~~~~~ \mp\, 
    \sqrt{1-\frac{4(w_2-w_1)(1+2w_1-w_2)\barOmega_1}{[1+w_2+(w_1-w_2)\barOmega_1]^2}}\,
    \vast]~. \nonumber \\
  \label{eq:GenwwEigenvaluesCaseI}
\end{eqnarray}
For all possible combinations of $\barOmega_1$, $w_1$, and $w_2$ within the ranges 
$0 < \barOmega_1\leq 1$ and $-1 < w_1 < w_2 < 1$, both of these eigenvalues 
are real and negative.  Thus, we conclude that for any combination of $w_1$ and $w_2$
which satisfies these conditions, the stasis solution is a local attractor.  

This general result is applicable to any pairwise stasis involving a Class~I pump.  
For example, this result implies that the stasis solution that we obtained for the matter/radiation 
system in Sect.~\ref{sec:MatterGamma} is a local attractor.  Indeed,  
for this realization of stasis
we have $w_1=0$ and $w_2=1/3$, whereupon 
we see that Eq.~(\ref{eq:GenwwEigenvaluesCaseI}) 
reduces to
\begin{equation}
  \lambda_\pm ~=~ -\frac{1}{2}\vast[
  \frac{4+\barOmega_M}{4-\barOmega_M}\mp
    \sqrt{1 -\frac{8\barOmega_M}{(4-\barOmega_M)^2}}\,
    \vast]~.
\end{equation}
Likewise, our results also imply that the stasis solution that we obtained for the 
vacuum-energy/radiation system in Sect.~\ref{sec:LambdaGamma} is a
local attractor.  For this realization of stasis, we have
$w_1=w$ and $w_2=1/3$, whereupon we see that
Eq.~(\ref{eq:GenwwEigenvaluesCaseI}) reduces to  
\begin{eqnarray}
  \lambda_\pm ~&=&~ -\frac{1}{2}
    \vast[
    \frac{4-(3w-1)\barOmega_\Lambda}{4+(3w-1)\barOmega_\Lambda}
    \nonumber \\ & & ~~~~~~~ \mp\,
    \sqrt{1 -\frac{8(1-9w^2)\barOmega_\Lambda}{[4+(3w-1)\barOmega_\Lambda]^2}}\,
    \vast]~.~~~~~~
\end{eqnarray}

\subsubsection{Local attractor behavior: Class~II transitions}

For a pairwise stasis involving Class~II transitions, as discussed above, 
the contribution $\Omega_\ell^{(1)}$ from each $\phi_\ell$ to the overall 
abundance $\Omega_1$ associated with the component with equation-of-state 
parameter $w_1$ is transferred to the component with equation-of-state
parameter $w_2$ when $H$ drops below a particular threshold $\hat{H}$.  
In the approximation that this transition occurs sharply at the time $t$
at which $H(t) = \hat{H}$, this contribution takes the form 
\begin{equation}
    \Omega^{(1)}_\ell(t) ~=~ \Omega^{(0)}_\ell \,
      h_1(t^{(0)},t)\,\Theta\big(H(t) -\Hat{H}_\ell\big)~,
\end{equation}
For a universe involving only two components, with $\Omega_1 + \Omega_2 = 1$, 
the time-derivative of this expression may be written in the form
\begin{eqnarray}
  \frac{d\Omega_\ell^{(1)}}{dt} &~=~& \Omega_\ell^{(0)}
    \,h_1(t^{(0)},t) \,\delta(H - \hat{H}) \,\frac{dH}{dt} \nonumber \\ 
    & & ~~ + 3H\Omega_\ell^{(1)}(w_2 - w_1)(1 -\Omega_1)~.~~~
  \label{eq:dOmegaelldtHIndivid}
\end{eqnarray}

The rate of change of the total abundance $\Omega_1$ is simply the direct sum of the 
individual contributions in Eq.~(\ref{eq:dOmegaelldtHIndivid}).  In order to 
evaluate this sum, we shall pass to the continuum limit in which we express
$\ell$ in terms of the corresponding transition scale $\hat{H}_\ell$ and 
then treat $\hat{H}$ as a continuous parameter.  In other words, we shall replace
\begin{equation}
  \sum_\ell \to \int d\hat{H} \, n_{\hat{H}}(\hat{H})~,
  \label{eq:HhatContLim}
\end{equation}
where $n_{\hat{H}}(\hat{H})$ denotes the density of states within the tower per unit 
$\hat{H}$, evaluated at the location within the tower for which the transition scale 
is $\hat{H}$.  For a scaling relation between $\hat{H}$ and $m$ of the form given in 
Eq.~(\ref{class2defn}), this density of states takes the form
\begin{equation}
  n_{\hat{H}}(\hat{H}) ~\equiv~ \left| \frac{d\ell}{d\hat{H}}\right| 
    ~=~ \frac{1}{\hat{\gamma}\delta\hat{H}_0}
    \left(\frac{m_0}{\Delta m}\right)^{1/\delta}
    \left(\frac{\hat{H}}{\hat{H}_0}\right)^{1/(\hat{\gamma}\delta)-1}~.
  \label{eq:nHhat}
\end{equation}

In analogy with the quantities $\Omega_{\tau}(\tau;t)$ and $\Omega_{\hat{t}}(\hat{t};t)$
that we defined in Sect.~\ref{sec:TripleStasis}, we shall find it convenient to define
$\Omega_{\hat{H}}(\hat{H};t)$ to represent the abundance --- evaluated at time $t$ ---
of that particular $\phi$-field whose transition threshold is $\hat{H}$.  For a 
scaling relation between $\hat{H}$ and $m$ of the form given in Eq.~(\ref{class2defn}), 
this abundance is given by 
\begin{equation}
  \Omega^{(1)}_{\hat{H}}(\hat{H};t) ~=~ \Omega_0^{(0)}
    \left(\frac{\hat{H}}{H_0}\right)^{\alpha/\hat{\gamma}} h_1(t^{(0)},t)
    \, \Theta\big(H(t) - \hat{H}\big)~.
  \label{eq:OmegaHhatExpr}
\end{equation}

The rate of change of the total abundance $\Omega_1$, evaluated in the continuum limit, is
\begin{equation}
  \frac{d\Omega_1}{dt} ~=~ \int_{\hat{H}_0}^{\hat{H}_{N-1}} 
    d\hat{H}\,\frac{d\Omega^{(1)}_{\hat{H}}(\hat{H};t)}{dt}~,
\end{equation}
which yields an expression of the general form given in Eq.~(\ref{gendiffeqs}).  
However, we see from Eq.~(\ref{eq:dOmegaelldtHIndivid}) that the pump term $P_{12}$ 
for a Class~II transition is in general given by
\begin{equation}
  P_{12} ~=~ n_{\hat{H}}(H)\,\Omega^{(1)}_{\hat{H}}(H;t)\, \frac{dH}{dt}~,
\end{equation}
For the particular set of transitions we are considering here, with 
$n_{\hat{H}}(\hat{H})$ and $\Omega^{(1)}_{\hat{H}}(\hat{H};t) $ given by 
Eqs.~(\ref{eq:nHhat}) and~(\ref{eq:OmegaHhatExpr}), respectively, we have
\begin{equation}
    P_{12} ~=~ \frac{3H_0\Omega_0^{(0)}}{\kappa\hat{\gamma}\delta}
    \left(\frac{m_0}{\Delta m}\right)^{1/\delta}
    \left(\frac{H}{H_0}\right)^{1+\eta/\hat{\gamma}} h_1(t^{(0)},t)~,
  \label{eq:PumpLambdaMNotInStasis}
\end{equation}
with $\kappa$ given by Eq.~(\ref{kappaseesaw}).

The total abundance of the component with equation-of-state parameter $w_1$ in this case, 
evaluated in the continuum limit, is
\begin{equation}
  \Omega_1~=~ \int_{\hat{H}_{N-1}}^{\hat{H}_0} d\hat{H} \,
    n_{\hat{H}}(\hat{H})\,\Omega_{\hat{H}}^{(1)}(\hat{H};t)~.  
\end{equation}
Explicit integration yields
\begin{eqnarray}
    \Omega_1 \,&=&\, \frac{\Omega_0^{(0)}}{\hat{\gamma}\delta}\! 
       \left(\frac{m_0}{\Delta m}\right)^{1/\delta} H_0^{-\eta/\hat{\gamma}}\,
        h_1(t^{(0)},t) \int_{\hat{H}_0}^{H(t)} d\hat{H} \hat{H}^{\eta/\hat{\gamma}-1}
         \nonumber \\ 
      \,&\approx&\, \frac{\Omega_0^{(0)}}{\delta \eta} \!
        \left(\frac{m_0}{\Delta m}\right)^{1/\delta}
        \left(\frac{H}{H_0}\right)^{\eta/\hat{\gamma}} \, h_1(t^{(0)},t)~, 
\end{eqnarray}
where in going from the first to the second line we have used the fact that 
$H \gg \hat{H}_0$ at times well before the energy density associated with the 
lightest tower state is transferred to the energy component with equation-of-state
parameter $w_2$.  Comparing this expression to the expression for $P_{12}$ in 
Eq.~(\ref{eq:PumpLambdaMNotInStasis}), we observe that
\begin{equation}
  P_{12} ~=~ \frac{\eta}{\hat{\gamma}}\,\Omega_1 \left(\frac{3}{\kappa} H\right)~.
  \label{eq:PumpGenwwClassII}
\end{equation}
For $\hat{\gamma}=1$, we find that this result reduces to Eq.~(\ref{pumpLambdaMatter}) 
during stasis. However, we now see that Eq.~(\ref{eq:PumpGenwwClassII}) holds even without 
the assumption of stasis. Moreover, comparing this result to the stasis expectation 
in Eq.~(\ref{pumpsoln}), we observe that $p = \eta/\hat{\gamma}$ for a Class~II transition of 
this sort with arbitrary $\hat{\gamma} > 0$.

It follows from the result in Eq.~(\ref{eq:PumpGenwwClassII}) that the time-evolution 
of $\Omega_1$ and $\mathcal{H}$ in this case is described by the equations
\begin{eqnarray}
  \frac{d\Omega_1}{d\log t} ~&=&~ - \frac{\barkappa}{2} p \mathcal{H} 
    \Omega_1 \Bigl[1+w_2 + (w_1-w_2)\Omega_1\Bigr] \nonumber \\
    & & ~~~~~~~~~+ \barkappa (w_2-w_1) \mathcal{H} \Omega_1 (1-\Omega_1)
    \nonumber \\
  \frac{d\mathcal{H}}{d\log t} ~&=&~ \mathcal{H}
    -\frac{\barkappa}{2}\mathcal{H}^2\Bigl[ 1 + w_2 + (w_1-w_2)\Omega_1\Bigr]~.~~~~~
  \label{eq:StabDiffEqsGenClassIIStasis}
\end{eqnarray}
The only equilibrium solution for this system with 
non-vanishing $\Omega_1$ is the stasis solution in which $\mathcal{H} = 1$ and 
in which $\Omega_1=\barOmega_1$, where $\barOmega_1$ given by 
Eq.~(\ref{eq:GeneralwsOmega1}) with $p=\eta/\hat{\gamma}$.

Using Eq.~(\ref{Hubble}) in order to eliminate $\barkappa$ in favor of 
$\barOmega_1$, we find that the eigenvalues of the Jacobian matrix in this case are
\begin{eqnarray}
  \lambda_+ ~&=&~ \frac{2(1+w_1)(w_1-w_2)\barOmega_1}{[1+w_2+(w_1-w_2)\barOmega_1]^2} 
  \nonumber \\
  \lambda_- ~&=&~ -1~.
  \label{eq:GenwwEigenvaluesCaseII}
\end{eqnarray}
For all possible combinations of $\barOmega_1$, $w_1$, and $w_2$ within the ranges 
$0 < \barOmega_1\leq 1$ and $-1 < w_1 < w_2 < 1$, both of these eigenvalues 
are real and negative.  Thus, we conclude that for any combination of $w_1$ and $w_2$
which satisfies these conditions, the stasis solution is a local attractor.

This general result is applicable to any pairwise stasis involving a Class~II pump.  
For example, it implies that the stasis solution that we obtained for the vacuum-energy/matter 
system that we examined in Sect.~\ref{sec:LambdaGamma} is a local attractor.  Indeed,  
for this realization of stasis, wherein an energy component with equation-of-state parameter 
$w_1=w$ transfers its energy density to an energy component with $w_2 = 0$ via a transition 
from overdamped to underdamped oscillation,  Eq.~(\ref{eq:GenwwEigenvaluesCaseII}) reduces to
\begin{eqnarray}
  \lambda_+ ~&=&~ \frac{2w(1+w)\barOmega_\Lambda}{(1+w\barOmega_\Lambda)^2} \nonumber \\
  \lambda_- ~&=&~ -1~.
\end{eqnarray}


\begin{figure*}[t]
\centering
\includegraphics[width=0.32\linewidth]{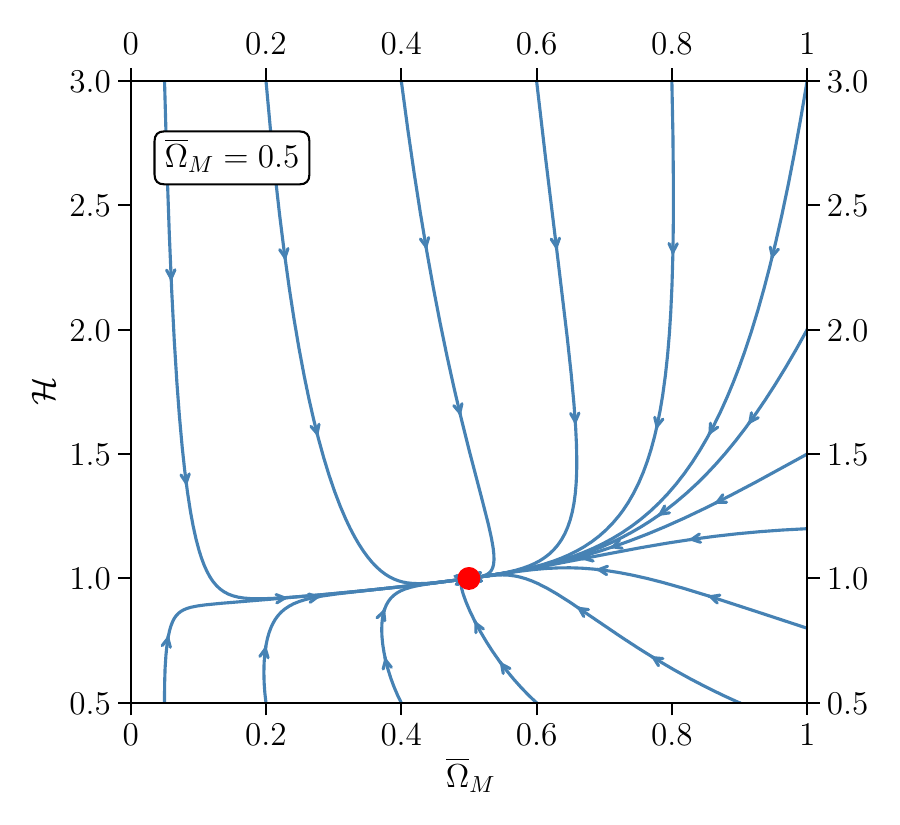}
\includegraphics[width=0.32\linewidth]{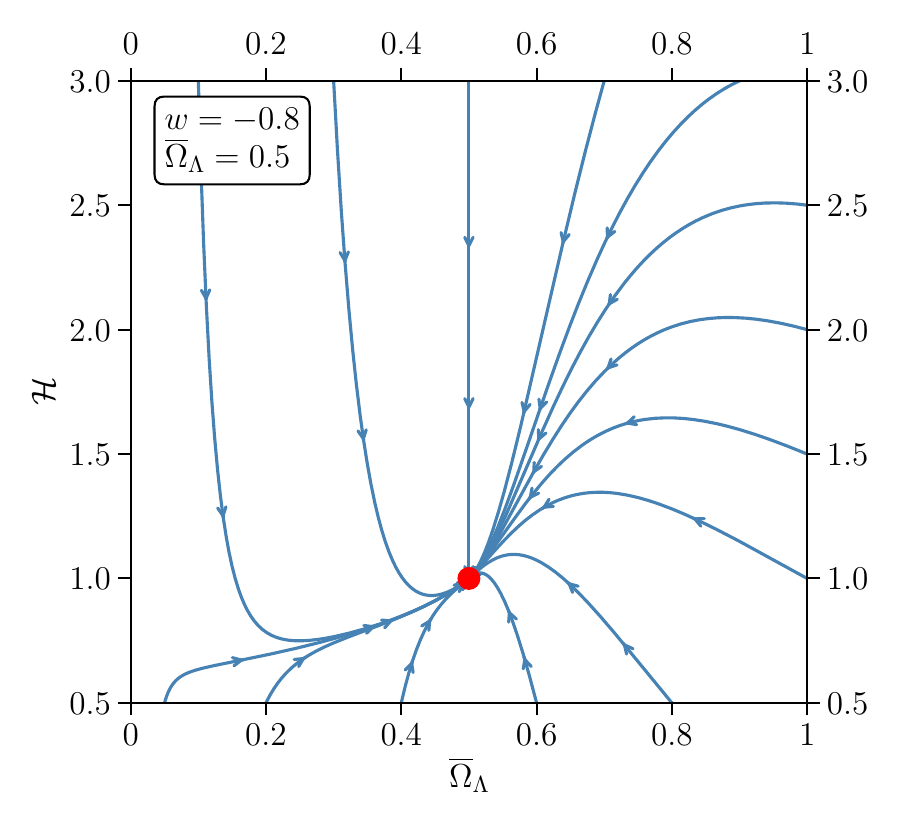}
\includegraphics[width=0.32\linewidth]{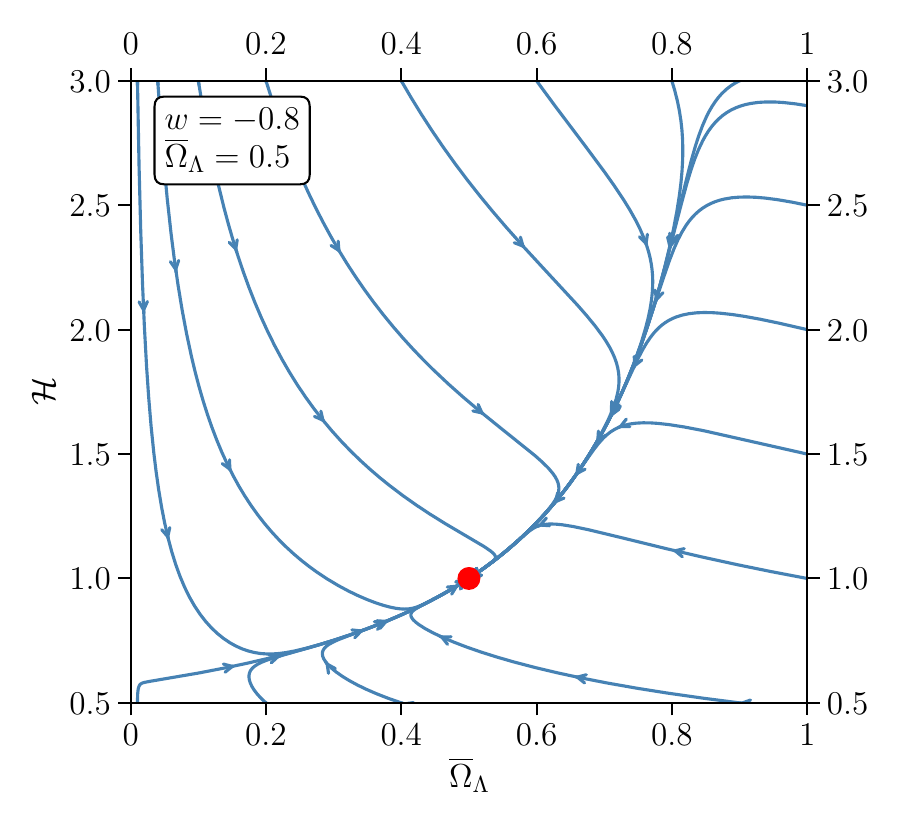}
\caption{
Attractor behavior for $M\gamma$ stasis (left panel), $\Lambda M$ 
stasis (middle panel), and $\Lambda \gamma$ stasis (right panel).  Shown in each 
case are the trajectories (blue curves) within the ($\Omega_1$,$\mathcal{H}$)-plane, where 
$\Omega_1$ in each case represents the abundance of the cosmological energy component with 
the smaller equation-of-state parameter.  These trajectories correspond respectively to the dynamical 
system described in Eq.~(\ref{eq:StabDiffEqsGenClassIStasis}) with $w_1 = 0$ and $w_2 = 1/3$,
the dynamical system described in Eq.~(\ref{eq:StabDiffEqsGenClassIIStasis}) with $w_1 = w = -0.8$
and $w_2 = 0$, and the dynamical system described in Eq.~(\ref{eq:StabDiffEqsGenClassIStasis})
with $w_1 = w = -0.8$ and $w_2 = 1/3$. 
The red dot in each case indicates the corresponding attractor stasis solution.
We have chosen $p$ in each case such that $\barOmega_1=0.5$, and we have used the 
instantaneous-decay approximation in obtaining the results shown in the left and right
panels.  We emphasize that while we have taken $w = -0.8$ for each pairwise stasis 
involving vacuum energy, similar results are obtained for other values of $w$.
\label{fig:PairwiseAttractors}}
\end{figure*}


\subsubsection{Global attractor behavior}

We have thus far demonstrated that each of our pairwise stasis solutions is a local 
attractor.  In order to assess whether these solutions are also {\it global}\/ 
attractors, we map the trajectories along which the system 
evolves in the ($\barOmega_1$,$\mathcal{H}$)-plane for different initial combinations 
of $\barOmega_1$ and $\mathcal{H}$, where in each case $\barOmega_1$ once again represents 
the abundance of the cosmological energy component with the smaller equation-of-state 
parameter.  In Fig.~\ref{fig:PairwiseAttractors}, the left panel shows a number of such 
trajectories for the matter/radiation system obtained by taking $w_1=0$ and $w_2=1/3$ in 
Eq.~(\ref{eq:StabDiffEqsGenClassIStasis}).  The middle panel shows 
trajectories for the vacuum-energy/matter system obtained by taking $w_1 = w = -0.8$ and
$w_2 = 0$ in Eq.~(\ref{eq:StabDiffEqsGenClassIIStasis}).  The right panel shows trajectories 
for the vacuum-energy/radiation system obtained by taking $w_1 = w = -0.8$ and $w_2 = 1/3$ 
in Eqs.~(\ref{eq:StabDiffEqsGenClassIStasis}).  The point within the 
($\barOmega_1$,$\mathcal{H}$)-plane which corresponds to the stasis solution in each
panel is indicated with a red dot.  
The value of $p$ in each case is chosen such that $\barOmega_1 = 0.5$.

In all three panels of Fig.~\ref{fig:PairwiseAttractors}, we see that our trajectories 
ultimately flow toward the stasis solution.  Indeed, this 
remains true even if we 
consider other values of $\barOmega_i$, other values of $w$ (when relevant), and 
even regions of the $(\barOmega_1,\mathcal{H})$ plane with values of $\mathcal{H}$ beyond 
those shown.  We therefore conclude that the stasis solution is not only a local attractor for 
all three kinds of pairwise stasis we have considered here, but a global attractor as well.

\FloatBarrier
\subsection{Triple stasis}

\begin{figure*}[htb]
\includegraphics[width=\linewidth]{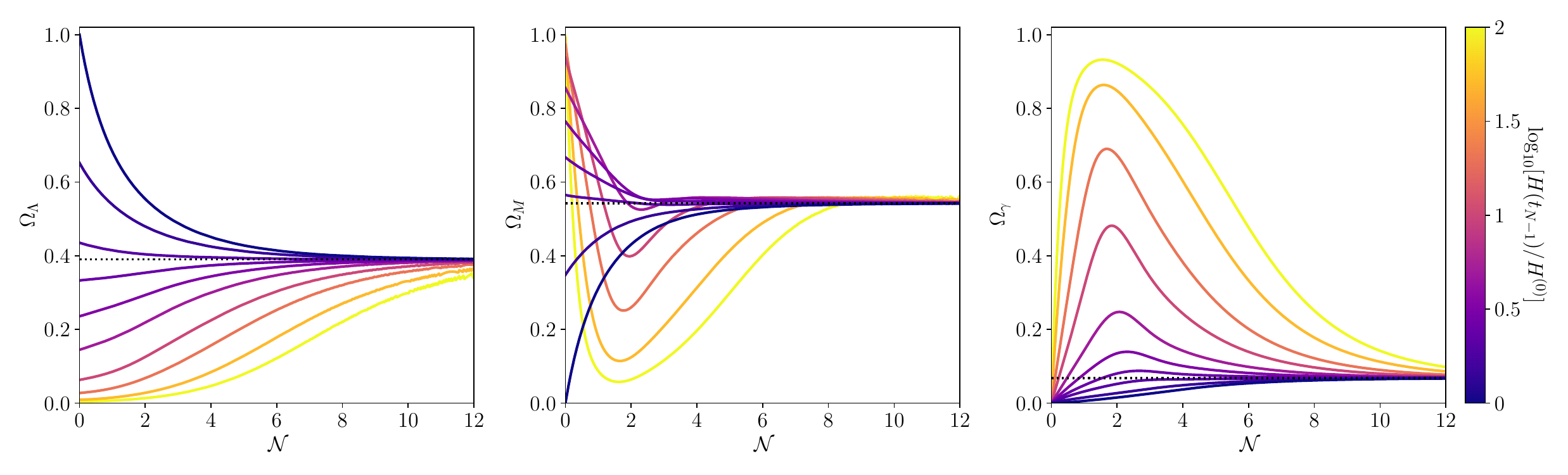}
\caption{The abundances of vacuum energy (left panel), matter (center panel), and
  radiation (right panel) in a three-component system exhibiting a triple-stasis 
  solution, plotted as functions of the number $\calN$ of $e$-folds of cosmological expansion  
  since the initial production time $t^{(0)}$.  The different 
  curves in each panel correspond to different values of the ratio $H(t_{N-1})/H^{(0)}$, where
  $t_{N-1}$ denotes the time at which the heaviest of the $\phi_\ell$ fields becomes underdamped and begins oscillating.
  These results were calculated within the full exponential-decay framework and correspond to the parameter choices 
  $\alpha=0.95$, $\delta = 4$, $w = -0.7$, $\xi = 20$, and $\Delta m/m_0=1$.
  For all values of $H(t_{N-1})/H^{(0)}$, we observe that 
  $\Omega_\Lambda(t)$, $\Omega_M(t)$, and $\Omega_\gamma(t)$ all evolve toward their stasis values.
  We also note that it takes longer for the universe to settle into its asymptotic stasis 
  state for some values of $H(t_{N-1})/H^{(0)}$ than for others. 
\label{fig:triple_converge_H0s}}
\includegraphics[width=\linewidth]{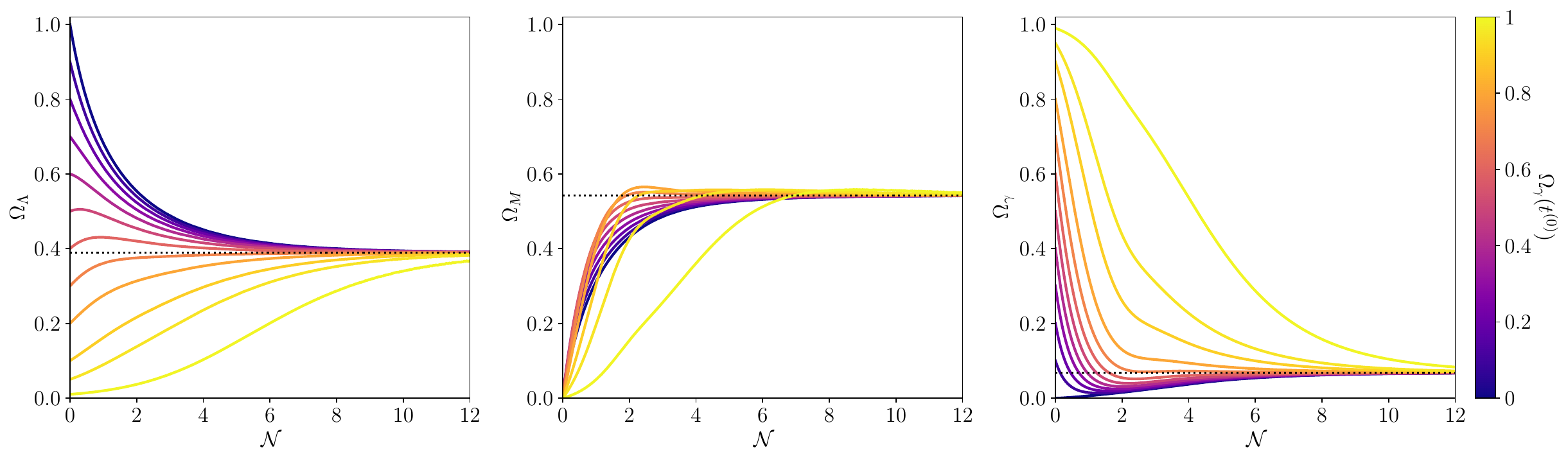}
\caption{Same as Fig.~\ref{fig:triple_converge_H0s}, except that
the different
  curves now correspond to different values for the initial vacuum-energy and radiation abundances $\Omega_\Lambda^{(0)}$ and $\Omega_\gamma^{(0)}=1-\Omega_\Lambda^{(0)}$, with $\Omega_M^{(0)}=0$ held fixed. 
  These results were calculated within the full exponential-decay framework and correspond to the parameter choices 
  $\alpha=0.95$, $\delta = 4$, $w = -0.7$, $\xi = 20$, $\Delta m/m_0=1$, and
  $3H^{(0)}/(2m_{N-1})=1$.
  For all values of $\Omega_{\Lambda}^{(0)}$ and $\Omega_\gamma^{(0)}$,
  we see that $\Omega_\Lambda(t)$, $\Omega_M(t)$, and $\Omega_\gamma(t)$ 
  all ultimately evolve toward their stasis values, with the universe taking longer to settle into its asymptotic stasis state for some values of    $\Omega_{\Lambda}^{(0)}$ and
  $\Omega_{\gamma}^{(0)}$ than for others.
\label{fig:triple_converge_Omegas}}
\end{figure*}

We now turn to consider whether the triple-stasis solution which emerged from the 
three-component system in Sect.~\ref{sec:TripleStasis} is likewise 
an attractor.  The analysis in this case is significantly more complicated than it 
is for the two-component systems that we considered in the previous subsection --- not 
merely because this system involves a larger number of dependent variables, but 
also because the coupled equations which describe the time-evolution of these variables 
are not differential equations, but rather integro-differential equations.  As a result,
we cannot determine whether the triple-stasis solution is a local attractor analytically using the
methods we employed when analyzing our two-component systems.  

For this reason, we instead 
investigate the attractor behavior of our triple-stasis system numerically by 
varying the initial conditions for the system at $t^{(0)}$.
In Fig.~\ref{fig:triple_converge_H0s}, we illustrate the effect on the abundances of 
varying the ratio $H(t_{N-1})/H^{(0)}$
where $t_{N-1}$ is the time at which the
heaviest field in the ensemble becomes underdamped and begins oscillating 
and where $H^{(0)}$ is the Hubble parameter at the initial production time
$t^{(0)}$.  The abundances $\Omega_\Lambda$, $\Omega_M$, and $\Omega_\gamma$
are plotted in the left, middle, and right panels of the figure, respectively, 
as functions of the number  $\mathcal{N}$ of $e$-folds of expansion since $t^{(0)}$.
The different curves appearing in each panel correspond to different values 
of $H(t_{N-1})/H^{(0)}$ within the range $0.01 \leq H^{(0)}/H(t_{N-1}) \leq 1$. 
The dotted horizontal line in each 
panel indicates the stasis value given in Eq.~(\ref{finalstasisabundances}) 
for the 
corresponding abundance.  In each case we 
have taken $\Omega_\Lambda^{(0)}=1$ and $\Omega^{(0)}_M=\Omega_\gamma^{(0)}=0$.

For all values of $H(t_{N-1})/H^{(0)}$, we observe that $\Omega_\Lambda$,
$\Omega_M$, and $\Omega_\gamma$ all evolve toward their stasis values.  Thus, we 
may conclude that the emergence of a triple stasis is not predicated on a particular
choice of $H(t_{N-1})/H^{(0)}$.
We also note that it takes the universe longer to settle into its asymptotic stasis 
state for some values of $H(t_{N-1})/H^{(0)}$ than it does for others.  

One might also ask whether the emergence of a stasis epoch depends sensitively on the
initial values $\Omega_{\Lambda}^{(0)}$, $\Omega_{M}^{(0)}$,
and $\Omega_{\gamma}^{(0)}$ of our three abundances.
However, given the assumptions inherent in our model, we do not have the freedom to vary these three abundances arbitrarily.
Indeed, our model comprises only the $\phi_\ell$ fields and one or
more effectively massless fields which behave as radiation throughout the entirety of the 
stasis epoch.   Likewise, the individual masses and abundances of the $\phi_\ell$ fields are assumed
to scale across the tower according to the relations in Eqs.~(\ref{MGscalings}) 
and~\eqref{MGmassform}, respectively.  Given these assumptions, it is not possible 
to adjust the relationship between $\Omega_\Lambda^{(0)}$ 
and $\Omega_M^{(0)}$ arbitrarily without introducing additional spectator fields which 
behave as either matter or vacuum energy.  That said, we {\it do}\/ have the freedom 
to adjust $\Omega_\gamma^{(0)}$ arbitrarily, provided that we compensate for this
adjustment by shifting the values of both $\Omega^{(0)}_\Lambda$ and $\Omega^{(0)}_M$ 
such that $\Omega^{(0)}_\Lambda + \Omega^{(0)}_M +\Omega^{(0)}_\gamma =1$ and
the appropriate relationship between $\Omega^{(0)}_\Lambda$ and $\Omega^{(0)}_M$ is 
maintained.  Thus, in what follows, we shall continue to take 
$\Omega^{(0)}_M=0$ and focus on the effect of varying 
$\Omega_\Lambda^{(0)}$ and $\Omega_\gamma^{(0)}$.

In Fig.~\ref{fig:triple_converge_Omegas} we illustrate the effect on the abundances that emerges upon varying $\Omega_{\Lambda}^{(0)}$ and $\Omega_{\gamma}^{(0)}$
subject to the constraints that $\Omega_\gamma^{(0)}=1-\Omega_\Lambda^{(0)}$ and $\Omega_M^{(0)}=0$.
As in Fig.~\ref{fig:triple_converge_H0s}, the three abundances 
$\Omega_{\Lambda}(t)$, $\Omega_{M}(t)$, and $\Omega_{\gamma}(t)$ 
are plotted as functions of $\calN$ in the left, middle, and right panels, respectively.   
In each panel, the dotted horizontal line once again indicates the 
stasis value given in Eqs.~(\ref{finalstasisabundances}) and (\ref{abundance_avalues})
for the corresponding abundance.

In each case, we observe that $\Omega_\Lambda(t)$, $\Omega_M(t)$, 
and $\Omega_\gamma(t)$ all evolve toward their stasis values.  Thus, we 
may further conclude that the emergence of triple stasis is not predicated on the universe 
being 
fully vacuum-energy dominated at the initial time $t^{(0)}$, much less devoid of radiation.
Indeed, we see that stasis emerges {\it regardless}\/ of the admixture of vacuum energy and radiation at the initial time.
However, we also observe that it takes the universe longer to settle into 
its asymptotic stasis state for some admixtures than others.

Taken together, the empirical results shown in  Figs.~\ref{fig:triple_converge_H0s} and~\ref{fig:triple_converge_Omegas} 
strongly suggest that our triple-stasis solution in Sect.~\ref{sec:TripleStasis}
is not only a local attractor but also a global attractor, pulling our system towards the stasis state for a wide range of dynamical parameters and initial configurations.

\FloatBarrier

\section{A phase diagram for stasis \label{sec:phase_diagram}}

It is interesting to synthesize the results of this paper thus far
by investigating how our different forms of stasis relate to each other.   
Collecting our results from Sects.~\ref{sec:MatterGamma}, \ref{sec:LambdaMatter}, \ref{sec:LambdaGamma}, and \ref{sec:TripleStasis}, we see that our different stases have 
the constraint equations
\beqn
     M\gamma~{\rm Stasis}:~~~~~
      \frac{\eta}{\gamma} &=& 2-\barkappa\nonumber\\
    \Lambda M~{\rm Stasis}:~~~~\,
      \eta &=& 2-(1+w)\barkappa 
       \nonumber\\ 
 {\Lambda}\gamma~{\rm Stasis}:~~~~~
      \frac{\eta}{\gamma} &
       = &
     2 - (1+w)\barkappa  
      \nonumber\\ 
  {\rm Triple~Stasis}:~~~~~
      \eta &=&  2-(1+w)\barkappa ~,~~ \gamma=1
       \nonumber\\ 
\label{synthesiscomparison}
\eeqn
Note that in the case of triple stasis, we are restricting our attention to Branch~A because Branch~B does not yield a full stasis.

The constraint equations in Eq.~(\ref{synthesiscomparison}) are all very similar to each other. Indeed, if we temporarily disregard the absence of the $w$-term within the constraint for $M\gamma$ stasis, {\it we see that these equations all become identical if $\gamma=1$.  }

\begin{figure*}
\centering
\includegraphics[keepaspectratio, width=0.52\textwidth]{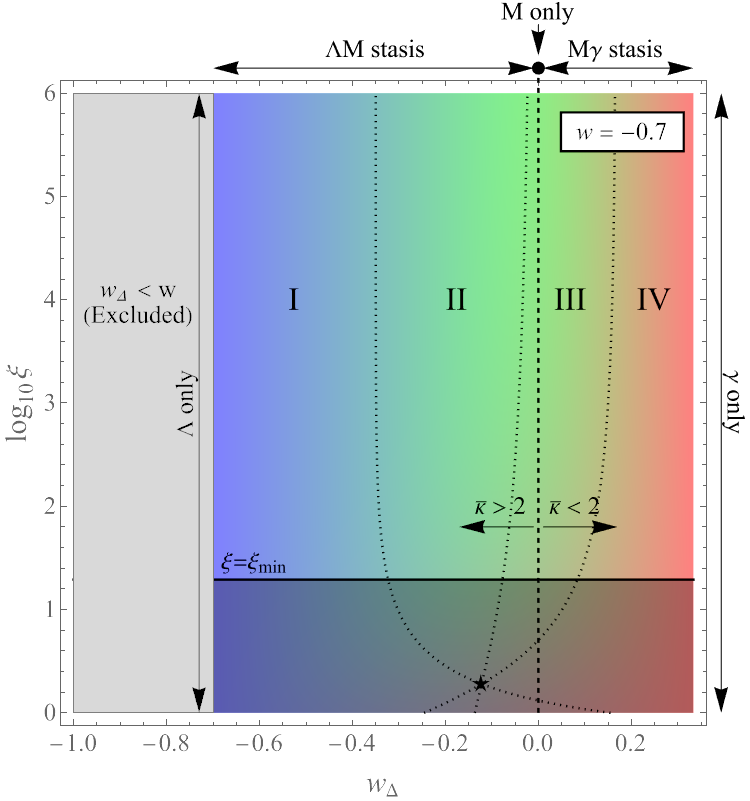}~
\hskip 0.2 truein
\raisebox{0.667\height}{\includegraphics[keepaspectratio, width=0.25\textwidth]{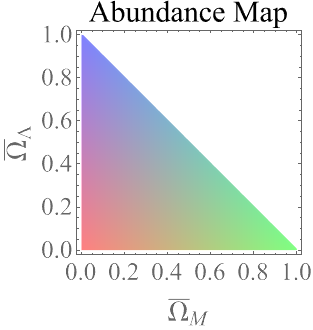}}\\
\vskip 0.3 truein
\includegraphics[keepaspectratio, width=0.32\textwidth]{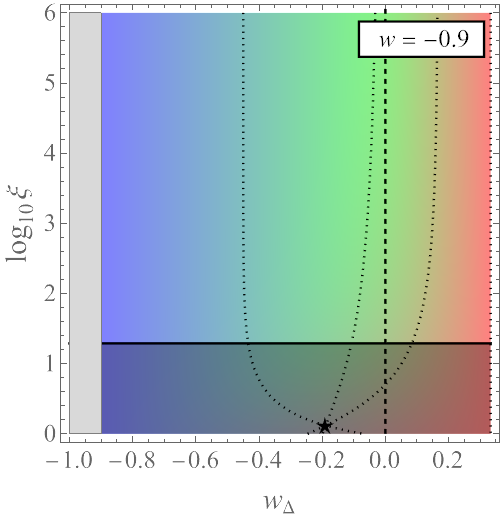}~
\includegraphics[keepaspectratio, width=0.32\textwidth]{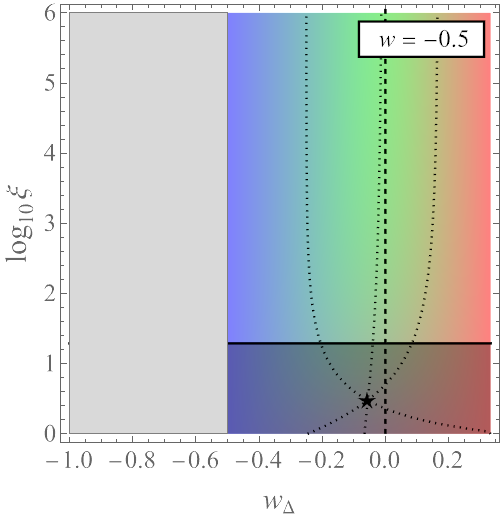}~
\includegraphics[keepaspectratio, width=0.32\textwidth]{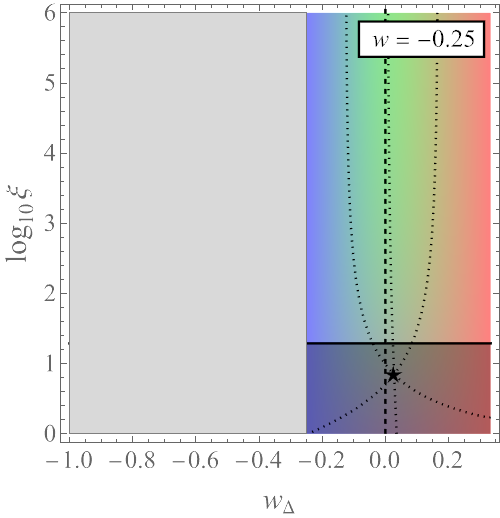}~
\caption{``Phase diagrams'' for triple stasis within the $(w_\triangle, \xi)$-plane for
$w=-0.7$ (top panel) and for $w = -0.9$, $w=-0.5$, and $w=-0.25$
(corresponding to the three panels along the lower row).
Working within the full exponential-decay framework, we have assigned the color of each 
point within the $(w_\triangle,\xi)$-plane according 
to the abundance-map palette shown on the top right, with the relative 
levels of blue, green, and red indicating the values of the stasis 
abundances $\barOmega_\Lambda$, $\barOmega_M$, and $\barOmega_\gamma$, respectively. 
The gray region on the left side of each panel is physically inaccessible, since 
self-consistency requires $w_\triangle > w$.  Likewise the shaded region 
with $\xi<\xi_{\rm min}$ lies outside our regime of validity for $\epsilon_{\rm dec}=0.05$.
The dashed black vertical line with $w_\triangle=0$ corresponds to universes with 
$\barkappa=2$, as shown in Fig.~\ref{fig:seesaw}, while points to the left 
(respectively right) of this line correspond to universes with $\barkappa>2$ 
(respectively $\barkappa<2$), as shown in the upper (respectively lower) panel of 
Fig.~\ref{fig:seesaw2}.  The points within Regions I through IV exhibit triple stases with
$\barOmega_\Lambda>\barOmega_M>\barOmega_\gamma$,
$\barOmega_M>\barOmega_\Lambda>\barOmega_\gamma$,
$\barOmega_M>\barOmega_\gamma>\barOmega_\Lambda$, and
$\barOmega_\gamma>\barOmega_M>\barOmega_\Lambda$, respectively.
By contrast, along the edges of this plane, our triple-stasis solutions reduce to simpler 
stasis solutions:  either to a pairwise $\Lambda M$ stasis (as discussed in 
Sect.~\ref{sec:LambdaMatter} and shown along the upper edge with $w_\triangle<0$), 
or to a pairwise $M\gamma$ stasis (as discussed in Sect.~\ref{sec:MatterGamma} 
and shown along the upper edge with $w_\triangle >0$), or to universes with only 
vacuum energy (as shown along the left edge), radiation (as shown along the right edge), 
or matter (as indicated at the top point with $w_\triangle=0$).  All universes along 
the vertical $w_\triangle=0$ line are effectively matter-dominated (as illustrated in 
Fig.~\ref{fig:seesaw}), with counterbalancing abundances of vacuum energy and radiation;  
however, as we move up this line towards greater values of $\xi$, these other abundances 
maintain their ratio but shrink to zero, leaving behind a universe consisting only of 
matter as $\xi\to\infty$.  The star represents the location of the ``triple point'' 
at which $\barOmega_\Lambda = \barOmega_M = \barOmega_\gamma$.
This figure therefore encapsulates and illustrates the 
relationships between the different versions of stasis discussed in this paper.  
\label{fig:PhaseDiagramPanels}}
\end{figure*}

There are also other commonalities between these different forms of stasis.
In general, a matter-dominated universe has $\barkappa=2$ while a 
radiation-dominated universe has $\barkappa=3/2$.   By contrast, within 
the context of our general-$w$ model for vacuum energy, a universe 
dominated by vacuum energy has $\barkappa= 2/(1+w)$.  It therefore 
follows that  our different mixed-component stases  have restricted 
ranges for $\barkappa$ given by
\beqn
M\gamma~{\rm Stasis}:~~~
       && 3/2 <\barkappa< 2     \nonumber\\
    \Lambda M~{\rm Stasis}:~~~
      && 2 <\barkappa< 2/(1+w)
       \nonumber\\ 
 {\Lambda}\gamma~{\rm Stasis}:~~~
      && 3/2 < \barkappa < 2/(1+w)
      \nonumber\\ 
    {\rm Triple~Stasis}:~~~
      && 3/2 < \barkappa < 2/(1+w)~.~~~~
\label{synthesiscomparison2}
\eeqn
However, substituting these results into the corresponding equations in 
Eq.~(\ref{synthesiscomparison}) we find in each case that 
\beq
\eta ~>~0~.
\label{alphadeltapositive}
\eeq
Of course, this condition has already been stated throughout this paper on the basis of other consistency constraints [see, \eg, Eqs.~(\ref{MGrange}), (\ref{rangge}), (\ref{LGrange}), and
(\ref{eq:poseta})].   However, the condition in Eq.~(\ref{alphadeltapositive}) is yet another commonality between our different forms of stasis.

The fact that all of our stases have the same basic constraints on their fundamental 
parameters suggests that they all populate different limiting regions of a common 
``phase space''.   We shall now demonstrate that this expectation is correct.

To do this, let us begin by considering the largest and most comprehensive of our stases, 
namely the triple stasis of Sect.~\ref{sec:TripleStasis}.~
In Fig.~\ref{fig:PhaseDiagramPanels}, we display the phase diagram for triple stasis within 
the $(w_\triangle, \xi)$-plane for several different values of $w$.  The results in the top
panel correspond to the choice $w=-0.7$, while the results shown in the three of panels at
the bottom of the figure correspond to the choices $w = -0.9$ (left panel), $w=-0.5$ 
(middle panel), and $w=-0.25$ (right panel).  The color of each point within each 
panel has been assigned according to the abundance-map palette shown on the far right, 
with the relative levels of blue, green, and red indicating the values of the stasis 
abundances $\barOmega_\Lambda$, $\barOmega_M$, and $\barOmega_\gamma$, respectively. 
The gray region on the left side of each panel is physically inaccessible, since 
self-consistency requires $w_\triangle > w$.  Likewise the shaded region 
with $\xi<\xi_{\rm min}$ lies outside our regime of validity for $\epsilon_{\rm dec}=0.05$.
The dashed black vertical line with $w_\triangle=0$ corresponds to universes with 
$\barkappa=2$, as shown in Fig.~\ref{fig:seesaw}.  Points to the right 
of this line correspond to universes with $\barkappa<2$, whereas points to the left 
correspond to universes with $\barkappa>2$.  Such universes respectively correspond to the 
situations illustrated in the upper and lower panels of Fig.~\ref{fig:seesaw2}.

The points within the Regions~I through~IV in each panel of Fig.~\ref{fig:PhaseDiagramPanels} 
exhibit triple stases with
$\barOmega_\Lambda>\barOmega_M>\barOmega_\gamma$, with
$\barOmega_M>\barOmega_\Lambda>\barOmega_\gamma$, with
$\barOmega_M>\barOmega_\gamma>\barOmega_\Lambda$, or with
$\barOmega_\gamma>\barOmega_M>\barOmega_\Lambda$, respectively.
By contrast, along the edges of this plane, our triple-stasis solutions reduce to simpler 
stasis solutions:  either to a pairwise $\Lambda M$ stasis (as discussed in 
Sect.~\ref{sec:LambdaMatter} and shown along the upper edge with $w_\triangle<0$), 
or to a pairwise $M\gamma$ stasis (as discussed in Sect.~\ref{sec:MatterGamma} 
and shown along the upper edge with $w_\triangle >0$), or to universes with only 
vacuum energy (as shown along the left edge), radiation (as shown along the right edge), 
or matter (as indicated at the top point with $w_\triangle=0$).  All universes along 
the vertical $w_\triangle=0$ line are effectively matter-dominated (as illustrated in 
Fig.~\ref{fig:seesaw}), with counterbalancing abundances of vacuum energy and radiation;  
however, as we move up this line towards greater values of $\xi$, these other abundances 
maintain their ratio but shrink to zero, leaving behind a universe consisting only of 
matter as $\xi\to\infty$.  Thus, as anticipated, we see that this figure 
encapsulates and illustrates the 
relationships between the different versions of stasis discussed in this paper.

As we remarked at the beginning of this section, the constraint equation for $M\gamma$ stasis in Eq.~(\ref{synthesiscomparison}) is slightly different from the others in that it does not depend 
on $w$.  This, of course, makes sense since $w$ is the equation-of-state parameter for the 
vacuum energy, and $M\gamma$ stasis does not involve vacuum energy.   However, at a purely 
algebraic level, we observe that the form of the $M\gamma$ constraint equation does match the 
others but has an ``effective'' $w=0$.  Requiring $w=0$ in turn implies that $w_\triangle\geq 0$, 
and we see from Fig.~\ref{fig:PhaseDiagramPanels} 
that this is indeed precisely the region to which the $M\gamma$ stasis is restricted.   

Finally, we note that each of the phase diagrams shown in Fig.~\ref{fig:PhaseDiagramPanels}  
exhibits a ``triple point'' --- \ie, a point at which the abundances of vacuum energy, matter, 
and radiation are equal during stasis.  The location of this triple point in each panel 
is indicated by a star.  Indeed, we observe that regardless of the form of 
$X$, the conditions under which the expressions for $\barOmega_\Lambda$, $\barOmega_M$, 
and $\barOmega_\gamma$ in Eqs.~(\ref{finalstasisabundances}) and~(\ref{abundance_avalues}) 
coincide are 
\begin{equation}
  w_\triangle ~=~ \frac{1+3w}{9}~,~~~~~~ 
  X ~=~ \frac{4 - 6w}{10 + 3w}~.
\end{equation}
The latter condition can be solved for any given form of $X$ in order to obtain the 
value of $\xi$ at the triple point.  This value of $\xi$ increases monotonically with $w$,
as can be seen by comparing the results shown in the different panels of 
Fig.~\ref{fig:PhaseDiagramPanels}, and can become as large as $\xi \sim \mathcal{O}(50)$ as 
$w$ approaches zero for both the form of $X$ in Eq.~(\ref{eq:XMDef}) and the form of 
$X$ in Eq.~(\ref{eq:XDefImproved}).

\FloatBarrier

\section{Beyond stasis \label{sec:variants}}

In addition to the stasis phenomenon discussed in previous sections,
there are also {\it variants}\/ of this phenomenon in which only some --- but not all --- of 
the features associated with stasis are retained.  Depending on which features are retained, 
we can obtain a variety of scenarios which may be exceedingly interesting in their own rights 
on both theoretical and phenomenological grounds.   In general, it is critical to study such 
variants because there exist many real-world effects which may push our system beyond some 
of the assumptions we have made when constructing our above models of stasis.   Understanding 
how robust the stasis phenomenon is when faced with such perturbations is therefore of 
profound importance for understanding the emergence of stasis within realistic models of physics.

\subsection{Quasi-stasis}
\label{subsect:Quasi-stasis}

Stasis, of course, refers to an epoch in which our abundances remain absolutely fixed as 
functions of time.  However, it is possible to obtain a {\it quasi-stasis}\/ situation in 
which our abundances do experience a non-zero time-dependence, but in which this time-dependence 
is extremely suppressed.  Within such scenarios, all of the leading power-law growth that would 
appear in the usual cosmology is still absent --- just as in ordinary stasis --- but a weak, 
quasi-logarithmic time-dependence remains. Depending on the rate of change associated with this 
residual time-evolution, such quasi-stasis solutions may effectively serve as (and in fact be 
phenomenologically indistinguishable from) true stasis solutions over relevant cosmological 
timescales.

It is easy to see how such a quasi-stasis might arise.  As we have seen, each of our pairwise 
stases is essentially unavoidable:   for any choices of fundamental scaling exponents 
$(\alpha,\gamma,\delta,w)$ within the specified ranges, our system necessarily evolves into a stasis configuration, with 
the relevant stasis abundances remaining absolutely constant as functions of time.   Indeed, 
only the values of these stasis abundances depend on our underlying parameter choices.

\begin{figure}[b]
\includegraphics[width=\linewidth]{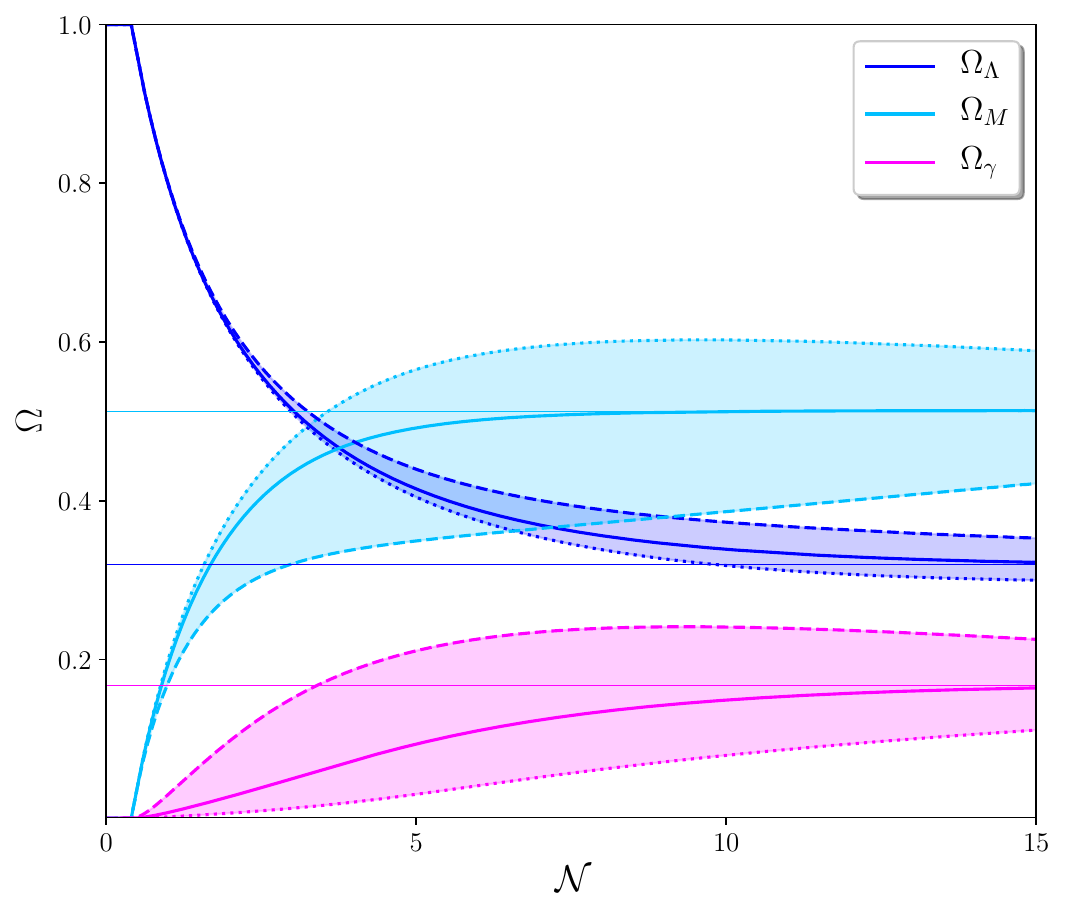}
\caption{Stasis versus quasi-stasis:  the abundances $\Omega_\Lambda$ (blue), $\Omega_M$ (cyan), and 
$\Omega_\gamma$ (magenta), plotted as functions of $\mathcal{N}$ for a true stasis (solid lines)
and two nearby quasi-stases, one with $\gamma = 0.95$ (dashed lines) and the other with 
$\gamma = 1.05$ (dotted lines). 
All curves are evaluated within  the framework of an exponential decay 
for the matter-radiation transition, with $(\alpha,\delta,w)= 
(1,2,-0.8)$ taken as benchmark values.
Corresponding curves for all other quasi-stases with 
$0.95 < \gamma < 1.05$ and the same $(\alpha,\delta,w)$ benchmark values
lie within the shaded bands.
In each case we see that variations in $\gamma$ induce 
deviations from true stasis, but in each case the abundances $\Omega_i$ nevertheless remain 
relatively constrained to lie near these stasis values, with time-evolutions that continue to 
be significantly suppressed.  The closer $\gamma$ is to unity, the more closely the $\Omega_i$ 
curves trace the true stasis values.}
\label{fig:quasistasis}
\end{figure}

For {\it triple}\/ stasis, by contrast, a new possibility opens up.  Because an additional 
constraint equation arises --- in particular, that in Eq.~(\ref{consteq}) which is needed 
in order to ensure that our two pumps are compatible with each other --- not every choice 
of $(\alpha,\gamma,\delta,w)$ 
leads to a successful triple stasis.  Indeed, as we have seen 
in Sect.~\ref{sec:TripleStasis}, only those choices which place our system along Branch~A 
in Fig.~\ref{fig:branches} lead to a full, triple stasis.

This observation implies that it is possible to choose values for $(\alpha,\gamma,\delta,w)$ 
[or equivalently for $(\gamma,\barkappa)$] for which we do {\it not}\/ obtain a triple stasis 
along Branch~A.~  Amongst these, however, there are two classes of special cases that are worthy 
of note.  Such special cases yield situations in which our abundances do not remain constant, 
but evolve exceedingly slowly.   (Indeed, this slow evolution replaces 
the stasis phenomenon itself, and exists independently of any features related to the {\it approach}\/ 
to stasis.)  The first class of such solutions 
consists of those lying along Branch~B in Fig.~\ref{fig:branches}.
Indeed, we recall from Sect.~\ref{sec:TripleStasis} that these solutions satisfy both of our 
scaling constraints in Eqs.~(\ref{constw1}) and (\ref{constw1prime}), but simply fail to satisfy 
the second constraint in Eq.~(\ref{eq:OMconds}) that ensures that these solutions avoid a logarithmic 
instability.  By contrast, the second class consists of solutions which do not lie along either 
Branch~A or Branch B, but whose underlying parameters place it {\it relatively close to Branch~A}\/.~ 
In such cases, our system does not satisfy our overall scaling relations.  However, our system does 
satisfy these relations {\it approximately}\/, and therefore we once again expect a highly 
suppressed time-evolution for the abundances.
This situation might easily arise, for example, if our system originally lay along 
Branch~A {\it at tree level}\/ --- and thus had $\gamma=1$ at tree level --- but then 
radiative corrections altered the scaling relations for our decay widths in such a way as to 
introduce a small correction for $\gamma$, pushing the effective value of this exponent 
slightly away from $\gamma=1$.

This quasi-stasis phenomenon is illustrated in Fig.~\ref{fig:quasistasis}.  In this figure, we 
plot the abundances $\Omega_\Lambda$ (blue), $\Omega_M$ (cyan), and $\Omega_\gamma$ (magenta) as 
functions of $\mathcal{N}$ for a true stasis (solid lines), a quasi-stasis with $\gamma = 0.95$ 
(dashed lines) and a quasi-stasis with $\gamma = 1.05$ (dotted lines). Working within the 
framework of an exponential decay for the matter-radiation transition, we have taken 
$(\alpha,\gamma,\delta,w) = (1,1,2,-0.8)$ for the true stasis.  Corresponding curves for all 
quasi-stases with $0.95 < \gamma < 1.05$ for the same choices of these other parameters lie 
entirely within the shaded bands delimited by the dashed and dotted curves.  Indeed, the 
closer $\gamma$ is to unity, the more closely the $\Omega_i$ curves track those obtained 
for a true stasis.  These results illustrate that stasis is robust against departures from 
the $\gamma = 1$ criterion, and that the universe indeed experiences a period of approximate 
stasis even when $\gamma$ is not exactly unity.

\begin{figure*}[htb]
 \includegraphics[width=0.32\linewidth]{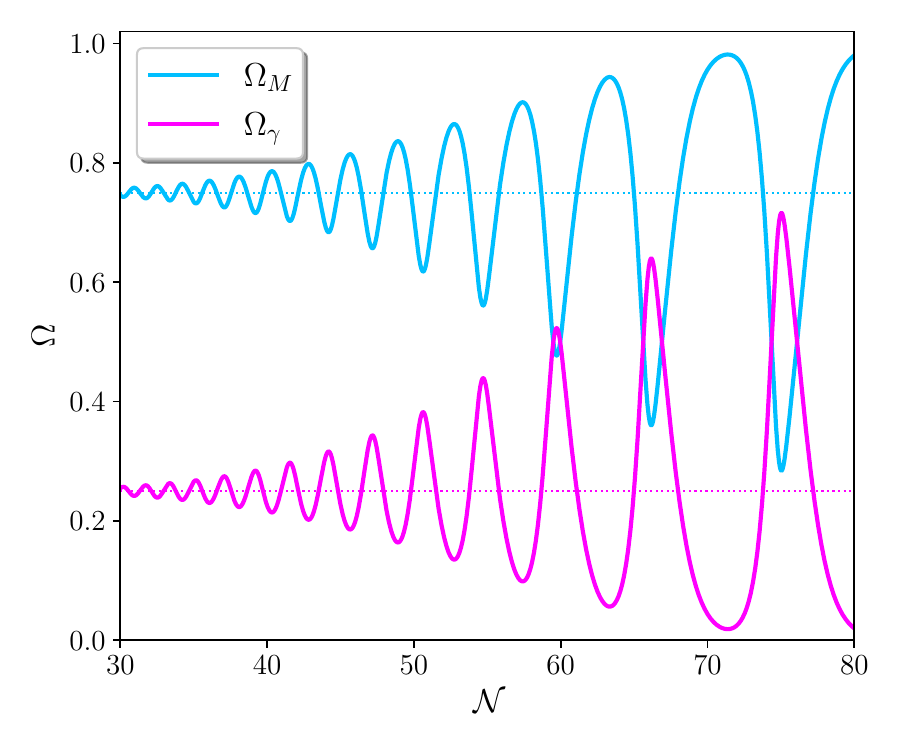}
\includegraphics[width=0.32\linewidth]{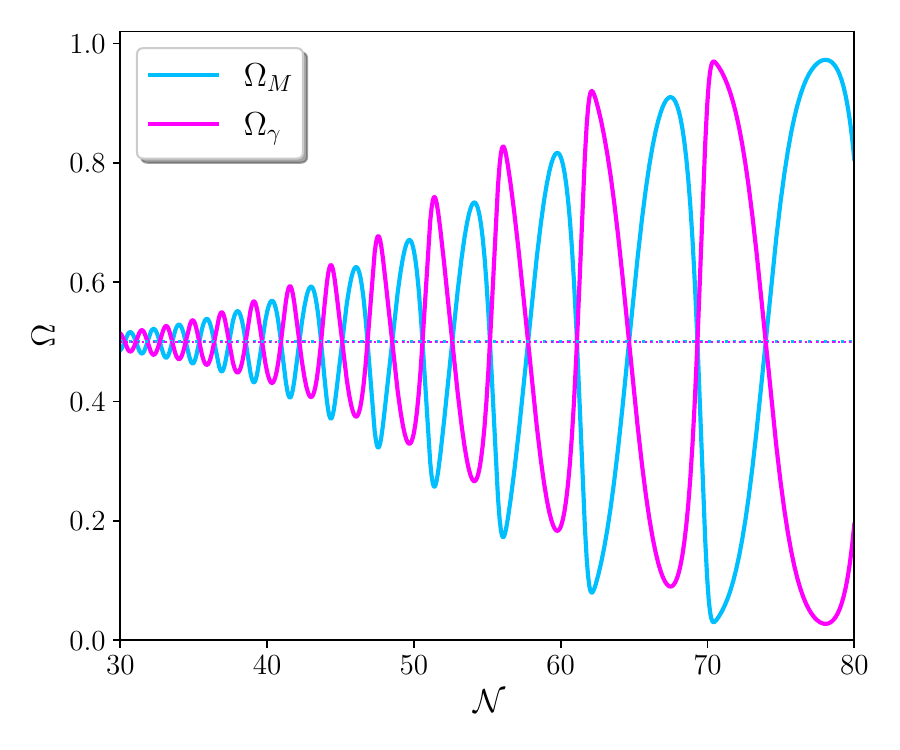}
\includegraphics[width=0.32\linewidth]{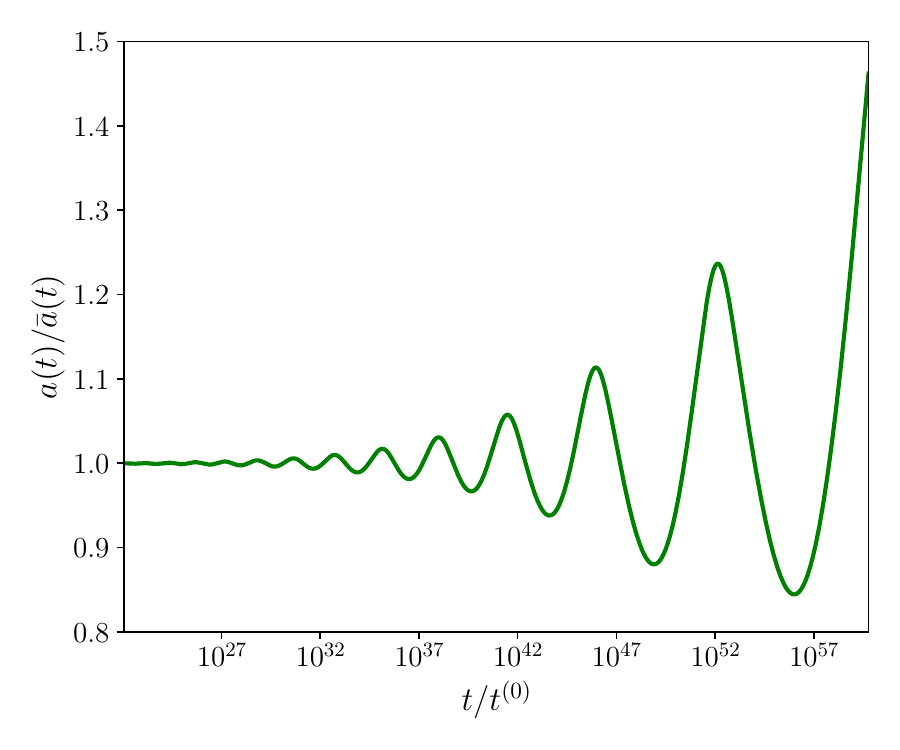}
\caption{Pairwise oscillatory stasis involving matter and radiation.  In the
  left panel, we plot $\Omega_M$ (solid cyan curve) and $\Omega_\gamma$ 
  (solid magenta curve) as functions of time for the parameter choices
  $\alpha\approx 1.13$, $\delta=4$, $\gamma=9$, which imply $\barOmega_M=3/4$ and $\barOmega_\gamma=1/4$.  The dotted horizontal cyan and magenta lines respectively indicate the 
  corresponding constant abundances $\barOmega_M$ and $\barOmega_\gamma$ 
  which would be obtained for a true stasis with the same values of $\alpha$, 
  $\delta$, and $\gamma$.  In the middle panel we show the abundance curves 
  obtained for the parameter choices $\alpha\approx 2.32$, $\delta=4$, $\gamma=9$, which imply
  $\barOmega_M=\barOmega_\gamma= 1/2$.
  These parameter choices  exemplify the special case in 
  which $\barOmega_M = \barOmega_\gamma $ and in which the abundances $\Omega_M$ and 
  $\Omega_\gamma$ exhibit a ``braiding'' phenomenon, mutually oscillating around 
  the same central value.  In the right panel, we plot as a function of time the ratio 
  of the scale factor $a(t)$
  to the corresponding value $\overline{a}(t)\sim t^{\barkappa/3}$ that the scale 
  factor would have in a true stasis for the same parameter choices as in the 
  middle panel.  We emphasize that similar oscillatory phenomena can also arise for 
  the other realizations of stasis that we have considered in this paper.
  \label{fig:M_gamma_braiding}}
\end{figure*}

\subsection{Oscillatory stasis and braiding}
\label{subsect:OscillatoryStasis}

In our discussions of stasis, we have implicitly assumed that $\Delta m$ --- the 
characteristic mass spacing between the states in our $\phi_\ell$ towers --- is 
sufficiently small that the continuum limit we have employed, \eg, in 
Eqs.~(\ref{sum-to-integral_tau}) and~(\ref{sum-to-integral_that}) remains a valid
approximation across the entire tower.
However, it is also interesting to consider what happens when $\Delta m$ is 
larger and discretization effects become important.   
For concreteness, in what follows we shall focus on the case of a matter/radiation stasis, 
though we emphasize that similar phenomena arise in the other realizations of 
stasis that we have discussed in this paper as well.

As a general rule of thumb, discretization effects become important in a 
matter/radiation stasis of the sort discussed in Sect.~\ref{sec:MatterGamma} when 
the timescale associate with the depletion of the abundance of a particular
state $\phi_n$ within the tower --- \ie, its lifetime $\tau_n$ --- is short in
comparison with the characteristic time interval $\tau_{n-1} -\tau_n$ between the 
lifetimes of successively decaying states within the tower 
at time $t = \tau_n$.
This the characteristic time interval between the lifetimes of successively 
decaying states in the tower is essentially the reciprocal of the density of 
states $n_\tau(\tau_n)$ per unit lifetime, evaluated at $\tau = \tau_n$.  
In general, this density of states is given by
\begin{equation}
    n_\tau(\tau) \,=\, \frac{1}{\gamma\delta}
      \left(\frac{m_0}{\Delta m}\right)^{1/\delta}\!
      (\Gamma_0 \tau)^{-1/\gamma}\Big[(\Gamma_0\tau)^{-1/\gamma}-1\Big]^{1/\delta-1}
      \frac{1}{\tau}~,
  \label{eq:DensOfStatestauNoApprox}
\end{equation}
which reduces to the approximate result in Eq.~(\ref{ns}) in the regime in which 
$\tau \ll \tau_0$.  Thus, we expect discretization effects to become important when
$n_\tau^{-1}(\tau) \gtrsim \tau$, or in other words, when
\begin{equation}
   \gamma\delta
      \left(\frac{\Delta m}{m_0}\right)^{1/\delta}\!
      (\Gamma_0 \tau)^{1/\gamma}\Big[(\Gamma_0\tau)^{-1/\gamma}-1\Big]^{1-1/\delta}
      ~\gtrsim~ 1~.
  \label{eq:OscStasisCondit}
\end{equation}

In general, the quantity on the left side of Eq.~(\ref{eq:OscStasisCondit})
depends non-trivially on $\tau$.  Thus, depending on the 
values of $\delta$, $\gamma$, $\Gamma_0$, and the ratio $\Delta m/m_0$,  
it is possible that discretization effects will be evident only within certain 
ranges of $\tau$.  Indeed, for sufficently small $\tau \ll \Gamma_0^{-1}$, 
the condition in Eq.~(\ref{eq:OscStasisCondit}) reduces to
\begin{equation}
   \gamma\delta
      \left(\frac{\Delta m}{m_0}\right)^{1/\delta}\!
      (\Gamma_0 \tau)^{1/\gamma\delta}
      ~\gtrsim~ 1~.
  \label{eq:OscStasisConditReduced}
\end{equation}
This implies that discretization effects typically become increasingly important at 
late times.  More specifically, they become important at timescales 
$\tau \gtrsim \tau_{\rm disc}$, where we have defined
\begin{equation}
  \tau_{\rm disc} ~\equiv~ \Gamma_0^{-1}(\gamma\delta)^{-\gamma\delta}
  \left(\frac{\Delta m}{m_0}\right)^{-\gamma}~.
  \label{eq:tauosc}
\end{equation}

Since we are assuming that $\gamma > 0$, we observe that $\tau_{\rm disc}$ 
decreases as $\Delta m$ increases.  Thus, as the characteristic scale of the 
mass splittings between the tower states increases, discretization effects
become important at earlier and earlier times.  Indeed, in situations in which 
$\tau_{N-1} \gtrsim \tau_{\rm disc}$, these effects are important  
throughout the entire time interval during which which the tower states are 
decaying.  Conversely, in situations in which $\tau_0 \ll \tau_{\rm disc}$,
the effects are nominally negligible throughout this entire time interval and the 
universe effectively remains in a true stasis.  We may therefore determine a 
rough threshold for $\Delta m$ above which above which discretization effects 
become important prior to the end of stasis by setting 
$\tau_{\rm disc} \rightarrow \tau_0$ in Eq.~(\ref{eq:tauosc}).  This threshold is 
\begin{equation}
  (\Delta m)_{\ast} ~\equiv~ (\gamma \delta)^{-\delta} m_0 ~.    
\end{equation}
In reality, however, we emphasize that discretization 
effects {\it always}\/ come into play at times $t\sim \tau_0$, when the longest-lived 
states in the tower are decaying and second term in the square brackets in 
Eq.~(\ref{eq:OscStasisCondit}) can no longer be neglected.  Indeed, such 
discretization effects are evident during the last few $e$-folds of the 
matter/radiation stasis illustrated in the left panel of Fig.~\ref{fig:review}.

In Fig.~\ref{fig:M_gamma_braiding}, we illustrate the impact that these discretization
effects have on the evolution of the abundances $\Omega_M$ and $\Omega_\gamma$, as well 
as on the expansion rate.  In the left panel, we plot these abundances (solid curves) 
as functions of time for the parameter choices $\alpha\approx 1.13$, $\delta=4$, $\gamma=9$, 
and $\barOmega_M=3/4$.  The corresponding constant abundances $\barOmega_M$ and 
$\barOmega_\gamma$ which would be obtained for a true stasis with the same values of
$\alpha$, $\delta$, and $\gamma$ are indicated by the dotted horizontal lines.
In the middle panel, we show the corresponding abundance curves for the parameter 
choices $\alpha\approx 2.32$, $\delta=4$, $\gamma=9$, and $\barOmega_M=1/2$. 

We see from the left and middle panels of Fig.~\ref{fig:M_gamma_braiding} that the net 
impact of discretization effects on the abundances of our cosmological components is 
to give rise to an quasi-oscillatory behavior wherein $\Omega_M$ and $\Omega_\gamma$ 
both vary around fixed central values --- values which correspond to the respective
stasis values $\barOmega_M$ and $\barOmega_\gamma$.  We shall refer to the variant of 
stasis wherein this phenomenon is manifest as ``oscillatory'' stasis in what follows.
The results shown in the left panel Fig.~\ref{fig:M_gamma_braiding} 
are representative of the more general case in which these central values  are 
different.  By contrast, the results shown in the middle panel exemplify the special
case in which $\barOmega_M = \barOmega_\gamma = 1/2$, with the abundance curves exhibiting 
a ``braiding'' phenomenon in which they mutually oscillate around the same central value.
The results shown in both panels also illustrate that the effective period 
$T_{\rm osc}(t)$ of the ``oscillations'' in an oscillatory stasis is in general not 
constant.  Indeed, at any given time $t$, this effective period is simply the time 
interval $T_{\rm osc}(t)\sim n_\tau^{-1}(t)$ between the decays of successive tower
states, and $n_\tau(t)$ is in general time-dependent.

The variation of the abundances during a period of oscillatory stasis implies
that $\kappa$ is also not constant during such a period; 
rather, $\kappa$ experiences a quasi-periodic oscillation as the universe expands.  
This behavior is illustrated in the right panel of Fig.~\ref{fig:M_gamma_braiding}, 
where we plot the ratio of the scale factor $a(t)$ during a period of oscillatory 
stasis to the corresponding value $\overline{a}(t) \sim t^{\barkappa/3}$ which 
would be obtained during a period of true stasis for the same parameter choices 
as in the middle panel.  This behavior implies that as the universe expands, 
its {\it average}\/ rate of growth remains consistent with a fixed time- and stasis-averaged 
value of $\langle w\rangle$.  However, the universe  ``reverberates'' as it expands, with 
pulsing periods of faster and slower expansion.  

The results shown in Fig.~\ref{fig:M_gamma_braiding} also indicate that oscillatory 
stasis, when it arises, can potentially persist for a significant number of $e$-folds 
of cosmic expansion.  In situations in which $\Delta m \gtrsim (\Delta m)_\ast$,
the expression for $\mathcal{N}_s$ in Eq.~(\ref{MGNN}) represents the total 
number of $e$-folds of expansion associated with both true stasis {\it and}\/ oscillatory
stasis.  For $\tau_{N-1} \gtrsim \tau_{\rm disc}$, as discussed above, true stasis
is never achieved and the number of $e$-folds $\mathcal{N}_{\rm osc}$ of oscillatory 
stasis is given by Eq.~(\ref{MGNN}) as well.
By contrast, for $\tau_{N-1} \lesssim \tau_{\rm disc} \lesssim \tau_0$,
oscillatory stasis begins roughly when $t \sim \tau_{\rm disc}$ and ends when 
the last state in the tower decays.  Thus, we have   
\begin{equation}
  \mathcal{N}_{\rm osc} ~\approx~
    \log\left(\frac{a(\tau_0)}{a(\tau_{\rm disc})}\right)~.
  \label{eq:Nosc}
\end{equation}
Approximating $a(\tau_0)$ and $a(\tau_{\rm disc})$ with their corresponding 
time-averaged values $\overline{a}\sim \tau_0^{\barkappa/3}$ and 
$\overline{a}\sim \tau_{\rm disc}^{\barkappa/3}$, and using
Eq.~(\ref{MGHubble}) in order to express $\barkappa$ in
terms of the constant matter abundance $\barOmega_M$ that would be obtained
for a true stasis with the same values of $\alpha$, $\gamma$, and $\delta$, we find that
\begin{equation}
  \mathcal{N}_{\rm osc} ~\approx~   
       \displaystyle \frac{2\gamma}{4-\barOmega_M} 
       \log \left(\frac{\Delta m}{(\Delta m)_{\ast}}\right)~.
  \label{eq:NoscSubbed}
\end{equation}
Once again, we emphasize that while this approximate expression for
$\mathcal{N}_{\rm osc}$ decreases to zero continuously as 
$\Delta m \rightarrow (\Delta m)_{\ast}$, discretization effects in 
fact always come into play at times $t\sim \tau_0$.

The quasi-oscillatory behavior we have described here may have observable 
consequences.  For example, this behavior could potentially affect the growth of 
both scalar and tensor perturbations in the early universe in distinctive
ways.  One particularly interesting possibility is that resonance effects
could arise as a consequence of these oscillations.  While
$T_{\rm osc}(t)$ is in general time-dependent during oscillatory stasis, an
alignment of this timescale with other relevant timescales
even during a single cycle of ``oscillation'' could potentially have
consequential effects.  Indeed, such single-cycle resonances are know to 
have a significant impact on cosmological dynamics in other contexts 
(see, \eg, Ref.~\cite{Dienes:2019chq}), and it is certainly conceivable 
that they could have observable consequences in the context of oscillatory
stasis as well.  We leave the investigation of such possibilities for future work.

\subsection{Stasis unrealized}
\label{subsect:stasis_unrealized}

Finally, there is another unique behavior which is potentially associated with 
stasis but which does not result in a stasis configuration.  Depending on the underlying 
model parameters, it may happen that our system begins heading {\it toward}\/ a stasis 
configuration as the result of the attractor behavior associated with stasis, but never fully reaches this destination because our level-by-level
transitions reach the bottom of the tower before the stasis is fully realized.
This can then result in abundances whose time-evolution begins to slow over many 
$e$-folds (as appropriate for the approach to stasis), but then grow again as another 
post-stasis dynamics comes into play.   This phenomenon may also have important 
phenomenological implications, and represents one of the few ways in which a 
pairwise stasis may ultimately be avoided.

\FloatBarrier

\section{Discussion and conclusions \label{sec:conclusions}}

Cosmic stasis --- a phenomenon in which the abundances of multiple cosmological energy 
components remain effectively constant across extended cosmological eras despite cosmological expansion --- arises naturally
in many extensions of the Standard Model.  In Ref.~\cite{Dienes:2021woi}, for example, it was 
shown that a pairwise stasis involving matter and radiation can arise from the decays of a 
tower of unstable particles with a broad spectrum of lifetimes and cosmological abundances.
In this paper, we have extended this prior analysis and demonstrated that cosmic stasis is a more general phenomenon which 
can also arise in the presence of other cosmological energy components with other equations of state. 
These include, as we have seen,
cosmological systems 
involving vacuum energy.  In such cases, the transition from overdamped to underdamped oscillation of a homogeneous scalar field 
provides a natural 
mechanism for the 
transfer of energy from vacuum energy to matter, thereby allowing
a tower of such scalar fields with a broad spectrum of masses to give rise to
a pairwise stasis involving vacuum energy and matter.  We have also
shown that a direct transfer of energy density from vacuum energy to radiation can, under certain
conditions, give rise to a pairwise stasis involving vacuum energy and radiation. 
Moreover, we have  shown that it is even possible for a {\it triple}\/ stasis to arise in which the abundances of 
vacuum energy, matter, and radiation all simultaneously remain 
constant despite cosmic expansion.
Indeed, this last result is highly non-trivial and does not emerge simply as the result of the existence of the previous pairwise stases.
We further demonstrated that all of
these types of stasis are dynamical attractors within their corresponding cosmological systems.  Thus, as 
long as these systems satisfy the basic conditions under which stasis can develop, all of these systems will 
ultimately flow toward the stasis state, irrespective of their initial conditions.

As indicated above, one of the keys to our analysis in this paper has been the use of the overdamping/underdamping transition as a means of transferring energy density from vacuum energy --- or more generally from a cosmological energy component with an effectively constant equation-of-state parameter $-1<w<0$  --- to matter. 
This in and of itself represents a significant broadening of the 
scope of cosmological scenarios which give rise to stasis to include those involving 
higher-dimensional axion or axion-like fields~\cite{Dienes:1999gw,Dienes:2012jb} as well as scenarios 
involving realizations of the string axiverse~\cite{Arvanitaki:2009fg}.
However, it is conceivable that a stasis epoch could naturally arise in other 
BSM scenarios as well.  These might include, for example, cosmologies involving a kination 
component with $w=1$, 
other cosmological energy components with $w > 1/3$, 
or even a cosmological energy component with a time-varying equation-of-state parameter 
$w(t)$.  Moreover, it is also conceivable that stasis could arise in cosmologies involving 
spatial curvature ($w=-1/3$), cosmic strings ($w=-1/3$), or domain walls ($w=-2/3$).   
Such cosmologies can also involve other energy components with effectively constant equation-of-state parameters within
the range $-1 < w < 0$, such as we have considered in this paper, but which transfer their 
energy density to other cosmological components via different mechanisms.
In order for this to occur, our analysis indicates that the  pump terms associated with 
these mechanisms would need to exhibit an appropriate $P\sim 1/t$ scaling behavior during 
stasis.  In some cases this may be non-trivial, especially if we further demand that such pump emerge naturally within the context of BSM physics. 

Given the possibility that cosmological energy components beyond vacuum energy, 
matter, and radiation could conceivably have constituted a significant fraction of the total 
energy density of the universe at early times, another obvious extension of our analysis 
would be to investigate whether and under what conditions a stasis epoch involving {\it more}\/ than 
three components might arise.  What additional conditions would then have to be satisfied?  In order to answer this question, history can be our guide.  
For each of the pairwise-stasis scenarios that we examined in 
Sects.~\ref{sec:MatterGamma},
\ref{sec:LambdaMatter}, and \ref{sec:LambdaGamma}, the system of equations that we obtained 
for our stasis abundances was over-constrained.  However, since some of these equations
happened to be redundant, we were nevertheless able to obtain self-consistent solutions for these
abundances.  By contrast, for the triple-stasis scenario that we examined in Sect.~\ref{sec:TripleStasis},  
no such redundancies arose within the corresponding system of equations.  We were therefore able to 
obtain a series of constraints that led to unique solutions for our stasis abundances.    

Given this pattern, it is natural to expect that
a {\it quadruple}\/ stasis would lead to an 
{\it under-constrained}\/ system of constraints.  We would then expect to find not a unique set 
of stasis abundances but rather a {\it line}\/ of possible solutions.  In such cases, the particular 
stasis abundances towards which the system evolves would presumably depend on initial conditions.
We would likewise expect this pattern of increasingly under-constrained solutions 
for the stasis abundances to continue as we introduce additional energy components into the mix.    Of course, these observations are predicated on the manner in which additional pumps for energy transfer 
are introduced as the number of additional energy components is increased.

In this connection, we  note [in complete analogy with the discussion below Eq.~(\ref{eq:weffdef})] 
that another way of introducing a fourth energy component would be to establish a triple stasis 
between three of the components and then ensure that the fourth component functions as a mere 
{\it spectator}\/ --- \ie, that it have an equation-of-state parameter which exactly matches the 
stasis average $\langle w\rangle$ determined by other three, and that it experience no energy 
transfers with the other components and thereby remain inert.  As a result, the dynamics of the 
underlying triple stasis would not be disturbed, and we would once again expect to obtain a line of 
potential solutions for the four stasis abundances in which the abundance of the fourth component is arbitrary and the abundances of the other three decrease to compensate but  remain in the same ratios as they had prior to the introduction of the fourth component. 

From a model-building perspective, the realizations of stasis that we have discussed in this 
paper all involve towers comprising large numbers of individual states.  However, as we have repeatedly emphasized, such towers arise naturally in many extensions of the SM.~  For example, scenarios 
involving additional, compactified spacetime dimensions naturally give rise to towers of 
Kaluza-Klein (KK) resonances.  Other examples of towers of states which emerge naturally in 
BSM scenarios include the towers of closed-string resonances which appear in Type~I string 
theories and the towers of hadron-like resonances which appear in theories involving confining 
hidden-sector gauge groups.  

Within the context of such BSM scenarios, there is often a straightforward 
relationship between the properties of the stasis epoch and the parameters
of the underlying particle-physics model which gives rise to it.  For example, in realizations of stasis 
in which the $\phi_\ell$ fields are the KK resonances associated with an extra spacetime dimension of radius $R$, the masses $m_\ell$ of the KK states span the range  
from $R^{-1}$ to some 
fundamental cutoff scale $\mu$ such as the string or GUT scale.  In such 
realizations of stasis, the range of lifetimes for the $\phi_\ell$ fields --- and therefore
the duration of the stasis epoch --- depends crucially on the hierarchy between
$R^{-1}$ and $\mu$.  As a result, in situations in which the stasis state has $\langle w\rangle < -1/3$ (causing the universe to
undergo accelerated expansion during stasis), the size of the comoving horizon today 
would in a very real sense be the manifestation of this hierarchy of energy scales.
Indeed, within such scenarios,
the large number of $e$-folds of expansion experienced by our universe across its history 
might actually be the manifestation 
of a hierarchy in the size of an otherwise unseen compactified dimension!

In all of our stasis realizations involving the decays of the $\phi_\ell$ states, the resulting cosmological dynamics is essentially determined by the corresponding masses $m_\ell$ and decay widths $\Gamma_\ell$.
Moreover, we have assumed that 
for all $\phi_\ell$ these widths are dominated by decays to effectively massless particles
outside the tower which behave like radiation throughout the stasis epoch.  However, within certain realizations of such stases there may be {\it multiple}\/ states $\phi_\ell^{(j)}$ at each level $\ell$ --- states which share the same mass $m_\ell$ and decay width $\Gamma_\ell$.
At tree level,
the resulting cosmological dynamics will be largely independent of the 
manner in which each energy density $\rho_\ell$ for each $\ell$ might be partitioned 
among such degenerate states.
Indeed, at tree level such a collection of degenerate states essentially functions as a single state whose total energy density $\rho_\ell$ is the sum of the contributions from its constituents.
At loop level, however, 
the cosmological dynamics 
is in general sensitive to such degeneracies. 
For example, the renormalization of the masses $m_\ell$ and couplings which give 
rise to each $\Gamma_\ell$ would depend on these degeneracies since all of these states can run independently within the loops.
Thus the scaling of the decay widths would ultimately depend on such degeneracies.

Moreover, in situations in which the $\phi_\ell^{(j)}$ states decay to the same species 
of radiation particle at tree level, loop-level diagrams will generically give rise to decay 
processes in which heavier $\phi_\ell$ states decay into lighter $\phi_{\ell'}$ states with 
$m_{\ell'}<m_\ell$.  For decay processes in which the resulting daughter particles are significantly 
lighter than the parents, these daughters will be produced with significant boosts as seen within 
the cosmological background frame.  These boosts will therefore result in non-trivial phase-space 
distributions for the lighter $\phi_{\ell'}$ particles.  The presence of potentially significant 
velocities for these states can also increase their  corresponding equation-of-state parameters 
from $w_\ell=0$ (consistent with our original assumption of cold matter) to anywhere within the 
range $0 < w_\ell < 1/3$.  These phase-space distributions will also thereafter evolve 
non-trivially during the stasis epoch due to cosmological redshifting effects~\cite{Dienes:2020bmn}.
Thus, if loop-level effects are significant, they could in principle have a non-negligible effect 
on the cosmological dynamics involved in establishing and sustaining stasis. 

Issues regarding the degeneracies of states at each mass level may be particularly relevant for 
string-theoretic realizations of stasis.  In general, perturbative string-theory spectra contain 
many kinds of states, including not only KK states (and winding states if the string is closed) 
but also string oscillator states.  All of these states come in infinite towers.
For example, the infinite towers of oscillator states have masses which 
scale as $\alpha' m_\ell^2\sim \ell$ where $\alpha'$ is the Regge slope, or equivalently 
$\sqrt{\alpha'} m_\ell\sim \ell^\delta$ with $\delta=1/2$.  However, at each level $\ell$ the 
number $n_\ell$ of oscillator  states grows {\it exponentially}\/: 
$n_\ell\sim e^{\sqrt{\ell}}\sim e^{\sqrt{\alpha'}m_\ell}$.   This can have a number of 
interesting consequences.  For example, in string theory these states will generally have 
different spins  (even though they share the same mass);   they will therefore couple 
differently to lighter states and potentially have different decay widths.
Second, this growth in the number of states at each mass level ultimately leads to the 
famous Hagedorn  phenomenon~\cite{Hagedorn:1965st} wherein various thermodynamic quantities 
experience divergences at a critical temperature $T_c$ --- a temperature beyond which the theory 
is believed to transition to a different phase with different underlying degrees of freedom.   
It may therefore be these new degrees of freedom that are relevant for understanding the 
dynamical properties of the early universe, at least for $T>T_c$.  But even when $T<T_c$, 
there remains the issue of how the total abundances $\Omega_\ell$ might scale as functions 
of $\ell$ across the tower, given that the densities of oscillator states are growing 
exponentially.  This issue was studied in some detail for one cosmological production 
mechanism in Ref.~\cite{Dienes:2016vei}, but many other production mechanisms are possible.
Indeed, like all predictions of abundances $\Omega_\ell$ across our tower, this is ultimately a 
model-dependent question which depends on the particular production mechanism envisaged.  
We also emphasize that in general, string oscillator states have mass splittings $\Delta m$ 
on the order of the string scale.  Such states can therefore potentially support a stasis at 
intermediate energy scales below the Planck scale only within the context of low-scale string 
theories, with $M_{\rm string}\equiv 1/\sqrt{\alpha'}\ll M_{\rm Planck}$. 

Of course, even if the string scale is situated near the Planck scale (implying that the 
excited oscillator states are therefore near the Planck scale as well), large-volume 
compactifications can lead to KK towers populating intermediate scales.  Such KK states 
can therefore support a stasis at intermediate scales, even though they emerge in a 
string context. Indeed, within such string theories, these KK states will likely be the 
only part of the perturbative string spectrum whose towers are lighter than the string scale.   
Moreover, their degeneracies will not experience exponential growth.

It is not only the abundances and decay widths of our states that might be affected by 
model-specific concerns;  the same may also be true of their {\it masses}\/.  For example, 
throughout this paper we have assumed that the fields $\phi_\ell$  have masses which 
scale according to Eq.~\eqref{MGmassform}. However, interactions of these fields with 
other fields in the theory may lead to radiative corrections for these masses.
For KK towers stemming from a single large flat extra dimension, such radiative corrections 
were studied in Refs.~\cite{Cheng:2002iz,Bauman:2011xf}.  In some cases, it was 
found~\cite{Bauman:2011xf}  that the overall scaling structure of these KK masses is 
actually preserved under such one-loop radiative corrections --- indeed, in these cases 
the radiative corrections can simply be bundled into ``renormalizations'' of the overall 
parameters $m_0$ and $\Delta m$ without changing the form of Eq.~(\ref{MGmassform}). 
By contrast,  in other cases, the one-loop radiative corrections were found to distort 
these scaling relations altogether.   In either case, however, one generally finds that 
these radiative corrections are exceedingly small.

Moreover, in particle-physics realizations of stasis in which the $\phi_\ell$ are 
coherently oscillating scalar fields, effects of this sort can affect not only 
the masses of these fields but also other aspects of the scalar 
potential~\cite{PhysRevD.7.1888}.  Indeed, at points in field space where the 
$\phi_\ell$ fields are significantly displaced from their vacua, loop corrections 
can have significant impacts on the shape of the potential.  For example, these impacts 
have been studied within the context of large-field inflation models, where the resulting 
modification of the potential can in turn lead to modified predictions for inflationary 
observables~\cite{Covi:1998jp, Covi:1998mb,Sloth:2006az, Sloth:2006nu, Senoguz:2008nok, 
Bostan:2019fvk, Heurtier:2019eou}.  One intriguing possibility that can arise in certain 
situations as a result of such corrections is that coherent oscillations of the $\phi_\ell$ 
fields can in fact behave at early times like matter or radiation --- or potentially even 
like a perfect fluid with $w > 1/3$.  Thus, in realizations of stasis involving fields 
of this sort, there might exist mechanisms which transfer energy density from matter or 
radiation to vacuum energy, rather than the other way around.  We leave the investigation 
of such possibilities for future work.

The existence of an early period of stasis throughout the cosmological timeline can have a 
number of phenomenological consequences.
Indeed, the modification of the expansion history alone relative to that of the standard 
cosmology can affect the evolution of density perturbations, the spectrum of gravitational waves,
and predictions for cosmic-microwave-background (CMB) observables.  Moreover, such a 
modification can also have an impact on 
a variety of out-of-equilibrium processes, including those which play a crucial role in the 
production of dark matter or the generation of a baryon asymmetry in many BSM scenarios.
Indeed, many of these possibilities were discussed in detail in Ref.~\cite{Dienes:2021woi}
within the context of cosmologies involving an epoch of pairwise matter/radiation stasis.
However, since these effects arise generically in any cosmological scenario 
involving a non-standard expansion history, they also pertain to cosmologies involving the 
alternative types of stasis that we have examined in this paper.

One of the main motivations for this paper was to show that an epoch of cosmic stasis can in principle arise 
in situations in which the vacuum-energy abundance is non-negligible.  This prompts the
question as to whether such a stasis epoch could potentially give rise to a period of
cosmic inflation~\cite{Starobinsky:1980te,Guth:1980zm,Linde:1981mu,Linde:1983gd,Mukhanov:1981xt} 
with a duration sufficient to address the horizon and flatness problems.  In typical 
inflationary scenarios, the extraordinary degree of large-scale homogeneity and isotropy that we 
observe in our universe is attributed to an epoch of rapid cosmological expansion wherein the 
energy density is dominated by a perfect fluid with equation-of-state parameter $w < -1/3$.  
However, such a period of rapid expansion could also be the result of an epoch of cosmic stasis 
in which a tower of states which behave like vacuum energy coexists with matter and/or radiation, 
provided that $\langle w\rangle < -1/3$.  Moreover, since such a 
stasis epoch concludes when the last of these vacuum-energy states transfers its energy to 
matter or radiation, a graceful exit from inflation occurs naturally in this context.  

In these respects, inflationary stasis scenarios of this sort have a great deal in common 
with so-called warm-inflation scenarios~\cite{Berera:1995wh,Berera:1995ie,Berera:1998px}
in which radiation is produced throughout the inflationary epoch due to dissipative effects and 
in which thermal fluctuations, rather than quantum fluctuations, represent the dominant 
contribution to primordial density perturbations~\cite{Taylor:2000ze,Hall:2003zp}.  Indeed, during
the slow-roll phase of warm inflation --- as in a vacuum-energy/radiation stasis ---  $\Omega_\gamma$
remains approximately constant due to a transfer of energy density from vacuum energy 
to radiation which counteracts the effect of cosmic expansion~\cite{Berera:2023liv,Kamali:2023lzq}.  
However, there are also fundamental differences between these scenarios.  In warm inflation, 
the vacuum-energy component vastly dominates the energy density of the universe during the 
slow-roll phase, and $H$ therefore remains approximately constant.  While vacuum energy is 
continually being transferred to radiation during this phase, it is not transferred at a rate that 
is sufficiently large in order to have a significant impact on $\Omega_\Lambda$.  
Thus, the back-reaction of $\rho_\gamma$ on $H$ plays no essential role and can be neglected.  
By contrast, during an epoch of vacuum-energy/radiation stasis, $\Omega_\Lambda$ and $\Omega_\gamma$ 
during inflation can {\it both}\/ be non-negligible --- and indeed even
$\mathcal{O}(1)$ --- and the back-reaction of $\rho_\gamma$ on $H$ is 
incorporated fully, with $H$ inversely proportional to $t$.

Of course, whether a given stasis scenario of this sort constitutes a viable model of inflation
ultimately depends on whether the spectrum of scalar and tensor perturbations that it yields
are consistent with observation (for a recent review, see, \eg, 
Ref.~\cite{Achucarro:2022qrl,Chou:2022luk}) --- 
and in particular with the properties of the CMB.~  The evolution of these perturbation spectra 
during an epoch of inflationary stasis would be highly non-trivial,
with significant roles potentially played not only by
the quantum fluctuations of each individual $\phi_\ell$ field which contributes to 
$\Omega_\Lambda$, but also by thermal fluctuations within the radiation bath.  Moreover, the collective behavior of the $\phi_\ell$ fields and the value of 
$\langle w\rangle$ to which they dynamically give rise during stasis would 
affect the manner in which both scalar and tensor perturbations evolve.
Investigating the perturbation spectra which can arise in inflationary stasis scenarios ---
spectra which may exhibit not only distinctive features at high frequencies but also  characteristic 
patterns of non-Gaussianities --- is the subject of ongoing work.

\bigskip

\begin{acknowledgments}

We are happy to thank J.~Kost for discussions.
The research activities of KRD are supported 
in part by the U.S.\ Department of Energy under Grant DE-FG02-13ER41976 / DE-SC0009913, and 
also by the U.S.\ National Science Foundation through its employee IR/D program.
The work of LH is supported in part by the U.K.\ Science and Technology Facilities Council (STFC) 
under Grant ST/P001246/1.  
The work of FH is supported in part by the Israel Science Foundation grant
1784/20, and by MINERVA grant 714123.
The work of TMPT is supported in part by the U.S.\ National Science Foundation under 
Grants PHY-1915005 and PHY-2210283.  The research activities of BT are supported in part by the 
U.S.\ National Science Foundation under Grants PHY-2014104 and PHY-2310622.  
 LH also acknowledges the hospitality and support provided by the 
Institut Pascal at Universit\'e Paris-Saclay during the Paris-Saclay Astroparticle Symposium 2022,  
which in turn was supported through the IN2P3 master projet UCMN, the P2IO Laboratory of Excellence 
(program “Investissements d’avenir” ANR-11-IDEX-0003-01 Paris-Saclay and ANR-10-LABX-0038), 
the P2I axis of the Graduate School Physics of Université Paris-Saclay, and
IJCLab, CEA, IPhT, APPEC,  and ANR-11-IDEX-0003-01 Paris-Saclay and ANR-10-LABX-0038.
FH also thanks ITP CAS for hospitality.
 The opinions and conclusions
expressed herein are those of the authors, and do not represent any funding agencies. 

\end{acknowledgments}

\vfill\eject

\bigskip
\bigskip

\bibliography{TheLiterature2}

\end{document}